\documentclass[11pt,a4paper]{book}

\usepackage[english]{babel}

\usepackage{amsmath}
\usepackage{a4,amssymb,amstext,theorem,enumerate}
\usepackage{amsfonts}
\usepackage{latexsym,epic,graphicx}

\usepackage{amsfonts}

\usepackage{fancyhdr}

\theoremstyle{break} 
\newtheorem{thm}{Theorem}[chapter]
\theoremstyle{plain}

\newtheorem{prop}[thm]{Proposition}
\newtheorem{cor}[thm]{Corollary}
\newtheorem{defi}[thm]{Definition}

\makeatletter

\@addtoreset{equation}{section}
\makeatother

\newcommand{\be}{\begin{eqnarray}}
\newcommand{\ee}{\end{eqnarray}}
\newcommand{\beg}{\begin{eqnarray*}}
\newcommand{\eeg}{\end{eqnarray*}}

\newcommand{\nn}{\nonumber}
\newcommand{\Spin}{\textit{Spin}}
\newcommand{\Diff}{\textit{Diff}}
\newcommand{\Symp}{\textit{Symp}}
\newcommand{\ka}{\mathfrak{k}}
\newcommand{\vp}{\varphi}

\newcommand{\diag}{\text{diag}}
\newcommand{\ad}{\mathbf{ad}}
\newcommand{\Inv}{\textit{Inv}}
\newcommand{\e}{\epsilon}
\newcommand{\ep}{\varepsilon}
\newcommand{\epr}{\ep'}
\newcommand{\psir}{\psi'}
\newcommand{\eps}{<\hspace{-11.6pt}-}
\newcommand{\epsr}{\eps'}
\newcommand{\chir}{\chi'}

\newcommand{\p}{\partial}

\newcommand{\G}{\Gamma}

\newcommand{\cP}{\mathcal{P}}

\newcommand{\cM}{\mathcal{M}}

\newcommand{\Z}{\mathbb{Z}}
\newcommand{\Lie}{\mathfrak{g}}
\newcommand{\Car}{\mathfrak{h}}

\newcommand{\g}{\gamma}

\newcommand{\pf}{\vspace{-\baselineskip} \paragraph*{Proof.}}

\newcommand{\qed}{\hfill $\square$}

\newcommand{\R}{\mathbb{R}} 
\newcommand{\N}{\mathbb{N}} 
\newcommand{\C}{\mathbb{C}} 
\newcommand{\id}{\mathbf{1\hspace{-2.9pt}l}}

\begin{document}

{\flushright ULB-TH/08-28\\[10mm]}

\thispagestyle{empty}

\begingroup

\parindent0pt

\begin{center}

\vspace*{1cm}

\begingroup\renewcommand{\bfdefault}{bx}\mathversion{bold}\Large\bfseries
$E_{7(7)}$ and $d=11$ supergravity\par\vspace{1cm}
\endgroup

\begingroup\LARGE\ttfamily
D I S S E R T A T I O N\par\vspace{1cm}
\endgroup

zur Erlangung des akademischen Grades\\
doctor rerum naturalium\\
(Dr. rer. nat.)\\
im Fach Physik\par\vspace{0.5cm}

eingereicht an der\\
Mathematisch-Naturwissenschaftlichen Fakult\"at I\\
der Humboldt-Universit\"at zu Berlin\par\vspace{1cm}

\begingroup\large\sffamily
von\\
Herrn Dipl.~Phys.~Christian Hillmann\\
geboren am 27.11.1979 in Verden/Aller\par
\endgroup
\vspace{1cm}

\end{center}
\vfill

Pr\"asident der Humboldt-Universit\"at zu Berlin\\[0.1cm]
Prof.~Dr.~Dr.~h.c.~Christoph Markschies\par\vspace{0.5cm}

Dekan der Mathematisch-Naturwissenschaftlichen Fakult\"at I\\[0.1cm]
Prof.~Dr.~Lutz-Helmut Sch\"on \par\vspace{0.5cm}

Gutachter:\\[0.3cm]
1.\quad Prof.~Dr.~Hermann Nicolai
\\[0.3cm]
2.\quad Prof.~Dr.~Bernard de~Wit
\\[0.3cm]
3.\quad Dr.~Axel Kleinschmidt
\par\vspace{0.5cm}

Tag der m\"undlichen Pr\"ufung: \quad 16.~Dezember 2008

\endgroup

\clearpage\thispagestyle{empty}\cleardoublepage

\pagenumbering{roman} \setcounter{page}{1}

\section*{Zusammenfassung}
Diese Dissertation ist den Symmetrien von Gravitationstheorien gewidmet. Nach einer allgemeinen Einf\"uhrung des Symmetriebegriffs f\"ur physikalische Theorien wird die Diffeomorphismensymmetrie der Gravitationsphysik diskutiert. Diese wird im zweiten Kapitel aus der Eichung der Poincar\'eisometriegruppe des Minkowskiraums konstruiert. Geleitet von dem Wunsch nach einer Quantisierungsvorschrift f\"ur Gravitationseffekte wird argumentiert, dass Supersymmetrie in diesem Kontext nicht zu vernachl\"assigen ist und wieso eine Diskussion von mehr als vier Raumzeitdimensionen von einem mathematischen Standpunkt aus interessant erscheint. Dies bildet den \"Ubergang zur Supergravitation in elf Raumzeitdimensionen.\\

Im dritten Kapitel analysiere ich die Methodik der \glqq Nichtlinearen Realisierung\grqq{} und deren Anwendung auf die Gravitationsphysik im Allgemeinen. Insbesondere setze ich mich mit dem Ansatz von Borisov \& Ogievetsky \cite{BO74} auseinander, der von West \cite{W00} auf die elfdimensionale Supergravitation angewendet wurde. Dies stellt den Ausgangspunkt f\"ur das eigentliche Ergebnis meiner Dissertation dar, das ich in Kapitel \ref{CHAP5} beschreibe.\\

Ich betrachte dort eine sechzigdimensionale Lorentzmannigfaltigkeit mit der zus\"atzlichen Struktur einer erhaltenen, degenerierten symplektischen Form von Kodimension vier und eines erhaltenen symmetrischen quartischen Tensors, was ich im Folgenden als exzeptionelle Geometrie bezeichne. Die daraus resultierende Einschr\"ankung der Diffeomorphismengruppe erm\"oglicht die konsistente Reduktion der Freiheitsgrade des Vielbeins in sechzig Dimensionen auf solch eine Weise, dass ein Untervielbein in sechsundf\"unfzig Dimensionen die Form einer $E_{7(7)}$-Matrix hat. Die darin enthaltenen bosonischen Freiheitsgrade identifiziere ich mit denen aus der elfdimensionalen Supergravitation, also dem uneingeschr\"ankten Vielbein und dem Dreiformpotential in elf Dimensionen. \\

In der exzeptionellen Geometrie ist es m\"oglich, eine Supersymmetrie mit 32 Superladungen konsistent zu definieren. Aufgrund der hohen Symmetrie ist deren Form jedoch stark eingeschr\"ankt. Um den Kontakt zur Supergravitation in elf Dimensionen herzustellen, m\"ussen folglich Abh\"angigkeiten der physikalischen Felder von den zus\"atzlichen Koordinaten ignoriert werden. Es ist nun m\"oglich, die Supersymmetrietransformationen der elfdimensionalen Supergravitation f\"ur den behandelten Sektor exakt aus denen der exzeptionellen Geometrie durch Vernachl\"assigen der \"ubrigen Koordinatenabh\"angigkeiten herzuleiten. Der interessante Aspekt ist, dass die Supersymmetrietransformationen im Gegensatz zur $d=11$ Supergravitation nicht durch die Dynamik, sondern rein durch gruppentheoretische \"Uberlegungen fixiert sind.\\

Ich beschlie\ss e meine Dissertation mit einem Ausblick auf die Dynamik im Rahmen der vollst\"andigen exzeptionellen Geometrie, deren spezielle L\"osungen mit denen der $d=11$ Supergravitation exakt \"ubereinstimmen d\"urften.

\section*{Summary and introduction}

Duality symmetries such as the electric-magnetic duality in electromagnetism have been of interest in theoretical physics for a long time. In the context of a potentially existing M-theory, U-duality \cite{HT95} is the most prominent example. The low energy effective action of M-theory is $d=11$ supergravity. In a reduction \`a la Kaluza--Klein of $d=11$ supergravity to $d=4$, the U-duality group is proposed to be a discrete version of $E_{7(7)}$.\\

The real Lie group $E_{7(7)}$ is a global symmetry of this $N=8$ supergravity in four dimensions. It is well known that the $Gl(7)\subset E_{7(7)}$ part of this global symmetry group is related to the diffeomorphism symmetry in $d=11$ \cite{CJS78}. This thesis investigates the question if $d=11$ supergravity can be lifted to 60 dimensions by restricting the geometry without introducing new fields. Thus, the entire Lie group $E_{7(7)}$ can be interpreted as a subgroup of the diffeomorphism group in the $60$-dimensional exceptional geometry.\\

I consider a sixty-dimensional Lorentzian manifold with preserved degenerate symplectic form $\Omega$ of codimension $4$ and a preserved totally symmetric quartic tensor $Q$. The resulting restriction of the diffeomorphisms provides the possibility to consistently reduce the degrees of freedom of the $60$-dimensional vielbein in such a way that a $56$-dimensional subvielbein $e^H$ is an $E_{7(7)}$ matrix. Therefore it is possible to completely parametrize the latter by degrees of freedom of the $d=11$ vielbein and the three-form potential of $d=11$ supergravity. Thus, I do not add additional fields to $d=11$ supergravity. The result of this thesis is that the $E_{7(7)}$-covariant supersymmetry variations\footnote{Following the definition of covariance from the sections \ref{SymmAct} and \ref{prim2b}, the $E_{7(7)}$ symmetry transformation of the coordinates induces an $SU(8)$ action on the fields.
}
\beg
{\left(e^{-H}\right)^{AB}}^\mu \underline{\delta}{\left(e^{H}\right)_{\mu}}^{CD}&=& \eps^{[A}\chi^{BCD]} + \frac{1}{4!}\e^{ABCDEFGH}\bar{\eps}_E\bar{\chi}_{FGH}\nn\\
\underline{\delta}\chi^{ABC} &=& \bar{\nabla}^{[AB}\eps^{C]}
\eeg
exactly reproduce the supersymmetry variations of $d=11$ supergravity, if the latter are restricted to the degrees of freedom that are covered by the $56$-dimensional subvielbein $e^H$ and the corresponding fermion $\chi^{ABC}$. The $56$-dimensional $E_{7(7)}$-covariant derivative $\bar{\nabla}$ acts on the $8$-dimensional spinor $\eps^C$ of the supersymmetry variation, which is linked to the $32$ supercharges of $N=8$ $d=4$ supergravity by the standard decomposition into $4\times 8$ used e.g. in \cite{CJ79,dWN86}. The gravitino $\psi_m^C$ is associated with the field $\chi^{ABC}$ in a way similar to \cite{dWN86}. Since I restricted the diffeomorphisms in $d=60$ to the ones that preserve the tensors $\Omega$ and $Q$, it is consistent to describe the fermions as $SU(8)\subset Spin(59,1)$ representations. The purpose of this dissertation is to explain the notation used in these $E_{7(7)}$-covariant supersymmetry variations, which I will state again at the end of chapter \ref{CHAP5}.\\

It was West's idea \cite{W00} to relate the construction of Borisov \& Ogievetsky \cite{BO74} in the context of non-linear realizations to symmetries of $d=11$ supergravity. In this dissertation, I follow this line of thought and devote chapter \ref{CHAP4} to the discussion of non-linear realizations. I start by reviewing the procedure of Borisov \& Ogievetsky and close by discussing the role of torsion in this context, which cannot be discarded for a supergravity theory \cite{FN76}.\\

Therefore, the adequate framework for the discussion of supergravity appears to be Einstein--Cartan theory. In chapter \ref{CHAP3}, I introduce this theory as the natural consequence of gauging the isometry group of flat Minkowski space \cite{SK61}. Before presenting the mathematical concepts that are necessary for this dissertation in chapter \ref{CHAP2}, I want to address some points that will not be covered in this thesis.\\

Most prominently, the question if the equations of motion of $d=11$ supergravity can be lifted to a $60$-dimensional theory without introducing new fields, is beyond the scope of this thesis. I will restrict myself to commenting on possible Lagrangians in the conclusion. Furthermore, I refrain from discussing the supersymmetry variations of the fields of $d=11$ supergravity that are not encoded in the $56$-dimensional subvielbein $e^H$ in detail. From the point of view of $d=4$ $N=8$ supergravity and global $E_{7(7)}$-covariance, it is very probable that these follow the same scheme as the ones $e^H$ comprises. The fate of the on-shell supersymmetry algebra of $d=11$ supergravity will not be part of this dissertation either. Since Cartan's theorem \cite{C09} rules out general $11$-dimensional diffeomorphisms, if both tensors $\Omega$ and $Q$ are preserved, it is a very interesting question in which way the supersymmetry algebra can close at all. It is not clear either, if this construction for the case of the symmetry groups $E_{8(8)}$ and $E_{9(9)}$ of $d=3$ and $d=2$ maximal supergravity, respectively, leads to similar results. I will also refrain from a detailed discussion of the candidates $E_{10(10)}$ \cite{DHN02} and $E_{11(11)}$ \cite{W01} for symmetries of M-theory, because only a complete theory in $d=60$ will allow for a relation. \\

From a string theory point of view, this construction is interesting for the following reasons. As this theory in $d=60$ dimensions is U-duality invariant by construction, it is expected to also contain $IIB$ supergravity. This is possible, because the choice of the eleven dimensions of $d=11$ supergravity in the $d=60$ theory is not canonical: choosing a different set of $10$ coordinate directions from the $60$ possible ones may allow an interpretation of the fields of this $d=60$ theory as the ones of $IIB$ supergravity. I want to conclude this introduction with the remark that requiring $E_{7(7)}$-covariance in $d=60$ and diffeomorphism invariance in $d=11$ may lead to a strong selection criterium for higher curvature corrections in M-theory.

\tableofcontents

\chapter{Symmetries and physical fields}\label{CHAP2}
\pagenumbering{arabic} \setcounter{page}{1}
I want to start with an introduction to the mathematical language that will be used to describe physical theories in the chapters \ref{CHAP3}, \ref{CHAP4} and \ref{CHAP5}. Since the investigation of symmetry structures in these theories has been my prime motivation to write the thesis, I will devote this chapter to the action of symmetries on physical fields. I will proceed in four steps that correspond to the four sections \ref{phys1}-\ref{prim2} of this chapter.
\begin{enumerate}
	\item Definition of a physical theory and of a \textbf{formal symmetry action} on physical fields. I will show that this naturally leads to a Lie structure.
	 \item Review of finite dimensional Lie structures, such as Lie groups, Lie algebras, their real forms and representations that will be relevant for this dissertation.
	 \item Discussion of the infinite dimensional Lie algebra of vector fields $\mathfrak{diff}_d$ and its finite dimensional subalgebras that generate $\mathfrak{diff}_d$ by Ogievetsky's theorem.
	 \item An application of this representation theory of real Lie groups allows to define an \textbf{explicit symmetry action} on physical fields. I close this chapter with the introduction of the vielbein frame.
\end{enumerate}
This thesis will be focussed on the symmetry group of coordinate transformations $\Diff(d)$ that corresponds to the Lie algebra of vector fields $\mathfrak{diff}_d$. $\Diff(d)$ is closely linked to finite dimensional groups: In chapter \ref{CHAP3}, I will show that gauging the finite dimensional Poincar\'e group leads to $\Diff(d)$-invariant theories. In chapter \ref{CHAP4}, I will review and reinterpret the result of Borisov \& Ogievetsky \cite{BO74} to construct a $\Diff(d)$-invariant theory by requiring invariance under its finite dimensional affine linear and conformal subgroups alone. And in chapter \ref{CHAP5}, I will show that $d=11$ supergravity seems to be extendable to a theory in $d=60$ dimensions, endowed with a particular geometry that contains a restriction of an infinite dimensional $\Diff(56)$ subgroup to a finite dimensional one. \\

Due to this close link between finite dimensional symmetry groups and infinite dimensional ones, I will discuss all symmetry groups on the same footing. I will also include a sketch of Cartan's classification of finite dimensional, complex, simple Lie algebras, because I want to highlight the analogy to Ogievetsky's theorem: in both cases, the algebras can be descibed by their simpler finite dimensional subalgebras with relations in between. These are the Serre relations in Cartan's case and the vector field representation in Ogievetsky's. I will provide further details in section \ref{Ogievetsky}.

\section{Symmetries in physical theories}\label{phys1}
\subsection{Structural remarks}\label{GeomStr}
In this dissertation, a physical theory will always denote a collection of fields $\Phi_1,\Phi_2,\dots$ with local dependence on space-time coordinates $x^\mu$ that locally parametrize a real manifold $\cM^d$ in $d$ dimensions, equipped with a non-degenerate metric $g$ of Lorentzian signature. The explicit dependence of the fields on the coordinates is called the dynamics of this field theory. It is prescribed by specific partial differential equations, which are called the equations of motion. Quite often, these can be derived from an action $S$ by a variational principle.\footnote{If the theory includes self-dual $p$-forms, it is not straightforward to state an action. An example for a case where this is possible nonetheless is provided in \cite{HT88}.} 
\\

In order to facilitate the analysis, I restrict myself to considering manifolds $\cM^d$ with analytic transition functions \cite{Wa83}. As experiments indicate that the concept of spinors is appropriate to describe physical processes, I further constrain the manifolds $\cM^d$ to have trivial first and second Stiefel--Whitney class \cite{Law,Mor} in order to be able to define a spin structure. This directly implies orientability of $\cM^d$.
\\

A theory consists of a collection of fields. In the language of differential geometry, these are sections of a tensor bundle\footnote{This is not restricted to tensor products of the tangent bundle, the spin bundle could also be involved.} over a manifold $\cM^d$. A simple example is a scalar field $\Phi$. This is a mapping from the manifold $\cM^d$ to the field of real numbers
\be\label{Scalar1}
\Phi : \cM^d \longrightarrow \R.
\ee
As the manifold is analytic, it is natural to require that $\Phi$ analytically depends on points $x\in \cM^d$. Its precise dependence however, is only determined by the dynamics, i.e. the equations of motion. More interesting examples are a vector field $A$ or a tensor field $g$
\begin{subequations}
\be
A:\cM^d &\longrightarrow & T\cM^d,\\
g:\cM^d &\longrightarrow & T^*\cM^d\otimes T^*\cM^d\label{metric5}.
\ee
\end{subequations}
To every position $x\in \cM^d$, this vector field $A$ associates a vector $A(x)$ in the tangent space $T_x\cM^d$ and $g$ a tensor $g(x)$ in the tensor product of two cotangent spaces. Examples for these objects are the electromagnetic potential $A$ and the metric $g$, if the tensor product $\otimes$ is symmetric. \newpage

The simplest example of a Lorentzian manifold $(\cM^d,g)$, i.e. a manifold $\cM^d$ with a non-degenerate metric $g$ of Lorentzian signature, is flat Minkowski space 
\be
(\cM^d,g) &\approx&(\R^d,\eta) \,=:\, \R^{d-1,1}\nn\\
\text{with}\quad \eta&:=&\text{diag}(-1,+1,\dots,+1).\label{eta}
\ee
In this special case, the manifold is a vector space. This implies that the linear span of an arbitrary choice of linearly independent basis vectors $\hat{P}_\mu$ with $\mu=0,\dots, d-1$ generates the manifold
\be\label{PDefi}
\cM^d \,\approx\,\R^{d-1,1}&=& \left\langle \hat{P}_0,\dots \hat{P}_{d-1}\right\rangle_\R.
\ee
Every point on $\R^{d-1,1}$ is hence uniquely determined by the coefficients $x^\mu$ with respect to this global basis:
\be\label{PDefi0}
x \,=\, x^\mu \hat{P}_\mu \in \cM^d \approx \R^{d-1,1}.
\ee
This defines a coordinate system on $\R^{d-1,1}$, which is globally defined. For a general manifold $\cM^d$, e.g. the $2$-sphere $S^2$, the situation is different: there is no distinguished set of coordinate systems as in the Minkowski case such that the entire manifold $\cM^d$ can be covered by a single coordinate system.\footnote{There also are curved manifolds that have this property, e.g. the $3$ sphere $S^3\approx SU(2)$. These are called parallelizable manifolds that are important for the teleparallel interpretation of general relativity addressed in section \ref{tele}.} However, there is a remnant of this structure: by the very definition of a differentiable manifold, for any point $x\in \cM^d$, there is an open neighbourhood $U$ with $x\in U\subset \cM^d$ which is diffeomorphic to a vector space. The important insight is that one can still use the coordinate basis introduced in (\ref{PDefi}), if the global validity is changed to a local one and if the metric $\eta$ is replaced by the Lorentzian metric $g$ (\ref{metric5}). Technically, this is done by substituting the open neighbourhood $U$ for $\cM^d$ in the definitions above and using the appropriate form of $g$ in local coordinates. This procedure will be of prime importance for the discussion of non-linear realizations in chapter \ref{CHAP4}, where I will work with a fixed coordinate system, i.e. with fixed vectors $\hat{P}_\mu$ (\ref{PDefi}) whose linear span is diffeomorphic to the open set $U\subset \cM^d$.

\subsection{Symmetries of a physical theory}\label{Symm1}
After these first structural remarks I want to introduce the notion of symmetry in a physical theory: a symmetry maps one set of physical fields to another one, with both sets satisfying the equations of motion. In this dissertation, I want to distinguish two possible ways a symmetry can act on physical fields. 

\begin{enumerate}
				\item \textbf{External symmetries}: \\
				An external symmetry acts on the coordinates and thus induces an action by the relevant bundle structure on the physical field. Even for a scalar field $\Phi$ (\ref{Scalar1}), this action is not trivial. This immediately follows from a Taylor expansion about the coordinate $x$. For a general manifold, these transformations are diffeomorphisms and illustrate only the arbitrariness of the choice which local coordinate system has been fixed. Sometimes, they are also referred to as space-time symmetries.
				\item \textbf{Internal symmetries}: \\
				An internal symmetry only acts on the fibres of the tensor bundle and not on the manifold directly. In particular, it does not act on the coordinates. A simple example is provided by the system of two scalar fields with identical dynamics and the symmetry would be a mixing of the two.
		\end{enumerate}
		For both kinds of symmetries, the following definition applies: if the same transformation is considered for all points $x\in \cM^d$, the symmetry is called \textbf{global}. If this is not the case, one speaks of a \textbf{local} or \textbf{soft} symmetry. The process of making a global symmetry local is known as \textbf{gauging} this global symmetry. This will be the main topic of chapter \ref{CHAP3}.\\
		
In chapters \ref{CHAP5} and \ref{CONCL}, I will address the question whether an (in this sense) internal symmetry of supergravity, the gauge symmetry of the $3$-form potential, is in fact an external symmetry in an extended picture with additional coordinates, on which these gauge transformations act in a non-trivial way. I want to emphasize that this interpretation is not in conflict with the Coleman-Mandula theorem \cite{HLS75}, because the geometry is enlarged. \\

Before discussing the symmetry action on physical fields in more detail, I want to mention a possible source of confusion: the discrimination between external and internal symmetries is not equivalent to the distinction between physical symmetries and gauge symmetries in the setting of general relativity. Gauge symmetries always are external symmetries, but physical ones can be of external or internal type.\footnote{An example for a gauge symmetry would be any isometry of the spacelike hypersurface $S^3$ in the de Sitter solution in $d=4$. As $S^3$ does not have a boundary, the Noether charges, e.g. the ADM-mass for asymptotically flat manifolds \cite{ADM62}, are not affected. Hence, the solution is not changed by the symmetry action. For a spacelike hypersurface with non-trivial boundary, however, the Noether charges would be affected, which may lead to a different solution. This is an example for a physical symmetry.} For this thesis however, these notions will not be important.

\subsection{Formal symmetry action on a physical field}\label{SymmAct}
Let $\Phi$ denote the physical field that solves the equations of motion. For this thesis, I want to define the action of a symmetry $\vp_A$ on $\Phi$ to depend analytically\footnote{More general symmetries, e.g. discrete symmetries, are interesting in their own right, but I will not treat them in this thesis.} on a (multi)label $A$
\be\label{symm56}
\vp_A: \Phi &\mapsto& \vp_A[\Phi]\\
\text{with }\quad\vp_{A=0}[\Phi] &=& \Phi.
\nn
\ee
In general, a physical theory contains more than one field. For these cases, $\Phi$ is an abbreviation for the set they create. Following the definition from section \ref{Symm1}, the physical field $\vp_A[\Phi]$ also solves the equations of motion. This defines an equivalence relation
\be
\Phi \sim \vp_A[\Phi].
\ee
The action of the same symmetry with a different (multi)label $B$ on $\vp_A[\Phi]$ results in
\be
\Phi \sim \vp_A[\Phi] \sim \vp_B\left[\vp_A[\Phi]\right].
\ee
Hence, I can define a third (multi)label $C$ for this symmetry by the concatenation of the two symmetry transformations
\be\label{Gruppe}
\vp_C &:=&\vp_B\circ \vp_A.
\ee
Thus, I have defined a product structure on the space of symmetry transformations. Due to the continuous dependence on the (multi)label $A$, there is a unity element $\vp_0=\id$, i.e. a trivial symmetry transformation. And as the two equivalence relations $\Phi_1\sim \Phi_2$ and $\Phi_2\sim \Phi_1$ should have the same meaning, there should also be an inverse symmetry transformation. In other words, for every (multi)label $A$ there is another label $B$ such that $\vp_C=\id$ in (\ref{Gruppe}). Finally, consider the chain of symmetry transformations linking the four solutions $\Phi_1,\dots, \Phi_4$:
\beg
\Phi_1\stackrel{\vp_A}{\sim} \Phi_2\stackrel{\vp_B}{\sim} \Phi_3\stackrel{\vp_D}{\sim} \Phi_4.
\eeg
Following (\ref{Gruppe}), this defines two symmetry transformations $\vp_C :=\vp_B\circ \vp_A$ and $\vp_E :=\vp_D\circ \vp_B$. Hence, there are two possibilities to define a symmetry transformation linking $\Phi_1$ to $\Phi_4$, either $\vp_D\circ \vp_C$ or $\vp_E\circ \vp_A$. Requiring a unique symmetry transformation that links the two solutions $\Phi_1$ and $\Phi_4$ is equivalent to demanding associativity of the product
\be\label{Assoc}
\left(\vp_D\circ \vp_B\right)\circ \vp_A  &=&\vp_D\circ \left(\vp_B\circ \vp_A\right).
\ee
Therefore, the space of analytic symmetry transformations exactly reproduces the definition of a \textbf{Lie group}. The set of all physical fields satisfying the equations of motion, forms an \textbf{orbit} of the Lie group action. The remaining part of this chapter is devoted to this topic. The reason for this is that symmetries play a crucial role in $d=11$ supergravity, whose structures I will investigate in chapter \ref{CHAP5}. In passing, I introduce the following definitions:

{\defi
Let $X,Y$ be objects, on which symmetry transformations can act, such as solutions of physical theories, their equations of motion or actions. 
\begin{itemize}
	\item $X$ is called \textbf{invariant} under a symmetry action $\vp_A$ if
\be\label{Invariant}
\vp_A[X] &=& X.
\ee
As an example may serve the Ricci scalar (\ref{Ricciscalar}) that is invariant under a general coordinate transformation $\vp_A$.
\item $Y$ is called \textbf{covariant} under a symmetry transformation if the induced action on $Y$ is (multi)linear. An example is the transformation of Einstein's equations of motion under a general coordinate transformation $x'=\vp_A(x)$: the induced action is a multiplication with the Jacobi matrix of the diffeomorphism $\vp_A$ of both free indices.\footnote{Sometimes, $Y$ is called a tensor, if it fulfills this property. I do not use this name at this point, because it is also used in contradistinction to ``spinor'' in the physics literature, which also is a covariant object. I will come back to this subtlety in section \ref{Representation}.}
\item A physical theory shall be called \textbf{invariant} under a symmetry transformation, if its equations of motion are covariant.\footnote{If an action $S$ exists, this is equivalent to an invariance of $S$ modulo a constant rescaling.}
\item A covariant derivative or connection $\nabla$ acting on a covariant object $Y$ is called \textbf{equivariant} under a symmetry transformation, if $\nabla Y$ also is covariant.
\end{itemize}
}
The last definition may require some further explanation: I have defined a physical theory by its equations of motion that are differential equations of the physical fields. Furthermore, I have already mentioned in section \ref{Symm1} that a symmetry transformation can be $x$-dependent, in particular if it is gauged. If I want both the physical fields and the equations of motion to be covariant objects, it is natural to replace the differential operator $d$ by a connection $\nabla$ that is equivariant under the symmetry transformation.

\section{Lie theory}\label{phys2}
\subsection{Lie groups and Lie algebras}\label{Lie4}
After this physical introduction to the Lie structure, I want to formalize this to make contact with the mathematical literature \cite{Wa83,FH91}.
{
\defi
A \textbf{Lie group} $G$ is a differentiable manifold which is also endowed with a group structure such that the map
\beg
G\times G&\rightarrow& G\\
(g,h)&\mapsto & g\cdot h^{-1}
\eeg
is differentiable.
}\\

As the group multiplication always is associative, a short comparison of this definition with section \ref{SymmAct} indeed shows that the symmetry structure of a physical theory is provided by a Lie group. The basic building blocks of finite dimensional Lie groups, i.e. of symmetry groups with finitely many symmetry generators, can be classified. This is done by the relation to Lie algebras:
{
\defi
Let $K$ be a field of characteristic $0$. A \textbf{Lie algebra} is a $K$-vector space $\Lie${} with a bilinear (\ref{(1)}), antisymmetric (\ref{(2)}) multiplication $[\cdot,\cdot]$ satisfying the Jacobi identity (\ref{(3)}):
\be\label{(0)}
	[\cdot,\cdot]:&&\Lie \times \Lie 	\longrightarrow \Lie.
\ee
	 For all $\hat{A},\hat{B},\hat{C}\in \Lie $ and for all $\alpha,\beta\in K$, this implies
\begin{subequations}\label{(g)}
\be
	&& \left[\alpha \hat{A}+\beta \hat{B},\hat{C}\right] = \alpha \left[\hat{A},\hat{C}\right] +\beta\left[\hat{B},\hat{C}\right]\label{(1)}\\
	&& \left[\hat{A},\hat{B}\right]=-\left[\hat{B},\hat{A}\right]\label{(2)}\\
	&& \left[\hat{A},\left[\hat{B},\hat{C}\right]\right] +\left[\hat{C},\left[\hat{A},\hat{B}\right]\right] +\left[\hat{B},\left[\hat{C},\hat{A}\right]\right]=0.
	\label{(3)}
\ee
\end{subequations}
}\\

Lie algebras $\Lie$ are linked to Lie groups $G$ by the exponential map
\be\label{exp}
\exp:\Lie &\rightarrow& G\\
\exp(0)&=& \id.
\nn
\ee
The local diffeomorphism $\exp$ is a homomorphism mapping the additive structure of the Lie algebra $\Lie$ to the multiplicative structure of the Lie group $G$.\footnote{The Lie product (\ref{(0)}) is linked to the group multiplication (\ref{Gruppe}) and the Jacobi identity (\ref{(3)}) to the associativity (\ref{Assoc}). Note that $\exp$ is not a global diffeomorphism in general. I will comment on this in section \ref{Representation}.} In the classical domain of physical theories, the real numbers $\R$ are the relevant choice for the field $K$. I will start the classification with the complex numbers $K=\C$ in the next section before refining to real forms in section \ref{realf}.

\subsection{The classification of simple, finite dimensional Lie algebras}\label{Class}
The main aim of this section is to motivate the way this classification of complex Lie algebras is proved \cite{FH91}. In chapter \ref{CHAP5}, I will need this argumentation to identify the geometric structures that are linked to $d=11$ supergravity. 

{
\defi\label{defiAd}
Let $\Lie$ be a finite dimensional, complex Lie algebra. There is a natural action of $\Lie$ on itself, defined by
\beg
	\mathbf{ad}:\Lie\times\Lie &\longrightarrow& \Lie\\
	(\hat{A},\hat{B})&\mapsto& \mathbf{ad}_{\hat{A}}(\hat{B}):=\left[\hat{A},\hat{B}\right].
\eeg
This is called the \textbf{adjoint action} of the Lie algebra $\Lie$ on the vector space $\Lie$. The \textbf{Cartan subalgebra} $\Car$ is defined to be the maximal Abelian diagonizable subalgebra of $\Lie$. Its dimension is called the rank of the Lie algebra $\Lie$.
}\\

In order to define the roots of a Lie algebra, I restrict the adjoint action to the Cartan subalgebra
\beg
	\left.\mathbf{ad}\right|_{\Car}:\Car\times\Lie &\longrightarrow& \Lie\\
	(\hat{H},\hat{G})&\mapsto& \mathbf{ad}_{\hat{H}}(\hat{G})=\left[\hat{H},\hat{G}\right].
\eeg
As all elements of the Cartan subalgebra $\Car$ commute, Schur's lemma and the Jacobi identity (\ref{(3)}) imply that all the maps $\mathbf{ad}_{\hat{H}}$ commute. Hence, there is a common eigenvector of $\mathbf{ad}|_{\Car}$. The eigenvalues however, do depend on the element $\hat{H}\in\Car$. This mapping is called the root of a Lie algebra:
{
\defi
For a given eigenvector $\hat{E}_\alpha\in \Lie$ of the adjoint action $\mathbf{ad}|_{\Car}$, the eigenvalues are given by the functional
\be
	\alpha:\Car&\longrightarrow&\C\nn\\
	\mathbf{ad}_{\hat{H}}\left(\hat{E}_\alpha\right)&=&\alpha(\hat{H})\hat{E}_\alpha.
	\label{Linear3}
\ee
For any eigenvector $\hat{E}_\alpha\in \Lie$, the non-zero functionals $\alpha$ are called \textbf{roots} or equivalently \textbf{weights of the adjoint representation}.
}\\

A further necessary definition is the one of a simple Lie algebra:
{
\defi \label{simple}
A \textbf{simple} Lie algebra $\Lie$ has no non-trivial ideal.\footnote{This is equivalent to stating that a simple $\Lie$ does not have a proper subspace $\ka\subset \Lie$ with the property $[\hat{G},\hat{K}]\in\ka$ for all $\hat{G}\in \Lie$ and all $\hat{K}\in \ka$.}
}\\

With a short calculation using the Jacobi identity (\ref{(3)}), one can prove that if $\alpha$ and $\beta$ are different roots, $\alpha+\beta$ also is a root, if the corresponding common eigenvector $\hat{E}_{\alpha+\beta}=\left[\hat{E}_{\alpha},\hat{E}_{\beta}\right]\in \Lie$ does not vanish. This is one of the most important insights in Lie theory. Together with the finite dimensionality, it results in the fact that there must be a highest root \cite{FH91}. Hence there are finitely many roots for all finite dimensional Lie algebras, which introduces a discrete lattice structure, the root lattice, for a Lie algebra. This is the starting point of the classification and the result is the following table:

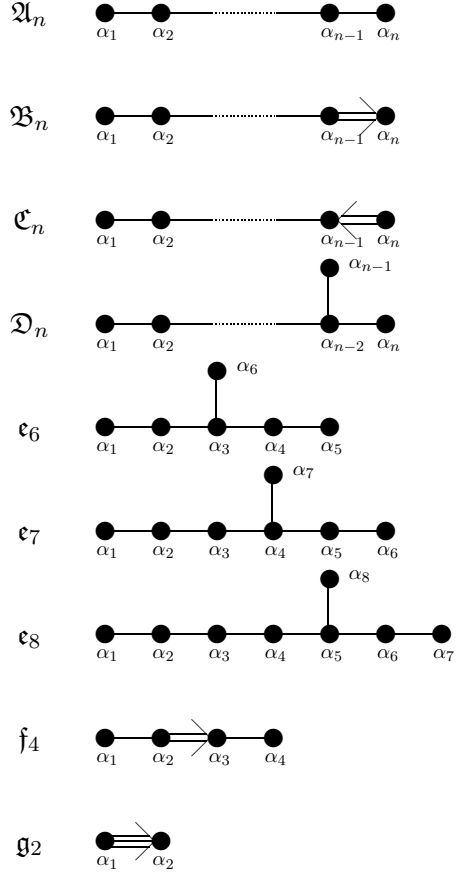
\begin{center}
\begin{tabular}{cl}
$\mathfrak{A}_n$
&
\scalebox{.7}{
\begin{picture}(160,50)(0,0)

\put(155,5){ \circle*{10}}

\put(154,-8){$\alpha_n$}

\put(125,5){ \circle*{10}}
\put(155,5){\line(-1,0){25}}
\put(124,-8){$\alpha_{n-1}$}

\put(125,5){\line(-1,0){25}}

\dottedline{2}(105,5)(60,5)

\put(35,5){ \circle*{10}}
\put(65,5){\line(-1,0){25}}
\put(34,-8){$\alpha_2$}

\put(5,5){ \circle*{10}}
\put(35,5){\line(-1,0){25}}
\put(4,-8){$\alpha_1$}
\end{picture}
}
\\
$\mathfrak{B}_n$
&
\scalebox{.7}{
\begin{picture}(160,50)(0,0)

\put(155,5){ \circle*{10}}

\put(154,-8){$\alpha_n$}

\put(125,5){ \circle*{10}}
\put(153,3){\line(-1,0){25}}
\put(153,7){\line(-1,0){25}}
\put(155,5){\line(-1,1){10}}
\put(155,5){\line(-1,-1){10}}
\put(124,-8){$\alpha_{n-1}$}

\put(125,5){\line(-1,0){25}}

\dottedline{2}(105,5)(60,5)

\put(35,5){ \circle*{10}}
\put(65,5){\line(-1,0){25}}
\put(34,-8){$\alpha_2$}

\put(5,5){ \circle*{10}}
\put(35,5){\line(-1,0){25}}
\put(4,-8){$\alpha_1$}
\end{picture}
}
\\
$\mathfrak{C}_n$
&
\scalebox{.7}{
\begin{picture}(160,50)(0,0)

\put(155,5){ \circle*{10}}

\put(154,-8){$\alpha_n$}

\put(125,5){ \circle*{10}}
\put(155,3){\line(-1,0){20}}
\put(155,7){\line(-1,0){20}}
\put(133,5){\line(1,1){10}}
\put(133,5){\line(1,-1){10}}
\put(124,-8){$\alpha_{n-1}$}

\put(125,5){\line(-1,0){25}}

\dottedline{2}(105,5)(60,5)

\put(35,5){ \circle*{10}}
\put(65,5){\line(-1,0){25}}
\put(34,-8){$\alpha_2$}

\put(5,5){ \circle*{10}}
\put(35,5){\line(-1,0){25}}
\put(4,-8){$\alpha_1$}
\end{picture}
}
\\
$\mathfrak{D}_n$
&
\scalebox{.7}{
\begin{picture}(160,50)(0,0)

\put(155,5){ \circle*{10}}

\put(154,-8){$\alpha_n$}

\put(125,5){ \circle*{10}}
\put(155,5){\line(-1,0){25}}
\put(124,-8){$\alpha_{n-2}$}

\put(125,35){ \circle*{10}}
\put(128,35){\line(0,-1){25}}
\put(139,35){$\alpha_{n-1}$}

\put(125,5){\line(-1,0){25}}

\dottedline{2}(105,5)(60,5)

\put(35,5){ \circle*{10}}
\put(65,5){\line(-1,0){25}}
\put(34,-8){$\alpha_2$}

\put(5,5){ \circle*{10}}
\put(35,5){\line(-1,0){25}}
\put(4,-8){$\alpha_1$}
\end{picture}
}
\\
$\mathfrak{e}_6$
&
\scalebox{.7}{
\begin{picture}(130,50)(0,0)

\put(125,5){ \circle*{10}}

\put(124,-8){$\alpha_5$}

\put(95,5){ \circle*{10}}
\put(125,5){\line(-1,0){25}}
\put(94,-8){$\alpha_4$}

\put(65,5){ \circle*{10}}
\put(95,5){\line(-1,0){25}}
\put(64,-8){$\alpha_3$}

\put(65,35){ \circle*{10}}
\put(68,35){\line(0,-1){25}}
\put(79,35){$\alpha_{6}$}

\put(35,5){ \circle*{10}}
\put(65,5){\line(-1,0){25}}
\put(34,-8){$\alpha_2$}

\put(5,5){ \circle*{10}}
\put(35,5){\line(-1,0){25}}
\put(4,-8){$\alpha_1$}
\end{picture}
}
\\
$\mathfrak{e}_7$
&
\scalebox{.7}{
\begin{picture}(160,50)(0,0)

\put(155,5){ \circle*{10}}

\put(154,-8){$\alpha_6$}

\put(125,5){ \circle*{10}}
\put(155,5){\line(-1,0){25}}
\put(124,-8){$\alpha_5$}

\put(95,5){ \circle*{10}}
\put(125,5){\line(-1,0){25}}
\put(94,-8){$\alpha_4$}

\put(95,35){ \circle*{10}}
\put(98,35){\line(0,-1){25}}
\put(109,35){$\alpha_{7}$}

\put(65,5){ \circle*{10}}
\put(95,5){\line(-1,0){25}}
\put(64,-8){$\alpha_3$}

\put(35,5){ \circle*{10}}
\put(65,5){\line(-1,0){25}}
\put(34,-8){$\alpha_2$}

\put(5,5){ \circle*{10}}
\put(35,5){\line(-1,0){25}}
\put(4,-8){$\alpha_1$}
\end{picture}
}
\\
$\mathfrak{e}_8$
&
\scalebox{.7}{
\begin{picture}(190,50)(0,0)

\put(185,5){ \circle*{10}}

\put(184,-8){$\alpha_7$}

\put(155,5){ \circle*{10}}
\put(185,5){\line(-1,0){25}}
\put(154,-8){$\alpha_6$}

\put(125,5){ \circle*{10}}
\put(155,5){\line(-1,0){25}}
\put(124,-8){$\alpha_5$}

\put(125,35){ \circle*{10}}
\put(128,35){\line(0,-1){25}}
\put(139,35){$\alpha_{8}$}

\put(95,5){ \circle*{10}}
\put(125,5){\line(-1,0){25}}
\put(94,-8){$\alpha_4$}

\put(65,5){ \circle*{10}}
\put(95,5){\line(-1,0){25}}
\put(64,-8){$\alpha_3$}

\put(35,5){ \circle*{10}}
\put(65,5){\line(-1,0){25}}
\put(34,-8){$\alpha_2$}

\put(5,5){ \circle*{10}}
\put(35,5){\line(-1,0){25}}
\put(4,-8){$\alpha_1$}
\end{picture}
}
\\
$\mathfrak{f}_4$
&
\scalebox{.7}{
\begin{picture}(100,50)(0,0)

\put(95,5){ \circle*{10}}

\put(94,-8){$\alpha_4$}

\put(65,5){ \circle*{10}}
\put(95,5){\line(-1,0){25}}
\put(64,-8){$\alpha_{3}$}

\put(35,5){ \circle*{10}}
\put(63,3){\line(-1,0){25}}
\put(63,7){\line(-1,0){25}}
\put(65,5){\line(-1,1){10}}
\put(65,5){\line(-1,-1){10}}
\put(34,-8){$\alpha_2$}

\put(5,5){ \circle*{10}}
\put(35,5){\line(-1,0){25}}
\put(4,-8){$\alpha_1$}
\end{picture}
}
\\
$\mathfrak{g}_2$
&
\scalebox{.7}{
\begin{picture}(40,50)(0,0)

\put(35,5){ \circle*{10}}

\put(34,-8){$\alpha_2$}

\put(5,5){ \circle*{10}}
\put(35,5){\line(-1,1){10}}
\put(35,5){\line(-1,-1){10}}
\put(35,5){\line(-1,0){25}}
\put(32,2){\line(-1,0){25}}
\put(32,8){\line(-1,0){25}}
\put(4,-8){$\alpha_1$}
\end{picture}
}
\end{tabular}
\end{center}
\begin{center}
\end{center}
Every node in these diagrams is labeled by a root $\alpha_i$ and corresponds to an $\mathfrak{A}_1$ algebra in the following sense: For the node with label $i=1,\dots,n$ with $n=\text{rank}(\Lie)$, one associates three vectors $(\hat{E}_{+\alpha_i},\hat{E}_{-\alpha_i},\hat{H}_{\alpha_i})$, a so-called Chevalley triple forming an $\mathfrak{A}_1$ subalgebra of $\Lie$. Its commutation relations are 
\begin{subequations}\label{A1C}
\beg
\left[\hat{E}_{+\alpha_i},\hat{E}_{-\alpha_j}\right]&=&\delta_{ij} \hat{H}_{\alpha_i}\\
\left[\hat{H}_{\alpha_i},\hat{E}_{\pm\alpha_j}\right]&=&\pm\alpha_{j}(\hat{H}_{\alpha_i}) \hat{E}_{\pm\alpha_j}.
\eeg
\end{subequations}
The $n$ commuting generators $\hat{H}_{\alpha_i}$ form the Cartan subalgebra $\Car$ of $\Lie$. The complete Lie algebra $\Lie$ is a subspace of the {\textbf{free algebra}}, which is the vector space generated by the $3n$ vectors $(\hat{E}_{+\alpha_i},\hat{E}_{-\alpha_i},\hat{H}_{\alpha_i})$ and arbitrary commutators thereof. The precise embedding of $\Lie$ in the free algebra is provided by the relation between the different $\mathfrak{A}_1$ subalgebras, i.e. the commutators of $E_{\alpha_i}$ with $E_{\alpha_j}$ for $i\neq j$. These are encoded in the Dynkin diagram or equivalently, in the Cartan matrix $A$:\footnote{For the cases discussed here, $A$ is an $n\times n$ matrix with $2$ on the diagonal. If the nodes $i$ and $j$ are linked with one line, $A_{ij}=A_{ji}=-1$. If there are two or three lines between the nodes $i$ and $j$ with the arrow pointing to $i$, $A_{ij}=-2$ or $A_{ij}=-3$ respectively, whereas one still has $A_{ji}=-1$. All the other elements are zero. For a generalization, consult \cite{K03}.}
\be
\alpha^i\left(\hat{H}_{\alpha_j}\right)&=&A_{ji}\label{CartanM}\\
	\left(\mathbf{ad}_{\hat{E}_{\alpha_i}}\right)^{1-A_{ij}}\hat{E}_{\alpha_j} &=&0.
	\label{Serre}
\ee
To sum up, every simple, finite dimensional, complex Lie algebra $\Lie$ is the quotient space of the corresponding free algebra modulo the Serre relations (\ref{Serre}), modulo the commutation relations $[\hat{H}_{\alpha_i},\hat{H}_{\alpha_j}]=0$ and modulo (\ref{A1C}). $\Lie$ is hence uniquely characterized by a Dynkin diagram or equivalently, by its commution relations.\\

This property is the starting point of Kac-Moody theory \cite{K95}. I have already mentioned above that the derivation of the discrete structure of Lie theory heavily relies upon the finite dimensionality of the algebras. In discussing infinite dimensional Lie algebras, Kac and Moody realized that postulating the Serre relations (\ref{Serre}) for an infinite dimensional Lie algebra also leads to a very interesting mathematical structure. This can also be encountered in physical theories. An example is the Geroch group \cite{BM87,G72}, which is a symmetry acting on the space of solutions of General Relativity with two commuting, independent Killing vectors.\footnote{Its corresponding Lie algebra $\mathfrak{A}_1^+$ \cite{J81} possesses two $\mathfrak{A}_1$ subalgebras, whose corresponding real groups are called Ehlers and Matzner-Misner group respectively. The Kramer-Neugebauer transformation from General Relativity \cite{ENKGM} corresponds to the automorphism of the Dynkin diagram of $\mathfrak{A}_1^+$.
}\\

As the Serre relations provide the same information as the Dynkin diagram, any Kac-Moody algebra is uniquely defined by its Dynkin diagram. However, note that in contradistinction to the finite dimensional case, this subset of infinite dimensional Lie algebras does not contain all physically interesting structures: the Virasoro algebra of $d=2$ conformal field theory and the algebra of all vector fields are two very prominent examples for infinite dimensional Lie algebr\ae{} of non-Kac Moody type, i.e. they do not have an associated Dynkin diagram.\\

\subsection{Representation theory}\label{Representation0}
I have defined physical fields to be sections of a tensor bundle over a $d$-dimensional manifold. This implies that for every point $x\in \cM^d$, the physical field can be written as a tensor product of vector spaces. Therefore, the induced action of a symmetry on these vector spaces has to be a linear one. Hence, one is led to the definition of a linear representation:

{
\defi
A linear representation $\mathbf{R}$ of a Lie algebra $\Lie$ is a map from the Lie algebra $\Lie$ to the endomorphisms $End(V)$ of a \textbf{representation space} $V$
\beg
	\mathbf{R}:\Lie&\rightarrow& End(V)\\
	  \hat{G} &\mapsto &\mathbf{R}_{\hat{G}}.
\eeg
Furthermore, this map must be compatible with all the properties of the Lie algebra (\ref{(g)}). In formul\ae:
	\begin{itemize}
	\item $\left[\mathbf{R}_{\hat{G}_1},\mathbf{R}_{\hat{G}_2}\right] = \mathbf{R}_{\left[\hat{G}_1,\hat{G}_2\right]}$
	\item $\mathbf{R}_{a\hat{G}_1+b\hat{G}_2}=a\mathbf{R}_{\hat{G}_1} + b\mathbf{R}_{\hat{G}_2}$
\end{itemize}
with $g_1,g_2\in \Lie$, $a,b\in \R$ or $\C$. The vector space $V$ is also called $\Lie$-\textbf{module}.
}\\

For the case $dim(V)<\infty$, the elements of $End(V)$ are matrices and the commutator is defined by the ordinary matrix product $\circ$
\beg
\left[\mathbf{R}_{\hat{G}_1},\mathbf{R}_{\hat{G}_2}\right]:=\mathbf{R}_{\hat{G}_1}\circ\mathbf{R}_{\hat{G}_2} -\mathbf{R}_{\hat{G}_2}\circ\mathbf{R}_{\hat{G}_1}.
\eeg
For the case of the diffeomorphism symmetry in section \ref{Diffeom}, the objects $\mathbf{R}_{\hat{G}}\in End(V)$ will be derivative operators and $\circ$ will be the concatenation of functions respecting the chain rule.\footnote{The reader will also encounter an example of a non-linear representation $\mathbf{R}$ in this context, the one of the conformal algebra $\mathfrak{so}_{(d,2)}$ on Minkowski space $\R^{d-1,1}$. The compatibility with all the properties of a Lie algebra still holds, but $\mathbf{R}_{\hat{G}}$ is not an endomorphism of the Minkowksi space.} To classify linear representations, the notion of irreducibility is useful.
{
\defi
A representation $\mathbf{R}$ of a Lie algebra $\Lie$ is called \textbf{irreducible} if the corresponding $\Lie$-module $V$ does not contain a proper submodule.\footnote{In other words, there is no proper subspace $\ka$ of $V$ on which a non-trivial representation $\tilde{\mathbf{R}}$ of $\Lie$ is defined $\tilde{\mathbf{R}}_{\Lie}:\ka\rightarrow\ka$. For example, $V=\mathfrak{gl}_d$ is not irreducible as a $\Lie=\mathfrak{so}_d$ representation, because $\ka=\mathfrak{so}_d$ is a proper subspace.}
}\\

I had already introduced an example for a representation before, the adjoint representation $\mathbf{R}=\mathbf{ad}$ (\ref{(0)}). Irreducibility for this representation is equivalent to the Lie algebra $\Lie$ being simple. This already hints at the fact that the classification of all admissible, irreducible, finite dimensional representations $\mathbf{R}$ of a Lie algebra and the classification of all finite dimensional, simple Lie algebras from section \ref{Class} are closely related: one simply has to replace $\mathbf{ad}$ by $\mathbf{R}$. Thus, the roots or weights of the adoint representation are replaced by the weights of a general representation. The finite dimensionality again introduces a discrete structure, the \textbf{weight lattice}. This leads to a classification of all inequivalent, irreducible, linear representations $\mathbf{R}$ of simple, finite dimensional, complex Lie algebras $\Lie$ \cite{FH91}.

\subsection{Real forms}\label{realf}
In this dissertation, real forms of the finite dimensional Lie algebras $\mathfrak{A}_n$, $\mathfrak{B}_n$, $\mathfrak{C}_n$, $\mathfrak{D}_n$ and of the exceptional one $\mathfrak{e}_7$ will play a prominent role. To define the notion of a real form, recall that all these complex algebras are uniquely determined by their Dynkin diagram or equivalently, by their non-degenerate Cartan-Killing matrices $A$. For the algebras under consideration, these $A$ can always be written as a matrix product of a diagonal matrix with a symmetric one \cite{K95}, from which it is possible to define a symmetric, non-degenerate $\C$-bilinear form
\beg
\langle\cdot,\cdot\rangle : \Car \times \Car &\rightarrow &\C.
\eeg
This form can be extended to the entire complex Lie algebra $\Lie$ by requiring \cite{K95}
\beg
\langle [\hat{A},\hat{B}],\hat{C}\rangle &=&\langle \hat{A},[\hat{B},\hat{C}]\rangle\\
\langle \hat{E}_{\alpha_i},\hat{E}_{-\alpha_j}\rangle&\propto&\delta_{ij}.
\eeg
Furthermore, there always exist sesquilinear endomorphisms $\omega$ on $\Lie$, so-called involutions with the property $\omega\circ \omega =\id_{\Lie}$. These induce Hermitean quadratic forms
\beg
(\cdot,\cdot)&:=&\langle\cdot,-\omega(\cdot)\rangle\in \R,
\eeg
which are equivalent to specifying a real norm on $\Lie$, the Killing norm. Due to the fact that, in contrast to $\C$, the field $\R$ is not algebraically closed, the signature of the Killing norm $(\cdot,\cdot)$ distinguishes different real forms of a single complex Lie algebra.\footnote{A complete list of all real forms can be found in \cite{C14}. Diagrammatically, they are distinguished by so-called Satake diagrams \cite{S60}.} The character $\chi$ of a real form is the difference between the numbers of linearly independent vectors in $\Lie$ with positive and negative Killing norm. In general, the character can have the values $-dim(\Lie)\leq \chi\leq rank(\Lie)$. The different real forms of $\Lie$ are then defined to be the fixed point sets of different choices of involutions $\omega$, e.g. for $\Lie=\mathfrak{e}_7$
\be
\mathfrak{e}_{7(\chi)} &:=& \left\{X\in \mathfrak{e}_{7}| \omega(X) = X\right\}.
\ee
In this dissertation, I will only use its split real form, which is simply obtained by taking real coefficients in the span of the free algebra from section \ref{Class}. It contains $63$ generators with negative Killing norm. As $dim_\R(\mathfrak{e}_{7(\chi)})=133$, there are $70$ generators with positive Killing norm and hence $\chi=70-63=7$. This real form $\mathfrak{e}_{7(7)}$ will be linked to $d=11$ supergravity in chapter \ref{CHAP5}.
\\

Generators with negative Killing norm are also referred to as compact generators, the ones with positive norm are called non-compact. This name results from the observation that in the matrix representation of $\mathfrak{A}_n$, the transposition of matrices together with a complex conjugation is an example for an involution $-\omega$. Since the corresponding Killing norm is the trace of the matrix product, the antihermitean matrices are called compact. This is consistent with the fact that the corresponding Lie group is compact as a topological manifold.\newpage

Another important real form for this thesis is the compact one. Following the name, all its generators are compact, i.e. $\chi =-dim(\Lie)$. The compact forms of the orthogonal algebras $\mathfrak{B}_n$ and $\mathfrak{D}_n$ have the particular name $\mathfrak{so}_{2n+1}$ and $\mathfrak{so}_{2n}$, respectively, because they have a representation as antisymmetric matrices $A$ in $d=2n+1,2n$ dimensions. The antisymmetry is defined with respect to a symmetric tensor $\eta_{\mu\rho}$ with $\mu,\rho=1,\dots,d$, the Euclidean metric $\eta=(+1,\dots,+1)$ and $\eta^{\nu\mu}\eta_{\mu\rho} = \delta_\rho^\nu$
\be\label{AntisyM}
{A_\mu}^\nu &=& -\eta_{\mu\rho}{A_\sigma}^\rho \eta^{\sigma\nu}.
\ee
I will also consider other real forms, which are conventionally denoted by $\mathfrak{so}_{(2n+1-p,p)}$ and $\mathfrak{so}_{(2n-p,p)}$ with $p=0,\dots n$, corresponding to metrics $\eta_{\mu\rho}$ with a different signature, i.e. $p$ minus signs. Prime examples are the Minkowski metric for the Lorentzian case $\mathfrak{so}_{(d-1,1)}$ and the split real form with $p=n$. \\

In chapter \ref{CHAP5}, the real forms of the symplectic algebra $\mathfrak{C}_n$ will be important. Their standard representation is provided by $2n\times 2n$ matrices $A$ that are subject to a similar constraint as (\ref{AntisyM})
\be\label{AntisyM2}
{A_\mu}^\nu &=& -\Omega_{\mu\rho}{A_\sigma}^\rho \Omega^{\sigma\nu}
\ee
with an antisymmetric, non-degenerate tensor $\Omega_{\mu\rho}$ and $\Omega^{\nu\mu}\Omega_{\mu\rho} = \delta_\rho^\nu$. I will denote its split real form by $\mathfrak{sp}_{2n}$ and its compact one by $\mathfrak{usp}_{2n}$. Finally, the split real form of $\mathfrak{A}_n$ has the name $\mathfrak{sl}_{n+1}$ and the compact one $\mathfrak{su}_{n+1}$. \\

To conclude, recall that the entire classification was aimed at simple Lie algebras, defined in \ref{simple}. An example of a non-simple algebra is $\mathfrak{gl}_{n+1}$, containing $\mathfrak{sl}_{n+1}$ and a one-dimensional Abelian ideal. This is the algebra of all real $(n+1)\times (n+1)$-matrices, which hence contains orthogonal and symplectic subalgebras. This fact also leads to my definition of the \textbf{maximal compact subalgebra} of a Lie algebra $\Lie$: it consists of the generators of $\Lie$ that are antisymmetric matrices in the embedding of $\Lie$ in $\mathfrak{gl}_{d}$ for $d$ as small as possible.\footnote{For $\Lie=Sp(2n)$, this deviates from the standard definition of a maximal compact subalgebra, because the dimension of the subalgebra $\mathfrak{usp}_n\oplus \mathfrak{usp}_n$ is bigger than the one of $\mathfrak{u}_n$, but the former does not consist of antisymmetric matrices in $d=2n$.} For the case of $\mathfrak{e}_{7(7)}$, I will explicitly show in chapter \ref{CHAP5} that the $63$ generators of $\mathfrak{e}_{7(7)}$ with negative Killing norm actually form the subalgebra $\mathfrak{su}_{8}$, which are antisymmetric matrices in $d=56$.

\subsection{Representations of Lie groups}\label{Representation}
It follows from the classification of irreducible representations in section \ref{Representation0} that every irreducible representation of $\mathfrak{sl}_{d}$ is a subrepresentation of a (multiple) tensor product of the $d$-dimensional vector representation $\mathbf{d}$ and its dual or contragredient representation $\overline{\mathbf{d}}$ \cite{FH91}. This is the reason why $\mathbf{d}$ and $\overline{\mathbf{d}}$ are referred to as \textbf{fundamental} and antifundamental representations of $\mathfrak{sl}_{d}$. Since the orthogonal algebra $\mathfrak{so}_{d}$ is a subalgebra of $\mathfrak{sl}_{d}$, every representation of $\mathfrak{sl}_{d}$ also is a representation of $\mathfrak{so}_{d}$. There are however representations of $\mathfrak{so}_{d}$ that are not representations of $\mathfrak{sl}_{d}$. These are the \textbf{spin} representations that are related to representations of the Clifford algebra \cite{FH91,dW02}, which will be important for the discussion of fermions in physical theories in chapters \ref{CHAP3} and \ref{CHAP5}.\\

For the representation of Lie groups, recall from section \ref{Lie4} that the homomorphism $\exp$ (\ref{exp}) locally maps the Lie algebra $\Lie$ to a corresponding group $G$. For the finite dimensional Lie algebras, $\exp$ is the standard matrix exponential. Thus, every representation $\mathbf{R}$ of a Lie algebra $\Lie$ induces a representation of a simply connected Lie group $\tilde{G}$ \cite{Mi83}. As $\exp$ merely is a local diffeomorphism, the Lie groups $G$ and $\tilde{G}$ may have different topologies. For $\Lie = \mathfrak{sl}_{d}$, it is standard to denote the simply connected Lie group $\tilde{G}$ by $Sl(d)$. For $\Lie = \mathfrak{so}_{(d-p,p)}$, this is not the case: $SO(d-p,p)$ is by definition a subgroup of $Sl(d)$. This is the reason why $SO(d-p,p)$ is not simply connected. Therefore, the spin representations are not representations of $SO(d-p,p)$, but representations of its simply connected covering group\footnote{The discussion of these discrete subgroups has often been disregarded in the literature in the past, e.g. in the context of possible symmetry groups $Sl(32)$ or $Spin(32)$ of M-theory. By referring to these discrete subgroups, these conjectures could be strongly constrained \cite{K03}.} 
\beg
\tilde{G} &=& Spin(d-p,p)\\
SO(d-p,p) &=& Spin(d-p,p)/\Z_2.
\eeg
It also is standard to reserve the notion \textbf{tensor representation} for $\mathfrak{sl}_d$ representations. $\mathfrak{so}_{(d-p,p)}$ or $Spin(d-p,p)$ representations that are not tensors in this sense are called \textbf{spinors}. This is the reason why I have not called $Y$ a tensor in the definitions of section \ref{SymmAct}. In other words, only representations with an even number of spinor indices are representations of the group $SO(d-p,p)$, because an even number of spinors can always be transformed into tensor representations by the Fierz identity. \\

This fact that spinors are not a representation of the Lorentz group $SO(3,1)$ may have led some authors to the conclusion that spinors must be described by infinite dimensional spin representations in the context of non-linear realizations \cite{K05}. Since all physical expectation values are of even degree in fermions, the Fierz identity argument applies, however. Hence, it is perfectly consistent to describe physical objects as $\Z_2$ equivalence classes of spinors, on which a Lorentz group action is well-defined. It is common in the physics literature not to mention this subtlety explicitly and just to talk about an induced Lorentz action on spinors instead. I will adopt this convention in the chapters \ref{CHAP3} and \ref{CHAP4}. \\

For the discussion of supergravity in chapter \ref{CHAP5}, this distinction will also be important. Therefore I want to add an easy prescription how to identify the topology of a subgroup. Let $K(G)$ be the maximal compact subgroup of a Lie group $G$. Then all admissible $K(G)$ representations arise from a decomposition of all admissible $G$ representations under $K(G)$. For the case $G=E_{7(7)}$, the fundamental representation $\mathbf{56}$ decomposes into the $\mathfrak{su}_8$ representations $\mathbf{28}\oplus \overline{\mathbf{28}}$. This implies that all the admissible representations of $K(E_{7(7)})$ have an even number of $SU(8)$ indices: the discrete group action that multiplies every index with a minus sign can hence be divided out. In the context of supergravity in chapter \ref{CHAP5}, this implies that $K(E_{7(7)})=SU(8)/\Z_2$ is the covariance group of the bosons, whereas its simply connected covering group $SU(8)$ is the covariance group of the fermions.

\section{Diffeomorphisms}\label{phys3}
I have already discussed in the context of Dynkin diagrams in section \ref{Class} that a Lie algebra $\Lie$ can often be described as the free algebra of $\mathfrak{A}_1$ subalgebras modulo some relations. Most prominent among these are the Serre relations (\ref{Serre}) for the case of (infinite dimensional) Kac-Moody algebras. This section will deal with a non-Kac-Moody algebra corresponding to a subgroup $\Diff(d)$ of the group of diffeomorphisms on a manifold. This can also be constructed from two finite dimensional subalgebras by Ogievetsky's theorem \cite{O73}, which can also be interpreted as defining relations between the two subalgebras with the help of their vector field representation $\mathbf{R}$.\\

I will start with the geometric origin of the symmetry group of diffeomorphisms in physical theories. Then, I will restrict this group to the subgroup $\Diff(d)$ that corresponds to the Lie algebra of analytic vector fields $\mathfrak{diff}_d$ in section \ref{Diffeom}. After discussing its relevant subgroups in \ref{Subalg}, I will prove Ogievetsky's theorem in \ref{Ogievetsky}.

\subsection{Diffeomorphisms in physical theories}\label{diffi}
I have defined physical fields in section \ref{GeomStr} to be sections of a tensor bundle over an analytic manifold $\cM^d$. By definition \cite{Wa83}, $\cM^d$ is equipped with an atlas of coordinate charts $(U_\alpha, x_\alpha)$ that consists of simply connected open sets $U_\alpha$ that cover $\cM^d$ and mappings 
\be\label{charts}
	x_\alpha: \quad U_\alpha &\rightarrow & x_\alpha (U_\alpha)\subset \R^d.
\ee
For every open set $U_\alpha$, the coordinate charts $x_\alpha$ provide analytic diffeomorphisms between $U_\alpha$ and an open set $x_\alpha (U_\alpha)$ in the flat vector space $\R^d$. Given an open set on $\cM^d$, it is obvious that there are many choices for a coordinate chart. The content of a physical theory should hence not depend on the arbitrariness of which coordinate chart has been chosen. This is referred to as \textbf{general coordinate invariance} of a physical theory, or \textbf{diffeomorphism symmetry}, equivalently. In formul\ae, for every point $p\in \cM^d$ and for two arbirary charts $(U_\alpha, x_\alpha)$ and $(U_\beta, x_\beta)$ with $p\in U_\alpha\cap U_\beta\subset \cM^d$, the diffeomorphism $\tilde{\vp}$ is a map from $x_\beta (U_\beta)\subset \R^d$ to $\R^d$
\be\label{diffeom56}
\tilde{\vp} \,:=\,x_\alpha\circ x^{-1}_\beta:\quad x_\beta (U_\beta)&\rightarrow & \R^d.
\ee

Hence, for the discussion of a physical theory with local dependence on $\cM^d$, it is no restriction to fix one coordinate chart $(U_\alpha, x_\alpha)$, if the equations of motion are covariant under the symmetry action of diffeomorphisms.\footnote{This covariance follows the definition of section \ref{SymmAct}. In other words, a solution to the equations of motion has to be identified with all solutions that are generated by the action of a diffeomorphism.} This implies that I can without loss of generality introduce basis vectors $\hat{P}_\mu$ (\ref{PDefi}) for the vector space $\R^d$, in which $x_\beta (U_\beta)$ is embedded. This open set can then be parametrized by coordinates $x^\mu\in \R$ with $\mu=0,\dots, d-1$ in the same way as the flat Minkowski space (\ref{PDefi0}).

\subsection{The group of diffeomorphisms $\Diff(d)$}\label{Diffeom}
The requirement of locality for a physical theory, which is standard for a theory that should be quantized \cite{HLS75}, has a further implication. Since the dependence on the point on the manifold $\cM^d$ is local, it is natural to ignore diffeomorphisms that link points on the manifold in a non-local way. Technically, this offers the possibility to use the same basis vectors $\hat{P}_\mu$ (\ref{PDefi}) for both open sets $x_\beta (U_\beta)\subset \R^d$ and $\tilde{\vp}\circ x_\beta (U_\beta)\subset \R^d$. Furthermore, I can restrict the group of diffeomorphisms to a subgroup $\Diff(d)$ whose elements $\vp_A$ are labelled by some (multi)label $A$ and are analytically connected to the identity map. In the basis provided by $\hat{P}_\mu$, the coordinates transform as follows
\be\label{GenDiff5}
\vp_A :x^\mu&\mapsto& {x'}^\mu \,:=\,\vp_A^\mu(x)\\
\vp_{A=0}:x^\mu&\mapsto& x^\mu\nn.
\ee
In this way, the group of diffeomorphisms of any $d$-dimensional manifold is restricted to a subgroup that does not depend on the global structure of $\cM^d$ any more. It is therefore consistent to simply denote it by $\Diff(d)$.\footnote{It may also be interesting to discuss the relevance of other diffeomorphism structures to physics, but I will refrain from doing so in this thesis.} \\

The definition of $\Diff(d)$ strongly reminds of the definition of a symmetry group in section \ref{SymmAct}. It turns out that one can exactly follow the line of argumentation used for finite dimensional Lie groups. The fact that any $\vp_A\in \Diff(d)$ is by definition analytically connected to the identity map, allows to prove that $\Diff(d)$ is locally diffeomorphic to a Lie algebra $\mathfrak{diff}_{d}$: 

{\thm\label{Satz1}
For any analytic manifold $\cM^d$, the subgroup $\Diff(d)$ (\ref{GenDiff5}) of diffeomorphisms on $\cM^d$ is a Lie group in the sense specified in the proof. Its corresponding Lie algebra is provided by
the vector space $\mathfrak{diff}_{d}$ of all analytic vector fields, endowed with the standard Lie bracket.
}

{\pf
At first, I would like to remind the reader of the definition of a vector field $X_A\in \mathfrak{diff}_{d}$
\beg
X_A: U &\rightarrow& TU\subset TM\\
x &\mapsto& X_A(x)\in T_x\cM^d\approx \R^{d-1,1},
\eeg
where $X_A$ may depend analytically on the point $x\in U\subset \cM^d$ and $A$ is some (multi)label, on which $X_A$ depends \textbf{linearly}. Since I fixed a particular chart $(U_\alpha,x_\alpha)$, it is sufficient to state the vector field in the basis $\hat{P}_\mu$ that I chose for the open set $x_\alpha(U)$ and that canonically induces a basis for the tangent vector space $T_x\cM^d$
\be\label{Vectorfields}
 x^\mu  &\mapsto& X_A^\mu(x),
\ee 
where the $d$ coefficients $X_A^\mu$ are analytic functions of the coordinates $x^0,\dots, x^{d-1}$. Thus, $\mathfrak{diff}_{d}$ is a vector space. To prove the theorem \ref{Satz1}, I have to define a Lie bracket on $\mathfrak{diff}_{d}$. This is a bilinear antisymmetric mapping 
\be\label{Lie7}
[\cdot,\cdot]: \qquad \mathfrak{diff}_d \times \mathfrak{diff}_d &\rightarrow & \mathfrak{diff}_d\nn\\
(X_A,X_B)&\mapsto&[X_A,X_B]
\ee
which has to satisfy the Jacobi identity (\ref{(3)}). The standard definition is by representing the basis vectors $\hat{P}_\mu$ of the tangent space $T_x\cM^d$, which are induced by the choice of coordinate vectors $\hat{P}_\mu\in\R^{d-1,1}\approx U\subset \cM^d$, as derivative operators 
\be\label{Lie9}
\mathbf{R}_{\hat{P}_\mu} &=&\p_\mu \,:=\,\frac{\p}{\p x^\mu}
\ee
that act on the $x$-dependent coefficients $X_A^\mu$ of the vector fields $X_A\in \mathfrak{diff}_d$. This implies that the coefficient $[X_A,X_B]^\mu$ of the vector field $[X_A,X_B]\in \mathfrak{diff}_d$ has the form
\be\label{Lie8}
 [X_A,X_B]^\mu &=&X_A^\nu \p_\nu X_B^\mu  -X_B^\nu \p_\nu X_A^\mu .
\ee
This endows the vector space $\mathfrak{diff}_{d}$ with the structure of a Lie algebra.
\\

Next, I have to make contact to the formula (\ref{GenDiff5}) for an arbitary analytic diffeomorphism $\vp_A\in \Diff(d)$ that depends analytically on the (multi)label $A$ and fulfills $\vp_{A=0}=\id$. Due to these properties, every $\vp_a\in \Diff(d)$ uniquely specifies a vector field $X_A\in \mathfrak{diff}_d$ by a Taylor expansion in $A$
\be\label{exp9}
\vp_A^\mu(x) &=& x^\mu + X_A^\mu(x) + \mathcal{O}(A^2)
\ee
in its domain of validity $x\in x_\alpha(U_\alpha)$. I prove in the appendix \ref{Pf1} that $\vp_A$ can be reconstructed from $X_A$ by the explicit formula 
\be\label{Diffeom2}
\vp_A^\mu(x) &=&\exp (X_A^\nu(x)\p_\nu) x^\mu.
\ee
Its evaluation is performed by an expansion of the exponential series taking into account the chain rule for differentiations. This defines the local homomorphism (\ref{exp}) for the Lie algebra $\mathfrak{diff}_d$ 
\be\label{Diffeom1}
\exp:\mathfrak{diff}_d&\rightarrow \Diff(d).
\ee
Thus, $\Diff(d)$ indeed is a Lie group, but there is a subtlety to observe that is related to the domain of validity $x_\alpha(U_\alpha)$ of the finite diffeomorphism $\vp_A$:

\begin{itemize}
\item Given any point $p\in \cM^d$, I can fix the chart such that $x_\alpha(p)=0$. For any fixed vector field $X_A\in \mathfrak{diff}_d$, there is an open neighbourhood $U\subset x_\alpha(U_\alpha)\subset \R^d$ of $x_\alpha(p)=0$ on which the formula (\ref{Diffeom2}) converges for all $x\in U$. 
\item In particular, it is not possible in general to multiply the Lie algebra element by an arbitrary finite number without risking to destroy the convergence of the series (\ref{Diffeom2}), which is in contrast to finite dimensional Lie groups. This is related to the question of local and global existence of a coordinate chart in an obvious way.
\item The concatenation $\vp_A\circ \vp_B$ of diffeomorphisms $\vp_A,\vp_B\in \Diff(d)$ only is an element in $\Diff(d)$, if the domain of validity $U_B$ of $\vp_B$ is restricted in such a way that the concatenation still converges.
\end{itemize}
\qed
}\\

Since I am discussing local physical theories in this thesis, I will not have to mention this subtlety in the sequel, unless it is explicitly necessary as for the case of the conformal diffeomorphism $\vp^{\mathfrak{c}}_A$ (\ref{SCTcoord}) in section \ref{Subalg}.\\

In concluding, I want to mention that $\vp_A\in \Diff(d)$ is also called the integral curve to a vector field $X_A\in \mathfrak{diff}_d$, which does not have to be globally defined \cite{Wa83}. Furthermore, it is crucial to observe that dropping all restrictions on the diffeomorphisms destroys the structure of a Lie group \cite{Mi83}.\\ 

Finally, observe that it is the representation $\mathbf{R}$ of the basis elements of the tangent space $\hat{P}_\mu$ as derivative operators (\ref{Lie9}) acting on the space-time coordinates $x^\mu$ in this chart that allowed to prove the Lie algebra properties of $\mathfrak{diff}_d$ and that is essential for the formula (\ref{Diffeom2}). I provide explicit examples for the evaluation of (\ref{Diffeom2}) in the next section.

\subsection{Subalgebras of $\mathfrak{diff}_{d}$}\label{Subalg}
In the preceding section, I have shown that the vector fields $\mathfrak{diff}_{d}$ form a Lie algebra. Next, I will construct two finite dimensional subalgebras that are defined by two particular forms of the vector fields (\ref{Vectorfields}), being the affine linear one (\ref{GLGen}) and the conformal one (\ref{KGen}). These are the ones whose closure in the vector field representation generates the infinite dimensional Lie algebra $\mathfrak{diff}_{d}$. This has become known as Ogievetsky's theorem \cite{O73} and I will explicitly prove it in section \ref{Ogievetsky}.

\subsubsection{Affine linear subalgebra $\mathfrak{a}_d$}
The first example for the evaluation of the formula (\ref{Diffeom2}) is the affine linear vector field $X^{\mathfrak{a}}_{(A,c)}\in \mathfrak{diff}_{d}$
\be\label{GLGen}
X^{\mathfrak{a}}_{(A,c)} &:=& \left({A_\nu}^\mu x^\nu +c^\mu\right)\p_\mu\qquad \forall A\in \mathfrak{gl}_d,\, \forall c\in \R^{d-1,1}.
\ee
The corresponding finite diffeomorphism hence is (\ref{Diffeom2})
\be\label{GLMatrix}
\left(\vp^{\mathfrak{a}}_{(A,c)}(x)\right)^\mu  &:=&\exp\left(\left.X^{\mathfrak{a}}_{(A,c)}\right.^\nu\p_\nu\right)x^\mu\\
&=&
\sum\limits_{n=0}^\infty \frac{1}{n!}{\left(A^n\right)_\nu}^\mu x^\nu +c^\mu\nn\\
&=&
{\left(e^A\right)_\nu}^\mu x^\nu +c^\mu ,
\nn
\ee
where	powers of $A$ are defined by ${(A^2)_\nu}^\mu = {A_\nu}^\rho{A_\rho}^\mu$ and $e^A$ is the standard matrix exponential series.\\

To prove that the transformations (\ref{GLGen}) form a subalgebra of $\mathfrak{diff}_{d}$, it is sufficient to show that the Lie bracket of two such transformations maps to a diffeomorphism with the same form as (\ref{GLGen}), but with another parameter, of course. A short computation of the commutator of vector fields (\ref{Lie8}) of the affine linear vector field (\ref{GLGen}) results in
\be\label{GLKom}
\left[X^{\mathfrak{a}}_{(A,c)},X^{\mathfrak{a}}_{(B,d)}\right] 
&=&
 X^{\mathfrak{a}}_{([A,B],Bc-Ad)}
\ee
with the matrix commutator ${[A,B]_\nu}^\mu = {A_\nu}^\sigma{B_\sigma}^\mu-{B_\nu}^\sigma{A_\sigma}^\mu$ and arbitrary $A,B \in \mathfrak{gl}_d$ and $c,d\in \R^{d}$. Hence, this is a finite dimensional subalgebra of $\mathfrak{diff}_{d}$, which is called the \textbf{affine algebra} $\mathfrak{a}_d$.\footnote{This affine algebra should not be confused with the notion of an affine Kac-Moody algebra or current algebra, as defined e.g. in \cite{K95}.}\\

It is standard to isolate the generators of the Lie algebra from the coefficients in (\ref{GLGen}) tantamount to the identification
\be
	X^{\mathfrak{a}}_{(A,c)} &=:&{A_\nu}^\mu {\mathbf{R}}_{{\left.\hat{M}\right.^{\nu}}_\mu} + c^\mu {\mathbf{R}}_{\hat{P}_\mu}\nn\\
		\Rightarrow \quad {\mathbf{R}}_{{\left.\hat{M}\right.^{\nu}}_\mu }\,=\,  x^\nu \p_\mu 
	&\text{and}&
	{\mathbf{R}}_{\left.\hat{P}\right._\mu }\,=\,  \p_\mu . \label{PRep}
\ee
It should be noted that the representation $\mathbf{R}$ of $\hat{P}_\mu$ as a derivative operator is the same one as in (\ref{Lie9}). With these definitions, the commutation relation (\ref{GLKom}) decomposes as follows:
\begin{subequations}\label{ComRel0}
\be
 \left[{\left.\hat{M}\right.^{\nu}}_\mu, {\left.\hat{M}\right.^{\tau}}_\sigma\right] &=& \delta_\mu^\tau {\left.\hat{M}\right.^{\nu}}_\sigma -\delta_\sigma^\nu {\left.\hat{M}\right.^{\tau}}_\mu \label{ComRel1}\\
   \left[{\left.\hat{M}\right.^{\nu}}_\mu, \left.\hat{P}\right._{\tau}\right] &=& -\delta^\nu_\tau   \left.\hat{P}\right._{\mu} \label{ComRel2}\\
     \left[\hat{P}_{\mu}, \hat{P}_{\tau}\right] &=& 0.
    \label{ComRel3}
\ee
\end{subequations}
The commutation relation (\ref{ComRel1}) reveals that the generators $\hat{M}$ generate a $\mathfrak{gl}_d$ subalgebra of $\mathfrak{a}_d$. The other relations show that the translations $\hat{P}_\mu$ form an ideal, which I denote by $\mathbf{d}$, because the $\hat{P}_\mu$ transform in the fundamental representation of $\mathfrak{gl}_d$. As a vector space, the affine algebra hence decomposes into
\be\label{affine1}
\mathfrak{a}_d=\mathfrak{gl}_d\oplus {\mathbf{d}}.
\ee
Note that the concept of the generators of a Lie algebra $\Lie$ always is an abstract one, determined by its commutation relations as discussed in section \ref{Class}. By the definition of a representation $\mathbf{R}$, any representation induces the commutation relations of the algebra and the normalization of the generators. In this thesis, I will always denote the vector field representation by $\mathbf{R}$. The normalization of an abstract algebra element $\hat{A}\in \mathfrak{gl}_d$ is also fixed by a comparison of the vector field representation (\ref{PRep}) with the definition of the affine linear vector field $X^{\mathfrak{a}}_{(A,c)}$ (\ref{GLGen}) to
\be\label{gldefi}
\hat{A}&=& {{A}_\nu}^\mu 
\hat{M}
{}^\nu{}_\mu .
\ee
The affine algebra $\mathfrak{a}_d$ (\ref{affine1}) corresponds to the affine group
\be\label{affine2}
A(d) &=& Gl(d) \ltimes \cP_d,
\ee
a finite dimensional subgroup of $\Diff(d)$, consisting of the semidirect product of the general linear group $Gl(d)$ and the Abelian group of constant translations $\cP_d$ in $d$ dimensions. There also are infinite dimensional subgroups of $\Diff(d)$. These are classified by Cartan's theorem \cite{C09}. One example is the group of symplectomorphisms that I will address again in the context of supergravity in section \ref{SympV1}.

\subsubsection*{Conformal isometries of the metric $\eta$}
In the section \ref{realf} on real forms of finite dimensional Lie algebras, I have already mentioned the concept of a metric 
\be\label{etaSkal}
\eta &=& \text{diag}(\pm 1,\dots ,\pm 1 )
\ee
of signature $p$, i.e. with $p$ minus signs in (\ref{etaSkal}). This defined the subalgebra $\mathfrak{so}_{(d-p,p)}$ of the matrix algebra $\mathfrak{gl}_{d}$ by (\ref{AntisyM}). The reader may recall that representations preserve the subalgebra property. This implies that the affine linear vector fields $X^{\mathfrak{a}}_{(A,c)}$ (\ref{GLGen}) with antisymmetric matrices $A$ (\ref{AntisyM}) form a vector field representation $\mathbf{R}$ of $\mathfrak{so}_{(d-p,p)}\oplus \mathbf{d}$. In this sense, $\mathfrak{so}_{(d-p,p)}\oplus \mathbf{d}$ also is a subalgebra of $\mathfrak{diff}_{d}$. The antisymmetry of $A$ induces the property of the finite diffeomorphisms $\vp_{(A,c)}^{\mathfrak{a}}$ (\ref{GLMatrix}) that they preserve the symmetric tensor $\eta$
\be\label{Isometry1}
x'&:=& \vp_{(A,c)}^{\mathfrak{a}}\quad\text{with antisymmetric } A\, (\ref{AntisyM}), \nn\\
\eta_{\mu\nu} &=& \frac{\p {x'}^\sigma}{\p x^\mu} \frac{\p {x'}^\tau}{\p x^\nu}\eta_{\sigma\tau}.
\ee
A diffeomorphism $x'=x'(x)$ with the property (\ref{Isometry1}) is called an \textbf{isometry} of the metric $\eta$. In the appendix \ref{Isometry} I prove that the subalgebra $\mathfrak{so}_{(d-p,p)}\oplus \mathbf{d}$ of diffeomorphisms $\mathfrak{diff}_{d}$ generates all diffeomorphisms with this property. For Lorentzian signature, i.e. $p=1$, this algebra is called the \textbf{Poincar\'e algebra} with the associated \textbf{Poincar\'e group} 
\be\label{Isometry8}
SO(d-1,1)\ltimes \cP_d &\subset & \Diff(d).
\ee
This group will be of crucial importance in chapter \ref{CHAP3}. Vector fields $X_A\in \mathfrak{diff}_d$ whose associated finite diffeomorphisms $\vp_A\in \Diff(d)$ preserve a metric tensor $g$, are called Killing vector fields of $g$. \textbf{Conformal Killing vector fields} $X^{\mathfrak{c}} \in \mathfrak{diff}_d$ are the ones whose associated integral curve $x'=\vp^{\mathfrak{c}}\in \Diff(d)$ preserve the metric tensor up to a scalar function. For the case of the constant metric $\eta$ (\ref{etaSkal}), the Poincar\'e algebra is enlarged by the following two classes of vector fields
\begin{subequations}
\be
X^{\mathfrak{a}}_{(b\id,0)} &=&b x^\mu \p_\mu \quad  \forall b\in \R,
\label{Ideal1}
\\
X^{\mathfrak{c}}_{a} &:=& \eta_{\tau\nu}x^\tau\left( x^\nu  a^\sigma -2  a^\nu  x^\sigma\right)\p_\sigma \quad \forall a\in \R^{d}.
\label{KGen}
\ee
\end{subequations}
The vector field $X^{\mathfrak{a}}_{(b\id,0)}$ is an element of the affine algebra $\mathfrak{a}_d$ (\ref{affine})\footnote{To be precise, it corresponds to the one-dimensional ideal of the algebra $\mathfrak{gl}_d$ that I have mentioned in section \ref{realf}.}, but the other one $X^{\mathfrak{c}}_{a}$ is not. This fact will be an important ingredient to prove Ogievetsky's theorem in \ref{Ogievetsky}.\\

In this context, I want to comment on the appearance of $\eta$ in the formula (\ref{KGen}). This does not restrict the manifolds under consideration to the ones that globally allow for a constant metric of the form $(\pm 1,\dots,\pm 1)$. Given a coordinate chart $(U_\alpha,x_\alpha)$ with local basis vectors $\hat{P}_\mu$, it is always possible to define a symmetric tensor $\eta$ of arbitrary signature in $\R^d$ that induces a symmetric tensor on one open set $U_\alpha\subset \cM^d$. However, this tensor cannot be consistently continued on the other open sets $U_\alpha$ covering $\cM^d$ in general.\footnote{As an example may serve the flat Euclidean metric on the compact three-sphere $S^3$.}  This is the crucial difference to the metric tensor $g$ defined on the entire manifold $\cM^d$, which I will start to discuss in section \ref{Gstructur}. The vector field $X^{\mathfrak{c}}_{a}$ should not be confused with the Weyl rescaling of a general metric $g$ either, which is not possible by a diffeomorphism action.\\

To sum up, the conformal Killing vector field $X^{\mathfrak{c}}_{a}$ merely is a special form of a general vector field $X_A\in \mathfrak{diff}_d$, because the latter can arbitrarily depend on the coordinates $x^0,\dots, x^{d-1}$ in an analytic way (\ref{Vectorfields}). $X^{\mathfrak{c}}_{a}$ also forms a subalgebra $\mathfrak{k}_{(d-p,p)}$ with the signature $p$ of $\eta$ (\ref{etaSkal}). Isolating the $d$ parameters $a^\mu\in \R$ from the generators $\hat{K}_\mu\in \mathfrak{k}_{(d-p,p)}$ as in (\ref{PRep}), I obtain
\be\label{KGen2}
X^{\mathfrak{c}}_{a} &=:& a^\mu {\mathbf{R}}_{\hat{K}_\mu}\\
 {\mathbf{R}}_{\hat{K}_\mu} &=& \eta_{\tau\nu}x^\tau\left( x^\nu  \delta_\mu^\sigma -2  \delta_\mu^\nu  x^\sigma\right)\p_\sigma.
  \label{KRep}
\ee
A short computation of the commutator of vector fields (\ref{Lie8}) induces the algebra relation
\be\label{KK}
\left[\hat{K}_\mu,\hat{K}_\nu\right] &=& 0.
\ee
Therefore, $\mathfrak{k}_{(d-p,p)}$ is an Abelian subalgebra of $\mathfrak{diff}_{d}$. The corresponding finite diffeomorphisms (\ref{Diffeom2})
\be
\left(\vp^{\mathfrak{c}}_{a}\right)^\mu (x) &:=&\exp\left(\left.X^{\mathfrak{c}}_{a}\right.^\nu\p_\nu\right)x^\mu
\label{SCTcoord4}
\ee
hence form an Abelian subgroup $K(d-p,p)\subset \Diff(d)$
\be\label{abelgroup}
\vp^{\mathfrak{c}}_{a}\circ \vp^{\mathfrak{c}}_{b} &=& \vp^{\mathfrak{c}}_{a+b}.
\ee

\subsubsection*{The subalgebra of conformal isometries $\mathfrak{so}_{(d-p+1,p+1)}$ of $\eta$}
The definition of a conformal Killing vector field stated that the corresponding finite diffeomorphism was an isometry (\ref{Isometry1}) of $\eta$ modulo a scalar function. This immediately implies that the Lie algebra of these vector fields must close. The resulting algebra is called the \textbf{conformal subalgebra} of $\mathfrak{diff}_d$. Since only the Poincar\'e subalgebra $\mathfrak{so}_{(d-p,p)}\oplus \mathbf{d}$ and the one-dimensional ideal (\ref{KGen2}) of $\mathfrak{gl}_d$ are the parts of the affine algebra $\mathfrak{a}_d$ (\ref{affine1}) with the conformal Killing property, it will prove convenient to introduce the following abbreviations:
\begin{subequations}\label{LDefi0}
\be\label{LDefi}
{\left.\hat{L}\right.^\nu}_\sigma &:=& \frac{1}{2}\left({\left.\hat{M}\right.^\nu}_\sigma -\eta_{\sigma\tau}{\left.\hat{M}\right.^\tau}_\mu {\big.\eta}^{\mu\nu}\right),\\
\hat{D} &:=& \delta_\nu^\sigma \hat{M}{}^\nu{}_\sigma\label{DDefi}.
\ee
\end{subequations}
A short calculation with the vector field representations $\mathbf{R}$ of the conformal generators (\ref{GLGen}) and (\ref{KGen}) fixes the remaining commutation relations of the conformal algebra to
\begin{subequations}\label{CFT3}
\be
\left[{\left.\hat{L}\right.^\nu}_\sigma ,\hat{K}_\mu\right]&=& -\frac{1}{2} \left(\delta^\nu_\mu \delta_\sigma^\tau - \eta_{\mu\sigma}\eta^{\nu\tau}\right)\hat{K}_\tau
\\
\left[\hat{D},\hat{K}_\mu\right]&=&   \hat{K}_\mu
\\
\left[\hat{K}_\mu,\hat{P}_\nu\right]&=& 2 \eta_{\mu\nu}\hat{D} +4 \eta_{\mu\tau}{\left.\hat{L}\right.^\tau}_\nu.
\ee
\end{subequations}
It is standard \cite{SS69} to show that the $d$-dimensional conformal algebra is isomorphic to the orthogonal algebra $\mathfrak{so}_{(d-p+1,p+1)}$ in $d+2$ dimensions. Following my definitions of section \ref{realf}, the latter is provided by antisymmetric matrices with respect to a metric $\eta$ with $p+1$ minus signs (\ref{AntisyM}). With the identifications 
\begin{subequations}\label{dplus2}
\be
\eta_{d+1,d+1}&=&+1 \,=\, -\eta_{d+2,d+2}\\
{\left.\hat{L}\right.^{d+1}}_{d+2} &=& \frac{1}{2}\hat{D}\\
{\left.\hat{L}\right.^{d+1}}_{\mu} &=& \frac{1}{4}\left(\hat{P}_\mu-\hat{K}_\mu\right)\\
{\left.\hat{L}\right.^{d+2}}_{\mu} &=& -\frac{1}{4}\left(\hat{P}_\mu+\hat{K}_\mu\right),
\ee
\end{subequations}
the commutation relations (\ref{ComRel1}) restricted to antisymmetric generators $\hat{L}$ in $d+2$ dimensions coincide with the ones of the conformal algebra. This implies with the discussion in section \ref{Representation} that the corresponding groups may only differ by a discrete group, which indeed is the case. I will address this point in the next section.

\subsubsection*{The conformal group}
Although the theorem \ref{Satz1} uniquely defines the finite diffeomorphism $\vp^{\mathfrak{c}}_{a}$ corresponding to the conformal vector field $X^{\mathfrak{c}}_{a}$, I have not stated its explicit formula in (\ref{SCTcoord4}). A direct computation as performed for the affine linear vector field $X^{\mathfrak{a}}_{(A,c)}$ in (\ref{GLMatrix}) is not the best way to obtain the result. The isomorphism of the conformal algebra to $\mathfrak{so}_{(d-p+1,p+1)}$ suggests to introduce $d+2$ light cone coordinates $(V^\mu,V^{d+1},V^{d+2})$ that are linked to the $d$ standard coordinates in the following way
\be\label{vCoord}
x^\mu &=:& \frac{V^\mu}{V_{d+1}-V_{d+2}}\\
0&=& \eta_{\mu\nu}V^\mu V^\nu +V^2_{d+1}-V_{d+2}^2.
\label{vConstr}
\ee
It should be noted that the vanishing of the bilinear symmetric form in the last line is a consistent constraint, because $\mathfrak{so}_{(d-p+1,p+1)}$ preserves this form by definition. The identification (\ref{dplus2}) allows to evaluate the action of $a^\mu\hat{K}_\mu$ in the representation of linear vector fields (\ref{PRep}) acting on the coordinates $V$ (\ref{vCoord}) in the standard way.\footnote{I provide more details in the appendix \ref{V1}. This procedure can also be found in \cite{SS69}.} Introducing the inversion
\be\label{Invdefi}
\Inv(x)^\nu := \frac{x^\nu}{x\cdot x}
\ee
that obviously has the property $\Inv\circ \Inv =\id_d$, I obtain
\be
\left(\vp^{\mathfrak{c}}_{a}\right)^\mu(x)
&=&
\left(\Inv\circ \vp^{\mathfrak{a}}_{(0,a)}\circ \Inv\right)^\mu (x).
\label{SCTcoord47}
\ee
This leads to the explicit form
\be
\left(\vp^{\mathfrak{c}}_{a}\right)^\mu(x)
&=&
\frac{x^\mu +a^\mu\,  x\cdot x }{1 +2 a\cdot x + a\cdot a\, x\cdot x}
\label{SCTcoord}
\\
\text{with }\quad a\cdot x&:=&\eta_{\mu\nu}a^\mu x^\nu.
\nn
\ee
I want to close this section with three comments
\begin{itemize}
	\item From the equation (\ref{SCTcoord}), it is obvious that the finite diffeomorphism $\vp^{\mathfrak{c}}_{a}$ is an example for the case of a restricted domain of validity $U$ as I explained at the end of section \ref{Diffeom}. For any parameter $a^\mu$ corresponding to the fixed vector field $X_A$, there is a domain of validity $0\in U\subset x_\alpha(U_\alpha)\subset\R^d$ for which the formul\ae{} (\ref{Diffeom2}) or (\ref{SCTcoord}) are well-defined.
	\item The fact that these diffeomorphisms $\vp^{\mathfrak{c}}_{a}$ form an Abelian group $K(d-p,p)$ (\ref{abelgroup}) also follows from (\ref{SCTcoord47}).
	\item The global structures of the conformal subgroup of $\Diff(d)$ and of the orthogonal group $SO(d-p+1,p+1)$ do not match. In even dimensions $d$, the conformal group is isomorphic to $SO(d-p+1,p+1)/\Z_2$. In odd dimensions, this is not the case, because the inversion $\Inv$ (\ref{Invdefi}) is an element of $SO(d-p+1,p+1)/\Z_2$. Since $Inv$ has a pole, it is not in $\Diff(d)$, however.\footnote{I exlain this subtlety in the appendix \ref{Global}. More details on the conformal group can be found in \cite{D1}.}
\end{itemize}

\subsection{Ogievetsky's theorem}\label{Ogievetsky}
I want to start this section with the observation that all generators in the conformal algebra $\mathfrak{so}_{(d-p+1,p+1)}$ that are not contained in the affine algebra $\mathfrak{a}_{d}$, are encoded in the Abelian algebra $\mathfrak{k}_{(d-p,p)}$
\beg
\mathfrak{so}_{(d-p+1,p+1)} &=& \mathfrak{k}_{(d-p,p)}\oplus \left( \mathfrak{so}_{(d-p+1,p+1)}\cap \mathfrak{a}_{d}\right).
\eeg
Hence, the closure of the algebras $\mathfrak{a}_{d}$ and $\mathfrak{so}_{(d-p+1,p+1)}$ in their vector field representation $\mathbf{R}$ is equivalent to the closure of the algebras $\mathfrak{a}_{d}$ and $\mathfrak{k}_{(d-p,p)}$ in their vector field representation $\mathbf{R}$. These algebras are parametrized by the affine linear vector fields (\ref{GLGen}) and the conformal ones (\ref{KGen}), respectively.\\

In order to show that this closure with respect to the Lie bracket of vector fields (\ref{Lie8}) indeed generates the entire algebra of analytic vector fields $\mathfrak{diff}_{d}$ (\ref{Vectorfields}), a basis of the vector space $\mathfrak{diff}_{d}$ has to be specified. It is a standard result that the closure of the vector space of all homogeneous monomials forms a basis of the space of all analytic functions. This implies that the vector space of analytic vector fields $\mathfrak{diff}_{d}$ is spanned by the closure of the vector fields
\be\label{poly6}
X^\mu(x)\p_\mu &=& \sum\limits_{{\mathbf{n}}=0}^\infty A^\mu_{\mathbf{n}} \mathbf{R}_{\hat{P}_{\mathbf{n},\mu}(x)}\\
\text{with }\quad\mathbf{n} &:=& (n_1,\dots,n_d),\nn\\
n&:=&n_1+\dots +n_d\nn.
\ee
The basis vectors $\hat{P}_{\mathbf{n},\mu}(x)$ of $\mathfrak{diff}_{d}$ have a vector field representation
\be\label{P2Defi}
\mathbf{R}_{\hat{P}_{\mathbf{n},\mu}(x)}&=& \big(x^1\big)^{n_1}\cdots \big(x^d\big)^{n_d}\p_\mu
\ee
as homogeneous monomials of degree $n$. The sum over ${\mathbf{n}}$ in (\ref{poly6}) is a sum over $n\in\N_0$, but it also contains a sum over the partitions of $n$ in $n_i\in \N_0$.\footnote{This is a Hamel basis, not a Schauder basis, i.e. for every analytic vector field $X_A\in \mathfrak{diff}_d$ there are coefficients $A^\mu_{\mathbf{n}}\in \R$ such that (\ref{poly6}) holds. Given $X_A\in \mathfrak{diff}_d$, the number of non-vanishing coefficients $A^\mu_{\mathbf{n}}\in \R$ does not have to be finite, however \cite{He92}.} This definition allows to state Ogievetsky's theorem \cite{O73} in the following way:

{\thm\label{Ogie2}
For $d>1$, the vector field representation $\mathbf{R}$ of any basis monomial $ \hat{P}_{\mathbf{n},\nu}(x)$ (\ref{P2Defi}) is a linear combination of the affine linear generators $(\hat{P}_\tau, \hat{M}{}^\mu{}_\nu)$ (\ref{PRep}), the conformal ones $\hat{K}_\nu$ (\ref{KRep}) and arbitrary commutators thereof in their representation $\mathbf{R}$ as vector fields.\\
}

{
\pf
Following Ogievetsky \cite{O73}, I prove this theorem in detail in the appendix \ref{OgieP}. It contains two parts: At first, I show that a general vector field (\ref{P2Defi}) of polynomial degree $n=2$ is contained in the closure. Then, I use this to construct vector fields $\hat{P}_{\mathbf{n},\mu}$ of arbitrary degree $n$ by induction.\qed
}\\

It is interesting to observe that not all conformal vector fields are necessary for the theorem to hold. This allows to refine the theorem by the following corollary that I prove in appendix \ref{Ogie6}.

{\cor\label{Coro2}
For $d>1$, fix $1<w\leq d$ arbitrary. Then the vector field representation $\mathbf{R}$ of any basis vector $ \hat{P}_{\mathbf{n},\mu}$ (\ref{poly6}) in $d$ dimensions is a linear combination of (commutators of) the affine linear generators $(\hat{P}_\tau, \hat{M}{}^\mu{}_\nu)$ (\ref{PRep}) in $d$ dimensions and the conformal ones $\hat{K}_\nu$ (\ref{KRep}) in $w$ dimensions in their representation $\mathbf{R}$ as vector fields.
}\\

Ogievetsky's theorem \ref{Ogie2} is the starting point of the procedure of Borisov \& Ogievetsky to generate a gravitational theory from the simultaneous non-linear realizations of the affine linear and the conformal symmetry group, which will be the topic of chapter \ref{CHAP4}. From the details of the proofs, I want to highlight the following points:
\begin{itemize}
		\item It is interesting to ask if the Lie algebra $\mathfrak{diff}_d$ is generated by the closure of the affine algebra $\mathfrak{a}_d$ with an arbitrary vector field whose dependence on the coordinates is quadratic or higher. This is not true: the vector field $\hat{X}^\mu = x^\mu x^\nu\p_\nu$ is quadratic in $x$, but it transforms as a vector representation under $\mathfrak{gl}_d$ and it also is Abelian. To generate $\mathfrak{diff}_d$, it is essential to include vector fields like $\hat{K}_\mu$ (\ref{KRep}) in the closure that do not transform as a linear representation of $\mathfrak{gl}_d$.
		\item The affine algebra $\mathfrak{a}_d$ does not preserve the structure defined by the symmetric form $\eta$ (\ref{etaSkal}), which could be interpreted as the reason why the commutator maps outside the conformal algebra $\mathfrak{so}_{(d-p+1,p+1)}$.
		\item This implies in particular that the signature $(d-p,p)$ of the flat metric $\eta$ had no effect on the proof of Ogievetsky's theorem. There is no connection to the signature of the metric $g$ on the Lorentzian manifold $\cM^d$.
		\item It is only for the construction of a gravitational theory \`a la Borisov \& Ogievetsky that it is necessary to endow the conformal vector field (\ref{KGen}) with the Lorentzian signature $p=1$.
\end{itemize}
To conclude, I want to close the loop of the analogy of Ogievetsky's theorem to the classification of simple, complex, finite dimensional Lie algebras from section \ref{Class}. In both cases, the precise way how the subalgebras are linked to each other is of crucial importance. As an example, the reader may consider the two Kac--Moody algebras $\mathfrak{sl}_3$ and $\mathfrak{sl}^+_2$: both have two $\mathfrak{sl}_2$ subalgebras, but $\mathfrak{sl}_3$ is finite dimensional and $\mathfrak{sl}_2^+$ is not. This is due to the fact that the two $\mathfrak{sl}_2$-subalgebras are linked by different Serre relations (\ref{Serre}). In the case of $\mathfrak{diff}_{d}$, the corresponding relation is provided by the vector field representation $\mathbf{R}$ of $\mathfrak{a}_d$ (\ref{PRep}) and $\mathfrak{so}_{(d-p+1,p+1)}$ (\ref{KRep}), whereas for the embedding of both algebras in e.g. $\mathfrak{gl}_{d+2}$, the relations would be different.\\
\smallskip\\

\section{Application to physical fields}\label{prim2}
\subsection{Explicit symmetry action on a physical field}\label{prim2b}
In the introduction to the section \ref{Representation0} on representation theory, I have already argued that the induced symmetry group action on physical fields always is a (multi)linear one. Passing to the corresponding Lie algebra, the induced action is linear. However, I have not explicitly specified how this linear action is evaluated in coordinates, even though it has to fulfill the criteria of a linear representation $\mathbf{R}$ of a Lie algebra $\Lie$. As the linear representation $\mathbf{R}$ by definition maps every algebra element $\hat{G}\in \Lie$ to a linear map $\mathbf{R}_{\hat{G}}\in End(V)$, a physical field $\psi$ can without loss of generality be parametrized by vectors $\hat{T}_\gamma$
\be\label{psiParam}
\psi(x) &=& \psi^\gamma(x) \hat{T}_\gamma
\ee
that span $V$, i.e. $\gamma=1,\dots, \dim(V)$. For every point $x\in \cM^d$, this vector space $V$ is isomorphic to the fibre of the tensor bundle at this point, keeping in mind that $\psi$ is a section of this tensor bundle. It is standard in the physics literature to call the $x$-dependent coefficients $\psi^\gamma$ physical fields, too.\\

Since the Lie algebra acts by an endomorhism $\mathbf{R}_{\hat{G}}\in End(V)$, the corresponding Lie group action on the vectors $\hat{T}_\gamma \in V$ is provided by the exponential series of matrices (\ref{exp})
\be\label{Tstrich}
\hat{T}'_\gamma &:=& \exp\left(\mathbf{R}_{\hat{G}}\right)\hat{T}_\gamma.
\ee
In other words, a symmetry action induces a change of the reference frame $\hat{T}_\gamma$ to $\hat{T}'_\gamma$ with respect to which the $x$-dependent coefficients $\psi^\gamma$ are defined. Their transformation hence is induced by (\ref{psiParam}), which results in the equality
\be\label{psistrich}
{\psi'}^\gamma \hat{T}'_\gamma &=& \psi^\gamma \hat{T}_\gamma.
\ee
The corresponding Lie algebra action $\delta_{\hat{\omega}}$ on the coefficients $\psi^\gamma$ is obtained from a linear expansion after substituting (\ref{Tstrich}) in (\ref{psistrich})
\be\label{SymmAct6}
\left(\delta_{\hat{G}} \psi^\gamma\right)\hat{T}_\gamma &:=& -\psi^\alpha \mathbf{R}_{\hat{G}}\left(\hat{T}_\alpha\right).
\ee
A simple, non-trivial example is a $\Lie=\mathfrak{gl}_{d}$ action on a $d$-dimensional vector field
\be\label{vector4}
V(x) &=& V^\mu(x)\p_\mu
\\
&\stackrel{(\ref{Lie9})}{=}& V^\mu(x)\mathbf{R}_{\hat{P}_\mu}.\nn
\ee
The discussion of real forms of Lie algebras from section \ref{realf} implies that any element $\hat{\omega}\in \mathfrak{gl}_{d}$ can be parametrized by a $d\times d$ matrix 
\beg
\hat{\omega} &=& {\omega_\mu}^\nu {\left.\hat{M}\right.^\mu}_\nu
\eeg
with $d^2$ real numbers ${\omega_\mu}^\nu\in \R$ and the abstract generators $\hat{M}\in \mathfrak{gl}_{d}$ (\ref{gldefi}). It is the vector field representation $\mathbf{R}$ that dictates the precise action on the vector field $V(x)$ (\ref{SymmAct6})
\beg
\left(\delta_{\hat{\omega}} V^\mu\right)\mathbf{R}_{\hat{P}_\mu}
 &=& -V^\tau \mathbf{R}_{\hat{\omega}}\left(\mathbf{R}_{\hat{P}_\tau}\right)\\
&=&
-V^\tau {\omega_\mu}^\nu \left[\mathbf{R}_{{\left.\hat{M}\right.^\mu}_\nu},\mathbf{R}_{\hat{P}_\tau}\right]\\
&\stackrel{(\ref{ComRel2})}{=}&
V^\tau {\omega_\mu}^\nu \delta_\tau^\mu\mathbf{R}_{\hat{P}_\nu}.
\eeg
Comparing the basis vectors fixes the action on the coefficients to
\be\label{GlAction}
\delta_{\hat{\omega}} V^\mu &=& {\omega_\tau}^\mu V^\tau.
\ee
This induced action by the commutator of vector fields (\ref{Lie7}) will be sufficient to describe all symmetry actions on physical fields. In section \ref{fermi}, I will explain in which way it also induces an action on spin representations.\\

In concluding this section, I want to highlight a subtlety. For \textbf{internal} symmetries, i.e. ${x'}^\mu =x^\mu$, it follows from their definition in section \ref{Symm1} that the symmetry group is the same Lie group that acts on the coefficients $\psi^\mu$. To phrase it differently, the induced action is the standard action by the symmetry Lie group. For \textbf{external} symmetries, this is not the case in general. As an example may serve the conformal subgroup of the diffeomorphism group $\Diff(d)$. Its action on the coordinates $x^\mu\mapsto {x'}^\mu$ is non-linear (\ref{SCTcoord}), but the induced action on e.g. a vector field is a linear one by the different Lie group $Gl(d)$
\beg
  {x'}^\mu &=& (\vp_a^{\mathfrak{c}})^\mu(x),\nn\\
  \Rightarrow \quad {V'}^\mu &=& \frac{\p {x'}^\mu}{\p x^\nu}V^\nu\nn.
\eeg
That the corresponding linear algebra action on a physical field is by $\mathfrak{gl}_d$ and not by $\mathfrak{so}_{(d-p+1,p+1)}$ is obvious from the fact that a $d$-dimensional vector field does not form a linear representation of $\mathfrak{so}_{(d-p+1,p+1)}$.\\

In this context, I also want to introduce the standard convention to denote transformed object by a prime. As the symmetry action always is induced by the action on the reference frame, e.g. for a diffeomorphism ${x'}^\mu=\vp_A^\mu(x)$
\beg
A &=& A_\mu dx^\mu \,=\, A'_\nu d{x'}^\nu,
\eeg
it is clear with respect to which frame the indices have to be contracted. Therefore, I do not adopt the convention to explicitly denote the $x$-dependence in the physical field, i.e. to write $A_\mu(x)$ and $A'_\mu(x')$, because the reference frame already is uniquely specified by one prime. I emphasize that the induced action $\delta_{\hat{\omega}}$ on a vector field $V^\mu(x)$ should not be confused with the Lie derivative of a vector field \cite{Wa83}. I will not need the concept of a Lie derivative in the sequel.

\subsection{The connection $\nabla$ on fields}\label{equivar}
For the discussion of local symmetry groups, the definition of an \textbf{equivariant connection} or covariant derivative $\nabla$ from section \ref{SymmAct} is of major importance. If the group action on a physical field $\psi^\gamma(x)$ is a \textbf{local} one, both the field $\psi$ and the equations of motion can only be covariant, if all partial derivatives $\p$ are replaced by equivariant connections $\nabla$. It immediately follows from the previous section that the covariance property is equivalent to $\psi$ transforming as a linear representation under the induced symmetry algebra action.\\

The case of prime importance for this thesis is the Lie algebra of vector fields $\mathfrak{diff}_d$ and its corresponding diffeomorphism group $\Diff(d)$. Its induced action on e.g. a vector field $V$ (\ref{vector4}) is a multiplication by the Jacobi matrix. To phrase this in other words: the symmetry $\Diff(d)$ induces a $G=Gl(d)$ action on the vector field $V$, whose coefficients are the coefficients of the Jacobi matrix that are $x$-dependent in most cases. For the $Gl(d)$ representation $V$, the connection $\nabla$ acting on the coefficients $V^\mu(x)$ in the basis $dx^\nu$ has the general form
\be\label{Conn56}
\nabla_\nu V^\mu(x)&:=&  \p_\nu V^\mu(x) - \delta_{\hat{\omega}_\nu} V^\mu(x)
\ee
with a $\Lie=\mathfrak{gl}_d$ action $\delta_{\hat{\omega}_\nu}$. Equivariance of this connection $\nabla$ (\ref{Conn56}) under $\Diff(d)$  is then equivalent to 
\beg
\nabla_\nu V^\mu(x) 
\eeg
being covariant, i.e. to transform as a $Gl(d)$ representation under a local $Gl(d)$ action induced by the diffeomorphism action on the coordinates
\beg
  {x'}^\mu &=& \vp^\mu_A(x),\nn\\
  \Rightarrow \quad {V'}^\mu &=& \frac{\p {x'}^\mu}{\p x^\nu}V^\nu\nn.
\eeg
To sum up, a connection $\nabla$ consists of a partial derivative and a $\Lie$-action on a $\Lie$-representation. It is straightforward from the definition (\ref{Conn56}) how to generalize this concept to arbitrary $Gl(d)$-representations. However, this does not cover all physical fields of interest. Since fermions correspond to the spin representations of the Lorentz group introduced in section \ref{Representation}, I will introduce the vielbein frame on a Lorentzian manifold in the next section. To describe fermions, this is the appropriate framework. Therefore, I will use it throughout this dissertation.

\subsection{The vielbein frame}\label{Gstructur}
I have already restricted the manifolds $\cM^d$ under discussion to be equipped with a non-degenerate metric $g$ (\ref{metric5}) of Lorentzian signature at the beginning of this chapter. For every point $x\in \cM^d$, the metric $g$ hence induces an invertible map
\be\label{canon1}
T_x\cM^d &\rightarrow & T_x^*\cM^d\\
v &\mapsto & g_x(v,\cdot).\nn
\ee
This canonical isomorphism between the tangent and the cotangent bundle allows to identify the dual vector spaces $T_x\cM^d$ and $T_x^*\cM^d$ for all points $x\in \cM^d$. Practically, this is the possibility to raise and lower indices consistently. I want to emphasize that this canonical ismorphism is not linked to the symmetry property of $g$ at all, but only to its non-degeneracy. This will be important in chapter \ref{CHAP5} where I will use a non-degenerate symplectic form $\Omega$ to establish the canonical isomorphism between tangent and cotangent spaces. \\

The symmetry property of $g$ is important for the result that for every point $x\in \cM^d$, there is a basis of the tangent space 
\be\label{viel2}
 T_x\cM^d &=& \langle \hat{e}_a(x)|a=0,\dots, d-1 \rangle_\R
\ee
with the property that every basis vector $\hat{e}_a(x)\in T_x\cM^d$ is mapped to its dual $\hat{e}^a(x)\in T^*_x\cM^d$ modulo a real constant by (\ref{canon1}). This is the spectral theorem of linear algebra. Since the map (\ref{canon1}) relates dual vector spaces, the basis can be refined such that the following equality holds
\be\label{gDefi}
g(x)&=& \eta_{ab}\hat{e}^a(x)\otimes \hat{e}^b(x)\,\in\, T^*_x\cM^d \otimes T^*_x\cM^d,
\ee
where $\eta$ is the flat Minkowski metric introduced in (\ref{eta}). From this relation, it is obvious that this choice of basis can only be unique up to an isometry of the Minkowski metric.\\

Next, recall from section \ref{Diffeom} that I have restricted the diffeomorphisms in such a way that allowed me to choose coordinates $x^\mu$ on some open set $U\subset \cM^d$ (\ref{GenDiff5}). These canonically induce another basis of the cotangent space
\beg
 T^*_x\cM^d &=& \langle dx^\mu|\mu=0,\dots, d-1 \rangle_\R.
\eeg
If I define the coefficients of the metric $g$ and of the dual basis elements $\hat{e}^a(x)\in T^*_x\cM^d$ with respect to this basis
\begin{subequations}
\be
g(x) &=:& g_{\mu\nu}(x)dx^\mu\otimes dx^\nu\\
\hat{e}^a(x) &=:& {e_\mu}^a(x) dx^\mu,
\label{vielbeindefi}
\ee
\end{subequations}
then the relation (\ref{gDefi}) implies for the coefficients of the basis vectors for every point $x\in U\subset \cM^d$
\be\label{Vielbeinequiv}
g_{\mu\nu} &=& \eta_{ab}{e_\mu}^a{e_\nu}^b.
\ee
The $d\times d$ matrices ${e_\mu}^a$ are called \textbf{rep\`ere mobile} or \textbf{vielbein}. I have already mentioned that given a metric $g(x)$, the corresponding vielbein matrix $e(x)$ is not unique. The possible choices of $e(x)$ decompose into disjoint $SO(d-1,1)$ orbits that are labelled by the different signatures of the eigenvalues of $e(x)$. For the rest of this thesis, I will restrict to the orbit $[e]$ of matrices with positive eigenvalues. Two vielbeine $e,f\in \R^{d\times d}$ are in the same orbit $[e]$, if they are linked by a Lorentz transformation in the following way
\beg
e,f\in [e]&:\Leftrightarrow & \exists\, O\in SO(d-1,1):\quad {f_\mu}^a \,=\, {e_\mu}^b {O_b}^a.
\eeg
To sum up, given any Lorentzian metric in coordinates $g_{\mu\nu}$, then for every point $x\in U\subset \cM^d$, there is a unique $SO(d-1,1)$ orbit of matrices $[e]$ with positive eigenvalues such that the relation (\ref{Vielbeinequiv}) is true. Since $g$ varies smoothly with $x$ by definition, the same holds for $e$, which proves the local existence of $e$ as a one-form. This also is true globally, because I have restricted all manifolds $\cM^d$ to the ones with trivial first and second Stiefel--Whitney class \cite{Law,Mor}. Therefore, the vielbein matrix ${e_\mu}^a$ defines $d$ one-forms $\hat{e}^a$ by (\ref{vielbeindefi}). This implies that the vielbein transforms by the standard pull-back under a general coordinate transformation $\vp_A\in \Diff(d)$
\be\label{vielbeintrafo0}
{x'}^\mu &=&\vp_A^\mu(x)\nn\\
{e'_\mu}^a &=& \frac{\p x^{\mu}}{\p {x'}^\nu}{e_\nu}^b {O_b}^a.
\ee
Recalling that for a given metric $g$, the vielbein is only defined as an $SO(d-1,1)$ orbit $[e]$, it is not possible to exclude an accompanying $O\in SO(d-1,1)$ action mixing the basis vectors $\hat{e}^a$ of the vielbein frame. In the next chapter, I will argue that it is sensible from a physical point of view to relate the action $O\in SO(d-1,1)$ to the diffeomorphism $\vp_A$. This will allow to define an induced action of diffeomorphisms $\vp_A\in \Diff(d)$ on the spinor representations that I have introduced in section \ref{Representation}. I want to conclude with some remarks.
\begin{itemize}
	\item The vielbein matrix $e$ is the matrix that links the coordinate induced frame $dx^\mu$ of the cotangent space to the one $\hat{e}^a$ that is preferred by the metric tensor (\ref{gDefi}). This is the reason, why the vielbein $e$ is sometimes called the soldering form.
	\item The frame $\hat{e}^a$ is not coordinate induced in general. In other words: for a Lorentzian manifold, the one-forms $\hat{e}^a$ that transform as $G=SO(d-1,1)$ vectors under coordinate transformations, are not integrable to coordinates. This is different for $G=Sp(d)$ in the case of a symplectic manifold due to Darboux's theorem \cite{McDS95}, which I will use for the discussion of supergravity in chapter \ref{CHAP5}.
	\item The vielbein frame is a special case of a $G$ structure as discussed in Joyce's book \cite{J00}.
\end{itemize}
In relating physical theories to experiments the flat Minkowski space $\R^{d-1,1}$ (\ref{PDefi}) plays a distinguished role. For this particular manifold there is a globally defined coordinate chart such that the vielbein frame and the coordinate induced frame coincide. Hence, the vielbein $e$ is the identity matrix in this case. This situation is the starting point to relate the mathematical concepts of this chapter to interesting physical theories in the next ones.

\chapter{The dynamics and why $d=11$ supergravity?}\label{CHAP3}
In this chapter, I will introduce the dynamics of the theories relevant for this dissertation. I have closed the previous chapter with the remark that the flat Minkowski space $\R^{d-1,1}$ (\ref{eta}) plays a prominent role in relating physical theories to experiments. I will start by providing an example for such a physical theory in the first part: the gauging of the global, internal $U(1)$ symmetry of a free Dirac field and its quantization that leads to quantum electrodynamics. The big advantage of describing an observable force by a gauge theory lies in the possibility to use the concept of a perturbation expansion in its domain of validity to relate the theory to experiments.\footnote{The restricted domain of validity of the perturbation expansion emphasizes the interpretation of quantum electrodynamics as an effective field theory: neither for very large energies nor to infinite accuracy, it is adequate to use the tool of perturbation expansion, which is an asymptotic series in general \cite{ZJ02}.}\\

This is the reason to devote the second part of this chapter to the gauging of the global, external Poincar\'e isometry group of Minkowski space. This procedure naturally leads to a gravitational theory with torsion \cite{SK61}, a so-called Einstein--Cartan theory such as supergravity. Thus, it restores the property of the metric being a dynamical object, which was dropped in establishing the contact to experimental physics to begin with. Where the role of torsion is concerned, I would like to suggest the point of view that it might be inconsistent for a theory of quantum gravity to require vanishing torsion on the level of the operators. In analogy to the Gupta--Bleuler formalism of quantum electrodynamics, it could be possible that this only is a consistent constraint on the physical states. If this interpretation was adopted, supersymmetry could be interpreted as a natural symmetry of the free $d=4$ gravity theory linking the torsion degrees of freedom to the vielbein ones. This is my personal motivation to study supergravity theories from a physical point of view.\\

I will conclude this chapter with an argument why it might be interesting to discuss higher dimensional supergravities in order to quantize the gravitational interaction in $d=4$. Taking the requirement of maximal supersymmetry as a guideline, one is led to discuss supergravity in $d=11$ dimensions \cite{CJS78}. This theory will be further investigated in chapter \ref{CHAP5}.

\section{A brief excursion to quantum electrodynamics}
\subsection{Gauging a symmetry}\label{u1s}
A physical theory that is quite well understood, adequately tested and verified by experiment is the classical Dirac theory that describes fermions such as electrons or quarks in Minkowski space $\cM^d=\R^{d-1,1}$. As long as the gravitational interaction can be neglected, it is expected that the restriction of the manifold to flat Minkowski space is sufficient to describe experiments.\\

Since the reason for discussing this topic is of mere motivational nature, I will stick to the simpler case of electrons, which are described by Gra{\ss}mann valued, i.e. anticommuting, sections of the spin bundle over the manifold $\cM^d=\R^{d-1,1}$. The corresponding dynamics are provided by the action
\be\label{Dirac}
S &=&\int\limits_{\R^{d-1,1}}d^dx \,\mathcal{L}_{\text{Dirac}}(x)\\
\mathcal{L}_{\text{Dirac}}(x)&=&\bar{\psi}\left(i\g^\mu\p_\mu -m\right)\psi.
\ee
Although there is no experimental evidence for a dimension $d\neq 4$, I have kept the dimension $d$ arbitrary in order to emphasize that the entire argumentation is independent of the value of $d$. The $\g$-matrices fulfill the standard Clifford property (\ref{Clifford}) and $\bar{\psi}=\left(\psi^t\right)^*\g^0$ denotes the Dirac conjugation, i.e. transposition, complex conjugation and multiplication by $\g^0$. Due to the latter, the Lagrangian $\mathcal{L}_{\text{Dirac}}$ (\ref{Dirac}) is invariant under the following transformation of the spinors
\be\label{u1}
\psi'(x) &= & e^{i\lambda}\psi(x)
\ee
with $\lambda\in \R$. As the action on the coordinates is trivial, this symmetry transformation is an internal symmetry according to the classification of section \ref{Symm1}. Furthermore, as $\lambda$ does not depend on $x$, it is a global symmetry transformation. Since the corresponding symmetry Lie group $U(1)$ is one-dimensional, I refrain from introducing basis vectors $\hat{T}_\gamma$ (\ref{psiParam}) for the $U(1)$ representation space $\psi$ explicitly.\\

I have already mentioned in the formal introduction to symmetry groups of section \ref{Symm1} that every global symmetry transformation can be enhanced to a bigger symmetry by making it local. For the case of the $U(1)$ symmetry, this corresponds to allowing $\lambda$ in (\ref{u1}) to become $x$-dependent. Following the definitions of section \ref{SymmAct}, the physical theory will only be invariant under this enlarged symmetry group, if the partial derivative $\p_\mu$ is replaced by an equivariant connection $\nabla_\mu$
\be\label{conn12}
\nabla_\mu \psi &=& \p_\mu \psi-\delta_{A_\mu}\psi
\ee
with the action $\delta_{A_\mu}$ of the Lie algebra $\mathfrak{u}_1$. The requirement of equivariance under the local version of the symmetry (\ref{u1}) implies the identity
\be\label{u1trafo}
A'_\mu &=& A_\mu +i\p_\mu \lambda(x),
\ee
where I have used the standard abbreviation from section \ref{prim2} to denote the transformed objects by a prime. Note that both the global and the local symmetry group are internal by the definition from section \ref{SymmAct}, because they do not act on the coordinates.\\
 
The process of transforming a global continuous symmetry into a local one is known as gauging the global symmetry group. To complete it, all partial derivatives $\p_\mu$ in the Lagrangian $\mathcal{L}_{\text{Dirac}}$ (\ref{Dirac}) have to be replaced by covariant ones $\nabla_\mu$. This implies that the gauged Lagrangian contains the term
\be\label{coupling}
A_\mu \left(
\bar{\psi}\g^\mu \psi
\right)
.
\ee
with the Noether current being the term in the bracket. In deriving the equations of motion from the gauged action, $A_\mu(x)$ is then treated as an independent physical field, the \textbf{gauge field}, that couples to the fermions $\psi$ at least by the term (\ref{coupling}). In order to relate this gauged theory to experiments, other terms containing the gauge field $A_\mu$ can be added to the Lagrangian. These terms have to be constructed in such a way that the resulting theory is invariant under the local version of the symmetry action (\ref{u1}). Therefore, these terms may only implicitly depend on $A_\mu$ by the connection $\nabla$. For an internal symmetry, this is equivalent to stating that all admissible terms must contain the field strength tensor
\be\label{FTen}
F_{\mu\nu}&=& \left[\nabla_\mu,\nabla_\nu\right].
\ee
Since the Lagrangian is a scalar quantity, one has to build a scalar from (\ref{FTen}). One choice for a gauged Dirac Lagrangian is
\be\label{Dirac2}
\mathcal{L}(x) &=& \bar{\psi}\left(i\g^\mu\nabla_\mu -m\right)\psi - F_{\mu\nu}F^{\mu\nu}
\ee
where indices have been raised with the canonical isomorphism (\ref{canon1}) provided by the Minkowski metric $g=\eta$ of $\cM^d=\R^{d-1,1}$. Before commenting on the physical interpretation of this Lagrangian, I want to recapitulate the recipe for gauging a global symmetry.
\begin{enumerate}
 	\item Start with a given theory that describes physical objects. 
 	\item Observe that the corresponding action is invariant under a global continuous symmetry group transformation.
 	\item Gauge the symmetry by introducing a connection $\nabla$.
 	\item Construct a gauged Lagrangian in such a way that the gauged theory is invariant.
\end{enumerate}

\subsection{Relating the gauged theory to physics}\label{gauged2}
It is obvious from the recipe I provided in the previous section that the resulting gauged Lagrangian is by no means unique. There are many mathematically well-defined gauged Lagrangians that could lead to physical observations. In order to find out which one indeed leads to a physically interesting theory, one has to establish some correspondence between an observable object in nature and the gauge field. \\

For the case of the Dirac theory, this relation is provided by electromagnetism. The observable electric and magnetic field strengths $\vec{E}$ and $\vec{B}$ are in one-to-one correspondence with the field strengh $F_{\mu\nu}$ (\ref{FTen}) in $d=4$. As electrons are charged objects, this identification allows to test if there is a gauged Lagrangian that corresponds to the interaction of electrons under the electromagnetic force. It is one of the most exact results in physics that there is a Lagrangian that fulfills this property, which is provided by the example I have stated in (\ref{Dirac2}).\\

The gauge field $A_\mu$ is usually referred to as the electromagnetic potential. In contrast to the field strength $F_{\mu\nu}$, it is not in one-to-one correspondence to observable objects, because it is only defined modulo the gauge transformation (\ref{u1trafo}). This allows to set the exact part of the one-form $A$ to zero classically, which implies $\p\cdot A =0$.\footnote{In classical electrodynamics, this is called the Lorenz gauge. This choice is not unique, however: the Coulomb gauge is a prominent different gauge.} This arbitrariness corresponds to the so-called horizontal polarizations of a photon. \\

In the introduction to this chapter, I mentioned that I will discuss quantum aspects at some point. The standard quantization procedure  of the Lagrangian (\ref{Dirac2}) has led to remarkable agreement with experiments \cite{PS95}. Therefore, I want to highlight two facts of this procedure.
\begin{enumerate}
\item \textit{The objects to be quantized are the modes parametrizing the space of solutions of the free field equation for the gauge potential $A$.}
\item \textit{It is inconsistent to impose $\p\cdot A =0$ as an operator relation. It has to be imposed on physical states.}
\end{enumerate}
The second point is referred to as the Gupta-Bleuler formalism in quantum electrodynamics. To phrase this in other words: it appears to be inconsistent for the quantization of a gauge theory to fix a symmetry prior to quantization, even though this would not make a difference for the classical theory. It is important to keep this in mind for the discussion of gravity in the next section. I close the discussion of quantum electrodynamics with three comments.
\begin{itemize}
	\item I want to emphasize that the definition (\ref{conn12}) is not a restriction, because the explicit form of the Lagrangian (\ref{Dirac2}) was not fixed a priori. Only in this case, the relation (\ref{conn12}) should be referred to as minimal coupling. Any non-minimal coupling is hence equivalent to choosing a different Lagrangian (\ref{Dirac2}) and keeping (\ref{conn12}).
	\item For all the other real finite dimensional symmetry Lie groups that follow the classification in the sections \ref{Class} and \ref{realf}, the procedure works in the same way, keeping in mind the results of section \ref{prim2} for more complicated groups. As an example may serve the quarks $\psi = \psi^\gamma (x)\hat{T}_\gamma $ of quantum chromodynamics that transform as non-trivial representations of $SU(3)$ (\ref{psiParam}). The connection has the form
	\be\label{conn13}
	\nabla_\mu \psi^\gamma (x)&=& \p_\mu \psi^\gamma (x)-\delta_{\hat{A}_\mu} \psi^\gamma (x)
	\ee
	with an action $\delta_{\hat{A}_\mu}$ of the Lie algebra $\mathfrak{su}_3$ on the coefficients $\psi^\gamma$. The gluons of the strong interaction are set into correspondence to the $\mathfrak{su}_3$ valued gauge field $\hat{A}_\mu$. 
	\item If the gauge group $G$ is not simple, it is convenient to introduce gauge fields for all simple parts and ideals separately. This is no restriction, because the gauge field always takes values in the corresponding Lie algebra $\Lie$, which is a vector space and can hence be decomposed. This will be the case in the next section.
	 		 \end{itemize}

\section{Gravity as a gauge theory}\label{gravmatter}
All physical theories that are related to observations are formulated in flat Minkowski space $\R^{d-1,1}$, unless they involve the gravitational interaction. This implies that they all have a common external symmetry group, the Poincar\'e isometry group $SO(d-1,1)\ltimes \cP_d$ of flat Minkowski space $\R^{d-1,1}$ (\ref{Isometry8}). Since all physical fields are expected to couple to gravity, it is hence natural to apply the gauging recipe provided in section \ref{u1s} to the Poincar\'e group and try to relate the result to the gravitational interaction afterwards again.

\subsection{Gauging an external symmetry}\label{gravmatter2}
To perform the gauging process, I have to introduce a connection $\nabla$ that is equivariant under a local version of the Poincar\'e group action on physical fields $\psi$. Since the Poincar\'e algebra $\mathfrak{so}_{(d-1,1)}\oplus \mathbf{d}$ decomposes into the Lorentz algebra $\mathfrak{so}_{(d-1,1)}$ and the algebra of translations $\mathbf{d}$, I introduce one gauge field $\omega$ for the Lorentz subgroup $SO(d-1,1)$ and another one $e$ for the subgroup of translations. Recalling that the generators of translations $\hat{P}$ only act on the sections in their representation $\mathbf{R}$ as derivative operators (\ref{Lie9})
\beg
\mathbf{R}_{\hat{P}_\mu}&=&\p_\mu,
\eeg
the natural way to introduce the gauge field $e$ is by defining a new coordinate frame
\be\label{ehneu}
\p_a &:=& {e_a}^\mu (x)\p_\mu,
\ee
which has to transform in the vector representation $\mathbf{d}$ of the Lorentz group under local, invertible translations:
\be\label{xTrafo4}
x^\mu \,\mapsto & {x'}^\mu(x) &=\, x^\mu +c^\mu(x)\\
\Rightarrow \quad \,\,\p_\mu \,\,\mapsto &\p'_\mu &=\,\, \frac{\p x^\rho}{\p {x'}^\mu} \p_\rho\nn\\
\Rightarrow \quad {e_a}^\mu\,\,\mapsto &{e'_a}^\mu &=\,\, {O_a}^b{e_b}^\nu\frac{\p {x'}^\mu}{\p {x}^\nu}. \label{vielbein3}
\ee
I want to stress that introducing the gauge field $e$ as an $SO(d-1,1)$ representation $\mathbf{d}$ only restricts the general form of the induced action on $e$ to (\ref{vielbein3}). The precise dependence of a possible, induced Lorentz action $O\in SO(d-1,1)$ on the translation $c^\mu(x)$ (\ref{xTrafo4}) is not fixed. In particular, the trivial action $O=\id$ is not an inconsistent choice at this stage.\\

Since $e$ describes a change of frames for the tangent space, it must be non-degenerate as a matrix. It is conventional to denote the inverse by switching the indices
\beg
{e_a}^\mu {e_\mu}^b &=& \delta_a^b.
\eeg 
In this notation, the cotangent space is spanned by the dual frame of one-forms
\be\label{dualf}
dx^a &:=& {e_\mu}^a dx^\mu.
\ee
The reader may have noticed that the gauge field of translations $e$ appears to be related to the vielbein frame that I have introduced in section \ref{Gstructur} and for which I have used the same symbol $e$. Before explaining why the gauge field of translations $e$ can indeed be identified with the vielbein $e$ in the next section, I want to address the other gauge field $\omega$, the one of the Lorentz part $SO(d-1,1)$ of the Poincar\'e group. \\

Where $SO(d-1,1)$ is concerned, it is important to recall from the discussion of representations of Lie groups in section \ref{Representation} that every physical field $\psi$ in Minkowski space transforms as a Lorentz representation. Therefore, I can apply the procedure of section \ref{prim2} to separate the $x$-dependent field coefficients $\psi^\gamma(x)$ from the constant vectors $\hat{T}_\gamma$ that span the representation space $V$: the identity (\ref{psistrich})
\beg
{\psi'}^\gamma \hat{T}'_\gamma &=& \psi^\gamma \hat{T}_\gamma
\eeg
with the transformation (\ref{Tstrich})
\be\label{ODefi}
\hat{T}'_\gamma &=& \exp\left({o_b}^c(x) \mathbf{R}_{\hat{L}{}^b{}_c}\right)\hat{T}_\gamma
\nn\\
&=:&{\left(O^{-1}\right)_\gamma}^\beta(x) \hat{T}_\beta,
\ee
generated by the Lorentz generators $\hat{L}$ (\ref{LDefi}) in the appropriate representation $\mathbf{R}$,
induces the $SO(d-1,1)$ action on the coefficients
\be\label{psiTraf}
{\psi'}^\gamma &=&{O_\beta}^\gamma (x)\psi^\beta\\
 &=&\psi^\gamma + {o_b}^c(x) \delta_{{\left.\hat{L}\right.^b}_c}\psi^\gamma +\mathcal{O}(o^2)\nn.
\ee
In complete analogy to the case of an internal symmetry (\ref{conn13}), the connection $\nabla$ that is equivariant under gauged Poincar\'e transformations hence has the general form 
\be\label{conn45}
\nabla_a \psi^\gamma&=& {e_a}^\mu\p_\mu \psi^\gamma- 
{{\omega_a}_b}^c\delta_{{\left.\hat{L}\right.^b}_c}\psi^\gamma.
\ee
The index $a$ indicates that I have used the dual frame (\ref{dualf}), because this is the one that transforms as a Lorentz tensor under a gauged Poincar\'e action. The abstract generators $\hat{L}$ of the Lorentz algebra $\mathfrak{so}_{d-1,1}$ defined in (\ref{LDefi}) parametrize the Lorentz action $\delta_{{\left.\hat{L}\right.^b}_c}$ on the physical field $\psi^\gamma$.\footnote{Due to the convention to distinguish the different frames of the (co)tangent space only by different indices (\ref{ehneu}), it may be necessary to emphasize that the names of the indices in (\ref{LDefi}) were completely arbitrary. The important fact is that the antisymmetry of the generators $\hat{L}$ is to be understood with respect to the flat Minkowski metric $\eta$ throughout this thesis, whatever symbols of the indices are used. If different symbols for indices actually denote a different object, e.g. the definition (\ref{ehneu}), I will always refer to the appropriate definition. Furthermore, I want to remind the reader that the action of the abstract generators $\hat{L}$ on $\psi^\gamma$ is defined by the representation space spanned by the vectors $\hat{T}_\gamma$ (\ref{Tstrich}) and not by the vector field representation (\ref{PRep}).}\\

Thus, the transformation of the gauge fields $e$ and $\omega$ under a local Lorentz transformation (\ref{psiTraf}) is fixed to
\be\label{etrafo}
{{e'}_a}^\mu &=& {\left(O^{-1}\right)_a}^b{{e}_b}^\mu \\
\label{omtrafo}
{\omega'_{ab}}^c  
&=& {\left(O^{-1}\right)_a}^d {\left(O^{-1}\right)_b}^e \left( {\omega_{de}}^f {O_f}^c
+\p_d{O_e}^c\right)
\ee
with ${(O^{-1})_a}^b$ defined by (\ref{ODefi}) with the generators $\hat{P}_b$ spanning the vector representation $\mathbf{d}$ of $SO(d-1,1)$, similarly to the example in section \ref{prim2}.\\

Before completing the gauging procedure by constructing invariant actions, I want to highlight a subtlety that is due to the semidirect product structure of the Poincar\'e group: local Lorentz actions always affect both gauge fields $e$ (\ref{etrafo}) and $\omega$ (\ref{omtrafo}), whereas local translations only affect $\omega$, if the possible Lorentz action ${O_a}^b$ in (\ref{vielbein3}) is non-trivial $O\neq \id$. Otherwise, $\omega$ is invariant under a local translation.\\

To construct invariant objects under gauged Poincar\'e transformations however, the argumentation from section \ref{u1s} still holds: these may only implicitly depend on the gauge fields via the connection. Keeping in mind the fact that the derivative operators transform as a representation of the Lorentz group, a quick calculation\footnote{To illustrate the procedure from section \ref{prim2}, I have included the explicit calculation in the appendix \ref{Expl1}.}
 with the commutation relations (\ref{ComRel0}) shows 
\be\label{curv3}
\left[\nabla_a,\nabla_d\right]
&=&{T_{ad}}^f \nabla_f
-{\left(R_{ad}\right)_e}^f\delta_{{\left.\hat{L}\right.^e}_f }
\ee
with the two abbreviations
\begin{subequations}\label{TRDefi}
\be\label{cont}
{T_{ad}}^f &:=&\left[ \p_a ,  \p_d \right]^f +2{\omega_{[ad]}}^f\\
{\left(R_{ad}\right)_e}^f &:=&2\p_{[a}  {\omega_{d]e}}^f 
-\left[ \p_a ,  \p_d \right]^c {\omega_{ce}}^f
+2{\omega_{[a|e|}}^g {\omega_{d]g}}^f.
\label{Riemann}
\ee
\end{subequations}
Recalling that the Lorentz algebra in $d$ dimensions $\mathfrak{so}_{(d-1,1)}$ is parametrized by antisymmetric matrices (\ref{AntisyM}) with respect to $\eta$, the Riemann tensor $R$ (\ref{Riemann}) prescribes an endomorphism of the vector space $\mathfrak{so}_{(d-1,1)}$. Its trace is the Ricci scalar
\be\label{Ricciscalar}
\tilde{R}\,\,:=& tr_{\mathfrak{so}_{(d-1,1)}}\left(R\right) &=\,\, \left(R_{ad}\right)^{ad}.
\ee
Since external symmetries also act on the coordinates, the curvature tensor $R$ (\ref{Riemann}) is not the only Lorentz tensor under local Poincar\'e transformations that arises from the commutator $[\nabla,\nabla]$ (\ref{curv3}). The torsion tensor $T$ (\ref{cont}) can also be used to construct a Lagrangian. \\

This is the last step in the gauging procedure, to construct a Lagrangian from gauge invariant objects. One possibility is  to add the terms
\be\label{EHKosm}
\mathcal{L}_{\text{EH}}(x) &=& \tilde{R} + \tilde{T} + \Lambda
\ee
to the matter Lagrangian $\mathcal{L}_{\text{matter}}$, where $\tilde{T}$ symbolizes a term constructed from the degrees of freedom of the tensor $T$ and a global constant $\Lambda\in \R$. A comment should be added concerning the latter. In the case of internal symmetries, a global constant would not change the theory, because it would drop out from the Lagrangian under a variation of the action in order to get the equations of motion. The reason for this is that no gauge field enters the measure of the action.\\

This is however inconsistent for gauging an external symmetry. In order for the action to be invariant under local Poincar\'e transformations, the measure has to be defined by the $d$ fold wedge product of the dual form $dx^a$ (\ref{dualf})
\be\label{measure}
d^d x \, \det(e) &=& \frac{1}{d!}\e_{a_1\dots a_d} dx^{a_1}\wedge \dots \wedge dx^{a_d}.
\ee
This implies that a global constant $\Lambda$ in the Lagrangian does lead to non-trivial dynamics provided by the action
\be\label{action2}
S &=& \int\limits_{\R^{d-1,1}}d^d x \, \det(e) \left(\mathcal{L}_{\text{EH}}(x) +\mathcal{L}_{\text{matter}}(x,\nabla)\right).
\ee
In the matter Lagrangian $\mathcal{L}_{\text{matter}}$, all partial derivatives have to be replaced by covariant ones $\nabla$ (\ref{conn45}) and all physical fields are contracted in the Lorentz covariant frame that is induced by the vielbein frame $\hat{e}_a$ (\ref{viel2}). The equations of motion can thus be deduced by the standard variational principle. This completes the gauging procedure of the Poincar\'e symmetry for any matter theory in Minkowski space.

\subsection{Conventions and diffeomorphisms}\label{convdiff}
I want to investigate the transformation under local, invertible translations (\ref{xTrafo4}) a bit more closely. Comparing this transformation with the action of a diffeomorphism $\vp_A\in \Diff(d)$ with its associated vector field $X_A\in \mathfrak{diff}_d$ (\ref{Diffeom2}) from section \ref{Diffeom}, it is obvious that the two transformations coincide in their common domain of validity
\beg
\exp (X_A^\nu(x)\p_\nu) x^\mu \,=&\vp^\mu_A(x) \,=\, {x'}^\mu(x) &=\, x^\mu +c^\mu(x).
\eeg
This implies that the gauge field of translations $e$ is nothing but the vielbein $e$ that solders the coordinate induced frame to the vielbein frame $\hat{e}_a$ as explained in section \ref{Gstructur}: its transformation under a diffeomorphism $\vp_A\in \Diff(d)$ (\ref{vielbeintrafo0}) is the same as the one of the gauge field (\ref{vielbein3}) with an up-to-now not specified action $O\in SO(d-1,1)$ \beg
{x'}^\mu &=&\vp_A^\mu(x)\nn\\
{e'_\mu}^a &=& \frac{\p x^{\mu}}{\p {x'}^\nu}{e_\nu}^b {O_b}^a.
\eeg
A short look at (\ref{etrafo}) reveals that the coefficients of the space-time metric $g$ in the coordinate induced frame (\ref{Vielbeinequiv})
\beg
g_{\mu\nu}&=&{e_\mu}^a{e_\nu}^b\eta_{ab}
\eeg
are invariant under local Lorentz transformations. Of course, $g_{\mu\nu}$ transforms as a symmetric $2$-tensor under local translations or diffeomorphisms, equivalently.\\

Next, I want to cast the definition of the Lorentz tensor $T$ (\ref{cont}) in a more convenient form for the discussion in chapter \ref{CHAP4}:
\be\label{omegaK}
{\omega_{ac}}^d  &=:&{\left(v^{\text{min}}_a\right)_c}^d -\left(\delta_c^k\delta_j^d -\eta_{cj}\eta^{dk}\right){\left(v_k\right)_a}^j
+{K_{ac}}^d
\ee
with the abbreviations\footnote{For this thesis, I use antisymmetrizations of strength one, i.e. $A_{[a} B_{b]}:=\frac{1}{2}\left(A_{a} B_{b} -A_{b} B_{a}\right)$. The abbreviations $v$ and $v^{\text{min}}$ (\ref{abbrev1}) naturally arise in the context of non-linear realizations. I will explain this in chapter \ref{CHAP4}.} after lowering the Lorentz index with the Minkowski metric $\eta$
\begin{subequations}\label{abbrev1}
\be
\left(v^{\text{min}}_a\right)_{cd} &:=& \eta_{b[d}\left({e_{c]}}^\mu\p_{a}{e_\mu}^b\right)\label{vmin2}\\
\left(v_a\right)_{cd} &:=& \eta_{b(d}\left({e_{c)}}^\mu\p_{a}{e_\mu}^b\right)\label{v2}\\
K_{acd}&:=&\frac{1}{2}\left(T_{acd} +T_{dac} -T_{cda}\right).\label{Kontorsion}
\ee
\end{subequations}
Having identified local translations with diffeomorphisms, the connection $\nabla$ (\ref{conn45}) with local Poincar\'e equivariance is the \textbf{Levi--Civita connection} of differential geometry modulo the \textbf{torsion} tensor $T$ in disguise: it is usually referred to as the spin connection with \textbf{contorsion} $K$. The Levi--Civita connection acts in the coordinate induced frame $dx^\mu$, whereas $\nabla$ acts in the vielbein frame $dx^a={e_\mu}^a dx^\mu$ (\ref{dualf}). This implies that the requirement of a covariantly constant metric $\nabla g=0$ is trivial, because in the vielbein frame, any metric has the form of the constant Minkowski metric $\eta_{ab}$ (\ref{gDefi}) by the very definition of the frame, which is annihilated by any Lorentz algebra valued connection $\nabla$.\newpage

Before linking the mathematical concept of gauging the Poincar\'e isometry group to a physically observable theory, I want to emphasize again that the relations (\ref{cont},\,\ref{omegaK}) are mere definitions of the Lorentz tensors $T$ and $K$. In the sections \ref{subtle} and \ref{tele}, I will provide two examples for constraints that establish a relation between the a priori independent gauge fields $e$ and $\omega$.

\subsection{General relativity}\label{ART}
Einstein proposed in 1915 that any gravitational coupling to a test particle can be interpreted as a force resulting from an appropriate change of the coordinate frame. To phrase it differently, for any matter system there is a coordinate system in which the gravitational coupling is absent. Any test particle in this frame is referred to as a freely falling particle. If the torsion tensor $T$ (\ref{cont}) is set to zero, the equations of motion of the action $S$ (\ref{action2}) are called the Einstein equations. This theory is referred to as General Relativity and it has led to remarkable agreement with experiment.
\\

An important point is that this theory can only be linked to experiments, if a holonomic or integrable frame is used: these are equivalent names for a frame $dx^\mu$ that is induced by coordinates $x^\mu$. And experiments always measure distances in these coordinates. This implies that the dynamical object of general relativity is the metric $g(x)$ (\ref{gDefi}). At every point $x\in\cM^d$, it is given by the $\frac{1}{2}d(d+1)$ independent components of a symmetric matrix $g_{\mu\nu}(x)$. Due to the associativity of matrix multiplication, the gauge field $e$ of translations can also be interpreted as a $d\times d$ matrix.\footnote{A local translation acts from the left and a local Lorentz rotation from the right. Due to associativity, the two actions commute.} Since there are no constraints on this matrix $e$, the degrees of freedom in equation (\ref{Vielbeinequiv})
\beg
g_{\mu\nu} &=& \eta_{ab}{e_\mu}^a{e_\nu}^b
\eeg
do not match. I have already addressed this fact in section \ref{Gstructur}: if the eigenvalues of the vielbein matrix $e$ are positive, then to every symmetric tensor $g_{\mu\nu}(x)$ there is a unique $SO(d-1,1)$ orbit $[e](x)$ of vielbein matrices $e(x)$.\footnote{From the gauge theoretic point of view, the requirement of positive eigenvalues of $e$ is natural due to the concept of a perturbation expansion $e=\id +h +\mathcal{O}(h^2)$.}\\

The important point is that the Lorentz group, whose orbit is $[e](x)$, acts in the same way as the Lorentz subgroup of the Poincar\'e group that I have gauged (\ref{etrafo}). This implies that it is consistent to fix any representative $e(x)$ in the equivalence class $[e](x)$, which is equivalent to fixing the matrix form of $e(x)$ for all $x\in \cM^d$. Thus, a local translation $\vp_A\in \Diff(d)$ with $x'=\vp_A(x)$ induces a compensating local Lorentz rotation $O(\vp_a,e)\in SO(d-1,1)$ such that the transformed matrix $e'$ has the same matrix form
\be\label{VielbeinTrafo}
{e'_\mu}^a &=& \frac{\p x^\nu}{\p {x'}^\mu}\,{e_\nu}^b \,{O(\vp_A,e)_b}^a.
\\\nn
\ee
It is this particular constraint of a fixed representative in the equivalence class $[e]$ that induces a diffeomorphism action on the vielbein frame $dx^a$ or in other words, that establishes the link from the tangent bundle to the spin bundle. Hence, physical fields in the spin bundle transform under a local translation of coordinates $\vp_A\in \Diff(d)$ by an induced Lorentz action $O(\vp_A,e)\in SO(d-1,1)$, which is well-defined because expectation values always are even in fermions as discussed in section \ref{Representation}. \\

I will adopt this choice (\ref{VielbeinTrafo}) for the induced Lorentz action of a diffeomorphism in the sequel. I want to emphasize again that this is a choice\footnote{I am grateful to T.~Damour for pointing this out to me.},but a canonical one, which is due to the following argument.\\

I have already mentioned in the sections \ref{Gstructur} and \ref{gravmatter2} that it would be consistent to fix the trivial Lorentz action $O=\id$ for the induced action on the vielbein (\ref{vielbein3}). This would imply that the gauge field $\omega$ would be invariant under a local translation or a diffeomorphism. In order to construct an action that is invariant under diffeomorphisms alone, it would hence be sufficient to consider any polynomial in $\omega$. Thus, even if I imposed vanishing torsion $T=0$, the Einstein--Hilbert action (\ref{EHKosm} with $\Lambda=0$) would not be the unique possibility of an action $S$ with diffeomorphism invariance any more that only depends on the vielbein matrix $e$ and is quadratic in derivatives. \\

This non-uniqueness is not a contradiction to results from differential geometry for one reason: the dependence of the action $S$ on $e$ would not in general be in a way that allows to rephrase $S$ as a function of the metric $g_{\mu\nu}$. Requiring this possibility is related to the symmetry of $g$ or to local Lorentz invariance of the action, equivalently. And as soon as the entire local Poincar\'e invariance is required, the uniqueness property of the Einstein--Hilbert action is restored, of course. Therefore, it is natural to require that every diffeomorphism $\vp_A\in \Diff(d)$ also induces a local Lorentz rotation $O(\vp_A,e)\in SO(d-1,1)$, because then it is sufficient to demand diffeomorphism invariance of the action alone to guarantee the uniqueness property of the Einstein--Hilbert action.

\subsection{Geometric interpretation}\label{subtle}
I have shown that the equations of motion of any theory for which the Poincar\'e symmetry is gauged, correspond to a theory that is invariant under the group of diffeomorphisms $\Diff(d)$, which is by definition the group of general coordinate transformations.\\

Since I defined a local physical theory by its analytic equations of motion, global properties like the topology of the underlying manifold do not influence the dynamics. As the gauging procedure is performed in local coordinates and as the Minkowski space is diffeomorphic to any open set on any Lorentzian manifold $(\cM^d,g)$ of equal dimension, I am free to start with an arbitrarily curved Lorentzian manifold and cover it with a coordinate chart $(U_\alpha,x_\alpha)$ containing open sets $U_\alpha$. For each open set $U_\alpha$, I can introduce basis vectors $\hat{P}_\mu$ (\ref{PDefi}) that span $x_\alpha(U_\alpha)\subset \R^d$. In the coordinate induced basis of the cotangent space $dx^\mu$, define a Minkowski metric $\eta=\eta_{\mu\nu}dx^\mu\otimes dx^\nu$ with respect to which the ungauged matter Lagrangian is specified.\footnote{The metric $g$ will of course have a different shape in these coordinates in general. Locally, $\eta$ is always well-defined, as I have already mentioned in the context of the conformal isometries (\ref{KGen}). I am grateful to D.~Giulini for pointing this out to me.} 
Gauging the Poincar\'e isometry of $\eta$ introduces gauge fields $e$ and $\omega$. As the gauged translations are equivalent to coordinate transformations, it is consistent to identify the gauge field of translations $e$ with the Lorentzian metric $g$ in this coordinate frame tantamount to the relation (\ref{Vielbeinequiv})
\beg
g_{\mu\nu} &=& {e_\mu}^a{e_\nu}^b \eta_{ab}.
\eeg
This procedure can be performed for any open neighbourhood on $(\cM^d,g)$ and for any coordinate chart $(U_\alpha,x_\alpha)$. The resulting theory is invariant under gauged translations or diffeomorphisms $\Diff(d)$-equivalently. The equations of motion of a local physical theory do not differ for different manifolds or for different open sets on the same manifold. They are universal in the sense that the abstract equations of motion for $g_{\mu\nu}(x)$ are the same for all points $x\in \cM^d$, whereas the metric tensor $g_{\mu\nu}(x)$ itself is different for different $x\in \cM^d$ in general. \\

For every open set, I could in principle define a separate action in some coordinate chart that provides these universal equations of motion after a variation. This is what is actually meant by integrating over a manifold $\cM^d$. In this spirit, the gauged version of the matter Lagrangian can be reinterpreted as a local physical theory on an arbitrary Lorentzian manifold $(\cM^d,g)$, whose universal equations of motion can be derived from the following action in the standard way:
\be\label{action3}
S &=& \int\limits_{\cM^d}d^d x \, \det(e) \left(\mathcal{L}_{\text{EH}}(x) +\mathcal{L}_{\text{matter}}(x,\nabla)\right).
\ee
This action is globally invariant under diffeomorphisms. Therefore, General Relativity can also be interpreted as follows: given the concept of a Lorentzian manifold, there is the arbitrariness which coordinate frame should be chosen to describe a physical theory. General Relativity claims that the physics is the same in every coordinate chart, i.e. the equations of motion are universal. Hence, having fixed a coordinate chart, the resulting equations of motion should be covariant under a general coordinate transformation, or equivalently under any diffeomorphism. The subtlety is that thus, non-local diffeomorphisms $\vp_a\not\in \Diff(d)$ would not be excluded. Hence, although I excluded them in section \ref{diffi}, the resulting theory (\ref{action3}) is invariant under their action.\\

To conclude this section, I want to stress that the argumentation from section \ref{gauged2} how to relate a gauge theory to physics applies to the Poincar\'e group in the same way as to any internal symmetry group. In particular, the gauged Lagrangian is not uniquely determined by the gauging procedure. Hence, it is again the \textbf{experiment} that has to fix the terms that have to be added to the matter Lagrangian, e.g. if higher curvature corrections have to be added or not. At the moment, experiments indicate that the Lagrangian $\mathcal{L}_{EH}$ (\ref{EHKosm}) is the right combination\footnote{A different possibility would e.g. be the untruncated MacDowell--Mansouri action (12) in \cite{MM77}, for which there is however no experimental evidence so far.}, in which $\Lambda>0$ is called the cosmological constant and torsion vanishes $K=T=\tilde{T}=0$. The latter constraint is the first example of a consistent constraint that establishes a relation between the independent gauge fields $e$ and $\omega$ (\ref{omegaK}).

\subsection{Teleparallel interpretation}\label{tele}
In his quest for a ``unified field theory'', Einstein \cite{E28} has also used a different interpretation of General Relativity that I briefly want to sketch. If some coordinate system is fixed, given any metric on an open set, it is always possible to choose the contorsion tensor $K$ (\ref{omegaK}) in the vielbein frame such that the Lorentz connection $\omega$ is trivial on this open set:
\be\label{omegaNull}
{\omega_{ac}}^d &=&0.
\ee
This implies that the curvature tensor $R$ (\ref{Riemann}) vanishes. This statement is not preserved under a change of the coordinate chart or equivalently a general transformation $\vp_A\in \Diff(d)$, but this is not important for the evaluation of experiments, because these are evaluated in a fixed coordinate chart anyway. Given some arbitrary metric, $\omega$ can be chosen to be trivial (\ref{omegaNull}) implying non-trivial torsion $T$ in general. Replacing $\omega$ by $T$ in the Einstein-Hilbert action (\ref{Riemann},\,\ref{action3}), one obtains a well-defined action that gives rise to the same dynamics as General Relativity in the geometric interpretation that I discussed in the previous section. Thus, the constraint of vanishing curvature is a well-defined procedure locally. It is the second example of a constraint that links the gauge fields $e$ and $\omega$.\\

The crucial ingredient to make this interpretation viable is the fixing of some coordinate system a priori. If the manifold $\cM^d$ is parallelizable, i.e. there is a global coordinate frame, as is the case for Lie groups and Minkowski space, this procedure is globally well-defined, too. If this is not the case, it is impossible to satisfy (\ref{omegaNull}) on the overlap of two coordinate charts, because (\ref{omegaNull}) is not preserved under a change of the coordinate chart with $K$ being a tensor in contrast to $\omega-K$. This may be the reason why the teleparallel interpretation of general relativity is often ignored in the physics literature.\footnote{A nice historical survey on this theory can be found in \cite{Go04}. Hehl has extended this idea in his work on metric affine geometry (MAG), in which the affine group $A(d)$ (\ref{affine2}) is gauged. More details and further references can be found in his review \cite{GH96}.}

\subsection{An aspect of quantization}\label{aspect}
In this section, I would like to push the analogy to quantizing internal gauge symmetries a bit further. To phrase the content of section \ref{ART} in a different way, physical observations do not exclude that in the same way as electromagnetic interactions can be interpreted as being mediated by a non-trivial gauge field $A=(\Phi,\vec{A})$, gravitational interactions can be interpreted as being mediated by the non-trivial gauge fields $(e,\omega)$ of the Poincar\'e group.\footnote{Non-triviality for the case of $e$ obviously means that the matrix ${e_\mu}^a$ is different from the identity matrix.} \\

I have already mentioned at the end of section \ref{subtle} that the model that is usually used to explain present observations assumes vanishing torsion $T$ and hence explains the gravitational interaction by a non-vanishing curvature of the universe. The teleparallel interpretation shows however that it is locally possible to choose a torsion such that the curvature vanishes without affecting the dynamics of general relativity. At least locally, there appears to be a symmetry linking the torsion to the curvature without affecting the dynamics. This arbitrariness is usually fixed by imposing a constraint on the connection $\nabla$ (\ref{conn45}) such that the torsion tensor $T$ (\ref{cont}) vanishes, what immediately constrains the two gauge fields $e$ and $\omega$ in the standard way (\ref{omegaK}) such that all independent degrees of freedom of $\omega$ are eliminated. \\

Next, recall from the discussion of quantizing the electromagnetic interaction in section \ref{gauged2} that it was the potential $A$ of the force that was quantized. This was however only determined up to a gauge symmetry. The fixing of this symmetry led to the constraint $\p\cdot A=0$. Finally, the argumentation showed that this constraint on the gauge field must be imposed on the physical states after having performed the quantization process, and not prior to it. Hence, if there was any symmetry relating the torsion to the vielbein, it could as well turn out to be inconsistent to impose vanishing torsion prior to quantization.\\

Therefore, I do not see any mathematical argument why one should not quantize the entire Einstein-Cartan theory, which is described by gauging the Poincar\'e group without the constraint of vanishing torsion being fixed on the operator level. For the physical states however, this constraint could be enforced in the same way as it is the case for the Lorenz gauge in quantum electrodynamics (Gupta-Bleuler formalism).\footnote{This analogy is not compelling, of course. Fixing the Lorenz gauge $\p\cdot A=0$ in electrodynamics breaks the local $U(1)$ gauge symmetry, whereas fixing the torsion $T=0$ does not break the local Poincar\'e symmetry. Only the additional symmetry linking the torsion to the vielbein is broken.}\\

It turns out that in $d=4$, a symmetry indeed exists that relates the torsion to the metric. Since I have restricted the class of manifolds under consideration to manifolds with vanishing first and second Stiefel--Whitney class \cite{Law,Mor}, it is possible to define a Gra\ss mann valued section $\chi$ of the tensor product of the cotangent bundle with the spin bundle. By the following identification, the field $\chi$ contains degrees of freedom of the torsion tensor
\be\label{gravitino}
{T_{ad}}^f&=&\frac14\chi^t_a\g^0\g^f\chi_d.
\ee
The coefficients $\chi_d$ of the section $\chi$ are denoted with respect to the vielbein frame (\ref{dualf}) of the cotangent space. The matrices $\g^f$ are representations of the Clifford algebra (\ref{Clifford}). In $d=4$, there is a Majorana representation of the Clifford algebra. Hence, the matrices $\g^f$ are real objects, which is the reason why no complex conjugation is involved in the definition (\ref{gravitino}). The spinor indices of $\chi_d$ are kept implicit as in section \ref{u1s}.\\
 
Since the torsion $T$ transforms as a Lorentz tensor under local Poincar\'e transformations, the section $\chi$ also transforms as a tensor modulo $\Z_2$, following the discussion from section \ref{Representation}. Hence any Lorentz scalar that is bilinear in $\chi$, is invariant under the diffeomorphism group $\Diff(d)$. \\

I want to construct a Lorentz scalar $\tilde{T}$ from the torsion degrees of freedom encoded in $\chi$. The standard choice is a Rarita--Schwinger term \cite{RS41} that can be added to the Einstein--Hilbert action (\ref{EHKosm})
\be\label{RS}
\tilde{T}&:=& \chi^t_a\g^0\g^{abc}\left(\nabla_b\chi_c -{K_{bc}}^d \chi_d\right)\\
\text{with}\quad \nabla_b\chi_c &=&\p_b\chi_c +{\omega_{bc}}^d\chi_d +\frac{1}{4}{\omega_{be}}^f{\g^e}_f \chi_c.
\nn
\ee
The covariant derivative $\nabla$ is the connection defined in (\ref{conn45}) with the contorsion tensor $K$ (\ref{Kontorsion}).\footnote{
In contradistinction to the original version of this theory in \cite{FN76}, $\g^f$ are real matrices and $\nabla$ is the standard connection with contorsion acting on the physical field $\psi$ as defined in section \ref{prim2}. The substitution $\psi_\mu:=\frac{1}{2}(1-i)\chi_\mu$ and $\e:=\frac{1}{2}(1-i)\ep$ would restore the original convention used in \cite{FN76} that was motivated from quantum field theory.}\\

The physical theory corresponding to the Lagrangian $\mathcal{L}_{\text{EH}}(x)$ (\ref{EHKosm}) with $\Lambda=0$ possesses a symmetry linking the vielbein $e$ to torsion. This is provided by the supersymmetry variation $\delta$ \cite{FN76} that relates solutions $e$ and $\chi$ of the equations of motion by a Gra\ss mann valued spinor $\ep$ parametrizing the symmetry
\begin{subequations}\label{vari1}
\be
\delta {e_\mu}^a &=& -\frac12\ep^t\g^0\g^a \chi_\mu\\
\delta \chi_\mu &=& \nabla_\mu\ep.
\ee 
\end{subequations}
The infinitesimal variation $\delta$ defines a physical symmetry in the sense of section \ref{SymmAct}, because it provides a Lie algebraic structure. A short calculation shows that the commutator of two supersymmetry variations 
\be\label{SUSY0}
\left[\delta_1,\delta_2\right] &=& \delta_3 + \delta_{\mathfrak{diff}_d} + \delta_{\mathfrak{so}_{(d-1,1)}}
\ee
produces a third supersymmetry variation $\delta_3$, a diffeomorphism $\delta_{\mathfrak{diff}_d}$ and a local Lorentz action $\delta_{\mathfrak{so}_{(d-1,1)}}$. The parameters of the three variations $\delta_3$, $\delta_{\mathfrak{diff}_d}$ and $\delta_{\mathfrak{so}_{(d-1,1)}}$ are the same for both solutions $e$ and $\chi$ of the equations of motion, but they explicitly contain the fields $e$ and $\chi$ \cite{FN76}. In other words, this algebra does not have structure constants like a finite dimensional Lie algebra. It has structure functions depending on the fields $(e,\chi)$ it acts on. Therefore, it is not a gauged version of a global symmetry algebra or superalgebra. \\
 
Chapter \ref{CHAP5} will deal with a gravitational theory that is defined by requiring maximal supersymmetry. Before presenting further details on this theory in section \ref{sugraDefi}, I want to close this part with some comments.

\begin{itemize}
	\item It is obvious from the definition of the gravitino $\chi$ (\ref{gravitino}) that not all $24$ degrees of freedom of the torsion tensor $T$ can be covered by the $16$ degrees of freedom of $\chi$. However, the vielbein $e$ in $d=4$ also has $16$ degrees of freedom prior to imposing the equations of motion and fixing the representative in the Lorentz orbit $[e]$. Hence, supersymmetry appears to be the maximal symmetry that can be introduced to link the torsion tensor $T$ to the vielbein $e$.\footnote{The degrees of freedom after imposing the equations of motion and fixing the gauges also coincide, of course, which follows from the closure of the supersymmetry algebra on the solutions $e$ and $\chi$ of the equations of motion.} 
	\item The $d=4$ $N=1$ supergravity \cite{FN76} provided by the Lagrangian $\mathcal{L}_{\text{EH}}$ (\ref{EHKosm}) with (\ref{gravitino},\,\ref{RS}) and $\Lambda=0$ possibly describes the free gravitational interaction that should be quantized. For the construction of propagators in such a quantum gravity, it is not important if a matter coupling to the system or $\Lambda\neq 0$ breaks the supersymmetry or not. This is in analogy to the conformal symmetry of the free Maxwell system that is quantized in quantum electrodynamics: the vertices break the conformal symmetry due to the coupling to massive Dirac fermions, but not the propagators of the Maxwell field. I want to emphasize that in the present construction, neither the matter system has to be supersymmetric as suggested in \cite{FN76}, nor that the Lorentzian manifold is elevated to some supermanifold. This would lead to a different theory \cite{IN92,OS78}. 
\item Since the connection $\nabla$ (\ref{conn45}) is Poincar\'e algebra valued and not $\mathfrak{diff}_d$ valued, I refrain from calling the diffeomorphism group $\Diff(d)$ the gauge group of general relativity. This distinction is important for the discussion of Goldstone bosons in the context of non-linear realizations in section \ref{Nonlinear2}, because with this interpretation, there only are finitely many Goldstone bosons in general relativity \cite{BO74}, and not infinitely many \cite{BK06}.
\item In the introduction to this chapter, I have already mentioned that the description of a physical force as a gauge theory offers the possibility to use the concept of a perturbation expansion in a coupling constant in its domain of validity to compare to experiments. For the gravitational interaction, the standard approach is an expansion of the gauge fields $(e,\omega)$ about the trivial case $(\id,0)$. Due to the semisimplicity of the Poincar\'e group, there is only one coupling constant for both gauge fields \cite{C76}. 
	\end{itemize}

\subsection{Comments on the literature}\label{origin}
Einstein orginally phrased General Relativity in the coordinate induced frame $dx^\mu$, and not in the one induced by the \textit{rep\`ere mobile} or \textit{vielbein} $e$ (\ref{gDefi}). The first interpretation of the gravitational interaction as a gauge group is due to Utiyama \cite{U54} in 1954, who only gauged the Lorentz subgroup of the Poincar\'e group: in order to obtain the dynamics of General Relativity he had to impose the symmetry of the Christoffel symbols by hand. Sciama and Kibble \cite{SK61} gauged the entire Poincar\'e group in 1961, which yields an Einstein-Cartan theory.\footnote{In 1976, Cho \cite{C76} claimed to have discovered an alternative approach to describe General Relativity as a gauge theory, by not identifying the translation operator with the derivative operator a priori and gauging the subgroup of translations alone. However, he requires the resulting theory to be independent of the particular coordinate frame chosen. This becomes clearer in Hehl's review \cite{GH96}, where it is referred to as ``soft Lorentz covariance''. In both cases, this is an additional constraint that is not provided by the gauge theory, as long as not the entire Poincar\'e group is gauged.} However, I am not aware of any proposal in the literature that elaborates a possibility to link the gauge fields $e$ and $\omega$ of the two subgroups by a (super)symmetry relation.

\section{Maximal supergravity}\label{sugraDefi}
\subsection{Quantization, higher dimensions and more supersymmetry}\label{QsugraDefi}
In section \ref{aspect}, I have argued that supersymmetry may be interesting for the quantization of the gravitational interaction in $d=4$ dimensions: as soon as one local coordinate system is fixed, there is a symmetry transformation {\bf in these coordinates} that generates torsion from the metric. Hence, it is possible that there is a symmetry action that maps curvature to torsion and vice versa in the fixed coordinate system. Global aspects of the manifold are by definition not important for the local physical theory which is why a restriction to parallizable manifolds is not compulsory. As for the discussion of quantum electrodynamics from section \ref{gauged2}, it is the experiment that will have to decide if this interpretation with non-trivial torsion degrees of freedom is relevant for a quantum theory of gravity or not.
	\\

In order to make contact to experiments, it looks promising to follow the analogy to quantum electrodynamics, i.e. to compute expectation values of a quantum theory of gravity by evaluating Feynman graphs. These consist of vertices and propagators. I want to focus on the propagators. In quantum electrodynamics, these correspond to the solutions in the zero coupling limit, which is the free field theory, described by the wave equation whose space of solutions is parametrized by the Fourier decomposition.\\

In the case of gravity, the free field theory is described by the Einstein equation, for which a parametrization of the space of solutions is problematic.\footnote{Of course, one could formally introduce an expansion $e=\delta +h +\mathcal{O}(h^2)$ and solve the wave equation for $h$. This is referred to as a gravitational wave. It would define a propagator for the linearized theory, and the infinitely many terms of higher polynomial order in $h$ could then be considered as interaction terms. This is the way quantum chromodynamics is handled, but there are only finitely many interaction terms in this case.} One reason for this is that these differential equations are weakly hyperbolic, which is related to the diffeomorphism invariance of the theory. \\

The problem one is confronted with is similar to the one Perelman solved in proving the Poincar\'e conjecture \cite{P02}. For this, Hamilton's concept of a Ricci flow was of crucial importance. I will not review it at this point, but I content myself with stating that this is a one parameter family of $d$-dimensional metrics fulfilling a weakly parabolic differential equation in $d+1$ dimensions. It is possible to establish a correspondence between $d$-dimensional diffeomorphisms and $(d+1)$-dimensional ones fulfilling a partial differential equation. This trick of a lift to a higher dimension allows to modify the dynamics to be strictly parabolic and hence solvable by introducing an additional scalar field $f$.\footnote{I am grateful to G.~Huisken for explaining the details.}\\

Although Perelman only discussed particular diffeomorphisms in $d=3$ and parabolic systems, the technical procedure could still be interesting for the present discussion. Perelman mentioned in \cite{P02} that the scalar field $f$ exactly is the dilaton used in string theories. From a gravitational point of view, this results from a reduction of a higher dimensional pure gravity theory on a torus \`a la Kaluza--Klein. Thus, the scalar field can be interpreted as a part of a metric in a higher dimensional space. I would like to suggest that this procedure of ``gauge fixing the diffeomorphism symmetry'' by adding more dimensions and thus more fields could also work for the Einstein equation in $d=4$, to transform them into strictly hyperbolic equations. This programme does of course not tell in which way the higher dimensional theory should be stated. \\

In this thesis, supersymmetry will serve as a guideline to construct a candidate for such a theory. It is well known that the maximal supergravity theory in $d=4$ has $N=8$ linearly independent supersymmetry variations. These link additional bosonic and fermionic degrees of freedom to the vielbein $e$ and the gravitino $\chi$ from section \ref{aspect}. That $N=8$ is indeed the maximal number of supersymmetries results from a perturbative argument: given a theory with more than $N=8$ supersymmetries in $d=4$, consider the linearized theory about Minkowksi background and constant transformation parameter $\ep$. In this setting, the supersymmetry algebra (\ref{SUSY0}) reduces to the Super-Poincar\'e algebra with well-defined, spin valued supercharges $Q$ that result from splitting the constant parameter $\ep$ from the supersymmetry variation $\delta =\bar{\ep} Q$ (\ref{vari1}) in complete analogy to extracting the generators $\hat{M}$ from the vector field in (\ref{PRep}). Hence, all fields of this theory build up a supermultiplet of this superalgebra and for $N>8$, the highest spin of the fields in this supermultiplet is greater than $2$. For this linearized theory, the argumentation from \cite{N77} is valid that rules out supergravity theories with $N>8$.\footnote{
I want to emphasize that this concept of a Super-Poincar\'e algebra fails as soon as the restriction to the linearized limit is lifted, which is obvious from equation (\ref{SUSY0}). Strictly speaking, supercharges $Q$ as operators mapping fields to fields are ill-defined in the non-linear domain of supergravity theories.
}\\

This is my personal motivation for studying $d=4$ $N=8$ supergravity and its possible relation to the gravitational interaction in $d=4$ dimensions. The additional fields of $N=8$ supergravity would hence be interpreted as some kind of Lagrangian multiplyers of the diffeomorphism symmetry and not as measurable excitations. In this context, it is also interesting to observe that $N=8$ supergravity is much better behaved in a perturbation expansion than was originally expected and may even be ultraviolet finite in perturbation theory \cite{BCFIJ08}. \\

I have suggested that adding more dimensions might help to improve the analytic behaviour of the Einstein equation in analogy to the discussion of the Ricci flow. The additional bosonic fields would follow from a higher dimensional vielbein by a reduction \`a la Kaluza--Klein. I will partly prove in chapter \ref{CHAP5} that this is indeed the case for $d=4$ $N=8$ supergravity: the bosonic fields can be combined in a $(d=60)$-dimensional vielbein, if the transition functions $\Diff(60)$ are restricted in a quite peculiar way. An intermediate step towards this goal is provided by $d=11$ $N=1$ supergravity \cite{CJS78} that I will discuss next.

\subsection{Supergravity in $d=11$}
A result of major importance for this thesis is due to Cremmer, Julia and Scherk \cite{CJS78}. They have shown that all solutions of $d=4$ $N=8$ supergravity are the solutions of $d=11$ $N=1$ supergravity with $7$ commuting Killing vectors. This is equivalent to stating that $d=4$ $N=8$ can be constructed from $d=11$ $N=1$ by a reduction \`a la Kaluza--Klein on a $7$-dimensional hypertorus $T^7$. This does not falsify my claim that it can also be derived from a $(d=60)$-dimensional theory, because $d=11$ supergravity has an additional bosonic degree of freedom apart from the vielbein $E$, a three-form potential $A$ which is only defined modulo the gauge transformation
\be\label{GaugeTrafo}
A'_{M_1\dots M_3}&=& A_{M_1\dots M_3} + 3\p_{[M_1}\lambda_{M_2M_3]}
\ee
with an arbitrary two-form $\lambda$. I am using the same letter $A$ as for the electromagnetic potential, because the gauge transformation is similar (\ref{u1trafo}) and as the equations of motion only depend on the corresponding field strength
\be\label{field11}
F_{M_1\dots M_4} &:=& 4\p_{[M_1}A_{M_2\dots M_4]}.
\ee
In contrast to the electromagnetic potential $A$, there is no obvious gauge group related to this transformation. In this thesis, I will follow West's argumentation \cite{W00} that this transformation should be considered on the same footing as the diffeomorphism symmetry of the vielbein. In chapter \ref{CHAP5}, I will prove that the supersymmetry variation is consistent with this interpretation for a $\mathbf{70}$-dimensional subsector of the bosonic degrees of freedom. This suggests that all the degrees of freedom of the vielbein in $d=11$ and of the three-form potential $A$ exactly parametrize a vielbein in $d=60$ dimensions in a restricted geometry, i.e. without any necessity to introduce new bosonic degrees of freedom. \\

The guideline how to restrict the sixty-dimensional geometry is provided by the hidden symmetries of $d=11$ $N=1$ supergravity. These are the global symmetries $E_{n(n)}$ of the theory, if the latter is reduced \`a la Kaluza--Klein on a hypertorus $T^n$ for $n=1,\dots,9$ \cite{J81,J83,N87}. For this thesis, I focus on $n=7$. After stating the action in the following section, I will briefly review the hidden symmetries in lower dimensions and their remnants in $d=11$, in particular the $Spin(3,1)\times SU(8)$-covariant formulation of $d=11$ supergravity \`a la de Wit--Nicolai \cite{dWN86}, which will serve as the starting point for the final result of this thesis, the $E_{7(7)}$-covariant supersymmetry variation presented in chapter \ref{CHAP5}.

\subsection{The $d=11$ supergravity action}\label{SUGRA1}
As far as the conventions are concerned, I follow Cremmer, Julia and Scherk \cite{CJS78}. The only difference to their original paper is that I use the signature $(-1,1,\dots,1)$ for the eleven-dimensional theory. This implies that I can choose a Majorana representation of the Clifford algebra (\ref{Clifford}) with real matrices $\tilde{\G}^A\in \R^{32\times 32}$. To simplify the notation, I have absorbed the gravitational constant $\kappa$ in the fields as suggested in \cite{CJ79}.\\

Since higher order terms in fermions will not be discussed in the sequel of this thesis, I refrain from stating them explicitly. In particular, the torsion terms that are quadratic in fermions will be neglected. Therefore, the covariant derivative $\nabla$ is the standard Levi--Civita connection. In the vielbein frame (\ref{ehneu}) the action hence has the form
\be\label{Action}
	S &=& \int\limits_{\cM^{11}}dx^{11}\det(E)\left(\frac{1}{4}\tilde{R} -\frac{i}{2}\bar{\psi}_B\tilde{\G}^{BCD}\nabla_C\psi_D - \frac{1}{48}F_{B_1\dots B_4}F^{B_1\dots B_4}\nn
	\right.\\
	&& -\frac{i}{96}\left( \bar{\psi}_{B_5}\tilde{\G}^{B_1\dots B_6}\psi_{B_6} +12\bar{\psi}^{B_1}\tilde{\G}^{B_2B_3}\psi^{B_4}\right)F_{B_1\dots B_4} \nn\\
	&&\left. +\frac{2}{12^4}\epsilon^{B_1\dots B_{11}}F_{B_1\dots B_4}F_{B_5\dots B_8}A_{B_9\dots B_{11}}\right).
\ee
The evaluation of the action of the Levi--Civita connection $\nabla$ (\ref{omegaK}) on the Gra\ss mann valued section $\psi$ of the tensor product of tangent and spin bundle follows the conventions from section \ref{prim2}.\footnote{I want to emphasize that $\nabla_{[C}\psi_{D]}$ differs from the ``covariant derivative'' $D_{[C}\psi_{D]}$ in \cite{CJS78} only by the torsion tensor that is of higher order in fermions and hence neglected. In contrast to the $(d=4)$-dimensional theory (\ref{RS}), I keep the more conventional form of the supergravity action with $\psi_M$ instead of $\psi_M=\frac{1}{2}(1-i)\chi_M$.} The Einstein--Hilbert term $\tilde{R}$ follows the definition (\ref{Riemann},\,\ref{Ricciscalar}) with the $11$-dimensional vielbein ${E_M}^A$. I have used the latter to transform the three-form potential $A$ and its corresponding field strength $F$ (\ref{field11}) into the vielbein frame. As before, I am using the convention to distinguish the different frames of the (co)tangent space by different names for the indices: $M,N,\ldots=0,\ldots,10$ correspond to the coordinate induced frame $dx^M$ and $A,B,\dots$ to the vielbein frame $dx^A$ (\ref{dualf}). Finally, the totally symmetric $\e$-tensor is normalized to $\e^{012345678910}=1$.\\

The independent physical fields of $d=11$ supergravity are the vielbein ${E_M}^A$, the gravitino $\psi_C$ with implicit spin indices and the three-form potential $A_{N_1\dots N_3}$. These are linked by the supersymmetry variations
\begin{subequations}\label{Trafo1}
\be
	\delta {E_M}^A &=& i\bar{\ep}\tilde{\G}^A\psi_M\\
	\delta \psi_M &=& \nabla_M\ep +\frac{1}{144}\left(\tilde{\G}{{}^{N_1\dots N_4}}_M-8\delta_M^{N_1}\tilde{\G}^{N_2\dots N_4}\right)\ep F_{N_1\dots N_4} \\
	\delta A_{N_1\dots N_3} &=& -\frac{3i}{2}\bar{\ep}\tilde{\G}_{[N_1N_2}\psi_{N_3]}.
\ee
\end{subequations}
The equations of motion of this theory follow from a variation of the action $S$ (\ref{Action}) in the standard way. For this thesis, only the equation of the three-form potential $A_{N_1\dots N_3}$ to zeroth order in fermions will be important:
\beg
\nabla_{B_1}F^{B_1\dots B_4} &=& -\frac{1}{24^2}\e^{B_2\dots B_4 A_1\dots A_8} F_{A_1\dots A_4}F_{A_5\dots A_8}.
\eeg
In a first order formalism, it is equivalent \cite{W00} to the two equations
\begin{subequations}\label{4FormG}
\be\label{4Form}
F^{B_1\dots B_4} &=& \frac{\alpha}{7!} \e^{B_1\dots B_4 C_1\dots C_7}F_{C_1\dots C_7}\\
F_{C_1\dots C_7} &=& 7\left(\nabla_{[C_1}A_{C_2\dots C_7]} + \frac{5}{\alpha}A_{[C_1\dots C_3}F_{C_4\dots C_7]}\right)
\label{7Form}
		\ee
		\end{subequations}
	with an arbitrary normalization constant $\alpha\in \R$ of the six-form potential $A_{C_1\dots C_6}$. It is standard to fix $\alpha=1$. The dual six-form potential $A_{C_1\dots C_6}$ obviously is only defined up to an arbitrary five-form $\lambda$ in complete analogy to the three-form potential $A_{C_1\dots C_3}$ (\ref{GaugeTrafo}). $A_{C_1\dots C_6}$ will be a crucial ingredient for the exceptional geometry of chapter \ref{CHAP5}.\\

Before reviewing hidden symmetries in the next section, I want to mention that $d=11$ supergravity is invariant under the gauge transformation (\ref{GaugeTrafo}) of the three-form potential. Since a gauge transformation of the action $S$ (\ref{Action}) results in a total derivative term, the equations of motion are invariant under this symmetry transformation, i.e. they only implicitly depend on $A$ through its gauge invariant field strength $F$ (\ref{field11}).

\subsection{Hidden symmetries}\label{hid1}
I have already mentioned that the guideline how to restrict the $(d=60)$-dimensional geometry is provided by the hidden symmetries of $d=11$ $N=1$ supergravity that are global internal symmetries. They result from a reduction \`a la Kaluza--Klein on a flat spacelike hypertorus $T^n$. I want to sketch briefly that this is equivalent to discussing the subspace of solutions of $d=11$ supergravity with $n$ independent commuting spacelike Killing vectors.\\

A reduction \`a la Kaluza--Klein on a flat spacelike hypertorus $T^n$ is based on the statement that $T^n$ endowed with the flat Euclidean metric $\eta$ can be embedded in the $11$-dimensional Lorentzian manifold $\cM^{11}$ 
\be\label{embed}
(T^n,\eta) &\hookrightarrow&  (\cM^{11},g)
\ee
using the same coordinate chart on both sides. Since I do not discuss non-local aspects in this thesis, it is sufficient to show that the tangent space $T_x\cM^{11}$ for every $x\in \cM^{11}$ allows for this embedding, because locally, the manifold is diffeomorphic to the tangent space. Next recall from the sections \ref{Subalg} and \ref{Gstructur} that the metric is flat if and only if the coordinate induced frame and the vielbein frame coincide and that this is the case if the coordinate induced basis vectors of the tangent space are independent Killing vectors. As coordinate induced basis vectors commute by definition, the presence of $n$ independent commuting spacelike Killing vectors restricts the possible metrics $g$ to the ones that allow for this embedding (\ref{embed}), which proves the equivalence.\qed\\

For a gravitational theory, the arbitrariness of the coordinate chart is encoded in the diffeomorphism symmetry. Having fixed a coordinate chart, the internal symmetry $\Diff(11)$ links one solution of the theory to another one. Given one solution that fulfills the constraint (\ref{embed}), it is obvious that a general diffeomorphism would not map it to another one that fulfills (\ref{embed}). Therefore, only a subgroup of $\Diff(11)$ is a symmetry of the space of solutions of $d=11$ supergravity with $n$ commuting Killing vectors.\\
 
The interpretation with $n$ commuting Killing vectors shows that the group of general coordinate transformations is restricted to $\Diff(11-n)\times A(n)$: in order to preserve the property of commutativity of Killing vector fields, the Jacobian matrix $\frac{\p x'}{\p x}$ of the diffeomorphism $x'=\vp_A(x)$ must be constant. And I have shown in the sections \ref{Diffeom} and \ref{Subalg} that this is the characterizing property of the affine group $A(n)$ (\ref{affine2}).\\

The symmetry group $\Diff(11-n)\times A(n)$ hence is the internal symmetry of the space of solutions of $d=11$ supergravity with $n$ independent commuting Killing spinors. Performing a Fourier expansion in the coordinates of the hypertorus $T^n\subset \cM^{11}$ and neglecting all non-trivial modes is known as a compactification on $T^n$. This procedure transforms the external symmetry group $\Diff(11-n)\times A(n)$ of an $11$-dimensional theory into an external part $\Diff(11-n)$ and a global internal part $A(n)$ of an $(11-n)$-dimensional theory. In accordance with the definitions of global internal symmetries from section \ref{Symm1}, $A(n)$ maps solutions to solutions without affecting the coordinates. It is however obvious from the constraint (\ref{embed}) that not all affine linear diffeomorphisms $A(n)$ map a given solution to a different one, because the isometry subgroup $SO(n)\ltimes \cP_n$ of $\eta$ will not affect a given solution. This implies that the $SO(10-n,1)$ scalars of the $(d=11-n)$-dimensional theory resulting from the eleven-dimensional metric, are parametrized by the coset 
\beg
A(n)/(SO(n)\ltimes \cP_n) &=& Gl(n)/SO(n).
\eeg

For $d=11$ supergravity, this global symmetry $Gl(n)$ is enlarged by the degrees of freedom of the three-form potential $A$ to $E_{n(n)}$.\footnote{For $n\geq 6$, the $(11-n)$-dimensional $\e$ tensor is used to combine the bosonic degrees of freedom in the appropriate representation of $E_{n(n)}$. The fermions transform as representations of the universal cover of the maximal compact subgroup $K(E_{n(n)})$ that I have introduced in the sections \ref{realf} and \ref{Representation}.} In the following table, I have listed the global symmetry groups $E_{n(n)}$, their maximal compact subgroups and the dimension of the cosets parametrizing the $SO(10-n,1)$ scalars of the $(d=11-n)$-dimensional theories \cite{J81,J83,N87}.\footnote{Note in particular that $\textit{K}(\textit{E}_{8(8)})=\Spin(16)/\Z_2$ is not diffeomorphic to $SO(16)$. Further information on these topics can be found in \cite{C07,NW89,N91,Pope,dW02}.}
 \vspace{-0.4cm}
\begin{center}
\scalebox{0.95}{
\begin{tabular}{c}
Reduction of  $D=11$ $N=1$ supergravity on $T^n$\\
\begin{tabular}{|c|c|c|c|}
\hline
 $n$ & Hidden symmetry $\textit{E}_{n(n)}$ & Compact subgroup $\textit{K}(\textit{E}_{n(n)})$& Dimension\\
\hline
 $1$
&
 $\textit{Gl}(1)$ &$\textit{SO}(1)$ &$1-0=1$
\\
 $2$
&
 $\textit{Sl}(2)\times \R$
 &$\textit{SO}(2)$&$4-1=3$
\\
 $3$&
 $\textit{Sl}(3)\times \textit{Sl}(2)$
 &$\textit{SO}(3)\times \textit{SO}(2)$
 &$11-4=7$
\\
 $4$&
 $\textit{Sl}(5)$
 &$\textit{SO}(5)$
 &$24-10=14$
\\
 $5$&
 $\Spin(5,5)$
 & $(\Spin(5)\times \Spin(5))/\Z_2$
 &$45-20=25$
\\
 $6$&
 $\textit{E}_{6(6)}$
 &$\textit{USp}(8)/\Z_2$
 &$78-36=42$
\\
 $7$&
 $\textit{E}_{7(7)}$
 &$\textit{SU}(8)/\Z_2$
 &$133-63=70$
\\
 $8$&
 $\textit{E}_{8(8)}$
 &$\Spin(16)/\Z_2$
 &$248-120=128$
\\
$9$&
 $\textit{E}_{9(9)}$
 &$\textit{K}(\textit{E}_{9(9)})$
 &$\infty-\infty=\infty$
\\\hline
\end{tabular}
\end{tabular}
}
\end{center}

I want to conclude this review with the remark that searching for higher dimensional supergravity theories that give rise to a coset model for a given semi-simple group $G$ is called group-disintegration or oxidation, because it is the inverse process to reduction \cite{CJLP99,J81,K02}. $d=11$ supergravity is commonly referred to as the ``oxidation end point'' of $N=8$ $D=4$ supergravity which is due to the following argument. If the diffeomorphism symmetry $\Diff(d)$ is not restricted, the parameter $\ep$ of the supersymmetry variation transforms as a spin representation of $Spin(d-1,1)$. For maximal supergravity with $N=8$ real, four component spinors in $d=4$ this leads to the restriction $d\leq 11$, because for any higher $d$, the dimension of the irreducible spin representation would be greater than $32$. However, this restriction can be circumvented if the diffeomorphism symmetry $\Diff(d)$ is restricted in such a way that the induced action is only by a proper subgroup of $Spin(d-1,1)$. Chapter \ref{CHAP5} will provide an example with $Spin(3,1)\times SU(8)\subset Spin(59,1)$. 

\subsection{Hidden symmetries in $d=11$}\label{Hid2}
It is still an up to date question in string theory to look for consistent compactifications of $d=11$ supergravity to $d=4$ physical dimensions. However, not on all manifolds with the correct dimensions, compactifications are admissible \cite{Pope}. Hence, $d=11$ supergravity must possess some information about preferred manifolds on which a compactification is possible. In proving the consistency of a compactification on the sphere $S^7$, de Wit and Nicolai showed that there is indeed some remnant of a hidden symmetry in the non-compactified theory:
\begin{center}
 \textit{$d=11$ supergravity has a local $Spin(3,1)\times SU(8)$-symmetry \cite{dWN86}.}
\end{center}

Guided by the $SO(8)$ gauged supergravity \cite{CJ79}, their ansatz reduced the local Lorentz symmetry $SO(10,1)$ to $SO(3,1)\times SO(7)$. Then, in order to discuss a linear problem, they focussed on the supersymmetry variations. They were able to write these in a $Spin(3,1)\times SU(8)$-covariant form in $d=11$ by combining the degrees of freedom of the vielbein $E$ and the three-form potential $A$ into $Spin(3,1)\times SU(8)$ representations with an even number of $SU(8)$ indices. Finally, they proved that this covariance can be extended to the equations of motion of $d=11$ supergravity.\footnote{There is no similar statement for $Spin(2,1)\times Spin(16)$ up to now, because Nicolai did not show the final step in \cite{N87b}, which is the covariance of the equations of motion under $Spin(2,1)\times Spin(16)$. There is no similar result for $Spin(4,1)\times USp(8)$ either, whereas $Spin(3,1)\times USp(8)$-covariance trivially follows from $Spin(3,1)\times SU(8)$ \cite{dW02}.}\\ 

The $Spin(3,1)\times SU(8)$-covariance serves as the starting point for my investigation of a $60$-dimensional exceptional geometry in chapter \ref{CHAP5}. Focussing on the bosons, the covariance group of the equations of motion of $d=11$ supergravity is $SU(8)/\Z_2$, which I have shown in section \ref{Representation} to be the maximal compact subgroup of $E_{7(7)}$. This nourishes the idea that a theory that is invariant under an $E_{7(7)}$ action on the coordinates may have the same field content as $d=11$ supergravity. Representation theory then fixes 
\beg
d=4+56=60
\eeg
as the lowest possible dimension in which such a theory can be defined. In chapter \ref{CHAP5}, I will explicitly show that for a $70$-dimensional subsector of $d=11$ supergravity, the supersymmetry variations (\ref{Trafo1}) can be consistently lifted to relations in $60$ dimensions, whose reduction to $11$ dimensions does not introduce additional degrees of freedom and that are strongly constrained by requiring an external $E_{7(7)}$-symmetry. In constructing this theory, I will use the concept of non-linear realizations, which will be the topic of the next chapter.

\chapter{Non-linear realizations}\label{CHAP4}
The aim of this thesis is to construct a sixty-dimensional exceptional geometry whose $60$-dimensional vielbein exactly comprises the bosonic fields of $N=8$ $d=4$ supergravity. The exceptional geometry corresponds to the restriction of the symmetry group $\Diff(60)$ to a subgroup that respects the fact that a $56$-dimensional subvielbein $e^H$ of the $60$-dimensional vielbein is an $E_{7(7)}$ matrix. The main result of chapter \ref{CHAP5} will be that it is possible to construct $E_{7(7)}$-covariant supersymmetry variations that are the ones of $d=11$ supergravity (\ref{Trafo1}) after a compactification of $49$ dimensions without adding further bosonic or fermionic fields, if the bosonic degrees of freedom are restricted to the ones described by the subvielbein $e^H$ and if only the corresponding $56$ dimensions are taken into account.\\

The crucial ingredient for this to be consistent at all is the restriction of the group of general coordinate transformations $\Diff(d)$ to a subgroup. An adequate description of this procedure is provided by the theory of non-linear realizations, which I briefly want to motivate.\\

I have shown in chapter \ref{CHAP3} that the connection $\nabla$ is the basic building block of a gravitational theory. In particular, $\nabla$ is sufficient to construct the Lagrangian $\mathcal{L}_{\text{EH}}$ (\ref{EHKosm}) of general relativity. A result of section \ref{gravmatter2} was that considering the vielbein $e$ as a gauge field implies that the Lagrangian may only depend on it through gauge covariant Lorentz tensors like torsion or curvature (\ref{TRDefi}). A corollary to this statement is that there does not exist a tensor that is of first order in derivatives and that only depends on the vielbein degrees of freedom, if full $\Diff(d)$-invariance for a theory is required. This is in particular true for a torsion tensor. Hence, a connection $\nabla$ with $\Diff(d)$-equivariance is uniquely determined to be the Levi--Civita connection (\ref{omegaK}) modulo a torsion tensor that cannot be linear in derivatives, if it exclusively depends on vielbein degrees of freedom - a statement that I will prove explicitly in this chapter.\\

Keeping in mind the necessity for the exceptional geometry to restrict the diffeomorphism group $\Diff(60)$ to a subgroup, this chapter will deal with the implications of such a restriction to this uniqueness result for the connection $\nabla$.\\ 

I will start by constructing the general connection $\nabla$ with equivariance under the affine subgroup $A(d)$ (\ref{affine2}) of the diffeomorphism group $\Diff(d)$. In section \ref{secC}, I will prove that requiring simultaneous equivariance under the abelian group $K(d-1,1)$ of conformal diffeomorphisms (\ref{SCTcoord4}) restores the uniqueness result for the connection $\nabla$. This has become known as the procedure of Borisov \& Ogievetsky \cite{BO74}.\\

A guideline for the proof is provided by Ogievetsky's theorem that I have quoted in section \ref{Ogievetsky}: any analytic vector field $X\in\mathfrak{diff}_d$ is in the closure of the vector space generated by the conformal Killing vector field of Minkowski space $X^{\mathfrak{c}}_{a}$ (\ref{KGen}), the affine vector field $X^{\mathfrak{a}}_{(A,c)}$ (\ref{GLGen}) and multiple commutators thereof. \\

I will conclude this chapter with highlighting the differences of my presentation to the orginal one by Borisov and Ogievetsky \cite{BO74} and to related work in the literature \cite{ISS71,K05,SS69}. I will also explain why I do not share their original motivation to consider the gravitational interaction on the same footing as internal symmetries in quantum field theory, to quote Ogievetsky \cite{O73}:
\smallskip\\
 \textit{``In deep analogy to the fact that pions are connected with non-linear realizations of the dynamical chiral $SU(2)\times SU(2)$ symmetry (see, e.g. \cite{W70}) gravity field proves to be connected with common non-linear realizations of the dynamical conformal and affine symmetries \cite{ISS71}.''}

\section{Connections with  affine linear equivariance}\label{clc}
In section \ref{Subalg}, I have defined the Poincar\'e group by the subgroup of $\Diff(d)$ that preserves the metric tensor $\eta$ (\ref{Isometry1}). This is the coordinate invariant, \textbf{geometric approach} to describe subgroups of diffeomorphisms which will be used to construct the exceptional geometry in chapter \ref{CHAP5}.\\

Since I am only interested in local properties of a theory in this thesis, it was sufficient to focus on an arbitrary simply connected open subset $U_\alpha\subset \cM^d$ (\ref{charts}) of the manifold. This allowed to fix a coordinate chart $x_\alpha$ and basis vectors $\hat{P}_\mu$ (\ref{PDefi}) in section \ref{diffi}. In this setting, the group of general coordinate transformations was further restricted to the analytic diffeomorphisms $\Diff(d)$ that form a Lie group. Its corresponding Lie algebra is the space of vector fields $\mathfrak{diff}_d$ (\ref{Vectorfields}). A subgroup of diffeomorphisms can hence also be defined by a subalgebra that is constructed from particular vector fields. It is this \textbf{algebraic approach} that I have used to define the affine linear diffeomorphisms in section \ref{Subalg} and that will be used in this chapter.\footnote{Due to the Lie property, every subgroup of $\Diff(d)$ can be defined by the algebraic approach, which is not obvious for the geometric approach.}

\subsection{Setup}\label{secA3}
The affine group $A(d)$ consists of the diffeomorphisms $\vp^{\mathfrak{a}}_{(A,c)}\in \Diff(d)$ that correspond to the affine linear vector field $X^{\mathfrak{a}}_{(A,c)}$ (\ref{GLGen}) by the exponential map (\ref{GLMatrix})
\be\label{GLMatrix2}
\big(\vp^{\mathfrak{a}}_{(A,c)}(x)\big)^\mu  &=&{\big(e^A\big)_\nu}^\mu {\big.x}^\nu +{\big.c}^\mu.
\ee
Having fixed a coordinate chart, the matrix $e^A$ has the same form for all points $x$ in the open set $U_\alpha\subset \cM^d$.\footnote{Changing the coordinate chart alters this property of course, but for any coordinate chart there exist diffeomorphisms in $\Diff(d)$ that have this form.} This implies that the Jacobian matrix
\be\label{lindefi}
\left.\frac{\p }{\p x^\mu}\left(\vp^{\mathfrak{a}}_{(A,c)}\right)^\nu
	\right|_{x}
  &=&{\left(e^{A}\right)_\mu}^\nu 
\ee
is constant for all $x$ in the domain of validity of the coordinate chart $U_\alpha\subset \cM^d$. Hence, the induced action of an affine linear diffeomorphism $x'=\vp^{\mathfrak{a}}_{(A,c)}(x)$ on the vielbein $e$ has the form (\ref{VielbeinTrafo})
\be\label{eTrafo}
	{{\big.e'}_\mu}^a &= & {\big(e^{-A}\big)_\mu}^\nu \,  {{\big.e}_\nu}^b \,{O\big(\vp^{\mathfrak{a}}_{(A,c)},e\big)_b}^a.
\ee
The matrix $O(\vp^{\mathfrak{a}}_{(A,c)},e)\in {SO}(d-1,1)$ is the compensating local Lorentz rotation that is needed to restore the arbitrary gauge fixing chosen for the vielbein $e$. Following the discussion from the sections \ref{gravmatter2} and \ref{ART}, the induced action of a diffeomorphism $\vp^{\mathfrak{a}}_{(A,c)}$ on the coefficients $\psi^\gamma$ (\ref{psiParam}) of physical fields $\psi$ is provided by (\ref{psiTraf})
\be\label{psiTrafo9}
	{\psi'}^\gamma &=&{O_\beta}^\gamma(\vp^{\mathfrak{a}}_{(A,c)},e)\psi^\beta\\
&=&\psi^\gamma + {o_b}^c(\vp^{\mathfrak{a}}_{(A,c)},e) \delta_{\hat{L}{}^b{}_c}\psi^\gamma +\mathcal{O}(o^2)\nn.	
\ee
I want to emphasize that the action on a fermion $\psi$ follows the argumentation of section \ref{Representation} and is hence well-defined, because all physical expectation values are of even degree in fermions.
\\

For the transformation of the vielbein (\ref{eTrafo}), I want to introduce the matrix notation that has already been suggested by the matrix commutator (\ref{GLKom})
\be\label{eTrafo2}
	e' &= & \left(e^{-A}\right) \cdot  e \cdot O(\vp^{\mathfrak{a}}_{(A,c)},e).
\ee
In general, the vielbein $e$ is an $x$-dependent $Gl(d)$ matrix, whereas $e^{-A}$ is an $x$-independent $Gl(d)$ matrix and $O(\vp^{\mathfrak{a}}_{(A,c)},e)$ is an $x$-dependent $SO(d-1,1)$ matrix. $O$ is needed to restore the arbitrary gauge fixing chosen for the vielbein. Therefore, the transformation $e\mapsto e'$ would correspond to a global left action with $e^{-A}\in GL(d)$ on an $x$-dependent coset $e\in Gl(d)/SO(d-1,1)$, if and only if the transformation $x\mapsto x'$ was also induced by this global left action. I will show in the next section that it is an affine coset $A(d)/SO(d-1,1)$ that exactly reproduces the (induced) action of an affine linear diffeomorphism on the coordinates $x^\mu$ (\ref{GLMatrix2}), the field coefficients $\psi^\gamma$ (\ref{psiTrafo9}) and the vielbein matrix $e$ (\ref{eTrafo2}).

\subsection{The affine coset}\label{secB}
To sum up the discussion at the end of the preceding section, the global left action on the coset must affect the coordinates. Hence, I have to extend the coset $Gl(d)/SO(d-1,1)$ with the coordinates in such a way that a global left action also changes them appropriately. The observation that coordinates $x^\mu$ and one-forms $dx^\mu$ transform with the same constant matrix $e^{A}$ under affine linear diffeomorphisms $\vp^{\mathfrak{a}}_{(A,c)}$ leads to the following ansatz for the coset
\be\label{paramCoset}
C&=& e^{x^\mu \hat{P}_\mu} e(x)\in A(d)/SO(d-1,1).
\ee
The exponentiation is the homomorphism $\exp$ (\ref{exp}) that links the affine group $A(d)$ (\ref{affine2}) to the affine algebra $\mathfrak{a}_d$ (\ref{affine1}). The coordinates $x^\mu$ are contracted with the generators of translation $\hat{P}_\mu$ (\ref{PRep}) that belong to the algebra $\mathfrak{a}_d$. Next, I will show that a global left action with the affine group element
\be\label{affine}
e^{c^\mu \hat{P}_\mu}e^{-A}\in A(d)
\ee
on the coset $C$ (\ref{paramCoset}) exactly reproduces the action of an affine linear diffeomorphism $\vp^{\mathfrak{a}}_{(A,c)}$ on the coordinates $x^\mu$ (\ref{GLMatrix2}) and on the vielbein $e$ (\ref{eTrafo2}).\\

In complete analogy to the Lorentz gauge fixing of the vielbein $e$ from section \ref{ART}, I want to fix a representative of the coset $Gl(d)/SO(d-1,1)$ by requiring that the matrix form of $e(x)$ (\ref{paramCoset}) be preserved under a global left action. This induces a right action on $C$ by a Lorentz group element. Thus, the complete transformation of the coset (\ref{paramCoset}) under the affine linear action (\ref{affine}) takes the form
\be\label{gltrafo}
C'&=&e^{c^\mu \hat{P}_\mu}e^{-A} \cdot C\cdot O(\vp^{\mathfrak{a}}_{(A,c)},e).
\ee
To evaluate the action of the affine group element on the coordinates, the following standard formula \cite{K95} is necessary
\be\label{formula2}
e^{X}Ye^{-X}=\exp\circ \ad_{X} (Y),
\ee
which is valid for any two Lie algebra elements $X,Y\in \Lie$. $\exp$ is the standard exponential series and $\ad$ is the adjoint action familiar from the Definition \ref{defiAd}. Using the same conventions for matrix products as in (\ref{GLMatrix}), I obtain the identity
\be\label{gltrafo3}
 e^{-A} e^{x^\tau \hat{P}_\tau} e^{A} &=&   e^{x^\tau e^{-A} \hat{P}_\tau  e^{A}}
 \nn\\
  &\stackrel{(\ref{formula2})}{=}& e^{x^\tau \exp\circ \ad_{-A} (\hat{P}_\tau)}
  \nn\\
  &\stackrel{(\ref{gldefi})}{=}& e^{x^\tau \sum\limits_{n=0}^\infty \frac{1}{n!} \left(-  {A_\nu}^\mu \mathbf{ad}_{{\left.\hat{M}\right.^\nu}_\mu}\right)^n (\hat{P}_\tau)}
  \nn\\
  &
  \stackrel{(\ref{ComRel2})}{=}& e^{x^\tau \sum\limits_{n=0}^\infty \frac{1}{n!} {(A^n)_\tau}^\nu  \hat{P}_\nu}\nn\\
  &=& e^{x^\tau {(e^{A})_\tau}^\nu  \hat{P}_\nu}.
\ee
This allows for a comparison of the transformation (\ref{gltrafo}) with the coset
\beg
C'&=& e^{{x'}^\mu \hat{P}_\mu} e'(x')
\eeg
and thus determines the transformations of the two parameters $x^\mu$ and $e(x)$ of the coset under the global left action by a general $A(d)$ element (\ref{affine}):
\begin{subequations}\label{gltrafo2}
\be
e'(x') & = & e^{-A}\cdot e(x)\cdot O(\vp^{\mathfrak{a}}_{(A,c)},e),
\\
		 {\big.x'}^\mu &=&{\big(e^{A}\big)_\nu}^\mu {\big.x}^\nu +{\big.c}^\mu.
\ee
\end{subequations}
These exactly are the equations (\ref{eTrafo2}) and (\ref{GLMatrix2}), which justifies the notation $O(\vp^{\mathfrak{a}}_{(A,c)},e)$ for the compensating right Lorentz action on the coset a posteriori.\\

The difference between the coset and the geometric point of view is the origin of the transformation. For the latter, a diffeomorphism action on the coordinates $x^\mu$ induces the action on the vielbein $e$, whereas the reverse logic applies to the coset picture: the action on the $Gl(d)$ matrix $e(x)$ induces the transformation of the coordinates $x^\mu$, because these form a $Gl(d)$ representation.\\

Where an induced action on physical fields $\psi$ is concerned, the same line of argumentation applies to both systems, because the action on $\psi$ is provided by the compensating Lorentz rotation (\ref{psiTrafo9}) that is common for both transformations.\footnote{In particular, the induced action of a global left $A(d)$ transformation on a fermion is well-defined, if the argumentation from section \ref{Representation} is taken into account that all physical expectation values are of even fermionic degree.}\\

The transformations induced by a general $A(d)$ action (\ref{affine}) on the coset $C$ (\ref{paramCoset}) and by an affine linear diffeomorphism $\vp^{\mathfrak{a}}_{(A,c)}$ (\ref{GLMatrix2}) completely coincide
\begin{enumerate}
	\item on the coordinates $x^\mu$,
	\item on the vielbein matrix $e(x)$ and
	\item on an arbitrary physical field $\psi$.
\end{enumerate}
Together with a glance at the definition of a connection $\nabla$ acting on phyical fields in section \ref{equivar}, this fact implies that any $\nabla$ that is built from the affine coset $C$ (\ref{paramCoset}) is an example for a connection that is equivariant under the affine subgroup $A(d)$ of $\Diff(d)$. This observation will be the most important ingredient for the construction of the general connection $\nabla$ with affine linear equivariance in the following two sections.

\subsection{The minimal connection of the affine coset}\label{ConnAff}
In the theory of non-linear realizations \cite{ISS71}, there is a natural candidate for a connection, which I am going to refer to as minimal connection $\nabla^{\text{min}}$. Given any non-linear $\sigma$-model built on the coset $C$ of Lie groups $G/H$, the connection $\nabla^{\text{min}}$ acting on $H$ representations $\psi$ is characterized by its $\Car$ action $\delta_{\hat{\omega}}$ (\ref{Conn56}). The algebra element $\hat{\omega}=\hat{v}^{\text{min}}\in \Car$ is defined as the projection of the Maurer--Cartan form $C^{-1}d C$ on the Lie algebra $\Car$ that corresponds to the Lie group $H$.\\

I will perform the construction of $\nabla^{\text{min}}$ for the affine coset $C$ (\ref{paramCoset}). With the formula for any algebra element $X\in \Lie$
\be\label{formula}
e^{-X}d e^X = \sum\limits_{j=0}^\infty \frac{(-1)^j}{(j+1)!}\ad_X^j dX,
\ee
the Maurer-Cartan form splits into two components 
\be\label{CdC}
C^{-1}d C &=& e(x)^{-1} \left(e^{-x^\mu \hat{P}_\mu} d\,e^{-x^\mu \hat{P}_\mu}\right) e(x) + e(x)^{-1} d\, e(x)\nn\\
&\stackrel{(\ref{ComRel3})}{=}&dx^\mu e(x)^{-1} \hat{P}_\mu e(x) + e(x)^{-1} d\, e(x)
\ee
Keeping in mind from section \ref{Gstructur} that all eigenvalues of the vielbein matrix are positive, it is consistent to associate to the $Gl(d)$ matrix $e(x)$ a corresponding Lie algebra element
\be\label{eMf}
e(x)&=:&e^{{{h}_\mu}^\nu {\hat{M}{}^\mu}_\nu}
\ee
in the basis spanned by the generators $\hat{M}$ (\ref{gldefi}) of the $\mathfrak{gl}_d$ algebra. It is important to note that the names of the indices are completely arbitary at this point. Only for the geometric picture, I defined the abbreviation that Greek indices corresponded to the coordinate induced frame $dx^\mu$ of the cotangent bundle whereas Latin indices were associated with the vielbein frame $dx^a$ (\ref{dualf}). If a different distinction for these basis vectors is used, the vielbein will simply be a matrix with arbitary names of the indices as in the present case.\\

The definition (\ref{eMf}) allows for an evaluation of both components of the Maurer--Cartan form (\ref{CdC}): 
\be\label{CdC2}
C^{-1}d C &=&dx^\mu {\left(e^{h}\right)_\mu}^\nu \hat{P}_\nu  + {\left(e^{-h}\right)_\mu}^\nu d{\left(e^{h}\right)_\nu}^\sigma {\left.\hat{M}\right.^\mu}_\sigma.
\ee
The first component follows the same calculation used in (\ref{gltrafo3}): it is an adjoint action of a $Gl(d)$ matrix on the Lie algebra generator $\hat{P}$ with $e^{A}$ being replaced by $e(x)$. The second components in (\ref{CdC}) and (\ref{CdC2}) coincide, because an action on an arbitrary $\mathfrak{gl}_d$ representation renders the same result.\\

In order to match the standard conventions in general relativity, I replace the top right Greek index of the matrix $e^h$ by a Latin one. I emphasize again that this is a mere relabeling from the coset point of view. I further adopt the convention from section \ref{gravmatter2} to distinguish the vielbein matrix from its inverse by a different position of the indices
\begin{subequations}\label{ehklein}
\be
{e_{\mu}}^a&=&{\left(e^h\right)_\mu}^a,\\
{e_{a}}^\mu&=&{\left(e^{-h}\right)_a}^\mu.
\ee
\end{subequations}
With the abbreviation $\p_a := {e_{a}}^\mu\p_\mu$ (\ref{ehneu}), the Maurer-Cartan form (\ref{CdC2}) takes the form
\beg
C^{-1}d C &=&dx^\mu {e_{\mu}}^a \left(\hat{P}_a  + {e_c}^\nu \p_a {e_\nu}^d \hat{M}{}^c{}_d\right).
\eeg
For the construction of the minimal connection $\nabla^{\text{min}}$, I have to split the $\Car=\mathfrak{so}_{(d-1,1)}$ generators $\hat{L}$ from the ones of $\mathfrak{gl}_{d}$. To do this, I define the symmetric generators $\hat{S}$ in analogy to the antisymmetric generators $\hat{L}$ (\ref{LDefi})\footnote{The generators $\hat{L}$ and $\hat{S}$ are always defined with respect to the flat metric $\eta$, whatever the names of the indices may be. I will always mention it explicitly, if different names of indices distinguish objects such as in the case of $dx^a$ (\ref{Leftinv}).}
\be\label{RDefi}
{\left.\hat{S}\right.^c}_d &:=& \frac{1}{2}\left({\left.\hat{M}\right.^c}_d +\eta_{df}{\left.\hat{M}\right.^f}_e \eta^{ce}\right).
\ee
Using the vielbein frame (\ref{dualf}) for the one-form indices
\be\label{Leftinv}
dx^a &=& {e_\mu}^a(x) dx^\mu,
\ee
I obtain for the Maurer-Cartan form
\be\label{MC}
	C^{-1}\p_a C =   \hat{P}_a + {\left({v_a}\right)_c}^d {\left.\hat{S}\right.^c}_d + 	{\left({v^{\text{min}}_a}\right)_c}^d {\left.\hat{L}\right.^c}_d.
\ee
The coefficients ${v}$ and ${v^{\text{min}}}$ are the same objects (\ref{abbrev1}) that I have already defined in the context of general relativity in section \ref{convdiff}
\begin{subequations}\label{omegaLin}
\be
	{\left(v_a\right)_c}^d &=& \left({e_g}^\nu\p_a {e_\nu}^{(d}\right)\eta^{f)g}\eta_{cf}
		\label{omega2},\\
			{\left(v^{\text{min}}_a\right)_c}^d &=& \left({e_g}^\nu\p_a {e_\nu}^{[d}\right)\eta^{f]g}\eta_{cf}
			\label{omega3}.
	\ee
\end{subequations}
Given any physical field $\psi$ in a $\Car=\mathfrak{so}_{(d-1,1)}$ representation, the minimal connection $\nabla^{\text{min}}$ acts on the coefficients $\psi^\gamma$ with respect to a basis $\hat{T}_\gamma$ of the representation space $V$ by (\ref{Conn56}) (in the vielbein frame $dx^a$ (\ref{dualf}))
\be\label{minimalCon}
	\nabla_a^{\text{min}}\psi^\gamma &=&\p_a\psi^\gamma - {\left(v^{\text{min}}_a\right)_c}^d \delta_{{\left.\hat{L}\right.^c}_d}\psi^\gamma.
	\ee
It is straightforward to show that this connection $\nabla_a^{\text{min}}$ indeed is equivariant under a global affine left action (\ref{affine}). The fact that physical fields $\psi$ transform as $\mathfrak{so}_{(d-1,1)}$ representations is the reason why I have chosen the vielbein frame (\ref{Leftinv}) for the one-form indices. Only in this frame, the coefficient $\nabla_a^{\text{min}}\psi^\gamma$ of the derivative field $\nabla\psi$ transforms as a Lorentz tensor with respect to the action that is induced by the affine action (\ref{affine}) on the coset $C$ (\ref{paramCoset}).
\\

To conclude, I want to summarize the two reasons why I had to include the coordinates $x^\mu\hat{P}_\mu$ in the coset (\ref{paramCoset}).
\begin{enumerate}
	\item One reason is to establish the complete equivalence of the coset transformation to the one of the coordinates $x^\mu$ under an affine linear diffeomorphisms and its induced action on the vielbein $e$. Without an inclusion of $x^\mu \hat{P}_\mu$ in the coset, the $x$-dependence of the vielbein would not be reproduced correctly: schematically, a global left action would map $e(x)$ to $e'(x)$ and not to $e'(x')$.
	\item The generators $\hat{P}_\mu$ provide the vielbein frame for the coset (\ref{Leftinv}), in which it is natural to denote the connection (\ref{minimalCon}). 
\end{enumerate}

\subsection{Connections with affine linear equivariance}\label{ConnAff2}
In this section, I will determine the possible dependences of connections $\nabla$ with affine linear equivariance on the vielbein $e$. At the end of section \ref{secB}, I have already argued that every connection that is built from the degrees of freedom of the affine coset $C$ (\ref{paramCoset}) provides an admissible choice. This is in particular true for the minimal connection $\nabla^{\text{min}}$ (\ref{minimalCon}). The statement that two connections may only differ by a tensor establishes the link between a general connection $\nabla$ with affine linear equivariance and $\nabla^{\text{min}}$:

{\prop\label{propo3}
Let $\psi$ be a physical field in a representation of $\mathfrak{so}_{(d-1,1)}$ on which a symmetry may act. Then any two connections $\nabla$ and $\tilde{\nabla}$ on $\psi$ that are equivariant under this symmetry transformation may only differ by a Lorentz tensor $X$ that is covariant with respect to the symmetry.
}
{
\pf
The proof starts by writing twice the definition of a connection on the coefficients $\psi^\gamma$ of a physical field following from section \ref{prim2}
\beg
\nabla \psi^\gamma &=& d\psi^\gamma - {\left.\omega\right._c}^d\delta_{{\left.\hat{L}\right.^c}_d}\psi^\gamma
\\
\tilde{\nabla} \psi^\gamma &=& d\psi^\gamma - {\left.\tilde{\omega}\right._c}^d\delta_{{\left.\hat{L}\right.^c}_d}\psi^\gamma.
\eeg
To simplify the notation, I have not fixed a basis of the cotangent space and included the one-forms $dx^\mu$ in the objects instead. To complete the proof, I have to show that the one-form valued object 
\be\label{XDefi2}
{X_c}^d&:=&{\left.\omega\right._c}^d - {\left.\tilde{\omega}\right._c}^d
\ee
is covariant. In other words, $X$ has to transform as a Lorentz tensor under the action induced by the symmetry. This is true, because the definition of equivariance under a symmetry transformation from section \ref{SymmAct} implies that $\nabla\psi$ is covariant with respect to it. Therefore, the difference
\beg
{\left.X\right._c}^d\delta_{{\left.\hat{L}\right.^c}_d}\psi^\gamma 
&=&
 \tilde{\nabla}\psi - \nabla\psi
\eeg
also is covariant and $X$ transforms as a Lorentz tensor under this symmetry transformation.\qed
}\\

The symmetry under consideration is the restriction of $\Diff(d)$ to its affine subgroup $A(d)$ (\ref{affine2}). Thus, any connection $\nabla$ with affine linear equivariance may only differ from the minimal connection $\nabla^{\text{min}}$ (\ref{minimalCon}) by a Lorentz tensor $X$. Using the frame (\ref{Leftinv}), I obtain the identity
\be\label{final1}
		\nabla_a \psi^\gamma &=& 
		\p_a\psi^\gamma 
		-\left({\left(v^{\text{min}}_a\right)_c}^d + {X_{ac}}^d \right)\delta_{{\left.\hat{L}\right.^c}_d}\psi^\gamma.
\ee
The class of all connections with affine linear equivariance is hence parametrized by all possible Lorentz tensors $X$ under affine linear diffeomorphisms $\vp^{\mathfrak{a}}_{(A,c)}$. It will be important for the construction of the exceptional geometry in chapter \ref{CHAP5} to elaborate the possible dependence of the tensor $X$ on the vielbein $e$. I summarize the answer in the following theorem whose proof can be found in appendix \ref{prop1beweis}.

{\thm\label{prop1}
Given locality of the theory, a Lorentz tensor $X$ that is covariant with respect to the affine subgroup $A(d)$ (\ref{affine2}) of the diffeomorphisms may only depend on the vielbein $e$ implicitly:
\be\label{prop12}
{X_{ac}}^d (e)&=& {X_{ac}}^d\left(v(e),\left.\nabla^{\text{min}}\cdot\right|\right).
\ee
$v(e)$ is defined in (\ref{omega2}) and the second dependence is to be understood in the following way: $X$ may depend on arbitrary powers of the minimal connection $\nabla^{\text{min}}$ (\ref{minimalCon}) acting on arbitrary Lorentz tensors that do not have to be constructed from the vielbein alone.
}\\

The tensors $X$ that are of first order in derivatives, will be of particular importance for supergravity in the context of exceptional geometry which is the topic of chapter \ref{CHAP5}. It follows from theorem \ref{prop1} that these tensors must be linear in $v$ (\ref{omega2}). From a purely group theoretic point of view however, there is no reason to restrict to tensors $X$ that are linear in derivatives. \\

Since the affine group $A(d)$ (\ref{affine2}) is a subgroup of the diffeomorphisms $\Diff(d)$, all connections with $\Diff(d)$-equivariance are of the form (\ref{final1}). The converse is not true, however: it is obvious that not all connections of the form (\ref{final1}) are $\Diff(d)$-equivariant. Requiring affine linear equivariance is only a necessary condition for $\Diff(d)$-equivariance. By Ogievetsky's theorem from section \ref{Ogievetsky}, the necessary and sufficient condition should be to demand simultaneous equivariance with respect to special conformal diffeomorphisms $\vp^{\mathfrak{c}}_{a}$ (\ref{SCTcoord4}). I will explicitly show this in the next section.

\section{Connections with $\Diff(d)$-equivariance}\label{secC}
The proof of Ogievetsky's theorem and the subsequent discussion from section \ref{Ogievetsky} led to the fact that the conformal Killing vector field $X^{\mathfrak{c}}_{a}$ (\ref{KGen}) is not the unique vector field with quadratic $x$-dependence that generates the algebra $\mathfrak{diff}_d$ of all vector fields in its Lie closure with the affine algebra $\mathfrak{a}_d$. Nevertheless, it is preferred due to its particular property: it forms the abelian subalgebra $\mathfrak{k}_{(d-1,1)}$ (\ref{KK}) of the Lie algebra $\mathfrak{so}_{(d,2)}$. This conformal algebra $\mathfrak{so}_{(d,2)}$ is spanned by the conformal Killing vector fields of the flat Minkowksi metric $\eta$.\\

I will follow the same procedure as in the preceding section. I start with introducing some conventions for the diffeomorphism under discussion, in this case the conformal one $\vp^{\mathfrak{c}}_{a}\in \Diff(d)$ (\ref{SCTcoord4}) corresponding to the conformal Killing vector fields $X^{\mathfrak{c}}_{a}$. Then, I will prove that the induced transformation of the vielbein is related to a global left action on a group coset $\bar{C}$. This will allow to construct a minimal connection $\bar{\nabla}$ as in section \ref{ConnAff}. An application of proposition \ref{propo3} to conformal equivariance implies that $\bar{\nabla}$ may only differ from any connection with $\Diff(d)$-equivariance by a tensor with respect to a conformal transformation. Together with theorem \ref{prop1}, this will prove the statement that the Levi--Civita connection is unique modulo a torsion tensor that cannot be linear in derivatives, if it exclusively depends on vielbein degrees of freedom.

\subsection{Properties of $\vp^{\mathfrak{c}}_{a}$}\label{Propert}
Following the definition from section \ref{Subalg}, a diffeomorphism with the conformal Killing property preserves the metric tensor $\eta$ up to a scalar function. A short calulation of the pullback of $\eta$ under the conformal diffeomorphism $x'=\vp^{\mathfrak{c}}_{a}$ (\ref{SCTcoord}) fixes this function to
 \begin{subequations}\label{CDiff}
 \be\label{Lorentzgroup}
 \frac{\p \left.x'\right.^\mu}{\p x^\nu}  \frac{\p \left.x'\right.^\rho}{\p x^\tau} \eta_{\mu\rho} &=& \frac{1}{J(x,a)^2}\eta_{\nu\tau}\\
  \text{with}\quad 
 J(x,a) &:=& 1+2a\cdot x +a\cdot a\,x\cdot x\label{SCTcoord3},
\ee
\end{subequations}
and with the abbreviation $a\cdot x$ for the contraction with the Minkowski metric from equation (\ref{SCTcoord}). This is equivalent to stating that for all $x\in U_\alpha\subset \cM^d$, the Jacobi matrix multiplied with the scalar function is orthogonal\footnote{Taking the determinant of (\ref{Lorentzgroup}) implies $\det(J(x,a)\frac{\p{x'}^\mu}{\p x^\rho})=\pm 1$. The determinant is $+1$, because $\det(J(x,0)\frac{\p \vp^{\mathfrak{c}}_{a}{}^\mu}{\p x^\rho})|_{a=0}=\det(\id)=1$ and $\vp^{\mathfrak{c}}_{a}$ is continuous in $a$. }
\be\label{SOelem}
J(x,a)\frac{\p \left.\vp^{\mathfrak{c}}_{a}\right.^\mu}{\p x^\nu}\in SO(d-1,1).
\ee
Using the matrix notation from section \ref{secA3}, the conformal diffeomorphism $x' = \vp^{\mathfrak{c}}_{a}(x)$ induces the vielbein transformation (\ref{VielbeinTrafo})
\be\label{zu3}
e' &=& \left(\frac{\p x}{\p x'}\right)\cdot e\cdot O(\vp^{\mathfrak{c}}_{a},e).
\ee
Keeping in mind the special property (\ref{SOelem}) of the conformal diffeomorphism $\vp^{\mathfrak{c}}_{a}$, it looks appealing to split this matrix equation in two by discussing the determinant part separately. With the definitions\footnote{The definition of $\sigma$ is the standard convention to discuss the determinant of the vielbein \cite{ISS71}. It is only valid for positive $J(x,a)$ (\ref{SCTcoord3}). This is no restriction due to the fact that the integral curve to the conformal Killing vector field (\ref{KGen}) is only defined for values of $a$ and $x$ such that $J$ is positive, which I have explained in section \ref{Diffeom}. In particular, there is no need for absolute value signs.}
\begin{subequations}
\be
\sigma &:=&\frac{1}{d}\log\circ \det\left(e\right),
\label{sigmaDefi}\\
\text{and}\quad \bar{e}&:=&e^{-\sigma}e\label{hbarD},
\ee
\end{subequations}
the transformation (\ref{zu3}) can be written in the following way
\begin{subequations}
\be
\sigma'(x')&=&\sigma(x) +\log(J(x,a))\label{sigmaTrafo},\\
\bar{e}'(x')
&=&
\left(J(x,a)\frac{\p \vp^{\mathfrak{c}}_{a}}{\p x} \right)^{-1}
\cdot \bar{e}(x) 
\cdot O(\vp^{\mathfrak{c}}_{a},e).
\label{zu4}
\ee
\end{subequations}
The induced action of a diffeomorphism $\vp^{\mathfrak{c}}_{a}$ on the unimodular vielbein $\bar{e}$ is hence a Lorentz action from the left (\ref{SOelem}) and from the right.

\subsection{The vielbein gauge}\label{suff}
In complete analogy to section \ref{secA3}, the induced action of a conformal diffeomorphism $\vp^{\mathfrak{c}}_{a}$ on the coefficients $\psi^\gamma$ (\ref{psiParam}) of physical fields $\psi$ is provided by the equation (\ref{psiTraf})
\be\label{PsiTrf}
	{\psi'}^\gamma &=&{O_\beta}^\gamma(\vp^{\mathfrak{c}}_{a},e)\psi^\beta.
\ee
The special orthogonal matrix $O(\vp^{\mathfrak{c}}_{a},e)$ is determined by the transformation of the unimodular vielbein $\bar{e}$ (\ref{zu4}). This is true for any arbitrary, but fixed gauge for the vielbein matrix $e$, as I have discussed in section \ref{ART}.\\

In this context, it is important to keep in mind that the entire argumentation from section \ref{clc} was indepedent of the particular choice of the vielbein gauge, too. In particular, the theorem \ref{prop1} comprising the restrictions on connections with affine linear equivariance is independent of the specific fixing of the local Lorentz gauge for the vielbein matrix $e$. It is natural to require this independence for a connection with general $\Diff(d)$-equivariance, too. This implies that fixing the vielbein gauge in some way must not disturb the $\Diff(d)$-equivariance property of a connection $\nabla$.\\
 
For a general fixing of the Lorentz gauge, the compensating Lorentz rotation $O(\vp^{\mathfrak{c}}_{a},e)$ (\ref{zu4}) strongly depends on the unimodular vielbein $\bar{e}$. However, if the Lorentz gauge for the vielbein is fixed in such a way that the matrix $\bar{e}$ transforms as a Lorentz representation under $\vp^{\mathfrak{c}}_{a}$, then the compensating Lorentz rotation does not depend on $\bar{e}$: it must have the form
\be\label{SpinRot}
	O\left(\vp^{\mathfrak{c}}_{a},e\right) &=& J(x,a)\frac{\p \vp^{\mathfrak{c}}_{a}}{\p x}.
\ee
There is only one choice for the vielbein gauge such that $\bar{e}$ transforms as a Lorentz tensor: the matrix $\bar{e}$ has to be symmetric with respect to the Minkowski metric $\eta$.\footnote{The (anti)symmetry with respect to $\eta$ follows the definition (\ref{AntisyM}). The polar decomposition of $Gl(d)$ guarantees that this choice of the vielbein gauge locally is possible and unique for Euclidean and Lorentzian signature \cite{ISS71}.} \\

In the sequel, I will show that for this explicit fixing of the local Lorentz gauge with $\bar{e}$ being a symmetric matrix, it is possible to extract specific connections $\nabla$ from the class of connections with mere affine linear equivariance (\ref{final1}) by a comparison with a conformal coset. These extracted connections $\nabla$ will have the characteristic property to also be equivariant under conformal diffeomorphisms $\vp^{\mathfrak{c}}_{a}$ and hence by Ogievetsky's theorem under all diffeomorphisms $\Diff(d)$. This extraction procedure will not lead to the empty set, because e.g. the $\Diff(d)$-equivariance of the Levi--Civita connection does not depend on the fixing of the vielbein gauge.

\subsection{The conformal coset}\label{Confcos}
Using the generators of the conformal group defined in section \ref{Subalg}, I parametrize the conformal coset $\bar{C}$ by the Bruhat decomposition \cite{D1,SS69}\footnote{The proper conformal group in even dimensions is $SO(d,2)/\Z_2$, which also is mentioned in section \ref{Subalg} and in \cite{SS69}. In order to keep the notation the least cumbersome, I will not emphasize this difference in the sequel any more. This also is consistent with the fact that I am only interested in local properties of diffeomorphisms close to $\id\in \Diff(d)$, which allows to use the parametrization (\ref{gaugefC}) \cite{D1}.}
\be\label{gaugefC}
	\bar{C} &=& e^{x^\mu \hat{P}_\mu}e^{\Phi^\nu(x) \hat{K}_\nu}e^{\sigma(x) \hat{D}} \in SO(d,2)/SO(d-1,1),
\ee
with space-time coordinates $x^\mu$, a vector field $\phi^\nu$ and a scalar field $\sigma$ both depending on $x^\mu$. Next, recall that the generators $\hat{P}$, $\hat{K}$ and $\hat{D}$ of the conformal Lie algebra can be interpreted as matrices in $\mathfrak{so}_{(d,2)}$ (\ref{dplus2}), i.e. as antisymmetric matrices with respect to the metric $\eta=\diag(\id_d,-\id_2)$ in $d+2$ dimensions. This implies that the parametrization (\ref{gaugefC}) fixes the matrix form of $\bar{C}\in SO(d,2)$: the coefficients of the $\mathfrak{so}_{(d-1,1)}$ generators $\hat{L}$ are set to zero.\footnote{A similar fixing of the gauge for the affine coset $C$ (\ref{paramCoset}) would consist of replacing the $\mathfrak{gl}_d$ generators $\hat{M}$ in (\ref{eMf}) by the symmetric generators $\hat{S}$ (\ref{RDefi}).}\\

A global left action, e.g. with $e^{a^\tau \hat{K}_\tau}\in K(d-1,1)\subset SO(d,2)$, will in general perturb this matrix form of the conformal coset $\bar{C}$ (\ref{gaugefC}).\footnote{For the construction of a connection, it is sufficient to discuss the effect of the subgroup $K(d-1,1)$ on the coset $\bar{C}$ (\ref{gaugefC}), because global left actions of the conformal generators $\hat{P}$, $\hat{L}$ and $\hat{D}$ either imply a constant compensating Lorentz rotation or none at all.} In order to restore it, one has to act by a compensating $SO(d-1,1)$ rotation $O(a,\bar{C})$ from the right in complete analogy to the discussion of the affine coset in section \ref{secB}. The effect of $K(d-1,1)$ on the coset parameters $(x^\mu,\Phi^\nu,\sigma)$ can hence be obtained by evaluating the equation
\be\label{conftrafo}
	\bar{C}' &=& e^{a^\nu \hat{K}_\nu} \cdot \bar{C}\cdot O(a,\bar{C}).
\ee
In contrast to the evaluation in the case of the affine coset, the translation operators $\hat{P}$ do not form a representation of the generators $\hat{K}$, which complicates the procedure. However, the trick from section \ref{Subalg} to lift the discussion to a $(d+2)$-dimensional space works in the same way as for the calculation of the integral curve to the conformal Killing vector field (\ref{SCTcoord}). This calculation with $(d+2)\times (d+2)$ matrices was pioneered in \cite{SS69}. With my conventions (\ref{dplus2}), it leads to the following transformations
\begin{subequations}\label{conftrafo2}
\be
{x'} &=& \vp^{\mathfrak{c}}_a(x)\label{conftrafo2a}\\
 \sigma'(x') &=&\sigma(x)+\log(J(x,a)) \label{conftrafo2b}\\
 {\Phi'}_\mu(x') &=&\frac{\p x^\nu}{\p {x'}^\mu}\left(\Phi_\nu +\frac{1}{2}\p_\nu \log \left(J(x,a)\right)\right).
 \label{conftrafo2c}
\ee
\end{subequations}
The function $J(x,a)$ was defined in (\ref{SCTcoord3}) and the indices of $\Phi^\mu$ were lowered with the Minkowski metric $\eta$ (\ref{etaSkal}).\\

I have used the same name for the determinant of the vielbein $\sigma$ (\ref{sigmaDefi}) in section \ref{Propert} and for the parameter $\sigma$ of the dilatation operator $\hat{D}$ in the coset $\bar{C}$ (\ref{gaugefC}). This identification is admissible, because both objects transform in the same way under a diffeomorphism $\vp_a^{\mathfrak{c}}$ (\ref{sigmaTrafo})  and a global left action $e^{a^\tau \hat{K}_\tau}$, respectively. In the appendix \ref{PrIdent56}, I prove that the induced local Lorentz rotations also agree, if and only if the vielbein gauge is fixed to be symmetric
\be\label{Ident56}
	O(\vp^{\mathfrak{c}}_a,e) &\stackrel{(\ref{SpinRot})}{=}&J(x,a)\frac{\p \vp_a^{\mathfrak{c}}}{\p x}\nn\\
	&=&O(a,\bar{C})
\ee

To sum up, a global left action with $e^{a^\tau \hat{K}_\tau}\in SO(d,2)$ on the conformal coset $\bar{C}\in SO(d,2)/SO(d-1,1)$ with the parametrization fixed by (\ref{gaugefC}) exactly reproduces the effect of a conformal diffeomorphism $ \vp_a^{\mathfrak{c}}\in \Diff(d)$
\begin{enumerate}
	\item on the coordinate $x^\mu$,
	\item on the determinant of the vielbein or equivalently $\sigma$ (\ref{sigmaDefi}) and
	\item on an arbitrary physical field $\psi$,
\end{enumerate}
if the Lorentz gauge is fixed in such a way that the vielbein is symmetric with respect to the Minkowski metric $\eta$.\footnote{As in the affine case, this statement strongly depends on the parametrization of the coset: e.g. $\bar{C}=e^{x^\mu \hat{P}_\mu +\phi^\nu \hat{K}_\nu +\sigma \hat{D}}$ would not lead to the same result.}\\

In the same way as the affine coset $A(d)/SO(d-1,1)$ defined a minimal connection $\nabla^{\text{min}}$ (\ref{minimalCon}) with $A(d)$-equivariance, the conformal coset $SO(d,2)/SO(d-1,1)$ defines a connection $\bar{\nabla}$ with $SO(d,2)$-equivariance. In the next section, I will construct $\bar{\nabla}$ and I will also explain a consistent reduction of the coset degrees of freedom, the so-called ``inverse Higgs effect'' that allows to replace the coset parameter $\Phi$ by the other coset parameters $x$ and $\sigma$.

\subsection{The minimal connection $\bar{\nabla}$ of $SO(d,2)/SO(d-1,1)$}\label{connbar}
In complete analogy to the affine case from section \ref{ConnAff}, I want to construct the minimal connection $\bar{\nabla}$ of the conformal coset $\bar{C}$ (\ref{gaugefC}). The formulas (\ref{formula2}) and (\ref{formula}) allow to elaborate the explicit dependence of the Maurer--Cartan form on the parameters $(x,\sigma,\Phi)$ (\ref{gaugefC}). Decomposing the $\mathfrak{so}_{(d,2)}$ valued one-form $\bar{C}^{-1} d \bar{C}$ with respect to the conformal generators $\hat{P}$, $\hat{K}$, $\hat{L}$ and $\hat{D}$ defined in (\ref{CFT3}) results in
\be\label{MC2}
\bar{C}^{-1} d \bar{C}&=& e^\sigma dx^\mu\left(\hat{P}_\mu +\left(\bar{v}_\mu\right)^\nu  \hat{K}_\nu +\left(\bar{v}_\mu\right) \hat{D} +
{\big(\bar{v}_\mu\big)_\nu}^\tau \hat{L}{\big.}^\nu{\big.}_{\tau}\right)
\ee
with the abbreviations
\begin{subequations}\label{omegaCalle}
\be
\left(\bar{v}_\mu\right)^\nu &=&\left(\p_\mu \Phi^\nu
	    -2\Phi_\mu\Phi^\nu +\Phi^\rho \Phi_\rho \delta_\mu^\nu\right)
	e^{-2\sigma},\label{omegaKC}\\
	\left(\bar{v}_\mu\right)&=& \left(\p_\mu \sigma -2  \Phi_\mu \right)e^{-\sigma},\label{omegaDC}\\
{\left(\bar{v}_\mu\right)_\tau}^\nu &=&-2\left(\delta_{\mu}^{\nu} \Phi_{\tau} -\eta_{\mu\tau} \Phi^{\nu}\right)e^{-\sigma}
\label{omegaLC}, 
\ee
\end{subequations}
where indices of $\Phi$ have been lowered with the Minkowski metric $\eta$ (\ref{etaSkal}).\newpage Physical fields $\psi$ transform as $\mathfrak{so}_{(d-1,1)}$ representations. Hence, it is natural to use basis vectors to span the cotangent space that are Lorentz tensors under the global left action (\ref{conftrafo}). It is obvious that the coordinate induced one-forms $dx^\mu$ do not have this property. As for the affine coset in section \ref{ConnAff}, it is the coefficient of the translation generator $\hat{P}_\mu$ in the Maurer-Cartan form (\ref{MC2}) that provides the Lorentz covariant basis. This one-form transforms under $x' = \vp^{\mathfrak{c}}_a(x)$ as follows:
\be\label{covector4}
e^{\sigma'}d{x'}^\mu &\stackrel{(\ref{conftrafo2b})}{=}& \left(J(x,a)\frac{\p \left.\vp^{\mathfrak{c}}_a\right.^\mu}{\p x^\nu}\right) e^{\sigma}dx^\nu.
\ee
The matrix $J(x,a)\frac{\p \vp^{\mathfrak{c}}_a}{\p x^\nu}$ indeed is in $SO(d-1,1)$ (\ref{SOelem}), which is the reason why I have defined the coefficients of the Maurer--Cartan form (\ref{MC2}) in this frame.\footnote{In contrast to the affine case in section \ref{ConnAff}, standard notations from general relativity do not provide a notational difference \`a la (\ref{ehklein}) between the coordinate frame spanned by $dx^\mu$ and the $SO(d-1,1)$-covariant frame for the conformal case. In other words, there are no ``flat'' indices for the conformal case. Hence, in discussing coefficients, one always has to pay attention to the relevant frame.} \\

Therefore, it is natural to define the minimal connection $\bar{\nabla}$ with conformal equivariance acting on physical fields $\psi^\gamma$ in the frame $e^\sigma dx^\mu$ (\ref{covector4}):
\be\label{minimalCon3}
	\bar{\nabla}_\mu \psi^\gamma &:=& e^{-\sigma}\p_\mu\psi^\gamma -{\left(\bar{v}_\mu\right)_\nu}^\tau \delta_{{\left.\hat{L}\right.^\nu}_{\tau}}\psi^\gamma .
\ee
Its definition follows the conventions from the sections \ref{equivar} and \ref{ConnAff}. This connection $\bar{\nabla}$ (\ref{minimalCon3}) is the last ingredient that is necessary to extract connections $\nabla$ with $\Diff(d)$-equivariance from the class of connections (\ref{final1}) with affine linear equivariance. I have already stated in which way the parameter $\sigma$ of the dilatation generator $\hat{D}$ is linked to gravity: it is identified with the determinant of the vielbein by the relation (\ref{sigmaDefi}). But what is the role of the coefficient $\Phi$ of the generator $\hat{K}$?

\subsubsection{The inverse Higgs effect}
To answer this question, I need to discuss the transformations of the objects (\ref{omegaCalle}) under a conformal left action (\ref{conftrafo}) in more detail. Before fixing the $SO(d,2)$ matrix form of $\bar{C}$, the Maurer--Cartan form $\bar{C}^{-1}d\bar{C}$ is invariant under a global left action by $K(d-1,1)$. If the matrix form is fixed by e.g. (\ref{gaugefC}), an in general $x$-dependent $SO(d-1,1)$ action from the right (\ref{conftrafo}) is induced. This spoils the covariance of the coefficient ${\left(\bar{v}_\mu\right)_\tau}^\nu$ of the Maurer--Cartan form along the Lorentz generators $\hat{L}$. Nonetheless, the objects $\left(\bar{v}_\mu\right)^\nu$ and $\left(\bar{v}_\mu\right)$ defined in (\ref{omegaKC},\,\ref{omegaDC}) transform as Lorentz tensors under a global left action. This implies that setting $\left(\bar{v}_\mu\right)=0$ initially is invariant under a global action. Thus, one can consistently reduce the degrees of freedom of the coset parametrization by replacing the vector $\Phi$ tantamount to the relation
\be\label{Ident1}
\frac{1}{2}e^\sigma\bar{v}_\mu &=& \Phi_\mu - \frac{1}{2}\p_\mu\sigma\,=\, 0.
\ee
This reduction is known as the inverse Higgs effect \cite{ISS71,IO75}.\footnote{Following Ivanov and Ogievetsky \cite{IO75}, this name comes about as follows. The standard Higgs effect can be interpreted in such a way that there is a possibility to choose the ``unitary'' gauge for a one-form $\Phi$ in which the Goldstone field $\sigma$ disappears from an invariant Lagrangian \cite{IO75}. The kinetic term for $\sigma$ turns into the mass term of $\Phi$. Hence, the Goldstone field is eliminated by absorption into $\Phi$. For the inverse effect, it is the other way round: by (\ref{Ident1}), the one-form $\Phi$ is replaced by the field $\sigma$, whose interpretation as a Goldstone field will be addressed in section \ref{Nonlinear2} again.} Substituting (\ref{Ident1}) in the formula for the minimal connection (\ref{minimalCon3}) with (\ref{omegaLC}) provides a covariant derivative $\bar{\nabla}$ that only depends on fields that are deducible from the vielbein $e$. This is the form in which I will use the connection $\bar{\nabla}$ to extract $\Diff(d)$-equivariant connections $\nabla$ from the class of connections with affine linear equivariance in the next section.\\

From a coset point of view, this reduction is not necessary. The field $\Phi$ has no natural interpretation in the context of general relativity, however. Since I only discussed the conformal coset $\bar{C}$ (\ref{gaugefC}) in order to construct the connection $\bar{\nabla}$, I am free to consistently restrict my parametrization of $\bar{C}$ in such a way that all degrees of freedom do have counterparts in gravity. Consistency means in this case that a global left action (\ref{conftrafo}) respects this reduction. By definition, this is true for (\ref{Ident1}). It is nice to observe that this reduction could also have been deduced by comparing the transformations of $x$ and $\sigma$ (\ref{conftrafo2a},\, \ref{conftrafo2b}) to the one of $\Phi$ (\ref{conftrafo2c}).

\subsection{Joint realizations of the symmetries}\label{secD}
In this section, I will finally single out the connections $\nabla$ with $\Diff(d)$-equivariance from the class of connections with affine linear equivariance. The latter consists of all connections $\nabla$ of the form (\ref{final1})
\beg
		\nabla_a \psi^\gamma &=& 
		\p_a\psi^\gamma - \left({\left(v^{\text{min}}_a\right)_c}^d + {X_{ac}}^d \right)\delta_{{\left.\hat{L}\right.^c}_d}\psi^\gamma
\eeg
with ${\left(v^{\text{min}}_a\right)_c}^d$ provided by (\ref{omega3}) and ${X_{ac}}^d$ constrained by theorem \ref{prop1}. As I had discussed in section \ref{suff}, all $\Diff(d)$-equivariant connections in this class must in particular be equivariant with respect to a conformal diffeomorphism $\vp^{\mathfrak{c}}_a$ for the symmetric gauge choice for the vielbein.\\

Comparing the Lorentz covariant frame under $A(d)$ transformations $dx^a$ of the affine coset (\ref{Leftinv}) with the one of the conformal coset (\ref{covector4}) shows that they differ by a multiplication with the unimodular vielbein $\bar{e}$. This however transforms as a Lorentz tensor (\ref{zu4},\,\ref{SpinRot}) under conformal diffeomorphisms $\vp^{\mathfrak{c}}_a$ for this particular choice of the vielbein gauge. This implies that the frame $dx^a$ inherits the property to transform covariantly under $\vp^{\mathfrak{c}}_a$ from the frame (\ref{covector4}), which is covariant under $\vp^{\mathfrak{c}}_a$ by construction. If this was not the case, the question to extract a connection $\nabla$ with $\vp^{\mathfrak{c}}_a$-equivariance from the class (\ref{final1}) would be ill-posed from the beginning.\\

The actual procedure of extraction consists of simply rewriting the general form of a connection (\ref{final1}) in a different way: splitting the determinant from the vielbein and using the definition of the minimal connection $\bar{\nabla}$ (\ref{minimalCon3}), I obtain
\beg
		\nabla_a \psi^\gamma &=& {\left.\bar{e}\right._{a}}^\mu 
		\left(
		\bar{\nabla}_\mu \psi^\gamma + {\left(\bar{v}_\mu\right)_\tau}^\nu \delta_{{\left.\hat{L}\right.^\nu}_{\tau}}\psi^\gamma
		\right)
\\
		&& -\left({\left(v^{\text{min}}_a\right)_c}^d + {X_{ac}}^d \right)\delta_{{\left.\hat{L}\right.^c}_d}\psi^\gamma
\eeg
The following chain of arguments is the crucial part of the entire procedure: 
\begin{enumerate}
	\item The connection $\bar{\nabla}$ (\ref{minimalCon3}) is by definition equivariant with respect to global left actions on the conformal coset $\bar{C}$.
	\item If and only if the vielbein gauge is fixed such that $\bar{e}$ transforms as a symmetric Lorentz tensor under a conformal diffeomorphism $\vp^{\mathfrak{c}}_a$, the induced action on a physical field $\psi$ of such a diffeomorphism and of a global left action on the coset $\bar{C}$ (\ref{conftrafo}) coincide.
	\item If furthermore the conformal coset $\bar{C}$ is restricted by the inverse Higgs effect (\ref{Ident1}), all parameters of the conformal coset can be consistently identified with objects in gravity.
	\item This in particular implies that the connection $\bar{\nabla}$ with the identification (\ref{Ident1}) and $\sigma$ identified with the determinant of the vielbein (\ref{sigmaDefi}) is a connection that is equivariant with respect to conformal diffeomorphisms.
	\item I want to extract connections $\nabla$ with this property from the class (\ref{final1}). 
	\item Since two connections with the same equivariance property may only differ by a tensor with respect to this transformation by proposition \ref{propo3}, the following object must transform as a Lorentz tensor under the action induced by a conformal diffeomorphism $\vp^{\mathfrak{c}}_a$:
\beg
  {\left.\bar{e}\right._{a}}^\mu {\left(\bar{v}_\mu\right)_\tau}^\nu \delta_{{\left.\hat{L}\right.^\nu}_{\tau}}\psi^\gamma
		 -\left({\left(v^{\text{min}}_a\right)_c}^d + {X_{ac}}^d \right)\delta_{{\left.\hat{L}\right.^c}_d}\psi^\gamma
\eeg
\end{enumerate}

\subsubsection{Evaluating the constraint}
In order to evaluate this constraint, I will first separate from it the coefficients $\psi^\gamma$ of the physical field. To do that, recall that I have introduced the vielbein $e$ as an $x$-dependent matrix ${e_\mu}^a$ mapping the coordinate induced frame $dx^\mu$ of the cotangent space to the Lorentz covariant one (\ref{vielbeindefi}). It is a convention in the physics literature to distinguish these two different sets of basis vectors only by their indices (\ref{Leftinv}). It is this convention that I will lift now: I introduce different names for the basis vectors and hence treat all indices on an equal footing, which is more natural from a linear algebra point of view. In doing so, I have to be careful with the abbreviations from section \ref{gravmatter}: I distinguished the inverse vielbein, i.e. the inverse matrix $e^{-1}$, from $e$ only by a different position of the indices. I am replacing this notation by defining for the unimodular vielbein $\bar{e}$ in analogy to (\ref{ehklein})
\begin{subequations}\label{ehklein2}
\be
{\left(e^{\bar{h}}\right)_\mu}^a &:=&{\left.\bar{e}\right._{\mu}}^a,\\
{\left(e^{-\bar{h}}\right)_a}^\mu &:=&{\left.\bar{e}\right._{a}}^\mu.
\ee
\end{subequations}
In the language of section \ref{ConnAff}, I could also interpret the left hand side as an exponentiated form of the Lie algebra element $\bar{h}\in \mathfrak{sl}_d$. Furthermore, I have to substitute the original expression ${e_a}^\mu \p_\mu$ for the abbreviation $\p_a$ (\ref{ehneu}). Then, Greek and Latin indices are on an equal footing. Both only denote a matrix multiplication.\\

After these redefinitions, a connection $\nabla$ from the class (\ref{final1}) is equivariant with respect to conformal diffeomorphisms $\vp^{\mathfrak{c}}_a$, if the following object transforms as a Lorentz tensor under $\vp^{\mathfrak{c}}_a$
\be\label{constr5}
{A_{ac}}^d  &:=& {\left(e^{-\bar{h}}\right)_{a}}^\mu 
				 e^{-\sigma}\left(\delta_{\mu}^{d} \delta^\rho_{c}-\eta_{\mu c} \eta^{d\rho}\right)\p_{\rho}\sigma
				 + {\left(v^{\text{min}}_a\right)_c}^d + {X_{ac}}^d.
\ee
In this formula, I have used the explicit form of ${\left(\bar{v}_\mu\right)_\tau}^\rho$ (\ref{omegaLC}) with the identification (\ref{Ident1}). I want to emphasize that both objects $\delta^\nu_c$ and $\eta_{\mu d}$ are well-defined and invariant under a Lorentz action, because all indices simply denote a matrix multiplication.\\

The question arises what could spoil the tensorial property of $A$. $v^{\text{min}}$ is completely determined by the vielbein degrees of freedom $(e^{\bar{h}},\sigma)$ and derivatives thereof (\ref{omega3}). The induced action of a conformal diffeomorphism $\vp^{\mathfrak{c}}_a$ on these shows that $e^{\bar{h}}$ transforms as a tensor, but $\sigma$ does not (\ref{sigmaTrafo}): $\sigma$ would have to be invariant under a conformal diffeomorphism in order to be a Lorentz tensor, because it is a scalar object. Hence, the dependence of $X$ on the vielbein $e$ has to be fixed in such way that the object $A$ only depends on Lorentz tensors, and in particular not on $\sigma$ explicitly. \\

Partial derivatives of tensors are not tensors in general, however. This implies that the partial derivative acting on the Lorentz tensor $e^{\bar{h}}$ in $v^{\text{min}}$ (\ref{omega3}) has to be replaced by a covariant one. The minimal connection $\bar{\nabla}$ (\ref{minimalCon3}) provides such a derivative. Its evaluation on the symmetric Lorentz tensor $e^{\bar{h}}$ follows the rules of section \ref{prim2}. The resulting object transforms as a Lorentz tensor under a conformal diffeomorphism $\vp^{\mathfrak{c}}_a$:
		\be\label{coveh}
	\bar{\nabla}_\nu {\left(e^{\bar{h}}\right)_\mu}^b
						&=&
									e^{-\sigma}\p_\nu {\left(e^{\bar{h}}\right)_\mu}^b
			-e^{-\sigma}{\left(e^{\bar{h}}\right)_\tau}^b\left(\delta_{\nu}^{\tau}\delta_{\mu}^{\rho} -\eta_{\nu\mu} \eta^{\tau\rho}   \right)\p_{\rho}\sigma
		  \nn\\
				&&
				+e^{-\sigma}{\left(e^{\bar{h}}\right)_\mu}^\pi\left(\delta_{\nu}^{b}\delta_{\pi}^{\rho} -\eta_{\nu\pi} \eta^{b\rho} \right)\p_{\rho}\sigma.
		\ee
		\newpage
		Substituting $v^{\text{min}}$ (\ref{omega3}) into the equation (\ref{constr5}) with the identity (\ref{coveh}) results in
				\be \label{ATra}
{A_{ac}}^d  &=&
		 {\left(e^{-\bar{h}}\right)_a}^\nu{\left(e^{-h}\right)_g}^\mu \bar{\nabla}_\nu \left(e^{h}\right)_\mu{}^{[d}\eta^{f]g}\eta_{cf} + {X_{ac}}^d\\
	&&
		+\frac{1}{2}\left\{
\delta_{a}^{d}
		 		  {\left(e^{-\bar{h}}\right)_c}^\rho
-\eta_{\nu\mu} \eta^{\tau\rho}{\left(e^{-\bar{h}}\right)_a}^\nu
		 		  {\left(e^{-\bar{h}}\right)_c}^\mu {\left(e^{\bar{h}}\right)_\tau}^d 
		 		  		 		   \right.\nn\\
		&&\left.
-\eta^{dg}\eta_{ca}
		 		  {\left(e^{-\bar{h}}\right)_g}^\rho
+\eta_{\nu\mu} \eta^{\tau\rho}\eta^{dg}\eta_{cf}{\left(e^{-\bar{h}}\right)_a}^\nu
		 		  {\left(e^{-\bar{h}}\right)_g}^\mu {\left(e^{\bar{h}}\right)_\tau}^f
	\right\}e^{-\sigma}\p_{\rho}\sigma	\nn
\ee

In order for $A$ to transform as a Lorentz tensor under a conformal diffeomorphism $\vp^{\mathfrak{c}}_a$, $X$ has to be chosen in such a way that the last two lines in (\ref{ATra}) are compensated for, because these do not transform covariantly on their own (\ref{sigmaTrafo}). As these non-covariant objects only depend on the vielbein degrees of freedom $(\bar{e},\sigma)$ and as they are of first order in derivatives, the theorem \ref{prop1} restricts the admissible compensating terms in $X_a$ to linear ones in $v$. The general ansatz for the $\mathfrak{so}_{(d-1,1)}$ valued Lorentz tensor $X_a$ is hence provided by the following decomposition into irreducible representations of the Lorentz algebra
\be\label{Xc}
	{X_{ac}}^d &=& {K_{ac}}^d\\
	&&+\left(\delta_c^k\delta_j^d -\eta_{cj}\eta^{dk}\right)\left(c_1{\left(v_k\right)_a}^j +c_2\delta_a^j {\left(v_k\right)_e}^e +c_3\delta_a^j{\left(v_e\right)_k}^e\right)\nn
\ee
with $K$ being an arbitrary Lorentz tensor that is covariant under both conformal diffeomorphisms $\vp^{\mathfrak{c}}_a$ and affine linear ones $\vp^{\mathfrak{a}}_{(A,c)}$. The constants $c_1,c_2,c_3$ are determined by substituting the general ansatz for $X$ (\ref{Xc}) into (\ref{ATra}) and by replacing the partial derivatives acting on the Lorentz tensor $\bar{e}$ by covariant ones $\bar{\nabla}\bar{e}$ (\ref{coveh}). The object $A$ (\ref{ATra}) only transforms as a Lorentz tensor under conformal diffeomorphisms $\vp^{\mathfrak{c}}_a$ for $c_1=-1$ and $c_2=c_3=0$, which is also shown in \cite{BO74}.\\

To sum up, the additional constraint of equivariance with respect to conformal diffeomorphisms in the symmetric vielbein gauge restricts the class of admissible connections (\ref{final1}) to
	\be\label{psiTrafo47}
		\nabla_a \psi^\gamma &=& 
		\p_a\psi 
		-{\omega_{ac}}^d \delta_{{\left.\hat{L}\right.^c}_d}\psi^\gamma\\
		\text{with}\quad {\omega_{ac}}^d  &=&{\left(v^{\text{min}}_a\right)_c}^d -\left(\delta_c^k\delta_j^d -\eta_{cj}\eta^{dk}\right){\left(v_k\right)_a}^j
		+{K_{ac}}^d
		\label{SpinCon}
\ee
and with a Lorentz tensor $K$ that is covariant under both affine linear and conformal diffeomorphisms. By Ogievetsky's theorem from section \ref{Ogievetsky}, $K$ is hence covariant under the diffeomorphisms $\Diff(d)$. \\

This is the same connection $\nabla$ as the one defined by (\ref{omegaK}), which I derived from gauging the Poincar\'e group in section \ref{gravmatter}. Since the tensor $K$ is in one-to-one correspondence to the torsion tensor $T$ of this connection, the $\Diff(d)$-equivariant connection $\nabla$ with vanishing torsion $T=K=0$ is the uniquely determined Levi--Civita connection (\ref{omegaK}).

\subsection{Summary and corollaries}\label{Rest}
Following the procedure pioneered by Borisov \& Ogievetsky, I started by constructing the general class of connections with affine linear equivariance (\ref{final1}). This was completely independent of the explicit matrix form of the vielbein $e$. In extracting the connections $\nabla$ with conformal equivariance from this class, I fixed the matrix form of the vielbein to be symmetric. This does not imply a priori that these $\nabla$ are equivariant under conformal diffeomorphisms $\vp_{a}^{\mathfrak{c}}$ for a different Lorentz gauge choice, too. However, since the result of the procedure is the same as the manifestly gauge invariant connection defined in (\ref{omegaK}), the $\Diff(d)$-equivariance of $\nabla$ is independent of the choice of the Lorentz gauge. To sum up, the fixing of the vielbein gauge was a technical procedure to construct this connection $\nabla$. The result is independent of the vielbein gauge and $\Diff(d)$-equivariant.\\

A corollary to theorem \ref{prop1} is the statement that the contorsion tensor $K$ (\ref{SpinCon}) cannot be linear in derivatives, if it exclusively depends on vielbein degrees of freedom: requiring conformal covariance rules out linear terms in $v$, which is obvious from equation (\ref{coveh}). I will show in chapter \ref{CHAP5} that this is not true any more, if the symmetry group $\Diff(d)$ is restricted to a subgroup as in the case appropriate for the discussion of $d=11$ supergravity.\\

In the context of this exceptional geometry, the restriction of the conformal diffeomorphisms to a $w$-dimensional submanifold with $1<w\leq d$ will be important. Where local questions are concerned, it is sufficient to discuss a $w$-dimensional open subset $W$ of the open set $x_\alpha(U_\alpha)\subset \R^d$, on which the conformal diffeomorphisms are well-defined as explained in section \ref{Diffeom}. Then, the corollary \ref{Coro2} to Ogievetsky's theorem from section \ref{Ogievetsky} is applicable. It implies that any connection $\nabla$ with affine linear equivariance in $d$ dimensions and conformal equivariance in $w$ dimensions must be $\Diff(d)$-equivariant.\\

\section{Other perspectives}\label{secE}
The necessity of the fixing of the Lorentz gauge for the vielbein matrix $e$ for the construction \`a la Borisov \& Ogievetsky may appear unnatural from a geometric point of view. Insisting on manifest Lorentz gauge covariance throughout the construction leads to a unified coset description of the diffeomeorphism group $\Diff(d)$. Its minimal connection is the Levi--Civita connection modulo a class of specific torsion tensors. Since the torsion tensor of supergravity theories (\ref{gravitino}) is not covered by this class, this description appears to be inappropriate for the discussion of supergravity, however.\\

After reviewing this unified coset description in the first section, I will comment on the original interpretation of Borisov \& Ogievetsky \cite{BO74} motivated from the discussion of internal symmetries. In particular, I will argue that the notion of Goldstone bosons may not be appropriate for this setting.\\

I will conclude with a brief survey of West's $E_{11(11)}$ conjecture that will serve as a guiding principle for the construction of the exceptional geometry of supergravity in chapter \ref{CHAP5}.

\subsection{A unified coset description}
To start with, I want to emphasize that I only used the affine coset and the conformal one as auxiliary objects in order to construct minimal connections $\nabla^{\text{min}}$ (\ref{minimalCon}) and $\bar{\nabla}$ (\ref{minimalCon3}). These proved useful to determine the general dependence of the $\Diff(d)$-equivariant connection $\nabla$ on the vielbein $e$. In particular, a dependence of $\nabla$ on an arbitrary torsion tensor $T$, which is not part of any coset, cannot be excluded.\\

A different approach would be to consider the coset as a fundamental object and to insist on manifest Lorentz covariance throughout the procedure of Borisov \& Ogievetsky. A quick look at the discussion in section \ref{Confcos} shows that the restriction to the symmetric gauge choice for the vielbein matrix is only due to the absence of the unimodular vielbein in the conformal coset $\bar{C}$ (\ref{gaugefC}). A global left action by $e^{a\hat{K}}\in K(d-1,1)$ on the modified coset
\beg
e^{x^\mu \hat{P}_\mu}e^{\Phi^\nu(x) \hat{K}_\nu}e^{\sigma(x) \hat{D}} \cdot \bar{e}
\eeg
would formally lead to the same compensating Lorentz rotation $O\left(\vp_a^{\mathfrak{c}},e\right)$ as a conformal diffeomorphism $\vp_a^{\mathfrak{c}}$ for any gauge choice of the vielbein (\ref{zu3}). Ogievetsky's theorem \ref{Ogie2} implies however that this modified coset does not close under a left action with all generators: it has to be enlarged to a parametrization of the entire diffeomorphism group coset $ \Diff(d)/SO(d-1,1)$. This introduces infinitely many new parameters for the generators $\hat{P}_{\bf{n},\mu}$ of $\Diff(d)$ with their representation $\mathbf{R}$ as vector fields (\ref{P2Defi})
\be\label{P2Defi3}
\mathbf{R}_{\hat{P}_{\mathbf{n},\mu}(x)}&=& \big(x^1\big)^{n_1}\cdots \big(x^d\big)^{n_d}\p_\mu.
\ee
The generalized coset would hence be of the form
\be\label{Coset5}
\tilde{C}&=& e^{x^\mu \hat{P}_\mu}\cdot e(x)\cdot \prod\limits_{n=2}^\infty e^{w_{(n)}^{\bf{n},\mu} \hat{P}_{\bf{n},\mu}}\in \Diff(d)/SO(d-1,1)
\ee
with the coefficients $w_{(n)}^{\bf{n},\mu}$ of the generators $\hat{P}_{\bf{n},\mu}$, whose vector field representation (\ref{P2Defi3}) is of polynomial degree $n\in \N$ and with the standard vielbein matrix $e(x)\in Gl(d)/SO(d-1,1)$ in an arbitrary Lorentz gauge.\\ 

Since the action of $\Diff(d)$ on the coset $\tilde{C}$ is a global left one, the corresponding minimal connection $\tilde{\nabla}^{\text{min}}$ acting on physical fields $\psi$ in $\mathfrak{so}_{(d-1,1)}$ representations is equivariant under $\Diff(d)$ by construction. Following the definition of the minimal connection for a coset from section \ref{ConnAff}, the $\mathfrak{so}_{(d-1,1)}$ part of the Maurer--Cartan form $\tilde{C}^{-1} d\tilde{C}$ is the essential ingredient of $\tilde{\nabla}^{\text{min}}$. A short calculation in the vector field representation (\ref{P2Defi3}) reveals that only two of the infinitely many parameters contribute
\beg
\left.\tilde{C}^{-1} d\tilde{C}\right|_{\mathfrak{so}_{(d-1,1)}} 
 &\stackrel{(\ref{formula2})}{=}&
 {e_\mu}^a dx^\mu \left\{
 -\left.
 w_{(2)}^{\bf{n},\mu} \left[  \hat{P}^{(2)}_{\bf{n},\mu}, \hat{P}_a\right]
 \right|_{\mathfrak{so}_{(d-1,1)}} 
  +
 {\left({v^{\text{min}}_a}\right)_c}^d {\left.\hat{L}\right.^c}_d\right\},
\eeg
where I have used the abbreviation $v^{\text{min}}$ (\ref{omega3}). Redefining the coefficient $w_{(2)}^{\bf{n},\mu}$ by
\be\label{redef}
w_{(2)}^{\bf{n},\mu} \mathbf{R}_{ \hat{P}^{(2)}_{\bf{n},\mu}} &=:& \frac12{w_{(ak)}}^j \,{e_\nu}^a{e_\rho}^k {e_j}^\mu\, \left(x^\nu x^\rho \p_\mu\right),
\ee
leads to the following form for the minimal connection
\be
\label{minimalCon5}
	\tilde{\nabla}_a^{\text{min}}\psi^\gamma &=&\p_a\psi^\gamma 
	-\left({\left(v^{\text{min}}_a\right)_c}^d
	 +\left(\delta_c^k\delta_j^d -\eta_{cj}\eta^{dk}\right){w_{(ak)}}^j\right) \delta_{{\left.\hat{L}\right.^c}_d}\psi^\gamma.
\ee
This connection is $\Diff(d)$-equivariant by construction, but it does not have the standard form familiar from general relativity (\ref{omegaK}). To achieve this, compare the conformal Killing vector field $X^{\mathfrak{c}}_\Phi$ (\ref{KGen}) with parameter $\Phi$ with the general vector field representation of polynomial degree two (\ref{redef}). Switching to the vielbein frame (\ref{ehneu}), I obtain the identity
\beg
\Phi^j = \eta^{ak}{w_{(ak)}}^j -2\eta^{ja}{w_{(ak)}}^k.
\eeg
Next recall that the inverse Higgs effect (\ref{Ident1}) related the parameter $\Phi$ of the quadratic generator $\hat{K}$ (\ref{KRep}) to the parameter of the linear generator $\hat{D}$ (\ref{PRep},\,\ref{DDefi}). This was possible by setting the Maurer-Cartan form along the linear generator $\hat{D}$ to zero. The generalization is straightforward: the coefficient of the Maurer--Cartan form along the linear symmetric generator $\hat{S}$ has to be set to zero. I want to introduce the following abbreviation for this coefficient
\beg
{e_\mu}^a dx^\mu \left\{
  {m_{ac}}^d \right\}{\left.\hat{S}\right.^c}_d
 &:=&\left.\tilde{C}^{-1} d\tilde{C}\right|_{\hat{S}} \\
 &=&
 {e_\mu}^a dx^\mu \left\{
 -\left.
 w_{(2)}^{\bf{n},\mu} \left[  \hat{P}^{(2)}_{\bf{n},\mu}, \hat{P}_a\right]
 \right|_{\hat{S}} 
  +
 {\left({v_a}\right)_c}^d {\left.\hat{S}\right.^c}_d\right\}.
 \eeg
Global left invariance of $\tilde{C}^{-1} d\tilde{C}$ implies that 
\be\label{torsionp}
{m_{ac}}^d &=&
  \left(\delta_c^k\delta_j^d +\eta_{cj}\eta^{dk}\right){w_{(ak)}}^j
  +{\left({v_a}\right)_c}^d
\ee
transforms as a Lorentz tensor under $\Diff(d)$. Setting ${m_{ac}}^d$ to zero is hence a consistent reduction of the coset degrees of freedom, which is tantamount to the identity
\beg
{w_{(ak)j}}&=& \frac{1}{2}\left({\left({v_k}\right)_{aj}} +{\left({v_a}\right)_{kj}} - {\left({v_j}\right)_{ak}}\right).
\eeg
A substitution of $w$ in the ansatz (\ref{Coset5}) proves that the minimal connection $\tilde{\nabla}^{\text{min}}$ (\ref{minimalCon5}) is the Levi--Civita connection, i.e. (\ref{SpinRot}) with vanishing torsion.\\

Furthermore, the inverse Higgs effect allows to parametrize the entire coset $\Diff(d)/SO(d-1,1)$ only with the coordinate $x$ and with vielbein degrees of freedom $e$ in a consistent way. To see this, first observe that the adjoint action of generators $\hat{P}_{\mathbf{n},\mu}$ (\ref{P2Defi3}) of degree $n$ for $n\geq 2$ on the translation generators $\hat{P}$ (\ref{Lie9}) maps to generators $\hat{P}_{\mathbf{n},\mu}$ of degree $n-1$. Therefore, the coefficients of the Maurer--Cartan form $\tilde{C}^{-1}d\tilde{C}$ along the generators of degree $n-1$ will have the form
\beg
{m_{ac_1\dots c_{n-1}}}^d &:=&
  {w^{(n)}_{(ac_1\dots c_{n-1})}}^d + \sum\limits_{l<n} {f_{ac_1\dots c_{n-1}}}^d\left(w^{(l)}\right)
\eeg
with an in general non-linear function $f$ of the components $w^{(l)}$ of degree $l<n$. In analogy to (\ref{torsionp}), $m$ transforms as a Lorentz tensor under $\Diff(d)$. Hence, $m$ can be consistently set to zero. This fixes all $w^{(n)}$ for $n\geq 3$ by induction.\\  

This approach to construct the Levi--Civita connection from a coset of the diffeomorphism group was pioneered by Kirsch \cite{K05}. In order to keep the analogy to non-linear realizations of internal symmetry groups \cite{ISS71}, he insisted on a symmetric gauge for the vielbein, however.

\subsection{What about torsion?}\label{tors1}
Since torsion is a possible part of the connection, it should be included in the coset $\Diff(d)/SO(d-1,1)$ in such a way that it affects the minimal connection. The equation (\ref{minimalCon5}) shows that it can only be a part of the coefficient ${w_{(ak)}}^j$ along the generator of degree $n=2$. As torsion is a tensor by definition, it would only change the tensorial part of ${w_{(ak)}}^j$. As $v$ is not a tensor under $\Diff(d)$, it is obvious from equation (\ref{torsionp}) that possible torsion in this setting corresponds to a non-vanishing ${m_{ac}}^d$. If I do not set ${m_{ac}}^d=0$, I can still substitute (\ref{torsionp}) in the formula for minimal connection $\tilde{\nabla}^{\text{min}}$ (\ref{minimalCon5}). The general ansatz for the coefficient
\beg
{w_{(ak)j}}&=& \frac{1}{2}\left({\left({v_k}\right)_{aj}} +{\left({v_a}\right)_{kj}} - {\left({v_j}\right)_{ak}}\right)
-\frac{1}{2}\left(m_{k(aj)} +m_{a(kj)} - m_{j(ak)}\right)
\eeg
leads to the general $\Diff(d)$-equivariant connection $\nabla$ (\ref{SpinRot}) with the following contorsion tensor 
\be\label{con7}
K_{a[df]}&=& m_{f(ad)}-m_{d(af)}
\\
\Leftrightarrow\quad T_{[ad]f}&=& K_{a[df]}- K_{d[af]}\,=\, m_{a(df)}-m_{d(af)}.\nn
\ee
This is however not a general expression for torsion, because it cannot be inverted: not for any $K$ exists an $m$ such that this equation holds. This is obvious from observing that any contorsion tensor of the form (\ref{con7}) fulfills the equation
\beg
K_{a[df]} + K_{f[ad]} +K_{d[fa]} &=&0,
\eeg
which is not true for a general contorsion tensor. In particular, the bifermionic torsion of $d=4$ supergravity (\ref{gravitino}) does not fulfill this constraint. I want to remark that Kirsch did not discuss any non-vanishing torsion in \cite{K05}.\footnote{He only conjectured that non-vanishing torsion might be connected to a non-commutativity of the space-time coordinates.} Hence, this result exceeds his work.
\\

To sum up, the unified coset model $\Diff(d)/SO(d-1,1)$ provides a nice way to discuss $\Diff(d)$-equivariant connections with a class of particular torsion terms. It is not general however. As the torsion of supergravity is not in this class, I will henceforth work with my interpretation that considers the two finite dimensional cosets as auxiliary objects and not as fundamental ones.\\

Concerning torsion, a comment on Borisov \& Ogievetsky's original presentation \cite{BO74} also is necessary. I proved in section \ref{Rest} that the contorsion tensor $K$ (\ref{SpinCon}) cannot be linear in derivatives, if it exclusively depends on vielbein degrees of freedom. An example for $K$ of third order in derivatives is provided by ${K_{ac}}^d = \eta^{dg}\nabla_{[c}{R_{g]fa}}^f$ with the Riemann tensor defined in (\ref{Riemann}). Only an additional constraint, such as a restriction on the power of derivatives, may exclude such torsion terms. Since Borisov and Ogievetsky did not state anything like that, their ansatz, eq. (28) in \cite{BO74}, is incomplete. This is the reason why they claim to obtain general relativity from $\Diff(d)$-covariance alone, without demanding vanishing torsion explicitly.\\

Torsion also is very interesting from another point of view: the requirement of vanishing torsion for the general connection $\nabla$ with affine linear equivariance (\ref{final1}) uniquely singles out the Levi Civita connection, without requiring the equivariance under conformal diffeomorphism $\vp^{\mathfrak{c}}_{a}$ at all.\footnote{To see this, observe that the non-vanishing torsion of the minimal connection $\nabla^{\text{min}}$ (\ref{minimalCon}) is proportional to $v$ (\ref{omega2}), which I show at the end of section \ref{prop1beweis} in the appendix.}

\subsection{Goldstone bosons for internal symmetry groups} \label{Nonlinear}
It's obvious from the introduction to this chapter that Borisov \& Ogievetsky's procedure \cite{BO74} was strongly inspired by the theory of non-linear realizations of internal symmetries \cite{ISS71,SS69a}. In particular, they linked their construction to the Goldstone phenomenon. In this section that closely follows Salam's and Strathdee's original work \cite{SS69a}, I want to recapitulate the original idea. In the following section \ref{Nonlinear2}, I will point out the problems with an extension of the Goldstone phenomenon to external symmetries because of which I am not going to follow this line of argumentation in this thesis.\\

To understand what is meant by a Goldstone boson, one first has to recall the definition of a spontaneously broken symmetry $G$, which is a symmetry of the Lagrangian, but not of the ground state of a quantum system. In order to translate this into the language used in my thesis, I want to quote the example from \cite{SS69a}:\\

Consider the chiral symmetry group $G=SU(2)\times SU(2)$ for $N_c=N_f=2$. The chiral Lagrangian contains pions $\pi$ only in the form $\p \pi$. Therefore, the spin-zero particles $\pi$ are massless. Hence, a vacuum state is indistinguishable from a state with any even number of zero frequency pions forming a representation of $H=SU(2)$ with isospin label $I=0$. In other words, the state with lowest energy (which is the ground state) is physically degenerate. Although the chiral Lagrangian is invariant under a $G$-action, the triplet of pions $\pi$ does not form a representation of $G$. It only parametrizes the coset $G/H$, which implies that a $G$-action on the pions is realized non-linearly. The degeneracy of the ground state together with the non-linear realization of the symmetry $G$ leads to a spontaneous break-down of the chiral symmetry $G$. The pions are the Goldstone bosons in this example.\\

Salam and Strathdee \cite{SS69a} generalized this procedure to arbitrary groups $G$. They introduced the notion of a preferred field. This is a field that transforms in a constrained way under the action of a global symmetry group $G$. The particles associated to these preferred fields are Goldstone bosons \cite{GSW62}. \\

Following Wigner's interpretation of the momentum four-vector as a Lorentz boost \cite{W39}, they identified these preferred fields as particular components of a coset $C=G/H$, the so-called reducing matrix. The action of an element $g$ of the symmetry group $G$ on the coset $C$ is then defined by a multiplication from the left by $g$ and a right multiplication with an element of the subgroup $H$
\be\label{Trafoh}
C' &=& g\cdot C\cdot h(g,C).
\ee
The latter is fixed by requiring that the transformed reducing matrix $C'$ or equivalently, the transformed preferred fields, again satisfy the constraint imposed on them. Thus, $h$ depends on $g$ in general which implies a non-linear transformation of the preferred fields under $G$. This constraint is nothing but the choice which representative in the $H$ orbit of $G/H$ has been fixed.\\

Let $\Lie$ and $\Car$ denote the Lie algebras corresponding to the Lie groups $G$ and $H$. Then the projection of the Maurer--Cartan form $C^{-1}dC$ on $\Lie\ominus \Car$ is invariant under a global left action (\ref{Trafoh}). Since the preferred fields parametrize the coset $C$, the coefficient of $C^{-1}dC$ along the vectors spanning $\Lie\ominus \Car$ defines a derivative of the preferred field, a ``covariant derivative''.\\

The quotation marks for ``covariant derivative'' were introduced in \cite{ISS71}, because the preferred field itself does not transform as an $H$-tensor under a global left $G$ action, which is essential for a covariant derivative in the setting of differential geometry or a connection defined in section \ref{SymmAct}. \\

In complete analogy to the gauging procedure I presented in chapter \ref{CHAP3}, these ``covariant derivatives'' can be added to a Lagrangian, which is invariant under a global $G$ action. If this global symmetry group $G$ is gauged, the preferred fields disappear from the Lagrangian, because they are included in the gauge fields of the local $G$ symmetry. This is the standard Higgs effect.

\subsection{Problems with a generalization to external symmetries}\label{Nonlinear2}
Neglecting gravitational interactions, Salam and Strathdee investigated in \cite{SS69} if the locally experimentally observed Lorentz isometry $H=SO(3,1)$ of our space time could result from a spontaneously broken conformal symmetry $G=SO(4,2)/\Z_2$ of space-time. Their result is that all the preferred fields parametrizing the coset $G/H$ turn out to be massive in Minkowski space and some of them have spin. Hence, the interpretation as ``Goldstone bosons'' is not as compelling as for the case of internal symmetries, which is why they also used quotation marks in \cite{SS69}. Only the effect of the linearized $G$ action on preferred fields is the characteristic one for Goldstone bosons: a constant displacement.\footnote{
Borisov \& Ogievetsky mentioned in a footnote in \cite{BO74} that strictly speaking, the preferred field $\sigma$ (\ref{gaugefC}) is the only independent one with this transformation behaviour. The other one $\Phi$ is a combination of $\sigma$ and a tensor field.
}\\

In a later paper \cite{ISS71}, they reinterpreted one of these preferred fields as part of the conformal Minkowski metric $g=e^{2\sigma}\eta$. For such a curved space-time however, the standard description of a Goldstone boson has to be treated with care: General coordinate covariance implies by definition that all coordinate systems are equivalent. If one associates to every coordinate system a vacuum state, the global symmetry group $G$ distinguishes these ground states, because $G$ only consists of the conformal Minkowski isometries, a subgroup of the diffeomorphisms. Taking e.g. the Einstein-Hilbert action for this particular metric $g=e^{2\sigma}\eta$ as an action, the Lagrangian is invariant under $G$, whereas the ground state is not. Hence, the diffeomorphism symmetry is spontaneously broken by the vacuum state, which merely is an explicit choice of a coordinate system. As this symmetry however affects the metric, it is not possible to measure the distance between these vacuum states in terms of energy as for the case of an internal symmetry. In other words, the question for the mass of this preferred field $\sigma$ in this setting is ill-posed.\\

To conclude, I want to summarize three problems in generalizing the notion of Goldstone fields to external symmetries:
\begin{enumerate}
	\item Salam \& Strathdee's definition of the reducing matrix $C$ assumes that it transforms as a tensor under a left $H\subset G$ action (eq. (2.3) in \cite{SS69a}). This is necessary for the preferred field to be an $H$ tensor under this transformation. In the present context, this is equivalent to fixing the symmetric gauge for the vielbein a priori, i.e. already in the affine coset $C\in A(d)/SO(d-1,1)$ (\ref{paramCoset}). This is not natural from a geometric point of view. 
	\item Borisov and Ogievetsky argue that the preferred field $h$ (\ref{eMf}) of the affine coset $C$ corresponds to a Goldstone boson, because $h$ must not appear explicitly in an invariant action. This refers to the case of internal symmetries again, where this is only true for the gauged version, i.e. $G(x)$, however. It is important to note that gauging the affine group produces a structure that is different from the diffeomorphism group $\Diff(d)$: although all diffeomorphisms induce a transformation of the vielbein $e$ by a matrix multiplication with the Jacobian, which is an in general $x$-dependent $Gl(d)$ matrix, not all $x$-dependent $Gl(d)$ matrices in $G(x)$ correspond to diffeomorphisms due to the integrability constraint.
	\item Most importantly, the question if the preferred field of the affine coset $C$ qualifies as a massless Goldstone boson is ill-posed, because different ground states cannot be compared, as there is no gauge invariant definition of energy in an arbitrarily curved space time.
\end{enumerate}
Hence, although the transformation behaviour of the preferred fields of the affine coset show the characteristic transformation behaviour of Goldstone bosons, I do not share the point of view of Borisov \& Ogievetsky that this formal association is compulsory. As far as experimental physics is concerned, there is no sign of a particle like the pion associated to these preferred fields $h$ so far.\\

For internal symmetries, non-linear models with spontaneously-broken symmetry often are non-renormalizable. This is taken as an indication that they only describe low energy phenomena \cite{IO75}. At higher temperatures or energies, more symmetry might be restored \cite{K05}. Due to the problematic definition of energy in the case of external symmetries, I do not agree with the conclusion in \cite{BK06,K05} either: the authors argued that the spontaneous breakdown of $\Diff(d)$ to $SO(d-1,1)$, with gravity corresponding only to the $Gl(d)$ part, should lead to a modification of gravity by introducing new interaction particles, just as in particle physics. They conjectured that these ``higher spin fields'' corresponded to the other preferred fields of the coset $\Diff(d)/SO(d-1,1)$ \cite{BK06}.

\subsection{Technical aspects}\label{Techn}
I concluded the section \ref{ConnAff} with the reasons why it was necessary to include the coordinates $x^\mu\hat{P}_\mu$ in the coset for the case of external symmetries. Since internal symmetries do not act on the coordinates by definition, there is no need to include them in a coset of an internal symmetry group. I emphasize this point, because for external symmetries both the coordinate indices and the gauge group indices transform in representations of the same group. This allowed to construct an antisymmetric Lorentz tensor from the symmetric Lorentz tensor $v$ (\ref{omega2}) that I could then add to the minimal connection $\nabla^{\text{min}}$ in (\ref{SpinCon}). It is important to note that this possibility does not exist for an internal symmetry: there is no way to modify the minimal connection by the extra part of the Maurer--Cartan form, because the indices do not transform under the same group.\\

The definition of a ``covariant derivative'' from the theory of non-linear realizations is problematic in the procedure of Borisov \& Ogievetsky. Considering $h$ (\ref{eMf}) as the preferred field, the definition reviewed in section \ref{Nonlinear} leads to the definition (eq. (26) in \cite{BO74})
\be\label{Kovh}
\nabla^{(BO)}_a {h_c}^d &:=& {\left(v_a\right)_c}^d
\ee
with the abbreviation ${\left(v_a\right)_c}^d$ as in (\ref{omega2}). Note that the important point is hidden in this definition: the $A(d)$-covariant Lorentz tensor ${\left(v_a\right)_c}^d$ is symmetric in $(cd)$ by construction independently of the gauge of the vielbein $e$, whereas ${h_c}^d$ is symmetric only for the special case of a symmetric vielbein gauge, which is necessary for the interpretation of $h$ as a preferred field. My first reason not to use this notation of a ``covariant derivative'' is that it obscures the fact that the procedure of Borisov \& Ogievetsky is independent of the explicit choice of the vielbein gauge.\footnote{This independence has already been observed by West \cite{W00}.}\\

The second reason is related to the extraction procedure of the $\Diff(d)$-equivariant connection from the class of connections with affine linear equivariance in section \ref{secD}. To do so, I had to replace the partial derivative $\p$ acting on the unimodular vielbein $\bar{e}$ by the minimal connection of the conformal coset $\bar{\nabla}$ (\ref{coveh}). This defines an action of the Lorentz connection $\bar{\nabla}$ on the symmetric traceless matrix $\bar{h}$ by the exponential series (\ref{eMf}). In complete analogy, the action of the Lorentz connections $\nabla^{\text{min}}$ (\ref{minimalCon}) or $\nabla$ (\ref{SpinCon}) on $\psi=h$ is defined, too.\footnote{
Strictly speaking, $h$ does not transform as a Lorentz tensor under the affine group $G=A(d)$ (\ref{affine2}), but under its Poincar\'e subgroup, it does. Any connection $\nabla$ with $G$-equivariance canonically induces by projection a connection with $H$-equivariance for any subgroup $H\subset G$.  It is however standard in the physics literature, not to distinguish these two connections by notation.
} This can be compared with the definition (\ref{Kovh}) by a direct evaluation of ${\left(v_a\right)_c}^d$ in terms of $h$ with the formula (\ref{formula})
\beg
e^{-h}d e^h = \sum\limits_{j=0}^\infty \frac{(-1)^j}{(j+1)!}\ad_h^j dh.
\eeg
At the first non-trivial order in $h$, the two objects disagree. Therefore, the definition (\ref{Kovh}) used in \cite{BO74} is ambiguous.

\subsection{Related work in the literature and $E_{11(11)}$} \label{Related}
Borisov \& Ogievetsky's procedure has been generalized by Ivanov and Niederle \cite{IN92} to superspace in the same quantum field theory inspired way. They applied it to the Ogievetsky-Sokatchev formulation \cite{OS78} of $N=1$ minimal Einstein supergravity by discussing finite dimensional superextensions of the affine group $A(4)$ and the conformal one $SO(4,2)/\Z_2$. \\

I will follow a different approach in this thesis. As I have already discussed in chapter \ref{CHAP3}, supergravity can be understood as a special case of Einstein--Cartan theory, which naturally arises from gauging the Poincar\'e isometry group of Minkowski space $\R^{d-1,1}$. In particular, there is no need to introduce a super Poincar\'e group on a superspace.\\

Where $d=11$ supergravity is concerned, it is the particular self-interaction of the three-form potential $A$ (\ref{GaugeTrafo}) by a Chern--Simons term (\ref{Action}) that together with the vielbein $e$ allows for a completion to a supersymmetric theory. Hence, it is very likely that this coupling is distinguished from others in some group theoretic way, which might also explain the hidden symmetries reviewed in section \ref{hid1}.\\

West was the first to connect $d=11$ supergravity to the procedure of Borisov \& Ogievetsky. In \cite{W00}, he proposed to extend the group $Gl(11)$ in the affine coset $C$ (\ref{paramCoset}) by a normal subgroup to $G_{11}$ in order to also include the degrees of freedom of the three-form potential $A$. In the sight of the chain of hidden symmetries, he further enlarged $G_{11}$ to the group corresponding to the infinite dimensional Kac--Moody algebra $\mathfrak{e}_{11(11)}$ \cite{W01}. To obtain a consistent global left action by this group on the coordinates in analogy to the coset description of section \ref{secB}, he had to generalize the $\mathfrak{gl}_{11}$ representation $\hat{P}_\mu$ of translations to a representation of $\mathfrak{e}_{11(11)}$. This formally introduces infinitely many additional coordinates that might be associated to brane charges \cite{W03}.\\ 

West's algebraic approach will be the guiding principle for the construction of the exceptional geometry in the next chapter. I will also combine the degrees of freedom of the vielbein $e$ and of the three-form potential $A$ into a single coset. To establish the contact to the dynamics of $d=11$ supergravity, I will content myself with discussing the finite dimensional exceptional algrebra $\mathfrak{e}_{7(7)}$ instead of $\mathfrak{e}_{11(11)}$. Since the fundamental representation space of $\mathfrak{e}_{7(7)}$ is $56$-dimensional, I will also have to introduce additional coordinates to obtain a consistent left action by the group $E_{7(7)}$ on the coset.\\

In contradistinction to the algebraic approach used by West, I will follow a geometric approach, however. In analogy to the definition of the Poincar\'e subgroup of $\Diff(d)$ from section \ref{Subalg}, I will require the invariance of certain tensors. This will restrict the diffeomorphisms of a higher dimensional space-time to the appropriate subgroup that corresponds to the global left action on the coset. In this sense, the entire coset will have a natural interpretation as a higher dimensional vielbein in this restricted or exceptional geometry.

\chapter{Maximal supergravity in $d=60$}\label{CHAP5}
I have concluded chapter \ref{CHAP3} with the observation that a $(d=60)$-dimensional geometry appears to be adequate for the treatment of a theory that is invariant under an $E_{7(7)}$ action on the coordinates and that contains the dynamics of $d=11$ supergravity. The $\Diff(60)$ symmetry of this theory has to be restricted in some way for two reasons.
\begin{enumerate}
	\item The degrees of freedom of a sixty-dimensional vielbein without further restriction exceed the ones of $d=11$ supergravity. A Kaluza--Klein reduction of an unrestricted $d=60$ geometry would hence imply a coupling to additional physical fields.
	\item A supersymmetry variation in $d=60$ dimensions would lead to more than $N=8$ supersymmetries in a compactification to $d=4$ dimensions, which is inconsistent as explained in section \ref{sugraDefi}.
\end{enumerate}
This is the reason why I will restrict to an exceptional geometry by requiring the invariance of a degenerate symplectic form $\Omega$ and a totally symmetric quartic tensor $Q$ of codimension $4$ in the $(d=60)$-dimensional geometry. This reduces the symmetry group $\Diff(60)$ to a subgroup, which allows to restrict the degrees of freedom of the $d=60$ vielbein and of the supersymmetry parameter in an appropriate way.\\

To relate the $d=60$ vielbein to the bosonic degrees of freedom of $d=11$ supergravity, recall from section \ref{SUGRA1} that these consist of the $d=11$ vielbein $E$ and the three-form potential $A$ (\ref{GaugeTrafo}). West pioneered the idea \cite{W00} to put both fields $E$ and $A$ into a single coset of $G_{11}/SO(10,1)$, which I have already discussed in the final section \ref{Related} of the chapter on non-linear realizations.  \\

An important ingredient in West's approach is manifest Lorentz covariance in $d=11$ dimensions. This is where the present discussion differs. Following Nicolai and de Wit \cite{dWN86}, I reduce the covariance group to $SO(3,1)\times SO(7)$ by fixing a block-triangular form for the $d=11$ vielbein matrix $E$, where the $SO(3,1)$ part corresponds to the codimension $4$ of the tensors $\Omega$ and $Q$.\\ 

Since the calculations are quite involved, I have chosen to focus on the $56$-dimensional subsector of the exceptional geometry in this thesis, on which both tensors $\Omega$ and $Q$ are non-degenerate. This will correspond to the $SO(7)$-covariant part of $d=11$ supergravity. I will show in section \ref{MaxS} that imposing the invariance of $\Omega$ and $Q$ restricts the remaining symmetry group $\Diff(56)$ to $E_{7(7)}$ and $56$-dimensional translations in accordance with Cartan's theorem \cite{C09}. This fact allows to reduce the degrees of freedom of the $d=56$ subvielbein $e^H$ to the ones of an $E_{7(7)}$ matrix.\\

It also establishes the link to the algebraic approach used in chapter \ref{CHAP4} and by West in \cite{W00}. Since the resulting subgroup of $\Diff(56)$ is a subgroup of the affine group $A(56)$ (\ref{affine2}), the coset description of section \ref{secB} can be used without loss of generality. This description will be the most appropriate one to control the independent degrees of the vielbein $e^H$ in a comparison with $d=11$ supergravity.\\

Of further importance for the dynamics and for the construction of supersymmetry variations will be the $56$-dimensional connection $\nabla$, which is the basic building block of any physical theory with a symmetry. Therefore, I will follow the fate of the result from chapter \ref{CHAP4} that the Levi--Civita connection is the unique $\Diff(d)$-equivariant connection, whose contorsion $K$ is linear in derivative operators, if $K$ may only contain vielbein degrees of freedom. An important insight is that this result does not hold any more, if the equivariance property of $\nabla$ is restricted to the subgroup of $\Diff(56)$ respecting the invariance of the tensors $\Omega$ and $Q$. It is this observation that will allow to define a connection $\nabla$ in $56$ dimensions that can consistently act on $SU(8)\subset Spin(56)$ representations, such as the transformation parameter $\ep$ of supersymmetry. \\

I will proceed as follows. I will start by proving the restriction of $\Diff(56)$ to a subgroup, if the invariance of the tensors $\Omega$ and $Q$ is imposed. Furthermore, I will also discuss the effect on possible connections $\nabla$ in section \ref{MaxS}. Then, I will parametrize the $E_{7(7)}$ generators by $Gl(7)$ representations in section \ref{Decompo} that is tailored for a comparison of the $56$-dimensional vielbein $e^H$ and of the connections $\nabla$ with $d=11$ supergravity in the sections \ref{affE7} and \ref{Comp1}. Finally, I will introduce supersymmetry variations on the $56$-dimensional exceptional geometry in section \ref{Super} that exactly reproduce the ones of $d=11$ supergravity to the extent that can be expected in this restriction to the $56$-dimensional subsector.\\
 
I will conclude with remarks on the dynamics and why it looks probable that all solutions of $d=11$ supergravity exactly correspond to the ones of a theory in this sixty dimensional exceptional geometry with $49$ spacelike Killing vectors.

\section{Exceptional geometry}\label{MaxS}
\subsection{The symplectic form $\Omega$}\label{first}
In the sequel, I will focus on the $56$-dimensional Euclidean subsector of the exceptional geometry. In a first step, I restrict the symmetry group $\Diff(56)$ to the subgroup $\Symp(56)$ that is defined by preserving the antisymmetric, non-degenerate two-form $\Omega$ that I denote in coordinates as
\beg
\Omega &=&\Omega_{\mu\nu}dx^\mu\wedge dx^\nu \quad \text{with }\mu,\nu=1,\dots,56.
\eeg
Due to Darboux's theorem \cite{McDS95}, there always is a coordinate chart $(U_\alpha,x_\alpha)$ (\ref{charts}) on the manifold $\cM^{56}$ such that the symplectic form $\Omega$ has canonical form in the coordinate induced basis of the cotangent space $dx^\mu$
\be
\Omega&=&\Omega_{\mu\nu}dx^\mu\wedge dx^\nu\label{om0}\\
&=&
 2\Omega_{\alpha,\,\beta+28}dx^{\alpha}\wedge dx^{\beta+28}\nn
\ee
with $\alpha,\beta=1,\dots,28$, $\Omega^{\mu\sigma}\Omega_{\sigma\nu}:=\delta_\nu^\mu$ and the Kronecker $\delta$ in
\be\label{omr1}
\Omega_{\alpha,\,\beta+28} &=& \delta_{\alpha\beta}\\
\Rightarrow \quad \Omega^{\beta+28,\,\gamma} &=& \delta^{\gamma\beta}\nn.
\ee
Furthermore, the non-degeneracy of $\Omega$ induces a canonical isomorphism between the tangent and the cotangent space at every point $x$ in the domain of the coordinate chart $U_\alpha\subset \cM^{56}$
\be\label{canon2}
T_x\cM^d &\rightarrow & T_x^*\cM^d\\
v &\mapsto & \Omega_x(v,\cdot).\nn
\ee
As in the case of the non-degenerate metric tensor $g$ (\ref{canon1}), this isomorphism allows to identify the dual vector spaces $T_x\cM^d$ and $T_x^*\cM^d$ for all points $x\in \cM^d$. There is one big difference to the discussion of the non-degenerate metric tensor $g$ from section \ref{Gstructur}: the basis of the cotangent space in which $\Omega$ has canonical form, is \textbf{coordinate induced}. Since the coefficients $\Omega_{\mu\nu}$ (\ref{omr1}) are constant, it is hence possible to ``integrate'' the canonical isomorphism (\ref{canon2}) to the coordinates $x^\mu$ that parametrize the open set $x_\alpha(U_\alpha)\subset \R^{56}$ with $U_\alpha\subset \cM^{56}$. This leads to dual coordinates that are well-defined under coordinate transformations $\Symp(56)\subset \Diff(56)$ that preserve $\Omega$
\be\label{mapp65}
x^\mu\, \mapsto \,p_\mu\,:=\,\Omega_{\mu\nu}x^\nu.
\ee
Hence, both sets of $56$ parameters $x^\mu$ and $p_\mu$ parametrize the open subspace
$x_\alpha(U_\alpha)\subset \R^{56}$. Due to the particular form of $\Omega_{\mu\nu}$ (\ref{omr1}), the set of coordinates and dual ones $(x^\alpha,p_\beta)$ with $\alpha,\beta=1,\dots,28$ also have this property, with $p_\beta$ defined by
\be\label{omr2}
p_\beta&:=&\Omega_{\beta\nu}x^\nu\\
\Rightarrow \quad \frac{\p}{\p p_\alpha} &=& \Omega^{\mu\alpha}\p_\mu\nn.
\ee

The notation is of course borrowed from classical mechanics: the $p$ are the conjugate momenta to the coordinates $x$. The different position of the indices of $x^\alpha$ and $p_\beta$ is a reminder that these coordinates are dual to each other. The canonical transformations then correspond to the symplectic coordinate transformations $\Symp(56)\subset \Diff(56)$ on phase space. These will be the topic of the next section.

\subsection{Symplectic vector fields}\label{SympV1}
The subgroup $\Symp(56)$ of $\Diff(56)$ was defined by the coordinate transformations $\vp$ that preserve the symplectic form $\Omega$ (\ref{om0})
\be\label{om1}
\vp^*\Omega &=&\Omega.
\ee
In the present case of constant coefficients $\Omega_{\mu\nu}$ (\ref{omr1}), this is equivalent to the constraint on the corresponding vector field $X\in \mathfrak{diff}_d$
\be\label{om4}
0&=& 2\Omega_{\sigma[\nu}\p_{\mu]} X^{\sigma}.
\ee
For the special case of affine linear vector fields $X^{a}_{(A,c)} \in \mathfrak{diff}_d$ (\ref{GLGen}), equation (\ref{om4}) constrains the matrices $A\in \mathfrak{gl}_{56}$ to be antisymmetric with respect to the tensor $\Omega_{\mu\nu}$ (\ref{AntisyM2}), what I used for the definition of the symplectic Lie algebra $\mathfrak{sp}_{56}$ in section \ref{realf}.\\

In the appendix \ref{symp12}, I prove that the general solution of the constraint (\ref{om1}) is provided by the vector field $X$ with $X_\beta := \Omega_{\beta\mu}X^\mu$ (\ref{omr2}) and
\begin{subequations}\label{om9}
\be
X &=& X^\alpha \frac{\p}{\p x^\alpha} +X_{\beta} \frac{\p}{\p p_\beta}\\
\text{with}\quad X^{\alpha} &=&\frac{\p}{\p p_\alpha} H(x,p)\\
X_{\beta} &=&-\frac{\p}{\p x^\beta} H(x,p)
\ee
\end{subequations}
with an arbitrary analytic scalar function $H(x,p)$ of the $56$ real coordinates $x^\nu$ that I chose to parametrize by $(x,p)$. In the language of Classical Mechanics, $H(x,p)$ is the generating function of canonical transformations, the Hamiltonian. It is nice to observe that this special class of vector fields in $56$ dimensions contains general vector fields $Y(x)\in \mathfrak{diff}_{28}$ (\ref{Vectorfields}) in $28$ dimensions by setting $H=p_\alpha Y^\alpha(x)$. This implies that the subgroup $\Symp(56)$ of $\Diff(56)$ is infinite dimensional.\footnote{This feature of symplectic geometry is in sharp contrast to Riemannian geometry: asking for the isometries of e.g. the Minkowski metric $\eta$ (\ref{Isometry1}), i.e. diffeomorphisms $\vp_A\in \Diff(56)$ with $\vp_A^*\eta =\eta$, leads to the finite dimensional Poincar\'e group that I showed in appendix \ref{Isometry}.}\\

At the beginning of this thesis in \ref{GeomStr}, I defined physical fields $\psi$ as sections of a tensor bundle over a Lorentzian manifold. This implies that the action of a general coordinate transformation $\vp\in \Diff(d)$ on $\psi$ is provided by a multiplication with the Lorentz action $O(\vp,e)$ that is induced by the Jacobian matrix $\frac{\p \vp}{\p x}$ after having fixed an arbitrary Lorentz gauge for the vielbein (\ref{VielbeinTrafo}). For the $56$-dimensional subsector under consideration in this thesis, the signature is Euclidean. Therefore, the induced Lorentz action $O(\vp,e)$ is via the compact group $SO(56)$.\\

The additional structure of a conserved symplectic form $\Omega$ allows to restrict the induced action $O(\vp,e)$ to a proper subgroup of $SO(56)$. To show this, start with expanding the Jacobian of an arbitrary symplectomorphism $\Symp(56)\subset \Diff(56)$ to linear order in its associated vector field $X$. With the equations (\ref{om9}), it has the form
\be\label{Symp7}
\frac{\p X^\mu}{\p x^\nu} &\stackrel{(\ref{om9})}{=}&\left(
\begin{tabular}{cc}
$\delta_\nu^\gamma \delta_\alpha^\mu\frac{\p}{\p x^\gamma}\frac{\p}{\p p_\alpha} $ 
&
$-\delta_\nu^\gamma \Omega^{\mu \beta}\frac{\p}{\p x^\gamma}\frac{\p}{\p x^\beta}$\\
   $\Omega_{\delta\nu }\delta_\alpha^\mu \frac{\p}{\p p_\delta}\frac{\p}{\p p_\alpha}$
& 
$-\Omega_{\delta\nu }\Omega^{\mu \beta}\frac{\p}{\p p_\delta}\frac{\p}{\p x^\beta}$ 
\end{tabular}
\right)H(x,p).
\ee
It is obvious that not all entries in this matrix are independent.\\

Next, observe that the matrix $\frac{\p X}{\p x}$ is $\mathfrak{gl}_{56}$ valued for a general vector field in $\mathfrak{diff}_{56}$. Since the antisymmetric matrices $\mathfrak{so}_{56}$ form a subalgebra of $\mathfrak{gl}_{56}$ (\ref{AntisyM}), I can always decompose $\frac{\p X}{\p x}$ into compact generators $\hat{L}$ (\ref{LDefi}) and noncompact ones $\hat{S}$ (\ref{RDefi})
\beg
\frac{\p X^\mu}{\p x^\nu}{\left.\hat{M}\right.^\nu}_\mu &=& \frac{\p X^\mu}{\p x^\nu}\left({\left.\hat{L}\right.^\nu}_\mu+{\left.\hat{S}\right.^\nu}_\mu\right)
\eeg
with respect to the Euclidean metric $\eta=\id_{56}$. In order to better handle the independent components of the real matrix $\frac{\p X}{\p x}$ (\ref{Symp7}) in this splitting, I perform the following change of coordinates
\begin{subequations}\label{zNeu}
\be
z^\alpha &:=& x^\alpha +i \eta^{\alpha\beta}p_\beta\\
\bar{z}^\alpha &:=& x^\alpha -i \eta^{\alpha\beta}p_\beta
\ee
\end{subequations}
with $\Omega_{\alpha,\,\beta+28} = \eta_{\alpha\beta}$ (\ref{omr1}) and $i^2=-1$. Substituting these coordinates (\ref{zNeu}) in the definition of the symplectic form $\Omega$ (\ref{om0}), I obtain for the real tensor $\Omega$
\be\label{Omega6a}
\Omega &=&i\Omega_{\alpha,\,\beta+28}dz^{\alpha}\wedge d\bar{z}^{\beta}.
\ee
In analogy to the definition of dual coordinates $p_\alpha$ from the coordinates $x^{\alpha+28}$ by the non-degenerate form $\Omega$ (\ref{omr2}), I can lower the index of the coordinate $\bar{z}$ with $\Omega$
\be\label{lower5}
\bar{z}_\beta &:=& \Omega_{\beta,\,\alpha+28}\bar{z}^\alpha.
\ee
In these coordinates, the compact and the non-compact part of the matrix $\frac{\p X}{\p x}$ (\ref{Symp7}) take the simple form
\begin{subequations}\label{Unity}
\be
\frac{\p X^\mu}{\p x^\nu}{\left.\hat{L}\right.^\nu}_\mu &=:&
\left(\frac{2}{i}\frac{\p}{\p z^\alpha}\frac{\p}{\p \bar{z}_\beta}H(z,\bar{z})\right)
{\left.\hat{M}\right.^\alpha}_\beta
\\
\frac{\p X^\mu}{\p x^\nu}{\left.\hat{S}\right.^\nu}_\mu &=:&\left(
\frac{2}{i}\frac{\p}{\p \bar{z}_\alpha}\frac{\p}{\p \bar{z}_\beta}H(z,\bar{z})\right)\hat{S}_{\alpha\beta}
+\text{c.c.}.
\ee
\end{subequations}
This defines the vector field representation $\mathbf{R}$ of the generators $\hat{M}$ and $\hat{S}$ in the coordinates $(z,\bar{z})$ (\ref{zNeu})
 \begin{subequations}\label{MSRep}
 \be
\mathbf{R}_{{\left.\hat{M}\right.^\alpha}_\beta}
&=&
 z^\alpha \frac{\p}{\p z^\beta}
 -\bar{z}_\beta \frac{\p}{\p \bar{z}_\alpha} 
 \label{U28}\\
  \mathbf{R}_{\hat{S}_{\alpha\beta}}\label{SRep}
&=&
\frac{1}{2} \left(
 \bar{z}_\alpha\frac{\p}{\p z^\beta} + \bar{z}_\beta \frac{\p}{\p z^\alpha}\right)
 \ee
 \end{subequations}
The Hamiltonian $H(z,\bar{z}) =H(x,p)$ in (\ref{Unity}) is real and the abbreviation ``+c.c.'' implies that the complex conjugate has to be added. With the vector field representation (\ref{U28}), it is straightforward to prove the commutation relation
\be\label{CRU28}
 \left[{\left.\hat{M}\right.^\alpha}_\beta,{\left.\hat{M}\right.^\gamma}_\delta\right]
&=&
\delta_\beta^\gamma {\left.\hat{M}\right.^\alpha}_\delta
  -\delta_\delta^\alpha{\left.\hat{M}\right.^\gamma}_\beta .
  \ee
These generators are compact by definition (\ref{Unity}). Since they fulfill the commutation relation of the complex Lie algebra of all $28\times 28$ matrices, they span its real compact form $\mathfrak{u}_{28}\subset \mathfrak{so}_{56}$. This also is the reason why I chose the same symbol $\hat{M}$ for the generators of both real forms $\mathfrak{gl}$ and $\mathfrak{u}$ of the same Lie algebra. It should be clear from the context and from the different indices, which real form is referred to. Using the symplectic form $\Omega$ to raise and lower the indices $\Omega_{\alpha,\beta+28}=\eta_{\alpha\beta}$ (\ref{lower5}), the vector field representation $\mathbf{R}$ (\ref{U28}) implies that the generators $\hat{M}$ of $\mathfrak{u}_{28}$ are anti-hermitean
\be\label{Mcc}
{\left.\hat{M}\right.^\alpha}_\beta &=& -\eta_{\beta\delta}\overline{{\left.\hat{M}\right.^\delta}_\gamma}\eta^{\gamma\alpha}.
\ee
I want to emphasize that the appearance of the complex unity $i=\sqrt{-1}$ does not mean that I have complexified the coordinates $x^\nu$. The matrix equations in (\ref{Unity}) are still real valued. This is just a more efficient way to write the generators. To put it in other words, the compact generators of $Sp(56)$ form a real representation of the unitary group $U(28)$. This is most easily seen by a decomposition of the algebras $\mathfrak{sp}_{56}$ and $\mathfrak{so}_{56}$ in $\mathfrak{u}_{28}$ representations.
\begin{subequations}\label{spsou}
\be
\mathfrak{sp}_{56} &=& \mathfrak{u}_{28}\oplus  \langle\hat{S}_{(\alpha\beta)} +\text{c.c.}\rangle_\R
\oplus \langle i\hat{S}_{(\alpha\beta)} +\text{c.c.}\rangle_\R\label{SPU}\\
\mathfrak{so}_{56} &=& \mathfrak{u}_{28}\oplus \langle\hat{T}_{[\alpha\beta]} +\text{c.c.}\rangle_\R\oplus \langle i\hat{T}_{[\alpha\beta]} +\text{c.c.}\rangle_\R\label{SOU}\\
\Rightarrow \quad \mathfrak{u}_{28} &=& \mathfrak{sp}_{56}\cap \mathfrak{so}_{56}
\ee
\end{subequations}
Before explaining why the $O(\vp,e)$-action induced by a symplectomorphism $\vp\in \Symp(56)$ on a physical field $\psi$ can be chosen to be in $U(28)\subset SO(56)$, I want to add comments concerning the holomorphic coordinates $(z,\bar{z})$ (\ref{zNeu}).

\subsection{Comments on the holomorphic frame}\label{holom1}
The appearance of the complex coordinates $(z,\bar{z})$ (\ref{zNeu}) should not be confused with a complex structure on a manifold \cite{H00}. Their introduction only is a reparametrization of the coordinates that suits best the real vector fields of $\Symp(56)$. The underlying manifold is not complex, because the symplectomorphisms with non-compact Jacobian do not preserve $\eta$ and hence perturb a possible complex structure ${J_\rho}^\mu := \Omega_{\rho\nu}\eta^{\nu\mu}$. I emphasize that $\eta_{\mu\nu}$ is not the metric in the coordinate frame, it merely is the Euclidean form that distinguishes compact generators from non-compact ones.\\

Nonetheless, it is clear from the definition of the holomorphic frame (\ref{zNeu}) that the holomorphic one-forms $dz^\alpha$ and the antiholomorphic ones $d\bar{z}_\beta$ do span the real cotangent space to the $56$-dimensional real manifold
\be\label{holframe}
T^*\cM^{56} &=& \langle  dz^\alpha+\eta^{\alpha\beta}d\bar{z}_\beta\rangle_\R \oplus i\langle dz^\alpha-\eta^{\alpha\beta}d\bar{z}_\beta\rangle_\R.
\ee

A last comment concerns the normalization of the dual coordinate $\bar{z}_\beta$ (\ref{lower5}). From the point of view of symplectic geometry, the normalization of the dual coordinates should follow the one used for $p$ (\ref{omr2}), which was deduced from the explicit form of $\Omega$ (\ref{om0}). With the equation (\ref{Omega6a}), this would lead to
\beg
\bar{z}^{(sy)}_\beta &:=& \frac{i}{2}\Omega_{\beta,\,\alpha+28}\bar{z}^\alpha.
\eeg
From the point of view of complex geometry however, it appears to be natural to define the dual object of the holomorphic coordinate $z$ simply by a complex conjugation. This is the reason why I have used the normalization (\ref{lower5})
\beg
\bar{z}_\beta &:=& \Omega_{\beta,\,\alpha+28}\bar{z}^\alpha.
\eeg
This will explain the appearance of the factor $i$ in the relation of the vielbein $e^H$ to its inverse (\ref{vili2}).

\subsection{Connections with symplectic equivariance}\label{LCc3}
For any physical theory with a symmetry $G$, a connection $\nabla$ with $G$-equivariance is an essential building block. In this section, I will discuss the implications on $\nabla$ of softening the constraint of $\Diff(56)$-equivariance to $\Symp(56)$ or symplectic equivariance.\\

To commence, recall from section \ref{SympV1} that restricting the diffeomorphisms $\Diff(56)$ to symplectomorphisms $\Symp(56)$ implies that the Jacobian matrix is symplectic (\ref{om1}). The transformation of the vielbein $e$ (\ref{VielbeinTrafo})
\beg
{e'_\mu}^a &=& \frac{\p x^\nu}{\p {x'}^\mu}\,{e_\nu}^b \,{O(\vp,e)_b}^a\\
\text{with}\quad x'&=&\vp(x)\quad \text{and }\vp\in \Symp(56)
\eeg
reveals that it is consistent to restrict the vielbein matrix to be an $x$-dependent $Sp(56)$ element. The orthogonal right action that compensates for the left action by the Jacobian, must hence also be a an element in $Sp(56)$. Since it also has to be orthogonal, the comparison of the corresponding Lie algebra generators (\ref{spsou}) proves that it has to be unitary
\beg
U(28)&=&Sp(56)\cap SO(56).
\eeg
In section \ref{Representation}, I showed that a general physical field $\psi$ transforms as a representation of the Lorentz algebra. For the subsector under consideration, this implies that all fields $\psi$ must transform as $\mathfrak{so}_{56}$ representations. If the vielbein matrix $e$ is an element of $Sp(56)$ with arbitrary $x$-dependence, restricting $\Diff(56)$ to the symplectomorphisms $\Symp(56)$ implies that the induced action on $\psi$ is by the covering group of $U(28)\subset SO(56)$. Hence, a general symplectomorphism does not link irreducible $\mathfrak{u}_{28}$ representations to each other. It is therefore consistent to treat all $\mathfrak{u}_{28}$ representations as independent physical fields. This is the first important message of this section:
\begin{center}
\textit{ Restricting the diffeomorpisms $\Diff(56)$ to symplectomorphisms $\Symp(56)$ allows to restrict the vielbein matrix to be symplectic $e(x)\in Sp(56)$. Then, it is admissible to interpret $\mathfrak{u}_{28}$ representations as physical fields in $d=56$. These do not have to be $Spin(56)$ representations.}
\end{center}

If the vielbein $e$ is symplectic, the symplectic form $\Omega$ has the same constant coefficients (\ref{omr1}) in both the coordinate induced frame $dx^\mu$ and the vielbein frame $dx^a={e_\mu}^a dx^\mu$ (\ref{dualf}), because the two frames differ by a multiplication with a matrix in $Sp(56)$, whose definition is the invariance of $\Omega$ (\ref{AntisyM2}). Then, the definition of the holomorphic coordinates $(z^\alpha,\bar{z}_\beta)$ (\ref{zNeu}) induces a holomorphic vielbein frame
\begin{subequations}\label{holframe2}
\be
\underline{dz}^\alpha &:=& \left(\delta^\alpha_a +i \delta^{\alpha}_{a-28}\right) {e_\mu}^a dx^\mu\\
\underline{d\bar{z}}_\beta &:=& \Omega_{\beta\,\alpha+28}\left(\delta^\alpha_a -i \delta^{\alpha}_{a-28}\right) {e_\mu}^a dx^\mu
\ee
\end{subequations}
with the indices $\mu,a=1,\dots,56$ and $\alpha,\beta=1,\dots,28$ and the sympletic form $\Omega$ to change their position $\Omega_{\beta\,\alpha+28}=\delta_{\beta\alpha}$ (\ref{omr1}).  The frame $(\underline{dz}^\alpha,\underline{d\bar{z}}_\beta)$ spans the cotangent space in the same way as the coordinate induced frame (\ref{holframe}). I want to emphasize that this definition only is possible, if the vielbein is an $Sp(56)$ matrix.\\

In this frame, a general connection (\ref{Conn56}) has the form
\begin{subequations}\label{Ccomp3}
\be
\underline{dz}^\alpha \nabla_\alpha \psi^\gamma &=:& 
		\underline{dz}^\alpha\left(\frac{\p}{\p z^\alpha}\psi^\gamma 
		-{\Big(\omega_{\alpha}\Big)_ c}^d \delta_{{\left.\hat{L}\right.^c}_d}\psi^\gamma\right)
		\\
				\underline{d\bar{z}}_\beta \bar{\nabla}^\beta \psi^\gamma &=:& 
		\underline{d\bar{z}}_\beta\left(\frac{\p}{\p \bar{z}_\beta}\psi^\gamma
				-
				{\left(\bar{\omega}^\beta\right)_{ c}}^d \delta_{{\left.\hat{L}\right.^c}_d}\psi^\gamma\right).
\ee
\end{subequations}
Next, recall from chapter \ref{CHAP4} that the Levi--Civita connection in the vielbein frame (\ref{SpinCon}) is constructed from the Maurer--Cartan form (\ref{omegaLin}) involving the vielbein matrix $e(x)$. For the case of $e(x)\in Sp(56)$, the compact and the non-compact parts of the Lie algebra valued coset $e^{-1}de$ follow the same decomposition as the one used in (\ref{Unity}). Without loss of generality, I can fix the normalization for the coefficients of the $\mathfrak{u}_{28}$ generators $\hat{M}$ and of $\hat{S}_{\alpha\beta}$ (\ref{MSRep}) to the canonical ones by
\begin{subequations}\label{vRestr}
\be
{\left(v^{\text{min}}_a\right)_c}^d {\left.\hat{L}\right.^c}_d
&=:&
{\left(v^{\text{min}}_a\right)_\alpha}^\beta {\left.\hat{M}\right.^\alpha}_\beta
\\
 {\left(v_a\right)_c}^d {\left.\hat{S}\right.^c}_d
 &=:&
 \left(v_a\right)^{\alpha\beta} \hat{S}_{\alpha\beta} +\text{c.c.}.
\ee
\end{subequations}
With these normalizations, it is a straightforward computation to obtain the $\mathfrak{so}_{56}$ valued one-form $\omega$ of the Levi--Civita connection in the vielbein frame (\ref{SpinCon}) with $K=0$. It is important to observe that $\omega$ is not $\mathfrak{u}_{28}$ valued, as one might have guessed. Recalling the antisymmetric generators $\hat{T}_{\alpha\beta}$ from the decomposition of $\mathfrak{so}_{56}$ under $\mathfrak{u}_{28}$ (\ref{SOU}), whose vector field representation is normalized to
\be\label{TVF}
\mathbf{R}_{\hat{T}_{\alpha\beta}} &:=& \bar{z}_{[\alpha}\frac{\p}{\p z^{\beta]}},
\ee
$\omega$ has the form
\begin{subequations}\label{sympCon}
\be
{\Big(\omega_{\alpha }\Big)_c}^d {\left.\hat{L}\right.^c}_d 
&=& 
\left({\left(v^{\text{min}}_\alpha\right)_\e}^\rho +\left(\bar{v}^\rho\right)_{\e\alpha}\right){\left.\hat{M}\right.^\e}_\rho
   -2\left(v_{\kappa}\right)_{\e\alpha}\overline{\hat{T}^{\kappa\e}}
   \\
      {\left(\bar{\omega}^\beta\right)_{ c}}^d {\left.\hat{L}\right.^c}_d 
&=& 
\left({\left(\bar{v}_{\text{min}}^\beta\right)_\e}^\rho -\left(v_\e\right)^{\rho\beta}\right){\left.\hat{M}\right.^\e}_\rho
   -
   2\left(\bar{v}^{\kappa}\right)^{\e\beta}\hat{T}_{\kappa\e}
\ee
\end{subequations}
with the pair of forms $(v^{\text{min}}_\alpha,\bar{v}_{\text{min}}^\beta)$ and $(v_\alpha,\bar{v}^\beta)$ (\ref{vRestr}) in the frame $(\underline{dz}^\alpha,\underline{d\bar{z}}_\beta)$ (\ref{holframe2}).\footnote{I want to point out a subtlety in the notation: $\overline{(v_\kappa)_{\alpha\beta}} = (\bar{v}^\kappa)^{\alpha\beta} \neq(v_\kappa)^{\alpha\beta}$. The bar in $(\bar{v})$ only corresponds to the frame $(\underline{dz}^\alpha,\underline{d\bar{z}}_\beta)$ (\ref{holframe2}).}\\

The observation that $\mathfrak{u}_{28}$ representations may be treated as physical fields in $d=56$ also has an implication for the covariant derivative. Starting with a physical field $\psi$ which transforms as an $\mathfrak{so}_{56}$ representation, the action of a symplectomorphism on $\psi$ is by a $U(28)$ rotation in the vielbein frame. This implies that the partial derivative of the field $\p\psi$ can be covariantized by adding a $\mathfrak{u}_{28}$ action alone. The Levi--Civita connection for an $Sp(56)$ valued vielbein is not $\mathfrak{u}_{28}$ valued, because it also contains the generators $\hat{T}$ (\ref{sympCon}). On the other hand, the Levi--Civita connection is $\Diff(56)$-equivariant by construction, which implies in particular $\Symp(56)$-equivariance for $e(x)\in Sp(56)$. Therefore, the coefficients $(v_{[\kappa})_{\e]\alpha}$ and $(\bar{v}^{[\kappa})^{\e]\beta}$ of the generators $\hat{T}$ of $\mathfrak{so}_{56}\ominus\mathfrak{u}_{28}$ in (\ref{sympCon}) must transform as a tensor. This can also be shown explicitly by using the fact that the vielbein is a symplectic matrix.\\

Hence, $(v_{[\kappa})_{\e]\alpha}$ is a $U(28)$ tensor under the action induced by a symplectomorphism $\vp\in \Symp(56)$. It is of first order in derivatives and is constructed from vielbein degrees of freedom alone (\ref{vRestr}). Note that this is a first example for the case mentioned in section \ref{Rest} in which symplectic equivariance is strictly weaker than $\Diff(56)$-equivariance. It also implies that there is a one-parameter family of connections on $\mathfrak{so}_{56}$ representations $\psi$ that are characterized by their $\mathfrak{so}_{56}$ valued one-form $\omega$ and $c\in \R$
\begin{subequations}\label{sympCon2}
\be
{\Big(\omega_{\alpha }\Big)_c}^d {\left.\hat{L}\right.^c}_d 
&=& 
\left({\left(v^{\text{min}}_\alpha\right)_\e}^\rho +\left(\bar{v}^\rho\right)_{\e\alpha}\right){\left.\hat{M}\right.^\e}_\rho
   -2c\left(v_{\kappa}\right)_{\e\alpha}\overline{\hat{T}^{\kappa\e}}
   \\
      {\left(\bar{\omega}^\beta\right)_{ c}}^d {\left.\hat{L}\right.^c}_d 
&=& 
\left({\left(\bar{v}_{\text{min}}^\beta\right)_\e}^\rho -\left(v_\e\right)^{\rho\beta}\right){\left.\hat{M}\right.^\e}_\rho
   -
   2c\left(\bar{v}^{\kappa}\right)^{\e\beta}\hat{T}_{\kappa\e}.
\ee
\end{subequations}

For the definition of the supersymmetry transformations, it will be of vital importance to obtain a connection that does not have to act on $\mathfrak{so}_{56}$ representations. A first step is to restrict to $\mathfrak{u}_{28}$ representations. However, these can only be interpreted consistently as physical fields $\psi$, if the connection does not contain any generators in $\mathfrak{so}_{56}\ominus \mathfrak{u}_{28}$, because $\mathfrak{so}_{56}$ is a simple Lie algebra. Therefore, there is only one $\Symp(56)$-equivariant connection in the family (\ref{sympCon2}) that preserves $\mathfrak{u}_{28}$ representations, which is the one with $c=0$. Equivalently, it is the projection of the Levi--Civita connection in the vielbein frame on the generators in $\mathfrak{u}_{28}$
\be\label{projCon}
\nabla &:=&\text{pr}_{\mathfrak{u}_{28}}\nabla^{(LC)}.
\ee
With this connection $\nabla$, it is consistent to describe physical fields as $U(28)$ representations. It is not free of torsion, however. A calculation analogous to the one performed in appendix \ref{Expl1} with the definition (\ref{Ccomp3}) results in 
\begin{subequations}\label{TorCon3}
\be
\Big[\nabla_\alpha,\nabla_\beta\Big] &=&-2(v_{[\alpha})_{\beta]\e} \bar{\nabla}^\e
-
{{\Big(R_{\alpha\beta}\Big)}_\gamma}^\e\delta_{{\left.\hat{M}\right.^\gamma}_\e}
\\
\left[\nabla_\alpha,\bar{\nabla}^\beta\right] 
&=&
-{{\left({R_\alpha}^{\beta}\right)}_\gamma}^\e\delta_{{\left.\hat{M}\right.^\gamma}_\e}
\\
\left[\bar{\nabla}^\alpha,\bar{\nabla}^\beta\right] &=&-2(\bar{v}^{[\alpha})^{\beta]\e} \nabla_\e
-
{{\left(\bar{R}^{\alpha\beta}\right)}_e}^f\delta_{{\left.\hat{M}\right.^\gamma}_\e.}
\ee
\end{subequations}
The curvature tensors in this expression have to be $\mathfrak{u}_{28}$ valued, because they result from a commutator of one $\mathfrak{u}_{28}$ valued connection with another one, which again maps to $\mathfrak{u}_{28}$.\\

The standard way to describe the dynamics of a theory with $\Symp(56)$-invariance is to construct an action from an invariant scalar object. Following the discussion from section \ref{gravmatter2}, the curvature tensor can always be interpreted as an endomorphism of the vector space $\mathfrak{so}_{56}$. Since the three tensors $R$ in (\ref{TorCon3}) are $\mathfrak{u}_{28}$ valued, this map has the form
\beg
R:\quad \mathfrak{u}_{28} &\rightarrow & \mathfrak{so}_{56}.
\eeg
As in the case of $\Diff(d)$-invariance (\ref{Ricciscalar}), its trace is an invariant scalar 
\be\label{RicciN}
\tilde{R}&=&tr_{\mathfrak{so}_{56}}(R)\nn\\
&=&
{{\left({R_\alpha}^{\beta}\right)}_\gamma}^\e\left({\left.{\left.\hat{M}\right.^\alpha}_\beta,\,\hat{M}\right.^\gamma}_\e\right)
\ee
with the Killing norm $(\cdot,\cdot)$ introduced in section \ref{realf}. It is important to note that this scalar is not $\Diff(56)$-invariant, because it differs from the Ricci scalar (\ref{Ricciscalar}) by terms proportional to the $\Symp(56)$-covariant tensor $(v_{[\kappa})_{\e]\alpha}$ that is not $\Diff(56)$-covariant. Furthermore, the parts $\left(R_{\alpha\beta}\right)$ and its complex conjugate (\ref{TorCon3}) do not contribute to the invariant scalar $\tilde{R}$ in this trace. In contradistinction to the $\Diff(56)$-invariant case, there is hence no unique action for the vielbein that is of second order in derivatives in the restricted case. \\

Before discussing the final constraint on the geometry by imposing the invariance of a symmetric quartic tensor $Q$, I would like to address the question of symmetry enhancement. This is related to the algebraic point of view of chapter \ref{CHAP4}, but there are subtleties to observe: if the affine coset from section \ref{secB} is taken as a starting point with a symplectic vielbein $e(x)\in Sp(56)$, it would be difficult to obtain the Levi--Civita connection (\ref{sympCon}) directly, because the generator $\hat{T}$ (\ref{TVF}) is not present in the affine coset at all. Furthermore, the joint realization of conformal and affine symmetries in section \ref{secD} crucially depended on the fact that $Gl(d)$ was not simple: it was the determinant of the vielbein that was used to single out the connections with $\Diff(d)$-equivariance from the class of connections with affine linear equivariance (\ref{final1}). Since $Sp(56)$ is simple, an analogous procedure to the one of Borisov \& Ogievetsky that only discusses finite dimensional subgroups of $\Diff(d)$, is impossible. \\

Nevertheless, there is one important argumentation related to the procedure of Borisov \& Ogievetsky that still holds. There is a connection $\nabla$ acting on $\mathfrak{u}_{28}$ representations $\psi$ with equivariance under the finite dimensional group $Sp(56)\subset A(56)$ that allows for a symmetry enhancement. If and only if the constant relating the antisymmetric matrices $(v_\alpha^{\text{min}},v^\beta_{\text{min}})$ to the symmetric ones $(v_\alpha,\bar{v}^\beta)$ (\ref{vRestr}) is fixed as in (\ref{sympCon2}), then the equivariance group is enlarged to the infinite dimensional group $\Symp(56)$, which is analogous to the symmetry enhancement of $A(d)$ to $\Diff(d)$ in section \ref{secD}.\footnote{Another coupling than the one in (\ref{sympCon2}) can be ruled out, because the objects $(v_\e)^{\rho\alpha}$ and $(\bar{v}^\rho)_{\e\alpha}$ do not transform as $\Symp(56)$ tensors.}\\

It is this symmetry enhancement that is the reason why the connection $\nabla$ (\ref{projCon}) is in a way preferred to other connections with mere $Sp(56)$-equivariance. After imposing the invariance of the quartic tensor $Q$, a similar enlargement of the equivariance group will not be possible any more. Hence, there will be no preferred connection from a symmetry point of view in the exceptional geometry of the following sections.

\subsection{The quartic symmetric tensor $Q$}\label{ESy}
In the preceding sections, I restricted the Lie group $\Diff(56)$ to the subgroup $\Symp(56)$ by requiring that an antisymmetric two tensor $\Omega$ be conserved (\ref{om1}). In the same way, I want to restrict $\Symp(56)$ to a subgroup that preserves a totally symmetric quartic tensor 
\be\label{Q0}
Q&=& Q_{(\mu_1\dots \mu_4)}dx^{\mu_1}\otimes dx^{\mu_2} \otimes dx^{\mu_3}\otimes dx^{\mu_4}
\ee
in addition to the symplectic form $\Omega$. Before I state this tensor explicitly, I want to adapt the notation to this task. Recall that for the discussion of the symplectic two-form $\Omega$ (\ref{om0}), it was a good choice to split the $56$ coordinates $x^\mu$ into two sets of $28$ coordinates $(x^\alpha,p_\beta)$ in section \ref{first}. In the same way, I want to split both sets of $28$ coordinates into $7+21$ each $(x^m,x^{[mn]},p_m,p_{[mn]})$, which will turn out to be suitable for this quartic tensor. The identification is as follows
\be\label{xDefi8}
x^m &:=& \delta^m_\alpha x^\alpha \quad m=1,\dots,7\\
x^{mn}&:=& \delta^{[mn]}_{\alpha-7}x^\alpha \quad m,n=1,\dots,7\nn\\
\text{with }\quad \delta^{[1\ 2]}_{8-7}&=&1\quad \text{et cetera}\nn.
\ee
The same decomposition is used for the dual coordinate $p$. In this notation, the quartic tensor with the symmetric tensor product $\circ$ has the form\footnote{$\e$ is the completely antisymmetric tensor in $d=7$ dimensions with the normalization $\e^{1234567}=1$.}
\be\label{Q2k}
Q&=&
-dp_m \circ dx^m \circ dp_n \circ dx^n 
\\
&&
+2dp_m \circ dx^m \circ dp_{pq} \circ dx^{pq} 
+8dp_m \circ dx^{mn} \circ dp_{nq} \circ dx^{q} \nn\\
&&
 -\frac{\sqrt{2}}{6}\left(\e_{m_1\dots m_7}dx^{m_1}\circ dx^{m_2m_3}\circ dx^{m_4m_5}\circ dx^{m_6m_7}
\right.\nn\\
&&\left.
+ \e^{m_1\dots m_7}dp_{m_1}\circ dp_{m_2m_3}\circ dp_{m_4m_5}\circ dp_{m_6m_7}\right)\nn\\
&&
+dp_{cd}\circ dx^{cd} \circ dx^{ef}\circ dp_{ef} 
-
4dx^{ab} \circ dp_{be}\circ  dx^{ef}\circ dp_{fa}.
\nn
\ee
In constructing the class of connections with affine linear equivariance from the affine coset in the procedure of Borisov \& Ogievetsky of chapter \ref{CHAP4}, I needed to split compact from non-compact generators. This split was the reason for introducing the (anti)holomorphic coordinates $(z,\bar{z})$ (\ref{zNeu}) in section \ref{SympV1}. These naturally induce the definition of (anti)holomorphic coordinates for the real ones $(x,p)$ defined in (\ref{xDefi8})
\begin{subequations}\label{zNeu2}
\be
z^m &:=& x^m + i \eta^{mr}p_r,\\
z^{mn}&:=& x^{mn} + i \eta^{mr}\eta^{ns}p_{rs}.
\ee
\end{subequations}
It will prove sensible for a better understanding of $Q$ (\ref{Q2k}) to recombine these coordinates again by defining with $\tau_7,\tau_8\in \R\backslash\{0\}$ and $M_1,M_2=1,\dots,8$
\be\label{zNeu3}
z^{M_1M_2}&:=&\frac{1}{i\tau_7}{\G_m}^{M_1M_2}z^m +\frac{1}{\tau_8}{\G_{mn}}^{M_1M_2}z^{mn}.
\ee
The $8\times 8$ matrices $\G_m$ that I introduced in the appendix \ref{Cliff}, fulfill the Clifford property $\{\G_m,\G_n\}=2\eta_{mn}$. For a Euclidean signature of $\eta$ in $d=7$, the matrices $\G_m$ have to be purely imaginary. Therefore, the coefficients of $z^m$ and $z^{mn}$ in (\ref{zNeu3}) are real. Hence, the antiholomorphic coordinate $\bar{z}$ follows from complex conjugation:
\be\label{zNeu4}
\bar{z}^{M_1M_2}&:=&\frac{1}{i\tau_7}{\G^m}^{M_1M_2}\bar{z}_m +\frac{1}{\tau_8}{\G^{mn}}^{M_1M_2}\bar{z}_{mn}.
\ee
Next recall from the definition (\ref{lower5}) that the position of the $Gl(7)$ indices $m$ and $mn$ of $\bar{z}$ are lowered with the symplectic form $\Omega$. I also want to use $\Omega$ to lower the indices of $\bar{z}^{M_1M_2}$. For this to work, the symplectic form $\Omega$ (\ref{Omega6a}) has to have canonical form in this frame, which is with the Kronecker $\delta$
\be\label{Omega6}
\Omega &=&\frac{i\tau_7^2}{8}\delta_{[M_1M_2],\,[N_1N_2]}dz^{M_1M_2}\wedge d\bar{z}^{N_1N_2}.
\ee
Substituting the definition (\ref{zNeu3}) in $\Omega$ (\ref{Omega6a}) only leads to the form (\ref{Omega6}) with the Clifford completeness relation (\ref{complete}), if the coefficients $\tau_7$ and $\tau_8$ are related by
\be\label{chi8fix}
\tau_8 &=& \tau_7\sqrt{2}.
\ee
Hence, the constant $\tau_8$ is fixed by requiring that $\Omega$ has canonical form in the coordinates (\ref{zNeu3}).
This allows to define in complete analogy to the lowering of the indices of the antihomorphic coordinate $\bar{z}$ in (\ref{lower5})
\be\label{lower6}
\bar{z}_{M_1M_2}&:=&\delta_{[M_1M_2],\,[N_1N_2]}\bar{z}^{N_1N_2}.
\ee
In this frame, the Hamiltonian vector fields (\ref{om9}) take the form
\begin{subequations}\label{om9c}
\be
X &=& X^{M_1M_2} \frac{\p}{\p z^{M_1M_2}} +\bar{X}_{M_1M_2} \frac{\p}{\p \bar{z}_{M_1M_2}}\\
\text{with}\quad 
X^{M_1M_2} &=&-\frac{16i}{\tau_7^2}\frac{\p}{\p \bar{z}_{M_1M_2}}H(z,\bar{z})\\
\bar{X}_{M_1M_2}&=&\frac{16i}{\tau_7^2}\frac{\p}{\p z^{M_1M_2}}H(z,\bar{z}).
\ee
\end{subequations}
Before resuming the discussion of the quartic tensor $Q$, I want to comment on the role of the constant $\tau_7\in \R\backslash \{0\}$ introduced in (\ref{zNeu3}). It is a global factor that links the two holomorphic frames. I have already mentioned in section \ref{holom1} that there is a discrepancy between natural choices of complex geometry and of symplectic geometry. For this thesis, I chose the conventions of complex geometry. Therefore, dual indices of the tangent space differ from cotangent space indices by a constant factor that is proportional to the factor $i\tau_7^2/16$ relating the coefficients of the symplectic form $\Omega$ in the frames (\ref{om0}) and (\ref{Omega6}). This will be the factor relating the vielbein $e^H$ to its inverse in (\ref{vili2}). In order to match the standard conventions of $d=11$ supergravity and to obtain simple formulas for the supersymmetry variations, I will fix $\tau_7$ in section \ref{Super}.\\

The introduction of the coordinates (\ref{zNeu3}) also leads to a nicer form for the quartic tensor $Q$ (\ref{Q2k}): 
\be\label{Q2}
Q
&=&
Q_{A_1\dots A_8}dz^{A_1A_2}\circ dz^{A_3A_4}\circ dz^{A_5A_6}\circ dz^{A_7A_8}\nn\\
&&+6{Q_{A_1A_2\,B_1B_2}}^{C_1C_2\,D_1D_2}dz^{A_1A_2}\circ dz^{B_1B_2}\circ d\bar{z}_{C_1C_2}\circ d\bar{z}_{D_1D_2}
\nn\\
&&+\bar{Q}^{A_1\dots A_8}d\bar{z}_{A_1A_2}\circ d\bar{z}_{A_3A_4}\circ d\bar{z}_{A_5A_6}\circ d\bar{z}_{A_7A_8}\nn
\ee
with
\begin{subequations}\label{QDefi}
\be
Q_{A_1\dots A_8} &=& \frac{\tau_7^4}{2^6 96}\e_{A_1\dots A_8}\\
\bar{Q}^{A_1\dots A_8} &=& \frac{\tau_7^4}{2^6 96}\e^{A_1\dots A_8}\\
{Q_{A_1A_2\,B_1B_2}}^{C_1C_2\,D_1D_2}&=&\frac{\tau_7^4}{2^6 6}\big(\delta_{(A_1A_2}^{D_2C_1}\delta_{B_1B_2)}^{C_2D_1} -\frac{1}{4}\delta_{(A_1A_2}^{C_1C_2}\delta_{B_1B_2)}^{D_1D_2}\big).
\ee
\end{subequations}
In the last line, the symmetry implies the interchange of pairs. It is obvious that this tensor is real.\footnote{The normalization of the completely antisymmetric tensor $\e$ is $\e^{12345678}=1$.} In this form $Q$ may look more familiar: it is the invariant quartic tensor of the Lie group $E_{7(7)}$ as it is also stated in \cite{dW02}. This will become clear in the following two sections.

\subsection{Vector fields preserving $\Omega$ and $Q$}\label{VectF}
I want to determine all $\vp\in \Symp(56)$ that fulfill
\be\label{Qinv}
\vp^*Q &=& Q.
\ee
In complete analogy to the discussion of the symplectic form $\Omega$ (\ref{om4}) in section \ref{SympV1}, equation (\ref{Qinv}) leads the constraint
\be\label{Q3}
	4Q_{\rho(\mu_1\mu_2\mu_3}\p_{\mu_4)}X^\rho &=& 0
\ee
for the Hamiltonian vector field $X$. In the appendix \ref{ESympm}, I prove that the general solution is provided by the Hamiltonian
\be\label{Eom10}
H_{(\Lambda,\Sigma,c)}(z,\bar{z})&:=& \frac{i\tau_7^2}{8}{\Lambda_A}^B\left(\delta_{[D_2}^{A}\delta_{D_1]}^{[C_1}\delta_{B}^{C_2]}
-\frac{1}{8}\delta_{B}^{A}\delta_{D_1D_2}^{C_1C_2}\right)z^{D_1D_2}\bar{z}_{C_1C_2}\\
&& +\frac{i\tau_7^2}{32}\Sigma^{[C_1C_2C_3C_4]}\left(\bar{z}_{C_1C_2}\bar{z}_{C_3C_4}
-\frac{1}{4!}\e_{C_1\dots C_8}z^{C_5C_6}z^{C_7C_8}\right)\nn\\
&&+\frac{i\tau_7^2}{16}c^{M_1M_2}\bar{z}_{M_1M_2} -\frac{i\tau_7^2}{16}\bar{c}_{M_1M_2}z^{M_1M_2}\nn.
\ee
Since the Hamiltonian $H(z,\bar{z})=H(x,p)$ is real, the arbitrary complex coefficients $({\Lambda_{A}}^{B},\Sigma^{[C_1C_2C_3C_4]},c)$ are subject to the following constraints
\begin{subequations}\label{Reell4}
\be
\overline{{\Lambda_{A}}^{B}}&=& -\delta_{AC}\delta^{BD}{\Lambda_{D}}^{C}\\
\overline{\Sigma^{C_1\dots C_4}}&=& \frac{1}{4!}\delta^{C_1D_1}\cdots\delta^{C_4D_4}\e_{D_1\dots D_8}\Sigma^{D_5\dots D_8}\label{SigmaBar}
\ee
\end{subequations}
and the obvious relation between $c$ and $\bar{c}$. I want to emphasize that the symmetric object $\delta$ is the Kronecker delta, which is part of the symplectic form (\ref{omr1}) and not a metric: the indices are hence raised and lowered as defined in (\ref{lower6}). The computation of the corresponding vector field follows the general rule (\ref{om9c}):

\be\label{E7Vektor}
X_{(\Lambda,\Sigma,c)} &=& 2{\Lambda_{A}}^{B}\left(\delta_B^{P}\delta_{Q}^A-\frac{1}{8}\delta_B^{A}\delta_{Q}^{P}\right)\left(z^{SQ}\frac{\p}{\p z^{SP}} - \bar{z}_{SP}\frac{\p}{\p \bar{z}_{SQ}}
\right)\nn\\
&& + \Sigma^{M_1\dots M_4}\left(\bar{z}_{[M_1M_2}\frac{\p}{\p z^{M_3M_4]}} +\frac{1}{4!}\e_{M_1\dots M_8} z^{M_5M_6}\frac{\p}{\p \bar{z}_{M_7M_8}}\right)\nn\\
&& +c^{M_1M_2}\frac{\p}{\p z^{M_1M_2}} +\bar{c}_{M_1M_2}\frac{\p}{\p \bar{z}_{M_1M_2}}.
\ee
This is the general vector field that preserves the tensors $\Omega$ and $Q$. The striking fact is that it is linear in the coordinates. This implies that the subgroup of $\mathfrak{diff}_d$ preserving $\Omega$ and $Q$ is isomorphic to a subgroup of the affine group $A(56)$. In the next section, I will show that this subgroup is $\cP_{56}\rtimes E_{7(7)}$.

\subsection{The vector field representation of $\mathfrak{e}_{7(7)}$}\label{VectE4}
In chapter \ref{CHAP2}, I showed that simple, finite dimensional, complex Lie algebras are determined by their Dynkin diagrams or equivalently, by their commutation relations. In order to find out which real Lie group is provided by the vector fields (\ref{E7Vektor}), the first task is to extract the commutation relations that are induced by the ones of $\mathfrak{diff}_{56}$. Following the same logic as in section \ref{Subalg}, I separate the parameters $(\Lambda,\Sigma,c)$ from the algebra generators in the vector field representation $\mathbf{R}$ by defining
\be\label{E7ZerlVF}
X^{\mathfrak{a}}_{(\Lambda,\Sigma,c)} &=:& 
{\Lambda_{A}}^{B}\mathbf{R}_{{\left.\hat{M}\right.^{A}}_{B}} 
+ 
\Sigma^{M_1\dots M_4} \mathbf{R}_{\hat{S}_{M_1\dots M_4}}\\
&&
+c^{M_1M_2}\mathbf{R}_{\hat{Z}_{M_1M_2}} +\bar{c}_{M_1M_2}\mathbf{R}_{\hat{\bar{Z}}^{M_1M_2}}.
\nn
\ee
A comparison with the general vector field $X_{(\Lambda,\Sigma,c)}$ (\ref{E7Vektor}) preserving $\Omega$ and $Q$ implies
\begin{subequations}\label{PRep2}
\be
\label{MDefi2}
\mathbf{R}_{{\left.\hat{M}\right.^{A}}_{B}} 
&=&
2\left(\delta_B^{P}\delta_{Q}^A-\frac{1}{8}\delta_B^{A}\delta_{Q}^{P}\right)\left(z^{SQ}\frac{\p}{\p z^{SP}} - \bar{z}_{SP}\frac{\p}{\p \bar{z}_{SQ}}
\right)\\
\label{RDefi2}
\mathbf{R}_{\hat{S}_{M_1\dots M_4}} 
&=&
\bar{z}_{[M_1M_2}\frac{\p}{\p z^{M_3M_4]}} +\frac{1}{4!}\e_{M_1\dots M_8} z^{M_5M_6}\frac{\p}{\p \bar{z}_{M_7M_8}}
\\
\label{ZDefi2}
\mathbf{R}_{\hat{Z}_{N_1N_2}} &=&\frac{\p}{\p z^{N_1N_2}}
\ee
\end{subequations}
and the obvious identifications for the complex conjugated operators. Note that the reality constraints of the components $(\Lambda,\Sigma)$ (\ref{Reell4}) are also fulfilled for the vector field representation $\mathbf{R}$ of the generators $(\hat{M},\hat{S})$. Hence, it is natural to lift these to the abstract generators
\begin{subequations}\label{MRquer}
\be
\overline{{\left.\hat{M}\right.^P}_Q} &=& -\delta^{PR}\delta_{QS}{\left.\hat{M}\right.^S}_R\label{MquerM}\\
\overline{\hat{S}_{M_1\dots M_4}} &=& \frac{1}{4!}\e_{M_1\dots M_8}\delta^{M_5N_1}\cdots \delta^{M_8N_4}\hat{S}_{N_1\dots N_4}\label{RquerR}
\ee
\end{subequations}
with the Kronecker $\delta$ from the symplectic form $\Omega$ to raise and lower indices as in (\ref{lower6}). A short calculation with the vector field representations (\ref{PRep2}) provides the commutation relations
\begin{subequations}\label{ComRel8}
\be
\left[{\left.\hat{M}\right.^{P}}_{Q}, {\left.\hat{M}\right.^{V}}_{W}\right]
 &=& 
 \delta_{Q}^{V} {\left.\hat{M}\right.^{P}}_{W} -\delta_{W}^{P} {\left.\hat{M}\right.^{V}}_{Q}
 \label{MMCR}
 \\
 \left[{\left.\hat{M}\right.^{P}}_{Q}, \hat{S}_{M_1\dots M_4}\right]
&=& 
4\left(\delta^{P}_{[M_1} \hat{S}_{M_2\dots M_4]Q}+\frac{1}{8}\delta_Q^P\hat{S}_{M_1\dots M_4}\right)\label{MRCR}
\\
\left[\,\overline{\hat{S}^{N_1\dots N_4}}, \hat{S}_{M_1\dots M_4}\right]
&=&\frac{2}{3}\delta^{[N_1\dots N_3}_{[M_1\dots M_3}{\left. \hat{M}\right.^{N_4]}}_{M_4]}\label{RRCR}
\\
\left[\hat{S}_{M_1\dots M_4},\overline{\hat{Z}^{N_1N_2}}\,\right]
&=& 
 -\delta^{N_1N_2}_{[M_1M_2}\hat{Z}_{M_3M_4]}\label{RZqCR}
 \\
\left[{\left.\hat{M}\right.^{P}}_{Q},\hat{Z}_{M_1M_2}\right]
&=& 
-2\left(\delta_{[M_2}^{P}\hat{Z}_{M_1]Q}-\frac{1}{8}\delta_Q^P\hat{Z}_{M_1M_2}\right)\label{MZCR}.
\ee
\end{subequations}
The commutation relation (\ref{MMCR}) are the same ones as used for $\mathfrak{gl}_d$ in (\ref{ComRel1}). This is the reason why I have used the same letter $\hat{M}$ and the awkward normalization in (\ref{MDefi2}). \\

It is well known \cite{dW02} that the commutation relations (\ref{ComRel8}) uniquely determine the complex Lie algebra $\mathfrak{e}_7$ together with a normal subalgebra corresponding to the $56$-dimensional Abelian group of translations $\cP_{56}$. Since I have constructed this Lie algebra as a subalgebra of the matrix algebra $\mathfrak{sp}_{56}$, the $63$ generators $\hat{M}$ are compact and the $\binom{8}{4}=70$ generators $\hat{S}$ are not compact. Following the definition from section \ref{realf}, the real form has to be 
\beg
\mathfrak{e}_{7(70-63)} &=&\mathfrak{e}_{7(7)}.
\eeg
Comparing with the discussion from section \ref{Representation}, the subgroup of diffeomorphisms preserving the two tensors $\Omega$ and $Q$ is isomorphic to $\cP_{56}\rtimes E_{7(7)}$ with its compact subgroup $SU(8)/\Z_2$. I want to close this section with a comment on the invariant tensor $Q$ (\ref{Q0}). In a fixed coordinate chart, $Q$ is a quartic tensor of the $\mathbf{56}$-dimensional vector space on which $\mathfrak{e}_{7(7)}$ acts as a matrix representation. It should not be confused with a quartic tensor of the $\mathbf{133}$-dimensional vector space $\mathfrak{e}_{7(7)}$. In \cite{GKN01}, it is shown that the quartic tensor $Q$ of the fundamental representation $\mathbf{56}$ can be related to the Cartan--Killing form or quadratic Casimir invariant of $\mathfrak{e}_{7(7)}$.

\subsection{Connections $\nabla$ for supergravity}\label{LCcE}
This section discusses the implications on physical fields $\psi$ and on connections $\nabla$, which is due to the restriction of $\Symp(56)$ to the diffeomorphisms that also preserve the quartic tensor $Q$ (\ref{Q0}). The argumentation strongly follows the one for symplectic equivariance of section \ref{LCc3}. \\

The first observation is a technical one: for the choice $\tau_8=\tau_7\sqrt{2}$ (\ref{chi8fix}), the symplectic form has canonical shape (\ref{Omega6}) in the coordinates $(z^{M_1M_2},\bar{z}_{M_1M_2})$ (\ref{zNeu3}). This implies that it is consistent to lower the index of the antiholomorphic forms in the vielbein frame in the same way as for the purely symplectic case (\ref{holframe2}). In analogy to (\ref{holframe}), the cotangent space of the real $56$-dimensional manifold with preserved tensors $(\Omega,Q)$ is therefore spanned by the one-forms
\begin{subequations}\label{holframe3}
\be
\underline{dz}^{A_1A_2}&:=&\frac{1}{\tau_7}\left(\frac{1}{i}{\G_a}^{A_1A_2}d\underline{z}^a +\frac{1}{\sqrt{2}}{\G_{ab}}^{A_1A_2}\underline{dz}^{ab}\right)
\\
\underline{d\bar{z}}_{B_1B_2}&:=&\frac{1}{\tau_7}\left(\frac{1}{i}{\G^b}_{B_1B_2}\underline{d\bar{z}}_b +\frac{1}{\sqrt{2}}{\G^{ab}}_{B_1B_2}\underline{d\bar{z}}_{ab}\right).
\ee
\end{subequations}
In this frame, the general form of a connection $\nabla$ (\ref{Ccomp3}) acting on $\mathfrak{u}_{28}$ representations $\psi$ is
\beg
\underline{dz}^{A_1A_2} \nabla_\alpha \psi^\gamma &=:& 
		\underline{dz}^{A_1A_2}\left(\frac{\p}{\p z^{A_1A_2}}\psi^\gamma 
		-{\Big(\omega_{{A_1A_2}}\Big)_\e}^\rho \delta_{{\left.\hat{M}\right.^\e}_\rho}\psi^\gamma\right)
		\\
				\underline{d\bar{z}}_\beta \bar{\nabla}^{B_1B_2} \psi^\gamma &=:& 
		\underline{d\bar{z}}_{B_1B_2}\left(\frac{\p}{\p \bar{z}_{B_1B_2}}\psi^\gamma 
		-{\Big(\bar{\omega}^{B_1B_2}\Big)_{ \e}}^\rho \delta_{{\left.\hat{M}\right.^\e}_\rho}\psi^\gamma\right).\\
\eeg
\smallskip\\

As in section \ref{LCc3}, the restriction of the group of diffeomorphisms to a subgroup allows for the possibility to cast the vielbein in a particular matrix shape. For the case of the preserved quartic tensor $Q$ (\ref{Q0}), it is consistent to restrict the vielbein matrix $e$ in $d=56$ to be an $E_{7(7)}$ group element with arbitrary coordinate dependence.\\

This implies that the diffeomorphisms $\vp_A\in \cP_{56}\rtimes E_{7(7)}\subset \Symp(56)$ induce an $SU(8)/\Z_2 \subset U(28)\subset SO(56)$ rotation in order to restore the arbitrary vielbein gauge (\ref{VielbeinTrafo}). To phrase this in other words, a diffeomorphism $\vp_A\in \cP_{56}\rtimes E_{7(7)}$ does not link irreducible $\mathfrak{su}_8$ representations to each other. It is therefore consistent to interpret every irreducible $\mathfrak{su}_8$ representation as an independent physical field $\psi$ in the vielbein frame.\\

As for the symplectic case in section \ref{LCc3}, such a restriction of the physical fields is only possible, if the action of the connection $\nabla$ preserves this structure, i.e. if the connection form $\omega$ does not contain any generator that links different irreducible $\mathfrak{su}_8$ representations to each other.\footnote{The standard Levi--Civita connection did contain these generators $\hat{T}$ for the symplectic case (\ref{sympCon}).}\\

However, there is a big difference to the symplectic case presented in section \ref{LCc3}: for the latter, there is a connection $\nabla$ with $Sp(56)\subset A(56)$-equivariance that allows for a symmetry enhancement. By adding a tensor $v$ (\ref{omega2}) with respect to affine linear diffeomorphisms, the $Sp(56)$-equivariance of $\nabla^{\text{min}}$ was enhanced to the bigger symmetry group $\Symp(56)$. This is the same mechanism that is used in the procedure \`a la Borisov \& Ogievetsky in chapter \ref{CHAP4}: the affine linear equivariance, provided by the minimal connection $\nabla^{\text{min}}$ (\ref{minimalCon}) alone, was enhanced to $\Diff(d)$-equivariance of $\nabla$ by adding an $A(d)$-covariant Lorentz tensor $v$ (\ref{omega2}).\\

The situation is different in the present case. Since the subgroup $\cP_{56}\rtimes E_{7(7)}$ of $\Diff(56)$ preserving both $\Omega$ and $Q$ is a subgroup of the affine group $A(56)$, the possibility to enlarge the symmetry group does not exist. Hence, there is no connection $\nabla$ in the class (\ref{final1}) provided by the theorem \ref{prop1} that is preferred due to a symmetry enhancement. In particular, the minimal connection $\nabla^{\text{min}}$ (\ref{minimalCon}) is completely sufficient to guarantee equivariance under all diffeomorphisms that preserve $\Omega$ and $Q$.\\

For the comparison to the supersymmetry variations of $d=11$ supergravity in section \ref{Super}, it will be important construct the general connection $\nabla$ acting on $\mathfrak{u}_8$ representations $\psi$ in this exceptional geometry, which is linear in derivatives and only depends on vielbein degrees of freedom. For the application of theorem \ref{prop1}, I have to identify the components (\ref{omegaLin}) of the Maurer--Cartan form $e^{-1}de$ with an $E_{7(7)}$ valued vielbein matrix $e$. In analogy to (\ref{vRestr}), I obtain\footnote{Note that there is no need to add the complex conjugate to the symmetric generator due to the reality constraint (\ref{RquerR}).}
\begin{subequations}\label{vRestr2}
\be
 {\left(v^{\text{min}}_a\right)_c}^d {\left.\hat{L}\right.^c}_d
&=&
 {\left(v^{\text{min}}_a\right)_A}^B {\left.\hat{M}\right.^A}_B
\\
 {\left(v_a\right)_c}^d {\left.\hat{S}\right.^c}_d
 &=&
 \left(v_a\right)^{D_1\dots D_4} \hat{S}_{D_1\dots D_4},
\ee
\end{subequations}
where I have used the same normalizations of the generators $\hat{M}$ and $\hat{S}$ as in (\ref{E7ZerlVF}). In order to construct a connection $\nabla$ acting on $\mathfrak{u}_8$ representations $\psi$ that is linear in derivatives and only depends on vielbein degrees of freedom, I have to investigate which connection one-forms $\omega$ can be constructed from the $\mathfrak{su}_8$ tensor $v$ (\ref{vRestr2}).\\

I want to follow the argumentation that I used to derive the equation (\ref{Xc}) during the joint realization of symmetries of section \ref{secD}. It was the decomposition of the tensor product $\mathbf{d}\otimes \mathbf{\frac12d(d+1)}$ and the projection on possible antisymmetric tensors that led to the general form for the Lorentz tensor $X$ in (\ref{Xc}). In the present case, the tensor product is $\mathbf{56}\otimes \mathbf{70}$ and the decomposition should be performed into irreducible $\mathfrak{su}_8$ representations. Following the definition of a general connection (\ref{Conn56}) from section \ref{prim2}, the general form of $\nabla$ is hence provided by the one-form valued $\mathfrak{u}_8$ action $\delta_{\hat{\omega}}$ 
\be\label{sympConE}
\delta_{\hat{\omega}}
&=& 
\left\{
{\left(v^{\text{min}}_{C_1C_2}\right)_A}^B \delta_{{\left.\hat{M}\right.^A}_B}
+
c_1\left(\bar{v}^{CD}\right)_{CD C_1C_2}\delta_{\hat{T}}\right.\\
&&\left.
+c_2\left(\bar{v}^{CQ}\right)_{CP C_1C_2} \left(\delta^P_A\delta^B_Q-\frac{1}{8}\delta_A^B\delta_Q^P\right)\delta_{{\left.\hat{M}\right.^A}_B}
\right\}
\underline{dz}^{C_1C_2} +\text{c.c.}\nn
\ee
with a $\mathfrak{u}_1$ generator $\hat{T}$ whose action on a field $\psi$ corresponds to a simple multiplication. In section \ref{Super}, I will fix the real constants $c_1,c_2\in \R$ in the appropriate way for a comparison with $d=11$ supergravity.\\

As a side remark, I want to mention that the Levi--Civita connection $\nabla^{(LC)}$ in $d=56$ again defines a connection in the class (\ref{sympConE}) by projection onto $\mathfrak{su}_{8}$, in analogy to (\ref{projCon})
\be\label{projCon2}
\nabla^{(LC)}_{\mathfrak{su}_{8}} &:=& \text{pr}_{\mathfrak{su}_{8}}\nabla^{(LC)}.
\ee
Starting with the $U(28)$-covariant form of (\ref{sympCon2}), a comparison of the vector field representations (\ref{MSRep}) and (\ref{PRep2}) implies that the connection $\nabla^{(LC)}_{\mathfrak{su}_{8}}$ corresponds to the choice $c_1=0$ and $c_2=\frac{1}{3}$.
I want to emphasize again that this choice is not related to a symmetry enhancement. Therefore, there is no reason why it should be preferred in the context of exceptional geometry.\footnote{The equations (\ref{TorCon3}) indicate that the torsion does not vanish for any values $c_1,c_2\in \R$. Hence, this is not a suitable connection $\nabla$ to discuss the holonomy of the underlying manifold.} I want to conclude this section with a short summary of the results.

\begin{enumerate}
	\item In the context of exceptional geometry, the symmetry group $\Diff(56)$ is restricted to $\cP_{56}\rtimes E_{7(7)}$ by requiring the invariance of the symplectic form $\Omega$ (\ref{om1}) and the symmetric quartic tensor $Q$ (\ref{Q0}).
	\item Therefore, it is consistent to restrict the $56$-dimensional vielbein $e(x)$ to be an $E_{7(7)}$ valued matrix with arbitrary dependence on the $56$ coordinates.
	\item Having fixed the Lorentz gauge of the vielbein, the induced Lorentz action in this setting is an element of $SU(8)/\Z_2\subset SO(56)$.
	 \item It is consistent to consider irreducible $\mathfrak{su}_8$ representations as independent physical fields $\psi$ in the $56$-dimensional exceptional geometry, because a restricted diffeomorphism $\vp\in \cP_{56}\rtimes E_{7(7)}$ does not link these to each other. 
	 \item Since $\cP_{56}\rtimes E_{7(7)}$ is a subgroup of the affine linear group $A(56)$ (\ref{affine2}), it is not possible to modify the minimal connection $\nabla^{\text{min}}$ (\ref{minimalCon}) in such a way that the equivariance group is enlarged. In particular, the Levi--Civita connection is not preferred in the general class of connections $\nabla$ acting on $\mathfrak{u}_8$ representations $\psi$ (\ref{sympConE}).
\end{enumerate}

\section{Decomposition of $\mathfrak{e}_{7(7)}$ under $\mathfrak{gl}_7$}\label{Decompo}
\subsection{The algebra $\mathfrak{e}_{7(7)}$}\label{Decompo2}
After having defined the exceptional geometry, I need to establish a parametrization of $\mathfrak{e}_{7(7)}$ that is appropriate for the comparison with $d=11$ supergravity. This will be achieved by a decomposition of the generators of $\mathfrak{e}_{7(7)}$ into $\mathfrak{gl}_7$ representations, which is natural from the point of view of the $4+7$ split used by de Wit and Nicolai \cite{dWN86}. The compact subalgebra $\mathfrak{so}_7$ of $\mathfrak{gl}_7$ is the part of the Lorentz algebra $\mathfrak{so}_{(10,1)}$ of $d=11$ supergravity that is relevant for the $56$-dimensional sector of the $60$-dimensional exceptional geometry under consideration. \\

For the decomposition, I will use the $\mathfrak{gl}_7$ subalgebra of $\mathfrak{e}_{7(7)}$ whose vector field representation on the seven coordinates $x^m$ (\ref{xDefi8}) has the standard form (\ref{PRep}). In order to obtain this vector field representation, I start with the observation that the vector fields (\ref{PRep2}) preserving both tensors $\Omega$ and $Q$, are written in the holomorphic frame $(z^{AB},\bar{z}_{CD})$ (\ref{zNeu3}). This is related to the standard coordinates $(x,p)$ by Clifford matrices. These were defined in the appendix \ref{Cliff} to be the purely imaginary $8\times 8$ matrices $\G_a$ that fulfill the Clifford property $\{\G_a,\G_b\}=2\eta_{ab}$ (\ref{Clifford}) for the Euclidean metric $\eta$ in seven dimensions. It is a standard result that the $63$ traceless matrices 
\be\label{Independent2}
		\G_a,\, \G_{ab},\, \G_{abc}\quad \text{with }a,b,c=1,\dots,7
\ee 
that follow the definition (\ref{AntisymG}), span the complex vector space of traceless $8\times 8$ matrices.\footnote{To prove this, observe that the standard norm of the space of matrices is the trace and that the trace of the product of any two matrices in (\ref{Independent2}) vanishes, if and only if the two are different (\ref{Clifford},\,\ref{Gnq},\,\ref{G3G}).} The matrices $\G_a$ and $\G_{ab}$ are antisymmetric, whereas $\G_{abc}$ is symmetric.\\

I will show next that this parametrization (\ref{Independent2}) induces a decomposition of $\mathfrak{e}_{7(7)}$ into $\mathfrak{gl}_7$ representations, for which the vector field representation $\mathbf{R}$ of $\mathfrak{gl}_7$ has the standard form (\ref{PRep}). Since the vector field representations of the non-compact generators $\hat{S}$ and of the compact ones $\hat{M}$ (\ref{PRep2}) provide the one of $\mathfrak{e}_{7(7)}$, I have to decompose these $\mathfrak{su}_8$ representations into $\mathfrak{so}_7$ representations at first. Then, I will recombine them into $\mathfrak{gl}_7$ representations.\\

I start with the compact generator $\hat{M}$ (\ref{MDefi2}) of $\mathfrak{su}_8$. Due to its reality property (\ref{MquerM}) and its tracelessness, I can without loss of generality decompose $\hat{M}$ into $\mathfrak{so}_7$ representations $\hat{L}$ as follows
\be\label{defiSo7}
{\left.\hat{M}\right.^A}_B &=:&-\frac{\tau_1}{2!8}{\left.\hat{L}\right.^a}_b{{{\G_a}^b}^A}_B - \frac{\tau_2}{3!8}\hat{L}^{a_1\dots a_3}{{\G_{a_1\dots a_3}}^A}_B \nn\\
&&
+\frac{i\tau_3}{6!8}\hat{L}^{a_1\dots a_6}\e_{a_1\dots a_6a}{{\G^a}^A}_B.
\ee
The normalization parameters $\tau_1,\tau_2,\tau_3\in \R\backslash\{0\}$ will be fixed in the sequel, partly by a comparison with $d=11$ supergravity. The symmetry properties of the $\G$-matrices imply that the decomposition (\ref{defiSo7}) respects the fact that the generators $\hat{M}$ are antihermitean (\ref{MquerM}), keeping in mind that $\G_a$ is purely imaginary. Therefore, the generators $\hat{L}$ are real.\footnote{It is obvious from the relation (\ref{epsdefi7}) that $i\G_a$ could be replaced by $\G_{a_1\dots a_6}$. To keep the notation as simple as possible, I will keep the purely imaginary unit $i$, however.}\\

With the relations (\ref{Gnq}) from the appendix \ref{SomeG2}, it is easy to check that the following identities hold:
\begin{subequations}\label{MGa}
\be
	{\left.\hat{L}\right.^{a}}_b &=&\frac{1}{\tau_1}{{{\G^{a}}_b}^B}_A{\left.\hat{M}\right.^A}_B, 
	\\
		 \hat{L}^{abc}&=&\frac{1}{\tau_2}{{\G^{abc}}^B}_A{\left.\hat{M}\right.^A}_B,
		 \\
	\hat{L}^{a_1\dots a_6}&=&\frac{1}{i\tau_3}\e^{a_1\dots a_6c}{{\G_{c}}^B}_A{\left.\hat{M}\right.^A}_B .
\ee
\end{subequations}
This allows to construct the explicit vector field representation $\mathbf{R}$ of the $\mathfrak{so}_7$ subalgebra of $\mathfrak{e}_{7(7)}$ with the equations (\ref{zNeu3}) and (\ref{MDefi2}): 
\be\label{Vect5}
\mathbf{R}_{{\left.\hat{L}\right.^{p}}_q} 
&=& \frac{1}{\tau_1}{{{\G^{p}}_q}^B}_A\mathbf{R}_{{\left.\hat{M}\right.^A}_B}\\
&=&
-\frac{2}{\tau_1}\left[\delta_m^p\delta_q^r -\eta^{pr}\eta_{qm}\right]
\left(
\left[x^m\frac{\p}{\p x^{r}} + \eta^{mv}\eta_{rw}p_v \frac{\p}{\p p_{w}}\right]
\right.\nn\\
&&\left.
+2\left[x^{ms}\frac{\p}{\p x^{rs}} + \eta^{mv}\eta_{rw}p_{vs} \frac{\p}{\p p_{ws}}\right]
\right)\nn.
\ee
It is the advantage of the algebraic approach presented in chapter \ref{CHAP4} that I do not have to use the complicated expression (\ref{Vect5}) in order to compute the connection $\nabla$ (\ref{sympConE}) in the $\mathfrak{gl}_7$ decomposition suitable for $d=11$ supergravity. In analogy to the discussion of the affine coset from section \ref{secB}, I can work with the abstract generators that are simpler to handle. However, one should always keep in mind that all these generators correspond to real vector fields in $56$ dimensions, whose explicit action on the $56$ real coordinates $x^\mu$ can be constructed in complete analogy to (\ref{Vect5}). It is nice to observe that an evaluation of the explicit formula (\ref{Vect5}) on the $7$ coordinates $x^m$
\beg
\mathbf{R}_{{\left.\hat{L}\right.^{p}}_q} \left(x^m\right) &=& -\frac{2}{\tau_1}\left(x^p \p_q -\eta^{pr}\eta_{qs}x^s\p_r\right)\left(x^m\right)
\eeg
is the standard vector field representation of the $SO(7)$ algebra that has already been used in chapter \ref{CHAP2} (\ref{PRep},\, \ref{LDefi}). For the normalizations to coincide, I have to fix
\be\label{chi1fix}
\tau_1&=&-4.
\ee
This explains why I have used the letter $\hat{L}$ for this generator in the decomposition of the $\mathfrak{su}_8$ element $\hat{M}$ (\ref{defiSo7}). Since $\mathfrak{su}_8$ is a subalgebra of $\mathfrak{so}_{56}$, all generators $\hat{M}$ are compact: they are presented as antisymmetric matrices in $56$ dimensions. Therefore, it is natural to also use the same letter $\hat{L}$ for the generators $\hat{L}^{abc}, \hat{L}^{a_1\dots a_6}$ in (\ref{defiSo7}) and for ${\left.\hat{L}\right.^{\mu}}_\nu$ in (\ref{LDefi}).\\

Since the generators $\hat{L}$ (\ref{defiSo7}) merely are a different parametrization of the $\mathfrak{su}_{8}$ generators $\hat{M}$ (\ref{MMCR}), the commutation relations of these compact elements of $\mathfrak{e}_{7(7)}$ can be derived by substituting the parametrization (\ref{defiSo7}) in the ones of $\hat{M}$ (\ref{MMCR}).\\ 

As I am interested in the commutation relations of $\mathfrak{e}_{7(7)}$ in the $\mathfrak{gl}_7$ decomposition, I have to discuss the non-compact generators, too. These are provided by the $\mathbf{70}$-dimensional representation $\hat{S}_{[M_1\dots M_4]}$ of $\mathfrak{su}_8$, defined by its vector field representation $\mathbf{R}$ (\ref{RDefi2}) in section \ref{VectE4}. The letter $\hat{S}$ reminds of its origin, because the elements $\hat{S}_{[M_1\dots M_4]}$ are presented as symmetric matrices in their action on a $56$-dimensional vector space.\\

The fact that the vector space of traceless $8\times 8$ matrices is spanned by the $\G$-matrices (\ref{Independent2}) also induces an $\mathfrak{so}_7$ decomposition of the $\mathfrak{su}_8$ representation $\hat{S}_{ABCD}$. With the relation (\ref{Finale6b}) from the appendix \ref{SomeG2}, the symmetry properties of the $\G$-matrices imply that the complex vector space of antisymmetric four-tensors with $A,B,C,D=1,\dots,8$ is spanned by the following $\G$ matrices:
\be\label{basisanti}
	{\G_{(a}}_{[AB}{\G_{c)}}_{CD]},\, {\G_{ab}}_{[AB}{\G^{a}}_{CD]},\, {\G_{[ab}}_{[AB}{\G_{c]}}_{CD]}.
\ee
This leads to the $\mathfrak{so}_7$ decomposition of the $\mathfrak{su}_8$ representation $\hat{S}_{ABCD}$:
\be\label{defiSo7R}
\hat{S}_{ABCD} &=:&\frac{3\tau_4}{32}{\left.\hat{S}\right.^{a}}_b\left(\delta_a^d\delta^b_c +\frac{1}{7}\left(\tau_9-1\right)\delta_a^b\delta_c^d\right){\G_{d}}_{[AB}{\G^{c}}_{CD]}\\
&&
 -\frac{\tau_5}{32}\hat{S}^{a_1\dots a_3}{\G_{[a_1a_2}}_{[AB}{\G_{a_3]}}_{CD]}
 +\frac{i\tau_6}{6!192}\hat{S}^{a_1\dots a_6}\e_{a_1\dots a_6f}{\G^{af}}_{[AB}{\G_{a}}_{CD]}.
 \nn
\ee
The normalization constants $\tau_4,\tau_5,\tau_6,\tau_9\in \R\backslash\{0\}$ will be fixed in the sequel. I want to emphasize that the Clifford algebra of $\G$-matrices does not distinguish, if an index $A,B,\dots$ is lowered or raised, i.e. if the associated space is the vec- tor space or its dual. Since the non-degenerate symplectic form $\Omega$ provides a canonical isomorphism between the tangent space and its dual as used in (\ref{lower6}), it is obvious from the vector field representations (\ref{PRep2}) of the two $\mathfrak{e}_{7(7)}$ generators that the position of the indices of $\hat{M}$ and $\hat{S}$ is not important, if the bar denoting the complex conjugation is kept explicitly as in (\ref{MRquer}). In order to stress that complex conjugated $\mathfrak{su}_8$ representations are dual to the original ones, I adopt the convention to denote them with raised indices. The standard relations for $\G$ matrices\footnote{To prove them, recall that all $\G$-matrices have entries of norm $1$. Computing e.g. ${\G_{a}}^{56}{\G_{b}}^{78}$ and ${\G_{a}}^{12}{\G_{b}}^{34}$ for arbitrary $a,b=1,\dots,7$ shows that they have the same sign. Due to the normalization $\e^{12345678}=1$, the coefficients also coincide.}
\begin{subequations}\label{epstrafo2}
\be
{\G_{a}}_{[AB}{\G_{b}}_{CD]}&=&+\frac{1}{4!}\e_{ABCDEFGH}{\G_{a}}^{[EF}{\G_{b}}^{GH]}\\
{\G^{ac}}_{[AB}{\G_{a}}_{CD]}&=&+\frac{1}{4!}\e_{ABCDEFGH}{\G^{ac}}^{[EF}{\G_{a}}^{GH]}\\
{\G_{[ab}}_{[AB}{\G_{c]}}_{CD]}&=&-\frac{1}{4!}\e_{ABCDEFGH}  {\G_{[ab}}^{[EF}{\G_{c]}}^{GH]}
\ee
\end{subequations}
allow to evaluate the reality condition (\ref{RquerR}) of the non-compact generator $\hat{S}$. Keeping in mind that the matrices $\G_a$ are purely imaginary, the parametrization (\ref{defiSo7R}) implies 
\be\label{defiSo7R2}
\overline{\hat{S}^{ABCD}} &=&\frac{3\tau_4}{32}{\left.\hat{S}\right.^{a}}_b\left(\delta_a^d\delta^b_c +\frac{1}{7}\left(\tau_9-1\right)\delta_a^b\delta_c^d\right){\G_{d}}^{[AB}{\G^{c}}^{CD]}\\
&&
 +\frac{\tau_5}{32}\hat{S}^{a_1\dots a_3}{\G_{[a_1a_2}}^{[AB}{\G_{a_3]}}^{CD]}
 +\frac{i\tau_6}{6!192}\hat{S}^{a_1\dots a_6}\e_{a_1\dots a_6f}{\G^{af}}^{[AB}{\G_{a}}^{CD]}\nn.
\ee
I refrain from calling the real generators ${\left.\hat{S}\right.^{a}}_b,\,\hat{S}^{a_1\dots a_3}$ the self-dual part and $\hat{S}^{a_1\dots a_3}$ the anti self-dual part of the tensor $\hat{S}_{ABCD}$, because firstly, the $\e$ tensor links complex conjugated tensors in (\ref{RquerR}) and secondly, indices are raised and lowered with the symplectic form $\Omega$ and not with a metric.\footnote{As for the case of $\hat{M}$ in (\ref{defiSo7}), I do not eliminate the $i$ in the parametrization (\ref{defiSo7R}) in order to avoid a cumbersome notation.}\\

With the formul\ae{} from the appendix \ref{SomeG2}, I obtain the relations
\be\label{MGc}
{\left.\hat{S}\right.^{a}}_b&=&\frac{1}{\tau_4} \left({\G_b}^{AB}{\G^a}^{CD}-\frac{1}{7}\left(1-\frac{2}{9\tau_9}\right)\delta_b^a{\G_c}^{AB}{\G^c}^{CD}\right)\hat{S}_{ABCD}
\nn\\
\hat{S}^{b_1\dots b_3}&=& \frac{1}{\tau_5}{\G^{[b_1b_2}}^{[AB}{\G^{b_3]}}^{CD]}\hat{S}_{ABCD} 
\nn\\
\hat{S}^{b_1\dots b_6} &=&\frac{i}{\tau_6}\e^{b_1\dots b_6c}{\G_{bc}}^{[AB}{\G^{b}}^{CD]}\hat{S}_{ABCD}.
 \ee
These allow to deduce the explicit vector field representation $\mathbf{R}$ of the non-compact generators of $\mathfrak{e}_{7(7)}$ from (\ref{RDefi2}). In analogy to (\ref{Vect5}), I am interested in the part that is relevant for the $\mathfrak{gl}_7$ subalgebra, which is
\be\label{Vect6}
\mathbf{R}_{{\left.\hat{S}\right.^{a}}_b} 
&=& \frac{1}{\tau_4} \left({\G_b}^{AB}{\G^a}^{CD}-\frac{1}{7}\left(1-\frac{2}{9\tau_9}\right)\delta_b^a{\G_f}^{AB}{\G^f}^{CD}\right)\hat{S}_{ABCD}\nn\\
&=&-\frac{2}{3\tau_4}\left\{
\left( \delta^a_m\delta_b^r +\eta_{mb}\eta^{ar} 
 \right)
\left(
 \left[ x^m\frac{\p}{\p x^{r}} - \eta^{mv}\eta_{rs}p_v \frac{\p}{\p p_{s}}\right]
 \right.\right.\nn\\
&&\left.
+2 \left[x^{ms}\frac{\p}{\p x^{rs}} - \eta^{mv} \eta_{rt}p_{vn} \frac{\p}{\p p_{tn}}\right]
\right)\nn\\
 &&
 +\frac{2}{7}\delta^a_b
 \left(
 \left(-1+\frac{1}{\tau_9}\right)
 \left[ x^r\frac{\p}{\p x^{r}} - p_m \frac{\p}{\p p_{m}}\right]
 \right.\nn\\
&&\left.\left.
-\left(2+\frac{1}{3\tau_9}\right)
 \left[ x^{rs}\frac{\p}{\p x^{rs}} - p_{mn} \frac{\p}{\p p_{mn}}\right]
\right)
\right\}.
\ee
An evaluation of this formula on the $7$ coordinates $x^m$ 
\beg
\mathbf{R}_{{\left.\hat{S}\right.^{p}}_q} \left(x^m\right) 
&=&
-\frac{2}{3\tau_4}\left\{
\left( x^p\p_q +\eta_{qr}\eta^{ps}x^r\p_s  \right)
 +
 \frac{2}{7}\delta^p_q
 \left(\frac{1}{\tau_9}-1\right)
  x^r\p_r
\right\}\left(x^m\right) 
\eeg
is the standard vector field representation of symmetric generators $\hat{S}$ that I have already introduced in chapter \ref{CHAP4} (\ref{PRep},\,\ref{RDefi}), if and only if the constants $\tau_{4}$ and $\tau_9$ are fixed by
\begin{subequations}\label{chifix}
\be\label{chi4fix}
\tau_4&=&-\frac{4}{3},\\
\tau_9 &=&1
\label{chi17fix}
.
\ee
\end{subequations}
I want to add a remark concerning the constant $\tau_9$. As an $\mathfrak{so}_7$ representation, only the symmetric traceless matrices are irreducible. Since I did not separate the trace from the symmetric generators ${\left.\hat{S}\right.^{a}}_b$ in (\ref{defiSo7R}), I had to introduce the constant $\tau_9$ for the ansatz (\ref{defiSo7R}) to be general. $\tau_9$ has now been fixed by requiring that the vector field representation $\mathbf{R}$ of ${\left.\hat{S}\right.^{a}}_b$ in its evaluation on the $7$ coordinates $x^m$ has the standard form (\ref{PRep},\,\ref{RDefi}).\footnote{From an algebraic point of view, this constant $\tau_9$ is nothing but the different way, in which a $\mathfrak{gl}_7$ subalgebra can be embedded in $\mathfrak{e}_{7(7)}$. It is interesting to note that the so-called ``gravity subline'' of the Dynkin diagram, used e.g. in \cite{DHN02}, corresponds to a different choice for $\tau_9$.}\\

Hence, I have succeeded in finding a $\mathfrak{gl}_7$ subalgebra of $\mathfrak{e}_{7(7)}$, whose vector field representation on the first $7$ coordinates $x^m$ (\ref{xDefi8}) has the standard form (\ref{PRep}). I will denote its generators (\ref{LDefi},\,\ref{RDefi})
\be\label{GLdefi}
{\left.\hat{M}\right.^a}_b&=&{\left.\hat{S}\right.^a}_b+{\left.\hat{L}\right.^a}_b
\ee
with the letter $\hat{M}$ as usual. The different indices should be sufficient to distinguish the $\mathfrak{gl}_7$ generators from the ones of $\mathfrak{su}_8$.\\

For the relation to $d=11$ supergravity, it will be important to decompose the $\mathfrak{e}_{7(7)}$ algebra into $\mathfrak{gl}_7$ representations. Therefore, I have to reassemble the non-compact $\mathfrak{so}_7$ representations $\hat{S}^{a_1\dots a_3},\,\hat{S}^{a_1\dots a_6}$ and the compact ones $\hat{L}^{a_1\dots a_3},\,\hat{L}^{a_1\dots a_6}$ into $\mathfrak{gl}_7$ representations. The necessary ones for the comparison to $d=11$ supergravity will be
\begin{subequations}\label{EDefi}
\be\label{E3Defi}
\hat{E}^{abc}&:=&\hat{S}^{abc}+\hat{L}^{abc},\\
\label{E6Defi}
\hat{E}^{a_1\dots a_6}&:=&\hat{S}^{a_1\dots a_6}+\hat{L}^{a_1\dots a_6}.
\ee
\end{subequations}
In the context of $\mathfrak{so}_7$ representations, the position of the indices is not important, because the seven-metric $\eta^{ab}$ is a conserved tensor. For the $\mathfrak{gl}_7$ representations $\hat{E}$ (\ref{EDefi}), this is not the case: lowered indices correspond to the contragredient representation, which is not equivalent to the original one with raised indices. Since I insisted on the standard vector field representation $\mathbf{R}$ of the $\mathfrak{gl}_7$ generators $\hat{M}$ (\ref{PRep}), the position of the indices in (\ref{GLdefi}) is fixed. Requiring that the $\mathfrak{so}_7$ representations $\hat{E}$ (\ref{EDefi}) are $\mathfrak{gl}_7$ representations, uniquely determines the value of the normalization constants $\tau_5,\tau_6$ in (\ref{defiSo7R}) to
\begin{subequations}\label{Fixchi}
\be
\label{Fixchi5}
\tau_5&=&-\frac{\tau_2}{3},\\
\label{Fixchi6}
\tau_6 &=&-2\tau_3.
\ee
\end{subequations}
To obtain the explicit $\mathfrak{e}_{7(7)}$ commutation relations of the $\mathfrak{gl}_7$ representations $\hat{M}{}^a{}_b$ (\ref{GLdefi}) and $\hat{E}$ (\ref{EDefi}), I have to substitute the parametrizations of $\hat{M}{}^A{}_B$ (\ref{defiSo7}) and $\hat{S}_{ABCD}$ (\ref{defiSo7R}) into the commutation relations (\ref{ComRel8}). The result is
\begin{subequations}\label{M2E}
\be
	\left[ {\left.\hat{M}\right.^{e}}_f,\hat{E}^{abc}\right]\label{E2E3}
  &=&
 3 \left(
  \delta_f^{[a}\hat{E}^{bc]e}
  -
  \frac{1}{9}\delta_f^{e}\hat{E}^{abc}
  \right),\\
    \label{E2E6}
\left[ {\left.\hat{M}\right.^{a}}_b,\hat{E}^{e_1\dots e_6}\right]
	&=&
	6\left(\delta_b^{[e_6}\hat{E}^{e_1\dots e_5]a}-\frac{1}{9}\delta_b^{a}\hat{E}^{e_1\dots e_6}\right).
	\ee
\end{subequations}
In particular, note that the non-standard trace subtractions are uniquely determined by the present choice of $\tau_9=1$ (\ref{chi17fix}).\\

I have chosen the name $\hat{E}$ for the generators on purpose. These correspond to the Chevalley generators $\hat{E}_\alpha$ I have introduced in section \ref{Class}.\footnote{The corresponding dual generators $\hat{E}_{-\alpha}$ would have lowered $\mathfrak{gl}_7$ indices and they would correspond to the differences $\hat{F}_{abc}:=\hat{S}_{abc}-\hat{L}_{abc}$ and $\hat{F}_{a_1\dots a_6}:=\hat{S}_{a_1\dots a_6}-\hat{L}_{a_1\dots a_6}$. Together with the generators $\hat{E}$ (\ref{EDefi}) and with $\mathfrak{gl}_7$, these span the entire Lie algebra $\mathfrak{e}_{7(7)}$. The dual nilpotent generators $\hat{E}_{-\alpha}$ will not be important in the sequel.} In particular for the finite dimensional case considered here, the Serre relations (\ref{Serre}) imply that there is an integer number $n$ such that the $n$ fold adjoint action of any $\hat{E}\in \mathfrak{g}\ominus \mathfrak{h}$ on any other generator vanishes. That this is indeed the case, follows from the commutation relations
\begin{subequations}\label{EE}
\be
\left[ \hat{E}^{efg},\hat{E}^{abc}\right]\label{E3E3}
	 &=&
	 -\frac{4\tau_3}{\tau_2^2}
	  	 \hat{E}^{efgabc},\\
\label{E3E6}
\left[\hat{E}^{efg},\hat{E}^{a_1\dots a_6}\right] &=& 0,\\
\label{E6E6}
\left[\hat{E}^{e_1\dots e_6},\hat{E}^{a_1\dots a_6}\right]&=&0
\ee
\end{subequations}
that are uniquely determined by the relations (\ref{ComRel8}), too.\\

\subsection{Action of $\mathfrak{e}_{7(7)}$ on the coordinates}\label{summ1}
In this dissertation, I will only use the action of the $\mathfrak{gl}_7$ generator $\hat{M}{}^a{}_b$ (\ref{GLdefi}) and of the nilpotent generators $\hat{E}^{e_1\dots e_3}$ and $\hat{E}^{e_1\dots e_6}$ (\ref{EDefi}) on the $56$ coordinates $x^\mu$. To describe this action, recall that the vector field representation $\mathbf{R}$ of the $\mathfrak{gl}_7$ generator $\hat{M}{}^a{}_b\in  \mathfrak{e}_{7(7)}$ (\ref{GLdefi}) had the canonical shape (\ref{PRep}) in its action on the first $7$ coordinates $x^m$ of $x^\mu\approx (x^m,x^{mn},p_m,p_{mn})$ (\ref{xDefi8}).\\

Since I want to use the algebraic approach of chapter \ref{CHAP4} for a comparison of the connection $\nabla$ (\ref{sympConE}) to $d=11$ supergravity, it is natural to also split the dual generators $\hat{P}_\mu$ (\ref{Lie9}) accordingly. Their vector field representation $\mathbf{R}$ is tantamount to the derivative operator $\frac{\p}{\p x^\mu}$ (\ref{Lie9}) and hence, the split of $x^\mu$ into $(x^m,x^{mn},p_m,p_{mn})$ (\ref{xDefi8}) induces
\be\label{PDefi3}
\mathbf{R}_{\hat{P}_m}\,:=\,\frac{\p}{\p x^m}
&&\mathbf{R}_{\hat{P}_{mn}}\,:=\,\frac{\p}{\p x^{mn}}\nn\\
\mathbf{R}_{\hat{Q}^m}\,:=\,\frac{\p}{\p p_m}
&&\mathbf{R}_{\hat{Q}^{mn}}\,:=\,\frac{\p}{\p p_{mn}}.
\ee
I have used the letter $\hat{Q}$ in order to distinguish the coordinate vectors for the $28$ coordinates $(x^m,x^{mn})$ from the ones for the $28$ dual coordinates $(p_m,p_{mn})$. \\

A comparison of these vector field representations (\ref{PDefi3}) with the one of the holomorphic generator $\hat{Z}_{N_1N_2}$ (\ref{ZDefi2}) leads with the definition of the holomorphic coordinate $z^{N_1N_2}$ (\ref{zNeu2},\,\ref{zNeu3}) to the following identity:
\be\label{defiSo7P2}
\hat{Z}_{AB} &=&-\frac{i\tau_7}{8}\hat{Z}_a{\G^a}_{AB}+\frac{\tau_{7}\sqrt{2}}{2!8}\hat{Z}_{ab} {\G^{ab}}_{AB}
\ee
with
\begin{subequations}\label{dzNeu2b}
\be
\hat{Z}_{a} &:=&\frac{1}{2}\left(\hat{P}_{a} -i\eta_{ac} \hat{Q}^{c}\right),\\
\hat{Z}_{ab} &:=&\frac{1}{2}\left(\hat{P}_{ab} -i\eta_{ac} \eta_{bd} \hat{Q}^{cd}\right).
\ee
\end{subequations}
Next, I substitute the relation (\ref{defiSo7P2}) into the commutation relations (\ref{RZqCR}) and (\ref{MZCR}), together with the parametrizations of the generators $\hat{M}{}^A{}_B$ (\ref{defiSo7}) and $\hat{S}_{ABCD}$ (\ref{defiSo7R}) of $\mathfrak{e}_{7(7)}$. The resulting commutation relations uniquely determine the action of the $\mathfrak{gl}_7$ generators $\hat{M}{}^a{}_b$ (\ref{GLdefi}) and of $\hat{E}$ (\ref{EDefi}) on the generators $\hat{P}$ and $\hat{Q}$ (\ref{PDefi3}). The action of the nilpotent generators $\hat{E}$ follows the sequence:\footnote{I have listed the explicit commutation relations in the appendix \ref{CRAE2}.}

\begin{center}
\begin{tabular}{ccccccccc}
&&$\stackrel{\hat{E}^{e_1\dots e_6}}{\longrightarrow }$&&&&$\stackrel{\hat{E}^{e_1\dots e_6}}{\longrightarrow }$\\
&
$\nearrow$ &  & $\searrow$&&$\nearrow$ &  & $\searrow$
\\
$\hat{P}_e$& $\stackrel{\hat{E}^{abc}}{\longrightarrow }$&$ \hat{Q}^{gh} 
$&$ \stackrel{\hat{E}^{abc}}{\longrightarrow }  $&$ \hat{P}_{ij}
$&$ \stackrel{\hat{E}^{abc}}{\longrightarrow }  $&$ \hat{Q}^{k}
$&$ \stackrel{\hat{E}^{abc}}{\longrightarrow }  $&$ 0$\\
&&&$\searrow$ &  & $\nearrow$&&$\searrow$  $\nearrow$
\\
&&&&$\stackrel{\hat{E}^{e_1\dots e_6}}{\longrightarrow }$&&&$\stackrel{\hat{E}^{e_1\dots e_6}}{}$
\end{tabular}
\end{center}

Before parametrizing the $E_{7(7)}$ valued vielbein with the help of the generators $\hat{M}{}^a{}_b$ (\ref{GLdefi}) and $\hat{E}$ (\ref{EDefi}), I want to review the origin of the different coordinates I have used.
\begin{enumerate}
	\item I started by evaluating the effect of preserving the symplectic form $\Omega$ in section \ref{first}, which led to the restriction of the diffeomorphisms $\Diff(56)$ to symplectomorphisms $\Symp(56)$. To write these in a nice way, I split the $56$ coordinates $x^\mu$ into the pair $(x^\alpha,p_\beta)$ with $\alpha,\beta=1,\dots,28$.
	\item It was essential for the construction of the general connection with affine linear equivariance in chapter \ref{CHAP4} to separate symmetric generators $\hat{S}$ from antisymmetric ones $\hat{L}$. For the Euclidean case, this is equivalent to the separation into non-compact generators $\hat{S}$ and compact ones $\hat{L}$. This was the reason for substituting the complex coordinates $(z^\alpha,\bar{z}_\beta)$ for $(x^\alpha,p_\beta)$ in section \ref{SympV1}, because in these coordinates, the vector field representations of both the compact and the non-compact generators preserving $\Omega$ had a particularly simple form (\ref{Unity}).
	\item In section \ref{ESy}, I refined the original splitting of the $56$ real coordinates in $(x^\alpha,p_\beta)$ to $(x^m,x^{mn},p_m,p_{mn})$ with $m,n=1,\dots,7$. In these coordinates, I specified a real symmetric quartic tensor $Q$ (\ref{Q2k}) whose preservation is imposed as an additional constraint on the symplectomorphisms. This constraint is solved in the (anti)holomorphic frame $(z^{M_1M_2},\bar{z}_{N_1N_2})$ that I introduced in (\ref{zNeu3}).
	\item The class of diffeomorphisms that preserve the two tensors $\Omega$ and $Q$ is isomorphic to the subgroup  $\cP_{56}\rtimes E_{7(7)}$ of the affine group $A(56)$, what I proved in section \ref{VectE4}.
	\item Finally, I parametrized the Lie algebra of $\cP_{56}\rtimes E_{7(7)}$ by $\mathfrak{gl}_7$ representations and I evaluated its action on the $56$ coordinates $(x^m,x^{mn},p_m,p_{mn})$.
\end{enumerate}
I want to conclude with a group theoretic argument why labelling the $56$ coordinates in the form $(x^m,x^{mn},p_m,p_{mn})$ (\ref{xDefi8}) is in a way preferred in this exceptional geometry. I defined the latter by imposing the invariance of the two tensors $\Omega$ and $Q$. Therefore, the group of admissible coordinate transformations is restricted to $\cP_{56}\rtimes E_{7(7)}\subset \Diff(56)$. This statement is independent of the labelling of the coordinates.\\

Next observe that a theory of unrestricted Euclidean gravity in the formulation of chapter \ref{CHAP4} needs a vielbein matrix that is $Gl(n)/SO(n)$ valued. Since I restricted the $56$-dimensional vielbein to an $E_{7(7)}$ matrix, it is natural to ask for $Gl(n)$ subgroups of $E_{7(7)}$ that correspond to a lower dimensional geometry with general coordinate invariance $\Diff(n)$. The maximal one is $Gl(7)$, whose action on $7$ coordinates $x^m$ can be fixed in the canonical way. This induces the labelling $(x^m,x^{mn},p_m,p_{mn})$ (\ref{xDefi8}).\footnote{Even though $\mathfrak{sl}_8$ also is a subalgebra of $\mathfrak{e}_{7(7)}$, it cannot be extended to $\mathfrak{gl}_8$.}\\

Finally, it is important to recall that not all $Gl(n)$ subgroups of $E_{7(7)}$ are subgroups of $Gl(7)\subset  E_{7(7)}$. A very prominent counterexample is $Gl(6)\times Sl(2)\subset E_{7(7)}$. After having fixed the supersymmetry variations of the $56$-dimensional theory by a comparison to $d=11$ supergravity, it would be very interesting to decompose the $E_{7(7)}$ valued vielbein and the $\mathbf{56}$ coordinates $x^\mu$ with respect to this subgroup. It is likely that these coincide with the supersymmetry transformations of $IIB$ supergravity in ten dimensions for the common subsector of fields contained in the $56$-dimensional vielbein $e^H$.\\

This leads to the possibility that all solutions of both $IIB$ supergravity in $d=10$ and $d=11$ supergravity exactly correspond to the solutions of the complete theory in the $60$-dimensional exceptional geometry that do not depend on the $56-n$ dimensions, on which the corresponding $Gl(n)$ subgroup acts in a non-canonical way, with $n=6,7$ respectively. This investigation is beyond the scope of this thesis, however.

\section{Vielbein and connection in $Gl(7)$ decomposition}\label{affE7}
\subsection{The vielbein of the exceptional geometry}\label{affE71}
At the end of the preceding section \ref{summ1}, I explained that the $Gl(7)$ subgroup of $E_{7(7)}$ induces the split of the $56$ coordinates $x^\mu$ into $(x^m,p_{mn},x^{mn},p_m)$ (\ref{xDefi8}). This can also be phrased in other words: it corresponds to the decomposition of the $\mathfrak{e}_{7(7)}$ representation $\mathbf{56}$ into irreducible $\mathfrak{gl}_7$ subrepresentations
\be\label{egl7}
\mathbf{56} &=& \mathbf{7}\oplus\overline{\mathbf{21}}\oplus\mathbf{21}\oplus\overline{\mathbf{7}}.
\ee
The representations $\mathbf{7}$ and $\mathbf{21}$ are spanned by the generators $\hat{P}_m$ and $\hat{P}_{mn}$, respectively, whereas the contragredient or dual representations $\overline{\mathbf{7}}$ and $\overline{\mathbf{21}}$ correspond to the generators $\hat{Q}^m$ and $\hat{Q}^{mn}$ that are associated to the dual coordinates $p_m$ and $p_{mn}$ by (\ref{PDefi3}). \\

Locally, these $56$ vectors $(\hat{P}_m,\,\hat{Q}^{mn},\,\hat{P}_{mn},\,\hat{Q}^n)$ parametrize the open set $U_\alpha$ of the $56$-dimensional submanifold of $\cM^{60}$ under discussion. Therefore any point $x\in U_\alpha$ is associated to the coefficients $(x^m,p_{mn},x^{mn},p_m)$ (\ref{xDefi8}) by a relation analogous to (\ref{PDefi0})
\be\label{coord6}
x &=& x^\mu\hat{P}_\mu\in \mathbf{56}\quad \text{with } \mu=1,\dots, 56\\
&=&
 x^m\hat{P}_m + p_{mn}\hat{Q}^{mn} + x^{mn}\hat{P}_{mn} + p_m\hat{Q}^{m}\quad\text{with } m,n=1,\dots, 7\nn.
\ee
The obvious $E_{7(7)}\subset Gl(56)$-covariance of the first line is obscured to obvious $Gl(7)$-covariance and hidden $E_{7(7)}$-covariance in the second line by this parametrization. I emphasize that the $E_{7(7)}$-covariance is not broken by this parametrization.\\

Furthermore, recall two implications of the statement from section \ref{LCcE} that the external symmetry group $\Diff(56)$ is restricted to $\cP_{56}\rtimes E_{7(7)}$ by requiring the invariance of the tensors $\Omega$ and $Q$:
\begin{enumerate}
	\item It is consistent to restrict the vielbein matrix in $56$ dimensions to an $E_{7(7)}$ matrix. Then, the orthogonal matrix that restores an arbitrary fixing of the vielbein gauge, is in $SU(8)/\Z_2\subset SO(56)$.
	\item Since $\cP_{56}\rtimes E_{7(7)}$ is a subgroup of the affine group $A(56)$ (\ref{affine2}), the analysis from the sections \ref{secA3} and \ref{secB} reveals that it is sufficient to discuss the affine coset $A(56)/SO(56)$. In contradistinction to the unrestricted geometry of chapter \ref{CHAP4}, there is no further constraint, because there is no symmetry enhancement.
\end{enumerate}
Therefore, I can without loss of generality focus on the affine coset (\ref{paramCoset}) in $56$ dimensions
\beg
C&=& e^{x^\mu \hat{P}_\mu} e(x)\in  A(56)/SO(56),
\eeg
with a restricted vielbein matrix
\beg
e^H&:=& e(x) \in E_{7(7)}/(SU(8)/\Z_2).
\eeg
With the split of the coordinates (\ref{coord6}), this leads to
\be\label{cosetE7}
C&=& e^{x^m\hat{P}_m + p_{mn}\hat{Q}^{mn} + x^{mn}\hat{P}_{mn} + p_m\hat{Q}^{m}} e^H.
\ee
Since the coordinates commute (\ref{ComRel3}), the Maurer--Cartan form takes the form
\beg
C^{-1}dC &=& dx^m \left(e^H\right)_m +dp_{mn} \left(e^H\right)^{mn} +dx^{mn} \left(e^H\right)_{mn} +dp_{m} \left(e^H\right)^{m}\\
&&
+ e^{-H}d e^H
\eeg
with the definitions of the $\mathbf{56}$ valued expressions
\begin{subequations}\label{eHD1}
\be
\left(e^H\right)_m&:=& e^{-H}\cdot \hat{P}_m \cdot e^H \\
\left(e^H\right)^{mn}&:=& e^{-H}\cdot \hat{Q}^{mn} \cdot e^H \\
\left(e^H\right)_{mn}&:=& e^{-H}\cdot \hat{P}_{mn} \cdot e^H \\
\left(e^H\right)^{m}&:=& e^{-H}\cdot \hat{Q}^{m} \cdot e^H.
\ee
\end{subequations}
These are abbreviations for the adjoint Lie group action of $e^H\in E_{7(7)}$ on the basis vectors of the representation $\mathbf{56}$ (\ref{coord6}). \\

A comparison with the discussion of the affine coset in section \ref{ConnAff} reveals that the expressions (\ref{eHD1}) comprise the degrees of freedom of the vielbein in $56$ dimensions. In order to determine the entries of the vielbein matrix, I have to calculate the coefficients of the basis vectors that span $\mathbf{56}$. I will use the holomorphic generators $\hat{Z}_{AB}$ (\ref{defiSo7P2}) for this task:
\begin{subequations}\label{eHD2}
\be
\left(e^H\right)_m &=:&{\left(e^H\right)_m}^{AB}\hat{Z}_{AB} + \text{c.c.}\\
\left(e^H\right)^{mn} &=:&{\left(e^H\right)^{mn}}^{AB}\hat{Z}_{AB} + \text{c.c.}\\
\left(e^H\right)_{mn} &=:&{\left(e^H\right)_{mn}}^{AB}\hat{Z}_{AB} +\text{c.c.}\\
\left(e^H\right)^m &=:&{\left(e^H\right)^m}^{AB}\hat{Z}_{AB} + \text{c.c.}.
\ee
\end{subequations}

In the remaining part of this section, I will explain why this choice to use the holomorphic generators $\hat{Z}_{AB}$ (\ref{defiSo7P2}) is preferred. A general admissible diffeomorphism in the exceptional geometry induces a global left action by $E_{7(7)}$ on the vielbein matrix $e^H$. Having fixed the vielbein gauge in an arbitrary way, the global left action on $e^H$ induces an orthogonal right action by
\beg
SU(8)/\Z_2&=& SO(56)\cap E_{7(7)}.
\eeg
Since the vielbein matrix $e^H$ links the coordinate induced frame of the tangent space to the vielbein frame by definition, the induced action on the former is by $E_{7(7)}$, whereas the action on the latter is by $SU(8)/\Z_2$. The important fact is that the irreducible $\mathfrak{e}_{7(7)}$ representation $\mathbf{56}$ is reducible with respect to its $\mathfrak{su}_8$ subalgebra:
\beg
\mathbf{56} = \mathbf{28}\oplus \overline{\mathbf{28}}.
\eeg
It is obvious from the commutation relation (\ref{MZCR}) that the holomorphic generators $\hat{Z}_{AB}$ form this irreducible representation $\mathbf{28}$ of $SU(8)/\Z_2$. Hence, this is also true for their coefficients
\beg
dx^m {\left(e^H\right)_m}^{AB} + dp_{mn} {\left(e^H\right)^{mn}}^{AB} +dx^{mn} {\left(e^H\right)_{mn}}^{AB} +dp_{m}{\left(e^H\right)^{m}}^{AB},
\eeg
which is the reason why the choice of $\hat{Z}_{AB}$ for the basis of the tangent space in the vielbein frame is preferred. Since the coordinate induced frame transforms under $\mathfrak{e}_{7(7)}$ in an irreducible way, a similar decomposition of $dx^\mu$ into complex coordinates is not preserved.\footnote{This would only be true, if the manifold allowed for an integrable complex structure \cite{J00}, which is not the case in general.}

\subsection{The explicit form of the vielbein}\label{VielbE}
At the end of section \ref{Decompo2}, I mentioned that the Lie algebra $\mathfrak{e}_{7(7)}$ is spanned by the $\mathfrak{gl}_7$ generators $\hat{M}{}^a{}_b$ (\ref{GLdefi}), by the nilpotent generators $\hat{E}^{abc},\,\hat{E}^{a_1\dots a_6}$ (\ref{EDefi}) and their dual generators $\hat{F}_{abc},\,\hat{F}_{a_1\dots a_6}$.  This implies in particular that the vielbein coset 
\beg
e^H\in E_{7(7)}/(SU(8)/\Z_2)
\eeg
can be parametrized by these vectors in the same way, as the standard vielbein was parametrized by the $\mathfrak{gl}_{d}$ generators $\hat{M}{}^\mu{}_\nu$ in equation (\ref{eMf}) of section \ref{ConnAff}. As long as no particular matrix form for the vielbein $e(x)$ is fixed, the invariance of the theory under the $d$-dimensional Lorentz group is manifest.\\
 
Since the $SU(8)/\Z_2$-covariance of the bosonic part of $d=11$ supergravity is not manifest, I will partly fix the vielbein gauge freedom by requiring that the matrix $e^H$ be of the following form
\be
e^H&=:&e^{{h_a}^b{\left.\hat{M}\right.^a}_b}e^{A_{abc}\hat{E}^{abc}}e^{A_{a_1\dots a_6}\hat{E}^{a_1\dots a_6}}\in E_{7(7)}/(SU(8)/\Z_2).
\label{cosetE73}
\ee
This parametrization is consistent, because it follows the Iwasawa decomposition keeping in mind that the generators $\hat{E}$ are nilpotent generators as discussed in section \ref{Decompo2}. In the following section, I will show that this choice (\ref{cosetE73}) corresponds to a block-diagonal form of the vielbein matrix with the blocks corresponding to the irreducible $Gl(7)$ subrepresentations of $\mathbf{56}$ (\ref{egl7}).\\

Since I have parametrized the vielbein $e^H$ by $Gl(7)$ representations, the remaining vielbein gauge freedom has to be
\beg
SO(7)&=& SU(8)/\Z_2\cap Gl(7).
\eeg
It corresponds to the undetermined form of the $Gl(7)$ valued matrix
\be\label{tildeE}
\tilde{e} &:=& e^{{h_a}^b{\left.\hat{M}\right.^a}_b} \in Gl(7)/SO(7).
\ee
It is important to note that this parametrization does not break the $E_{7(7)}$ symmetry. It only hides it in the same way as it was obscured by the parametrization of the coordinates (\ref{coord6}). This implies that the theory is still invariant under $SU(8)/\Z_2$, but it is not a manifest invariance as before, because the $\mathfrak{gl}_7$ representations that I used for the parametrization (\ref{cosetE73}) are linked to each other by a general $E_{7(7)}$ action.\\

This is the same argumentation familiar from a Kaluza--Klein reduction of a $(d+1)$-dimensional theory on a circle: the theory is invariant under $\Diff(d+1)$, but the parametrization is by $\mathfrak{so}_{(d-1,1)}$ representations. Only if the dependences on the $d+1^{\text{st}}$ coordinate, i.e. the ``non-trivial modes'', are neglected, the symmetry is broken to $\Diff(d)$. As long as the dependences on all coordinates are kept, the symmetry is still the original one. In constructing the vielbein, I do keep all the dependences on the $56$ coordinates and hence the entire symmetry group.\\

The decomposition of the $56$-bein $e^H$ into $Gl(7)$ representations (\ref{cosetE73}) allows to state the explicit dependence of $e^H$ (\ref{eHD2}) on the parameters $\tilde{e}$, $A_{a_1\dots a_3}$ and $A_{a_1\dots a_6}$. With the formula (\ref{formula2}) and the commutators from appendix \ref{CRAE2}, an analogous computation to the one in section \ref{ConnAff} evaluates the adjoint actions of (\ref{eHD1}):  
\begin{subequations}\label{VielbeinE}
\be\label{eHQ1b}
{\left(e^H\right)^m}^{AB} &=&\frac{1}{\tau_7}
 {\left.\tilde{e}\right._c}^m  {\G^{c}}^{AB}
 \\
  {\left(e^H\right)_{mn}}^{AB} \label{eHP2b}
&=&
\frac{1}{\tau_{7}\sqrt{2}}e^{-\frac{7}{3}\tilde{\sigma}} 
 {\left.\tilde{e}\right._m}^c {\left.\tilde{e}\right._n}^d
 \left({\G_{cd}}^{AB}
 -\frac{24}{\tau_2}A_{cd z}{\G^{z}}^{AB}\right)
 \\
\label{eHQ2b}
{\left(e^H\right)^{mn}}^{AB}  &=& \frac{1}{\tau_{7}\sqrt{2}}e^{\frac{7}{3}\tilde{\sigma}} 
 {\left.\tilde{e}\right._c}^m{\left.\tilde{e}\right._d}^n  
 \\
  &&\left(i{\G^{cd}}^{AB} -\frac{2}{\tau_2} A_{a_1\dots a_3}\e^{cda_1\dots a_3jk}{\G_{jk}}^{AB}
-
16{U_{z}}_{-}^{cd}{\G^{z}}^{AB}\right)\nn\\
\label{eHP1b}
{\left(e^H\right)_m}^{AB} &=& \frac{1}{i\tau_7}
 {\left.\tilde{e}\right._m}^d \\
 &&\left({\G_{d}}^{AB} +\frac{12}{\tau_2} A_{abd} {\G^{ab}}^{AB}
 -8i{U_{d}}_{+}^{jk}{\G_{jk}}^{AB} -8iU_{ad}{\G^{a}}^{AB}\right).
 \nn
\ee
\end{subequations}

I introduced the abbreviations
\begin{subequations}\label{UDefi}
\be
{U_{d}}_{-}^{jk}
&:=&
-\frac{3}{2\tau_2^2}A_{abc}A_{ghd}\e^{jkghabc}
-
\frac{1}{2\tau_3}A_{a_1\dots a_6}\e^{a_1\dots a_6[j}\delta_d^{k]}
\\
{U_{d}}_{+}^{jk}
&:=&
-\frac{3}{2\tau_2^2}A_{abc}A_{ghd}\e^{jkghabc}
+\frac{1}{2\tau_3}A_{a_1\dots a_6}\e^{a_1\dots a_6[j}\delta_d^{k]}
\\
U_{ad}
&:=&
-\frac{12}{\tau_2\tau_3}A_{a_1\dots a_6}A_{abd}\e^{ba_1\dots a_6}
+\frac{12}{\tau_2^3} A_{ars}A_{ghi}A_{kld}\e^{rsghikl}
\\
e^{\tilde{\sigma}} &:=& \det\left(\tilde{e}\right)^{\frac{1}{7}}
\label{sigmaDefi2}
\ee
\end{subequations}
where the last definition follows the one of $\sigma$ in (\ref{sigmaDefi}). Furthermore, I used the standard convention for the residual vielbein $\tilde{e}$ (\ref{ehklein}) to separate it from its inverse only by a different naming of the indices.\footnote{This implies ${\left.\tilde{e}\right._c}^m {\left.\tilde{e}\right._m}^d=\delta_c^d$ etc.} I want to emphasize again that it is only this convention that associates different meanings to the same symbol $\tilde{e}$ with different indices.

\subsection{Matrix form and inverse vielbein}\label{Matrix}
The matrix form of the vielbein $e^H$ is obtained by spanning both vector spaces that are linked by the linear map $e^H$, with the $56$ standard basis vectors $\hat{P}_\mu$ (\ref{PDefi0}). In section \ref{affE71}, I showed that the $\mathfrak{gl}_7$ decomposition of the $\mathfrak{e}_{7(7)}$ representation $\mathbf{56}$ induces a relabelling of the vectors $\hat{P}_\mu$ by $(\hat{P}_m,\,\hat{Q}^{mn},\,\hat{P}_{mn},\,\hat{Q}^n)$. This leads to the block notation for the vielbein matrix $e^H$, in which the top left block corresponds to a $\mathbf{7}\times \mathbf{7}$ submatrix etc.
\beg
e^H=\left(
\scalebox{.78}{
\begin{tabular}{c|c|c|c}
${\left.\tilde{e}\right._m}^d$
&
$-\frac{12}{\tau_2}\sqrt{2}{\left.\tilde{e}\right._m}^d A_{abd}$
&
$-8\sqrt{2}{\left.\tilde{e}\right._m}^d {U_{d}}_{+}^{jk}$
&
$-8{\left.\tilde{e}\right._m}^d U_{ad}$\\
\hline
$0$&$e^{\frac{7}{3}\tilde{\sigma}} 
 {\left.\tilde{e}\right._a}^m{\left.\tilde{e}\right._b}^n$
  &
  $-\frac{2}{\tau_2}e^{\frac{7}{3}\tilde{\sigma}} 
 {\left.\tilde{e}\right._c}^m{\left.\tilde{e}\right._d}^n
 A_{a_1\dots a_3}\e^{cda_1\dots a_3jk}$
  &
  $-8\sqrt{2}e^{\frac{7}{3}\tilde{\sigma}} 
 {\left.\tilde{e}\right._c}^m{\left.\tilde{e}\right._d}^n
 {U_{a}}_{-}^{cd}$\\
\hline
$0$&$0$&$e^{-\frac{7}{3}\tilde{\sigma}} 
 {\left.\tilde{e}\right._m}^j {\left.\tilde{e}\right._n}^k$
 &$-\frac{12}{\tau_2}\sqrt{2}e^{-\frac{7}{3}\tilde{\sigma}} 
 {\left.\tilde{e}\right._m}^c {\left.\tilde{e}\right._n}^d A_{cd a}$\\
 \hline
$0$&$0$&$0$&${\left.\tilde{e}\right._a}^m$
\end{tabular}
}
\right)
\eeg
Obviously, this vielbein matrix is real. The appearance of the complex unity $i$ in the presentation (\ref{VielbeinE}) only is due to the holomorphic frame $\hat{Z}_{AB}$ (\ref{defiSo7P2}).\\

Apart from the $\mathfrak{gl}_7$ generators $\hat{M}{}^a{}_b$ (\ref{GLdefi}), I have only used nilpotent generators $\hat{E}$ (\ref{EDefi}) for the parametrization of the vielbein $e^H$ (\ref{cosetE73}). This may have already led to the expectation that the corresponding vielbein matrix is of block-triangular shape. Since these matrices form a group, its inverse also has this shape.\\

In the sections \ref{ConnAff} and \ref{ConnAff2}, it has become clear that I need the inverse vielbein matrix $e^{-H}$ for the construction of connections $\nabla$ in the vielbein frame. The easiest way to obtain $e^{-H}$ is to use the fact that the vielbein is a symplectic matrix $e^H\in E_{7(7)}\subset Sp(56)$, which leads to the relation
\beg
{\left(e^{-H}\right)_\alpha}^\mu  \Omega_{\mu\nu} &=& {\left(e^H\right)_\nu}^\beta\Omega_{\alpha\beta}  
\eeg
with $\mu,\nu,\alpha,\beta=1,\dots,56$.\\

Since the symplectic form $\Omega$ has the same coefficients (\ref{omr1}) in both the vielbein and the coordinate induced frame, a short calculation provides the identity
\begin{subequations}\label{vili2}
  \be
     {\left(e^{-H}\right)_{AB}}^m&=&\frac{i\tau_7^2}{16}\overline{{\left(e^H\right)^m}_{AB}}
     \\
    {\left(e^{-H}\right)_{AB}}^{mn}&=&\frac{i\tau_7^2}{16}\overline{{\left(e^H\right)^{mn}}_{AB}}
    \\
    {\left(e^{-H}\right)_{AB}}_{mn}&=&-\frac{i\tau^2_{7}}{16}\overline{{\left(e^H\right)_{mn}}_{AB}}
    \\
  {\left(e^{-H}\right)_{AB}}_m&=&-\frac{i\tau_7^2}{16}\overline{{\left(e^H\right)_m}_{AB}}
  \ee
  \end{subequations}
  where I used the canonical isomorphism between tangent and cotangent space again to change the position of the indices, which is provided by the symplectic form $\Omega$ (\ref{lower6}). The factor $i\tau_7^2/16$ in these relations is the one that I have already mentioned in section \ref{ESy}: it is due to the fact that I chose the canonical normalizations of complex geometry that are not canonical from the point of view of symplectic geometry. Changing from the holomorphic frame $\hat{Z}_{AB}$ (\ref{defiSo7P2}) to the standard coordinates (\ref{coord6}) provides the matrix form of $e^{-H}$.
 \beg
e^{-H}=\left(
\scalebox{.78}{
\begin{tabular}{c|c|c|c}
${\left.\tilde{e}\right._c}^m$
&
$\frac{12}{\tau_2}\sqrt{2}e^{-\frac{7}{3}\tilde{\sigma}} 
 {\left.\tilde{e}\right._m}^a {\left.\tilde{e}\right._n}^b A_{abc}$
&
$-8\sqrt{2}e^{\frac{7}{3}\tilde{\sigma}} 
 {\left.\tilde{e}\right._j}^m {\left.\tilde{e}\right._k}^n
 {U_{c}}_{-}^{jk}$
&
$8{\left.\tilde{e}\right._m}^a
  U_{ca}$\\
\hline
$0$&$e^{-\frac{7}{3}\tilde{\sigma}} 
 {\left.\tilde{e}\right._m}^c{\left.\tilde{e}\right._n}^d$
 &
 $\frac{2}{\tau_2}e^{\frac{7}{3}\tilde{\sigma}} 
 {\left.\tilde{e}\right._j}^m{\left.\tilde{e}\right._k}^n 
 A_{a_1\dots a_3}\e^{cda_1\dots a_3jk}$
  &
  $-8\sqrt{2}{\left.\tilde{e}\right._m}^a
 {U_{a}}_{+}^{cd}$\\
\hline
$0$&$0$&$e^{\frac{7}{3}\tilde{\sigma}} 
 {\left.\tilde{e}\right._c}^m {\left.\tilde{e}\right._d}^n$
 &
 $\frac{12}{\tau_2}\sqrt{2}{\left.\tilde{e}\right._m}^a A_{cd a}$\\
 \hline
$0$&$0$&$0$&${\left.\tilde{e}\right._m}^c$\\
\end{tabular}
}
\right)
\eeg
  
I needed the inverse vielbein $e^{-H}$ for the construction of connections $\nabla$ acting on Lorentz representations, because it provided a Lorentz covariant basis for the tangent space. Technically, I have to multiply the partial derivative $\frac{\p}{\p x^\mu}$ by $e^{-H}$ to switch to this basis (\ref{ehneu})
\be\label{pUd}
\underline{\p}_\alpha &:=&{\left(e^{-H}\right)_\alpha}^\mu\p_\mu
\ee
with $\alpha,\mu=1,\dots,56$. As in the definition of the holomorphic frame $\underline{dz}^{AB}$ (\ref{holframe3}), I have introduced the convention to distinguish the objects in the $56$-dimensional vielbein basis from the ones in the coordinate induced frame by an underline. I will explain next, why this is necessary to avoid a possible confusion.\\

At first, I switch to the holomorphic frame $\hat{Z}_{AB}$ (\ref{defiSo7P2}) and split the $56$ coordinates $x^\mu$ in the four $Gl(7)$ representations $(x^m,p_{mn},x^{mn},p_m)$ (\ref{coord6}). The definition of $\underline{\p}_\alpha$ (\ref{pUd}) is thus equivalent to defining the holomorphic derivative in the vielbein frame
\be\label{pDefi5}
\underline{\p}_{AB} &:=&{\left(e^{-H}\right)_{AB}}^m\frac{\p}{\p x^m} +{\left(e^{-H}\right)_{AB}}^{mn}\frac{\p}{\p x^{mn}}\nn\\
&&
+{\left(e^{-H}\right)_{AB}}_{mn}\frac{\p}{\p p_{mn}} +{\left(e^{-H}\right)_{AB}}_m\frac{\p}{\p p_m}.
\ee
A complex conjugation provides the antiholomorphic derivative operator in the vielbein frame $\overline{\underline{\p}}$, whose indices are raised with the symplectic form $\Omega$ (\ref{lower6}). \\

Next, I want to recall that the parametrization of the vielbein $e^H$ is explicitly stated in $Gl(7)$ representations (\ref{VielbeinE}). Therefore, it is natural to define the real valued abbreviations for derivative operators in the $56$-dimensional vielbein frame in analogy to the decomposition of $\hat{Z}_{AB}$ into the real generators $\hat{P}$ and $\hat{Q}$ (\ref{defiSo7P2},\,\ref{dzNeu2b}).
\begin{subequations}\label{pDefi}
\be
\underline{\p}_c&:=&\frac{1}{i\tau_7}{\G_c}^{AB}\underline{\p}_{AB} + \text{c.c.}
\\
\underline{\p}_{cd}&:=&\frac{1}{\tau_7\sqrt{2}}{\G_{cd}}^{AB}\underline{\p}_{AB} + \text{c.c.}\\
\underline{\p}^{cd}&:=&\frac{i}{\chi_7\sqrt{2}}{\G^{cd}}^{AB}\underline{\p}_{AB} + \text{c.c.}\\
\underline{\p}^{c}&:=&\frac{1}{\chi_7}{\G^{c}}^{AB}\underline{\p}_{AB} + \text{c.c.}
\ee
\end{subequations}
This definition finally allows to determine the dependences of the partial derivatives in the $56$-dimensional vielbein frame on the $Gl(7)$ representations $\tilde{e}$, $A_{a_1\dots a_3}$ and $A_{a_1\dots a_6}$ parametrizing the vielbein $e^H$ (\ref{cosetE73}), in an easy way. A standard matrix multiplication with $e^{-H}$ provides
\begin{subequations}\label{EMinusH}
\be
\underline{\p}_{c}
&=&
 {\left.\tilde{e}\right._c}^m  \frac{\p}{\p x^m}
 +\frac{12}{\tau_2}\sqrt{2}A_{ab c}e^{-\frac{7}{3}\tilde{\sigma}} 
 {\left.\tilde{e}\right._m}^a {\left.\tilde{e}\right._n}^b  \frac{\p}{\p p_{mn}}
 \nn\\
 && -8\sqrt{2}e^{\frac{7}{3}\tilde{\sigma}} 
 {\left.\tilde{e}\right._j}^m{\left.\tilde{e}\right._k}^n 
 {U_{c}}_{-}^{jk}
\frac{\p}{\p x^{mn}}
+8 {\left.\tilde{e}\right._m}^a
 U_{ca}
\frac{\p}{\p p_{m}}
\\
\underline{\p}^{cd}
&=&
 e^{-\frac{7}{3}\tilde{\sigma}} 
 {\left.\tilde{e}\right._m}^c {\left.\tilde{e}\right._n}^d
   \frac{\p}{\p p_{mn}}
-8\sqrt{2}{U_{a}}_{+}^{cd}
  {\left.\tilde{e}\right._m}^a 
  \frac{\p}{\p p_{m}}
  \nn\\
   &&
   +\frac{2}{\tau_2} A_{a_1\dots a_3}\e^{cda_1\dots a_3jk}e^{\frac{7}{3}\tilde{\sigma}} 
 {\left.\tilde{e}\right._j}^m{\left.\tilde{e}\right._k}^n  
 \frac{\p}{\p x^{mn}}
\\
\underline{\p}_{cd}
&=&
 e^{\frac{7}{3}\tilde{\sigma}} 
 {\left.\tilde{e}\right._c}^m{\left.\tilde{e}\right._d}^n
 \frac{\p}{\p x^{mn}}
+\frac{12}{\tau_2}\sqrt{2}A_{cda}
  {\left.\tilde{e}\right._m}^a
 \frac{\p}{\p p_{m}}
 \\
\underline{\p}^{c}
&=&
{\left.\tilde{e}\right._m}^c
\frac{\p}{\p p_{m}}.
\ee
\end{subequations}
It is the first line that could cause a problem. Since it is natural to consider $\tilde{e}$ as a vielbein in seven dimensions, the object $\underline{\p}_c$ might be confused with $\tilde{e}{}_c{}^m\frac{\p}{\p x^m}$, if the underline in the definitions of the partial derivative in the vielbein frame $\underline{\p}$ (\ref{pUd},\,\ref{pDefi5},\,\ref{pDefi}) was dropped. In particular for a comparison to $d=11$ supergravity, it is sensible to keep the underline, because the vielbein in eleven dimensions also induces an abbreviation $\p_a$.

\subsection{The Maurer--Cartan form $e^{-H}de^H$}\label{cII}
In section \ref{LCcE}, I stated the general form of a connection $\nabla$ (\ref{sympConE}) that is equivariant under all diffeomorphisms preserving the tensors $\Omega$ and $Q$, that entirely depends on vielbein degrees of freedom and that is linear in derivatives. It was uniquely determined up to two constants $c_1,c_2\in R$.\\

To relate this result to $d=11$ supergravity, I parametrized the $E_{7(7)}$ vielbein matrix $e^H$ in terms of $Gl(7)$ representations $\tilde{e}$, $A_{a_1\dots a_3}$ and $A_{a_1\dots a_6}$ (\ref{cosetE73}). Since the general form of the connection $\nabla$ was provided by the compact and the non-compact parts $(v^{\text{min}},v)$ of the Maurer--Cartan form $e^{-H}de^H$ (\ref{omegaLin}) in (\ref{sympConE}), it is necessary to decompose $e^{-H}de^H$ into $Gl(7)$ representations, too. With the general formul\ae{} (\ref{formula},\,\ref{formula2})
\beg
e^{-X}d e^X = \sum\limits_{j=0}^\infty \frac{(-1)^j}{(j+1)!}\ad_X^j dX,\\
e^{X}Ye^{-X}=\exp\circ \ad_{X} (Y)
\eeg
and the commutation relations (\ref{ComRel1},\,\ref{M2E},\,\ref{EE}), I obtain
\be\label{E7Coset}
 e^{-H}d e^H &=&dx^\mu\left[
 \left({\left(\tilde{v}^{\text{min}}_\mu\right)_{c}}^{d} + {\left(\tilde{v}_\mu\right)_{c}}^{d}\right){\left.\hat{M}\right.^c}_d \right.\\
&&\left.
+ \left(v_\mu\right)_{a_1\dots a_3}\hat{E}^{a_1\dots a_3}
+\left(v_\mu\right)_{a_1\dots a_6}\hat{E}^{a_1\dots a_6}\nn
\right]
\ee
with the abbreviations
\be\label{vSym}
{\left(\tilde{v}_\mu\right)_{c}}^{d}&:=&
{\left.\tilde{e}\right._{g}}^m \p_\mu {\left.\tilde{e}\right._m}^{(d}\eta^{f)g}\eta_{cf}
 \nn\\
{\left(\tilde{v}^{\text{min}}_\mu\right)_{c}}^{d}&:=&{\left.\tilde{e}\right._{g}}^m \p_\mu {\left.\tilde{e}\right._m}^{[d}\eta^{f]g}\eta_{cf}
\nn\\
\left(v_\mu\right)_{a_1\dots a_3}&:=&
\p_\mu A_{a_1\dots a_3} 
+3A_{f[a_1a_2}{\left.\tilde{e}\right._{a_3]}}^m \p_\mu {\left.\tilde{e}\right._m}^{f}
-\frac{7}{3}  A_{a_1\dots a_3} \p_\mu\tilde{\sigma}
\nn\\
\left(v_\mu\right)_{a_1\dots a_6}&:=& 
\p_\mu A_{a_1\dots a_6}
-6 A_{f[a_1\dots a_5}{\left.\tilde{e}\right._{a_6]}}^m \p_\mu {\left.\tilde{e}\right._m}^{f}
-\frac{14}{3} A_{a_1\dots a_6} \p_\mu\tilde{\sigma}
\nn
\\
&&+\frac{2\tau_3}{\tau_2^2}A_{[a_1\dots a_3}\left(v_\mu\right)_{a_4\dots a_6]}.
\ee
The components $(\tilde{v}^{\text{min}},\tilde{v})$ of the $\mathfrak{gl}_7$ generators $\hat{M}{}^a{}_b$ are similarly defined to the ones (\ref{omegaLin}) of chapter \ref{CHAP4}. They depend on the unrestricted subvielbein $\tilde{e}$ (\ref{tildeE}). This was expected, because the embedding of the $\mathfrak{gl}_7$ subalgebra in $\mathfrak{e}_{7(7)}$ was determined in section \ref{Decompo} by requiring the standard form $x^\mu\p_\nu$ (\ref{PRep}) for $\mathfrak{gl}_7$ action on the coordinates.\\

This requirement also was the reason to fix $\tau_9=1$ (\ref{chi17fix}), which is responsible for the non-standard form of the commutation relations in (\ref{M2E}) and hence for the occurence of the determinant $\tilde{\sigma}$ (\ref{sigmaDefi2}) of $\tilde{e}$ in the expressions $(v_\mu)_{a_1\dots a_3}$ and $(v_\mu)_{a_1\dots a_6}$ (\ref{vSym}).\\

To establish the contact to the general form of the connection $\nabla$ in (\ref{sympConE}), I have to decompose the Maurer--Cartan form $e^{-H}de^H$ into compact and non-compact parts $(v^{\text{min}},v)$. In section \ref{LCcE}, I explained that these are parametrized by the $\mathfrak{su}_8$ representations  $\hat{M}{}^A{}_B$ and $\hat{S}_{ABCD}$ (\ref{vRestr2}), respectively:
\be\label{E7Coset2}
 e^{-H}d e^H &=:&dx^\mu\left[ {\left(v^{\text{min}}_\mu\right)_{A}}^{B}{\left.\hat{M}\right.^A}_B 
 +
\left(v_\mu\right)^{ABCD}\hat{S}_{ABCD}
\right].
\ee
If I recall the decomposition of the $\mathfrak{gl}_7$ generators $\hat{M}{}^a{}_b$ (\ref{GLdefi}) and of $\hat{E}^{abc},\,\hat{E}^{a_1\dots a_6}$ (\ref{EDefi}) into compact and non-compact parts, I can identify the $\mathfrak{su}_8$ representations $(v^{\text{min}},v)$ in (\ref{E7Coset2}) by a comparison with (\ref{E7Coset}):
\beg
{\left(v^{\text{min}}_\mu\right)_{A}}^{B}{\left.\hat{M}\right.^A}_B &=&
{\left(\tilde{v}^{\text{min}}_\mu\right)_{c}}^{d}{\left.\hat{L}\right.^c}_d 
+
\left(v_\mu\right)_{a_1\dots a_3}\hat{L}^{a_1\dots a_3}
+
\left(v_\mu\right)_{a_1\dots a_6}\hat{L}^{a_1\dots a_6}
\\
&=&
{\left(\tilde{v}_\mu\right)_{c}}^{d}{\left.\hat{S}\right.^c}_d 
+ 
\left(v_\mu\right)_{a_1\dots a_3}\hat{S}^{a_1\dots a_3}
+
\left(v_\mu\right)_{a_1\dots a_6}\hat{S}^{a_1\dots a_6}.
\eeg
The decomposition of the generators $\hat{M}{}^A{}_B$ (\ref{MGa}) and $\hat{S}_{ABCD}$ (\ref{MGc}) into $\mathfrak{so}_7$ generators allows to relate the coefficients, taking into account the fixing of the parameters $\tau_1,\tau_4,\tau_5,\tau_6,\tau_9$ (\ref{chi1fix},\,\ref{chifix},\,\ref{Fixchi}) and the symmetry properties of the $\G$ matrices:
\begin{subequations}\label{vGL}
\be\label{vminE7}
{\left(v^{\text{min}}_\mu\right)_{A}}^{B}
&=&
\frac{1}{4}{\left(\tilde{v}^{\text{min}}_\mu\right)_{e}}^{f} {{{\G^e}_f}_A}^B
+\frac{1}{\tau_2}\left(v_\mu\right)_{a_1\dots a_3}{{\G^{a_1\dots a_3}}_A}^B
\nn\\
&&
+\frac{i}{\tau_3}\left(v_\mu\right)_{a_1\dots a_6}\e^{a_1\dots a_6c}{{\G_{c}}_A}^B
\\
\label{vE7}
\left(v_\mu\right)^{ABCD}
&=&
-\frac{3}{4}{\left(\tilde{v}_\mu\right)_{e}}^{f}
\left(\delta^e_g\delta_{f}^h-\frac{1}{9}\delta^e_f\delta_{g}^h \right)
{\G^g}^{[AB}{\G_h}^{CD]}\nn\\
&&
-\frac{3}{\tau_2}\left(v_\mu\right)_{a_1\dots a_3}
{\G^{[a_1a_2}}^{[AB}{\G^{a_3]}}^{CD]}
\nn\\
&&
-\frac{i}{2\tau_3}\left(v_\mu\right)_{a_1\dots a_6}
\e^{a_1\dots a_6c}
{\G_{bc}}^{[AB}{\G^{b}}^{CD]}.
\ee
\end{subequations}

In a final step, I switch to the holomorphic vielbein frame $\underline{dz}^{AB}$ (\ref{holframe3}) of the cotangent space, because this is the one that is used in the general formula for the connection $\nabla$ (\ref{sympConE}):
\begin{subequations}\label{frame4}
\be
dx^\mu {\left(v^{\text{min}}_\mu\right)_{A}}^{B}{\left.\hat{M}\right.^A}_B &=& \underline{dz}^{EF} {\left(v^{\text{min}}_{EF}\right)_{A}}^{B}{\left.\hat{M}\right.^A}_B +\text{c.c.}\\
dx^\mu \left(v_\mu\right)^{ABCD}\hat{S}_{ABCD} &=& \underline{dz}^{EF} \left(v_{EF}\right)^{ABCD}\hat{S}_{ABCD} +\text{c.c.}
\ee
\end{subequations}
The antiholomorphic objects follow from a complex conjugation
\beg
\underline{d\bar{z}}_{B_1B_2} {\left(\bar{v}_{\text{min}}^{B_1B_2}\right)_{A}}^{B}{\left.\hat{M}\right.^A}_B 
&:=&
 \overline{\underline{dz}^{B_1B_2} {\left(v^{\text{min}}_{B_1B_2}\right)_{A}}^{B}{\left.\hat{M}\right.^A}_B}
 \\
 \underline{d\bar{z}}_{B_1B_2}\left(\bar{v}^{B_1B_2}\right)^{A_1\dots A_4}\hat{S}_{A_1\dots A_4}
&:=&
\overline{\underline{dz}^{B_1B_2} \left(v_{B_1B_2}\right)^{A_1\dots A_4}\hat{S}_{A_1\dots A_4}}
\eeg
after adjusting the position of the indices with the symplectic form $\Omega$ as defined in (\ref{lower6}). I want to emphasize that the objects $(v^{\text{min}},v)$ in the $56$-dimensional vielbein frame (\ref{vGL}) contain the derivatives that are denoted with an underline $\underline{\p}$ (\ref{pUd}).\\

This completes the $Gl(7)$ decomposition of the vielbein $e^H$ in $56$ dimensions and of its Maurer--Cartan form $e^{-H}de^H$ that provides the ingredients to relate the general connection $\nabla$ (\ref{sympConE}) to $d=11$ supergravity. I will start with this task in the next section.\\

\section{A first comparison to $d=11$ supergravity}\label{Comp1}
\subsection{The bosonic fields}\label{boson1}
In the introduction to this chapter, I have restricted the exceptional geometry in $60$ dimensions to a $56$-dimensional subsector. This was defined by focusing on the part on which the symplectic form $\Omega$ was non-degenerate. Its codimension $4$ corresponds to the $4$ dimensions that are singled out in the $4+7$ split of de Wit \& Nicolai \cite{dWN86} reviewed in section \ref{Hid2}. Hence, only the seven dimensions are expected to be contained in the $56$-dimensional subsector.\\

This implies that the $Gl(7)$ submatrix $e$ of the vielbein matrix $E$ in $11$ dimensions has to be present in this subsector. Guided by the parametrization of the $56$-dimensional vielbein matrix $e^H$ (\ref{cosetE73}) 
\be
e^H&=&e^{{h_a}^b{\left.\hat{M}\right.^a}_b}e^{A_{abc}\hat{E}^{abc}}e^{A_{a_1\dots a_6}\hat{E}^{a_1\dots a_6}}\in E_{7(7)}/(SU(8)/\Z_2),
\label{cosetE732}
\ee
it is natural to relate the subvielbein $e$ to the $\mathfrak{gl}_7$ generators $\hat{M}{}^a{}_b$.\\

The next relation that catches the eye is connected to the three-form potential $A$ (\ref{GaugeTrafo}): it looks promising to identify the coefficients $A_{abc}$ in $e^H$ with the coefficents of $dx^a\wedge dx^b\wedge dx^c$ in the eleven-dimensional three-form $A$ with $a,b,c=4,\dots,10$.\\  

To complete the identification of the bosonic degrees of freedom contained in $e^H$ with the ones of $d=11$ supergravity, I have to establish a relation for the coefficient $A_{a_1\dots a_6}$ in $e^H$. The canonical candidate is the dual six-form potential of $d=11$ supergravity that I have introduced in equation (\ref{4FormG}):
\begin{subequations}\label{4Form2}
\be
F^{B_1\dots B_4} &=& \frac{1}{7!} \e^{B_1\dots B_4 C_1\dots C_7}F_{C_1\dots C_7}\\
F_{C_1\dots C_7} &=& 7\left(\nabla_{[C_1}A_{C_2\dots C_7]} + 5A_{[C_1\dots C_3}F_{C_4\dots C_7]}\right).
	\ee
	\end{subequations}
There is a subtlety to observe. By the very definition of exceptional geometry, all degrees of freedom are independent, if they are contained in the $60$-dimensional vielbein that is restricted by the invariance of the tensors $\Omega$ and $Q$.\footnote{An additional restriction on the vielbein would have to be accompanied by a further restriction of the geometry, e.g. by imposing the invariance of an additional tensor.} In particular, this is true for the coefficients $A_{abc}$ and $A_{a_1\dots a_6}$ in the subvielbein $e^H$. \\

In the context of $d=11$ supergravity, however, the three-form potential and the six-form potential are related by the duality (\ref{4Form2}). Hence, they do not correspond to independent degrees of freedom in general.\\

The striking fact is that I have not associated all degrees of freedom of the three-form potential $A$ to the vielbein $e^H$ so far. Only for the coefficients of $dx^a\wedge dx^b\wedge dx^c$ with $a,b,c=4,\dots,10$, this is the case. The other degrees of freedom could be associated to the six-form potential. In particular, it is possible to associate the degrees of freedom of $A_{\alpha\beta\gamma}$ with $\alpha,\beta,\gamma=0,\dots,3$ to $A_{a_1\dots a_6}$ in seven dimensions by the duality relation (\ref{4Form2}):
\begin{subequations}\label{4Form23}
\be
F^{\alpha\beta\gamma\delta} &=& \frac{1}{7!} \e^{\alpha\beta\gamma\delta a_1\dots a_7}F_{a_1\dots a_7}\label{4Form23a}
\\
F_{a_1\dots a_7} &=& 7\left(\nabla_{[a_1}A_{a_2\dots a_7]} + 5A_{[a_1\dots a_3}F_{a_4\dots a_7]}\right).
\label{4Form23b}
	\ee
\end{subequations}
Therefore, the degrees of freedom contained in $A_{abc}$ and in $A_{a_1\dots a_6}$ are independent indeed. Thus, the entire subvielbein $e^H$ in $56$ dimensions is associated to bosonic fields of $d=11$ supergravity. My restriction of the $(d=60)$-dimensional exceptional geometry to the subsector of fields comprised by $e^H$ hence implies that I set the remaining bosonic fields of $d=11$ supergravity to zero. In the complete theory, it is likely that they correspond to the degrees of freedom of the $d=60$ vielbein that are not contained in $e^H$.\\

To complete the identification, I have to fix the relation between the subvielbein $e$ of the $11$-dimensional vielbein $E$ and the degrees of freedom of $e^H$ that are associated to the $\mathfrak{gl}_7$ generator ${\hat{M}}^a{}_b$ (\ref{cosetE732}). This is achieved by recalling the definition of the four form field strength $F$ (\ref{field11}) in eleven dimensions
\beg
F_{M_1\dots M_4} &:=& 4\p_{[M_1}A_{M_2\dots M_4]}.
\eeg
Restricting the range of the indices to $4,\dots,10$ and switching to the vielbein frame with $\p_a := {e_a}^m\p_m$ (\ref{ehneu}) results in
\beg
F_{abcd} &=& 4\nabla_{[a}A_{bcd]}\\
&=&
4\left(\p_{[a}A_{bcd]} +3A_{f[bc}{e_{d}}^m\p_{a]} {e_m}^f\right).
\eeg
From the point of view of exceptional geometry, this is a derivative acting on the coefficient $A_{abc}$ of the coset $e^H$ (\ref{cosetE732}). For this derivative to be equivariant under a restricted diffeomorphism in $56$ dimensions, i.e. in particular under a left action by $E_{7(7)}$, it must be related to the invariant coefficient (\ref{vSym}) of the Maurer--Cartan form $e^{-H}de^H$
\beg
\left(v_\mu\right)_{a_1\dots a_3}
&=&
\p_\mu A_{a_1\dots a_3} 
+3A_{f[a_1a_2}{\left.\tilde{e}\right._{a_3]}}^m \p_\mu {\left.\tilde{e}\right._m}^{f}
-\frac{7}{3}  A_{a_1\dots a_3} \p_\mu\tilde{\sigma}\\
\text{with}\quad 
e^{\tilde{\sigma}} &=& \det\left(\tilde{e}\right)^{\frac{1}{7}}.
\eeg
This is possible, if I identify the $Gl(7)$ valued matrix $\tilde{e}$ (\ref{tildeE}) in $e^H$ with the subvielbein $e$ in the following way
\begin{subequations}\label{Weyl}
\be
{\left.\tilde{e}\right._m}^a &=& \det(e)^{\frac{1}{2}}{\left.e\right._m}^a 
\\
\Leftrightarrow\quad {\left.e\right._m}^a &=& \det(\tilde{e})^{-\frac{1}{9}} {\left.\tilde{e}\right._m}^a\\
\Rightarrow \quad \det(\tilde{e})&=& \det(e)^{\frac{9}{2}}\,=:\,e^{\frac{63}{2}\sigma}.
\ee
\end{subequations}
Then, I obtain the relation
\be\label{FDefi}
  F_{abcd}&=&4(v_{[a})_{bcd]},
    \ee
    if I use the abbreviation $\p_a={e_a}^m\p_m$ (\ref{ehneu}) in the definition of $(v_{a})_{bcd}$ (\ref{vSym}). It was to distinguish this notation $\p_a$ from the derivative $\underline{\p}_a$ (\ref{pDefi}) in the $56$-dimensional vielbein frame that I introduced the underline in (\ref{pUd}).

\newpage

    The fact that a derivative acting on coset degrees of freedom must be $E_{7(7)}$-equivariant also applies to the definition of the seven form field strength (\ref{4Form23b}):
    \beg
    F_{a_1\dots a_7} &=& 7\left(\nabla_{[a_1}A_{a_2\dots a_7]} + 5A_{[a_1\dots a_3}F_{a_4\dots a_7]}\right)\\
    &\stackrel{(\ref{FDefi})}{=}& 
    7\Big(\p_{[a_1} A_{a_2\dots a_7]} -6 A_{f[a_2\dots a_6}{e_{a_7}}^m \p_{a_1]} {e_m}^{f}
\\
&&+20A_{[a_1\dots a_3}\left(v_{a_4}\right)_{a_5\dots a_7]}\Big).
    \eeg
    The corresponding coefficient (\ref{vSym}) of the Maurer--Cartan form $e^{-H}de^H$ has the form after substituting the subvielbein $e$ for $\tilde{e}$ (\ref{Weyl})
    \beg
    \left(v_\mu\right)_{a_1\dots a_6}&=& 
\p_\mu A_{a_1\dots a_6}
-6 A_{f[a_1\dots a_5}{e_{a_6]}}^m \p_\mu {e_m}^{f}
\nn
\\
&&+\frac{2\tau_3}{\tau_2^2}A_{[a_1\dots a_3}\left(v_\mu\right)_{a_4\dots a_6]}.
    \eeg
  A comparison provides the relation
  \be\label{F7Defi}
     F_{a_1\dots a_7} &=& 7(v_{[a_1})_{a_2\dots a_7]},
   \ee
   if the normalization constant $\tau_3$ of the generator $\hat{L}^{a_1\dots a_6}$ (\ref{defiSo7}) is fixed by
   \be\label{tau3fix}
   \tau_3 &=& -10\tau_2^2.
   \ee
It is obvious from the first order form of the field equation for the three-form potential $A$ (\ref{4FormG}) that $\tau_3$ corresponds to the normalization constant $\alpha$ of the six-form potential $A_{C_1\dots C_6}$ in $11$ dimensions. It was fixed to $\alpha=1$ in order to write the equation (\ref{4Form}) in canonical form.\\

Since a seven form field strength in seven dimensions is equivalent to a scalar, I can without loss of generality introduce the notation
\be\label{F7Defi2}
F&:=&F_{a_1\dots a_7}\e^{a_1\dots a_7}\\
&=&
7\left(v_{a_1}\right)_{a_2\dots a_7}\e^{a_1\dots a_7},\nn
\ee
before I conclude this section with two comments on the rescaling (\ref{Weyl}) of the vielbein.

\begin{itemize}
	\item The Weyl rescaling with the determinant of the vielbein (\ref{Weyl}) exactly is the one used by de Wit and Nicolai in \cite{dWN86}.
	\item It corresponds to a different choice of the constant $\tau_9$ that fixed the embedding of $\mathfrak{gl}_7$ in $\mathfrak{e}_{7(7)}$. If I had worked with $\tau_9=\frac{2}{9}$ instead of $\tau_9=1$ (\ref{chi17fix}), I would have encorporated this rescaling in the $\mathfrak{gl}_7$ decomposition of $\mathfrak{e}_{7(7)}$. Setting $\tau_9=\frac{2}{9}$ corresponds to the so-called ``gravity subline'', used e.g. in \cite{DHN02}.
	
	I did not adopt this choice in section \ref{Decompo2}, because the $\mathfrak{gl}_7$ action on the coordinates would have a non-standard form, which would imply a non-standard form of the vielbein matrix $e^H$ in section \ref{Matrix}. 
\end{itemize}

\subsection{The connection}\label{conn32}
In the preceding section, I have identified fields of $d=11$ supergravity with $Gl(7)$ representations that parametrize the vielbein $e^H$ (\ref{cosetE73}). In order to extend this relation to the dynamics in an exceptional geometry, it has become clear in chapter \ref{CHAP3} that the covariant derivative $\nabla$ is the fundamental object to discuss. For any physical field $\psi$ that transforms as a representation of $\mathfrak{u}_8$, the general form of $\nabla$ is defined by (\ref{Conn56})
\be\label{conn9}
\nabla \psi^\gamma &:=&  d \psi^\gamma - \delta_{\hat{\omega}} \psi^\gamma
\ee
with the action of the $\mathfrak{u}_8$ valued one-form $\hat{\omega}$ (\ref{sympConE})
\beg
\delta_{\hat{\omega}}
&=& 
\underline{dz}^{C_1C_2}{\left(\omega_{C_1C_2}\right)_A}^B\left(\delta_{{\left.\hat{M}\right.^A}_B} +\frac{1}{8}\delta^A_B\delta_{\hat{T}}\right) +\text{c.c.}\nn
\eeg
and the abbreviation
\be\label{omegaDefiN}
{\left(\omega_{C_1C_2}\right)_A}^B &:=&{\left(v^{\text{min}}_{C_1C_2}\right)_A}^B 
+c_1\left(\bar{v}^{CD}\right)_{CD C_1C_2}\delta^B_A\nn\\
&&+c_2\left(\bar{v}^{CQ}\right)_{CP C_1C_2} \left(\delta^P_A\delta^B_Q-\frac{1}{8}\delta_A^B\delta_Q^P\right).
\ee
The $\mathfrak{su}_8$ representations $(v^{\text{min}},v)$ in this action are related to the fields of $d=11$ supergravity by the equations (\ref{vGL}) and (\ref{frame4}). It is this form of the connection in the $\mathfrak{su}_8$-covariant vielbein frame $\underline{dz}^{C_1C_2}$ (\ref{holframe3}) that will be used for a comparison to the supersymmetry transformations in section \ref{Super}. \\

Nevertheless, it also is interesting to state the connection $\nabla$ in the $\mathfrak{gl}_7$-covariant vielbein frame (\ref{pDefi}), because I decomposed $\mathfrak{e}_{7(7)}$ with respect to $\mathfrak{gl}_7$ to establish the contact to the fields of $d=11$ supergravity. This change of basis in the vielbein frame follows the one in (\ref{defiSo7P2}) and leads to the definition
\be\label{ConnDefi2}
{\left(\omega_{C_1C_2}\right)_A}^B &=:&-\frac{i\tau_7}{16}\left({\left(\omega_{a}\right)_A}^B -i\eta_{ac} {\left(\omega^{c}\right)_A}^B\right)
{\G^a}_{C_1C_2}\\
&&
+\frac{\tau_{7}\sqrt{2}}{32}
\left({\left(\omega_{ab}\right)_A}^B -i\eta_{ac} \eta_{bd} {\big(\omega^{cd}\big)_A}^B\right)
 {\G^{ab}}_{C_1C_2}\nn.
\ee
For a comparison to $d=11$ supergravity, it is sensible to assume trivial dependence of the fields ${e_m}^a$, $A_{abc}$ and $A_{a_1\dots a_6}$ on the $49$ coordinates $(p_{mn},\,x^{mn},\,p_m)$ (\ref{coord6}) of the exceptional geometry, because these directions do not exist in the $d=11$ theory. A brief look at the expressions for partial derivatives in the $56$-dimensional vielbein frame (\ref{pDefi}) reveals that these formul\ae{} take a very simple form in this truncation: only one derivative is non-trivial
\be\label{oneder}
\underline{\p}_c &\stackrel{(\ref{Weyl})}{=}&e^{-\frac72\sigma}{e_c}^m \frac{\p}{\p x^m}.
\ee
The covariant derivative $\nabla$ (\ref{conn9}) along the common directions $x^m$ of the $56$-dimensional sector of
 the exceptional geometry and of $d=11$ supergravity is provided by the $\mathfrak{u}_8$ representation ${(\omega_c)_A}^B$ (\ref{ConnDefi2}) in the vielbein frame
\be\label{Conn11}
\nabla_c \psi^\gamma &=& e^{-\frac72\sigma}{e_c}^m \frac{\p}{\p x^m}\psi^\gamma 
-
{(\omega_c)_A}^B\left(\delta_{{\left.\hat{M}\right.^A}_B} +\delta^A_B\delta_{\hat{T}}\right)\psi^\gamma.
\ee
For general constants $c_1,c_2\in \R$, the abbreviation ${(\omega_c)_A}^B$ is related to the fields ${e_m}^a$, $A_{abc}$ and $A_{a_1\dots a_6}$ by
  \be\label{omegaWN}
 {\left(\omega_{c}\right)_A}^B 
  &=&
  \frac{1}{4}e^{-\frac72\sigma}\left[
  {\left(\tilde{v}_c^{\text{min}}\right)_{e}}^{f} 
 -
 \frac{c_2}{2}{\left(\tilde{v}_e\right)_{c}}^{f}
      \right]{{{\G^e}_f}_A}^B\\
     &&
+\frac{1}{\tau_2}e^{-\frac72\sigma}\left[\left(v_c\right)_{a_1\dots a_3}
-\frac{3c_2}{4}\left(v_{a_1}\right)_{a_2 a_3c}\right]{{\G^{a_1\dots a_3}}_A}^B
\nn\\
&&
-\frac{i}{10\tau_2^2}e^{-\frac72\sigma}
\left(v_a\right)_{a_1\dots a_6}
\e^{a_1\dots a_6b}
\left[\frac{c_2}{4}\delta^a_b\delta^{k}_c +\left(1-\frac{c_2}{4}\right) \delta^a_c\delta^{k}_b\right]
{{\G_{k}}_A}^B.\nn
\ee
In order to establish the contact to $d=11$ supergravity, it seems to be natural to fix $c_2=4$, which leads with the definition of the field strengths $F$ (\ref{FDefi},\,\ref{F7Defi2}) and with the one of the torsionfree spin connection $\omega$ (\ref{omegaK}) to
\beg
 {\left(\omega_{c}\right)_A}^B 
  &=&
  e^{-\frac72\sigma}\left[\frac{1}{4}
  \left.\tilde{\omega}_{ce}\right.^{f} 
        {{{\G^e}_f}_A}^B
+\frac{1}{4\tau_2}F_{ca_1\dots a_3}{{\G^{a_1\dots a_3}}_A}^B
-\frac{i}{70\tau_2^2}F{{\G_{c}}_A}^B\right].
\eeg
This expression coincides with the covariant derivative defined in equation (3.32) of de Wit \& Nicolai's publication \cite{dWN86} in which they proved the $Spin(3,1)\times SU(8)$-covariance of $d=11$ supergravity.\\

I want to emphasize that $\left.\tilde{\omega}_{ce}\right.^{f}$ is the standard form of the Levi--Civita connection in the spin frame (\ref{omegaK}) for the \textbf{rescaled vielbein} $\tilde{e}=\det(e)^\frac{1}{2}e$ (\ref{Weyl}). It is obvious from the general form of a connection with $E_{7(7)}$-equivariance (\ref{omegaWN}) that it is impossible to obtain the spin connection for the unrescaled vielbein $e$ instead. Following the discussion of section \ref{boson1}, the approach to consider $\tilde{e}$ instead of $e$ (\ref{Weyl}) as the subvielbein of $d=11$ supergravity is in conflict with the definition of the field strengths $F$.\footnote{A different choice for $\tau_9$ in (\ref{chi17fix}) would not alter this fact, because the explicit formula (\ref{omegaWN}) would also have to be modified.}\\

De Wit \& Nicolai's $4+7$ split of $d=11$ supergravity \cite{dWN86} obscured the $\Diff(11)$ symmetry of $d=11$ supergravity, only its $\Diff(4)\times \Diff(7)$ subgroup was manifest in their formulation. This is in accordance with the fact that the derivatives of $A_{abc}$ and $A_{a_1\dots a_6}$ in ${\left(\omega_{c}\right)_A}^B $ appear as $\Diff(7)$-covariant field strengths $F$ (\ref{field11},\,\ref{7Form}). \\

However, the connection $\nabla$ (\ref{conn9}) is defined in the $56$-dimensional exceptional geometry. This implies that the covariant derivative $\nabla$ along the $49$ coordinates $(p_{mn},\,x^{mn},\,p_m)$ (\ref{coord6}) cannot be neglected. The assumption that the physical fields $\psi^\gamma$, ${e_m}^a$, $A_{abc}$ and $A_{a_1\dots a_6}$ do not depend on these coordinates only implies that the three $\mathfrak{u}_8$ representations ${(\omega^c)_A}^B$, ${(\omega_{cd})_A}^B$ and ${(\omega^{cd})_A}^B$ (\ref{ConnDefi2}) have to transform as tensors under the symmetry group.\\

It is obvious that this is not the case, if the symmetry group was $\Diff(7)$:   
\beg 
  {\left(\omega^{c}\right)_A}^B
  &=&
  e^{-\frac{7}{2}\sigma}\left[ic_1\left(\tilde{v}_a\right)^{ca}\delta_A^B
  +\frac{c_2}{40\tau_2^2}
\left(v_{b_3}\right)_{a_1\dots a_6}
{\e^{a_1\dots a_6}}_{b_2}
{{\G^{cb_2b_3}}_A}^B\right.
\\
&&\left.
-\frac{3ic_2}{2\tau_2}\left(v_a\right)^{kac} 
     {{\G_k}_A}^B\right]
  \\
{\left(\omega_{cd}\right)_A}^B
&=&
e^{-\frac{7}{2}\sigma}\left[
\frac{6ic_1\sqrt{2}}{\tau_2}{\left(v_a\right)_{cd}}^a\delta_A^B
    -\frac{c_2\sqrt{2}}{40\tau_2^2}
\left(v_a\right)_{a_1\dots a_6}
{\e^{a_1\dots a_6}}_{[c}
 {{{\G^a}_{d]}}_A}^B
\right.\\
   &&-\frac{c_2\sqrt{2}}{16\tau_2}\left(v_{b_3}\right)_{a_1\dots a_3}
{\e^{a_1\dots a_3}}_{cdb_1b_2}
 {{\G^{b_1b_2b_3}}_A}^B
 \\
  &&\left.
  -\frac{ic_2\sqrt{2}}{8}
 {\left(\tilde{v}_a\right)_{e}}^{f}
 \left(2\delta_{[c}^e\delta_{d]}^{[k}\delta_{f}^{a]} -\frac{1}{3}\delta_f^e\delta_{d}^{[k}\delta_{c}^{a]}
 \right)
{{\G_k}_A}^B
\right]
\\
{\big(\omega^{cd}\big)_A}^B
&=&e^{-\frac{7}{2}\sigma}\left[\frac{ic_1\sqrt{2}}{5\tau_2^2}\left(v_a\right)_{a_1\dots a_6}\eta^{a[c}\e^{d]a_1\dots a_6}\delta_A^B
     +\frac{3c_2\sqrt{2}}{4\tau_2}
     \left(v_a\right)^{cdg }
  {{{\G^a}_g}_A}^B
\right.\\
   &&+\frac{c_2\sqrt{2}}{8}
   {\left(\tilde{v}_{b_3}\right)_{e}}^{f}
 \delta^{[d}_{b_1}\left(\delta_f^{c]}\delta^e_{b_2} -\frac{1}{6}\delta^{c]}_{b_2}\delta_f^e
 \right)
 {{\G^{b_1b_2b_3}}_A}^B
 \\
  &&\left.
  -\frac{ic_2\sqrt{2}}{8\tau_2}
  \left(v_a\right)_{a_1\dots a_3} 
\e^{aa_1\dots a_3kcd}
 {{\G_k}_A}^B
 \right].
   \eeg
   Recapitulating the construction of the exceptional geometry in this chapter, this does not come as a surprise. The symmetry group of general coordinate transformations $\Diff(56)$ was restricted by imposing the invariance of the two tensors $\Omega$ and $Q$ to the finite dimensional symmetry group $\cP_{56}\rtimes E_{7(7)}$. Hence, $\Diff(7)$ is not a symmetry group of the exceptional geometry and there is no reason for the objects $(\omega^c)_A{}^B$, $(\omega_{cd})_A{}^B$ and $(\omega^{cd})_A{}^B$ (\ref{ConnDefi2}) to transform as tensors under $\Diff(7)$, if the dependence on the $49$ coordinates $(p_{mn},\,x^{mn},\,p_m)$ (\ref{coord6}) is neglected.\\
   
It is important to observe that this does not falsify the claim that all solutions of $d=11$ supergravity are special solutions of a theory in a $(d=60)$-dimensional exceptional geometry. Solutions are characterized by their equations of motion and not by the symmetry group of these equations. The dependence on the additional $49$ coordinates is similar to the dependence on the mass of a scalar field in flat Minkowski space that is subject to the massive version of the Klein--Gordon equation. All fields with the same mass form representations of the Poincar\'e isometry group. Only the special fields with vanishing mass allow for an extension of this symmetry group to include the scaling symmetry. In the present setting of $49$ truncated coordinates, the special fields allow for an extension of the $Gl(7)$ part of $E_{7(7)}$ to $\Diff(7)$.\footnote{In both cases, the equations of motion impose an additional constraint, of course.} \\

For the proof of $Spin(3,1)\times SU(8)$-covariance of $d=11$ supergravity in \cite{dWN86}, this subtlety is not of importance. De Wit and Nicolai did not claim that the $SU(8)$-covariance of the equations of motion is related to a higher dimensional exceptional geometry. They decoupled the right action on the vielbein from the left one and discussed a general $SU(8)$ action on physical fields $\psi$ that is not induced by a diffeomorphism. This fact may have led to the ``fixing'' of the $SU(8)$ gauge of the generalized inverse vielbein $e_{AB}{}^m$ that they defined in equation (3.5) of \cite{dWN86}. It is nice to observe that this object coincides with the corresponding part $(e^{-H})_{AB}{}^m$ (\ref{vili2}) of the vielbein in the exceptional geometry.

\section{Supersymmetry}\label{Super}
The analysis of section \ref{conn32} showed that it is not an appropriate guideline for fixing the constants $c_1,c_2\in \R$ in the general form (\ref{omegaDefiN}) of the connection $\nabla$ (\ref{conn9}) in the truncation to the seven coordinates $x^m$ to require abstract $\Diff(7)$-equivariance. In this section, I will show that it is possible to fix $c_1,c_2$ in such a way that a particular way to act with $\nabla$ on a $\mathfrak{u}_8$ representation $\eps$ is $\Diff(7)$-covariant. The result reproduces the supersymmetry variation of the fermion in $d=11$ supergravity that corresponds to the bosonic degrees of freedom $e$, $A_{abc}$, $A_{a_1\dots a_6}$ discussed in section \ref{boson1}.

\subsection{The equivariant supersymmetry variation}\label{deltau}
A characteristic feature of $d=11$ supergravity is its invariance under the supersymmetry transformation (\ref{Trafo1}). To lift this to the exceptional geometry, I have to define a supersymmetry variation $\underline{\delta}$ in $60$ dimensions with the following properties\footnote{As $\underline{\delta}$ does not have an index, there should not arise any confusion with the notation for the $Gl(7)$-covariant basis $\underline{\p}_a$ that I defined in (\ref{pDefi}).}
\begin{enumerate}
	\item $\underline{\delta}$ is a derivation, i.e.
	\begin{subequations}\label{Deriv1}
	\be
		\underline{\delta}(A\cdot B) &=& (\underline{\delta}A) \cdot B + A\cdot \underline{\delta}B\\
		\underline{\delta}(\alpha A)&=&\alpha \underline{\delta}A\quad \forall \alpha\in \R
	\ee
	\end{subequations}
	with arbitrary $\mathfrak{u}_8$ representations $A$ and $B$.
	\item $\underline{\delta}$ is equivariant under the diffeomorphisms that preserve the tensors $\Omega$ and $Q$ defined in section \ref{MaxS}.
\end{enumerate}
Since it is beyond the scope of this thesis to discuss the entire theory in $d=60$, I focussed on the subsector that corresponds to the $56\times 56$ submatrix $e^H$ of the vielbein in $d=60$. Next recall from section \ref{affE71} that the independent degrees of freedom of $e^H$ can be parametrized in the holomorphic frame $\hat{Z}_{AB}$ (\ref{defiSo7P2}) by $(e^H)_\mu{}^{AB}$ (\ref{eHD2}) with $\mu=1,\dots,56$. Therefore, it is sufficient to discuss the two objects
\begin{subequations}\label{susy1}
\be
&{\left(e^{-H}\right)_{AB}}^\mu \underline{\delta}{\left(e^{H}\right)_{\mu}}^{CD}&\\
\text{and }&{\left(e^{-H}\right)^{AB}}^\mu \underline{\delta}{\left(e^{H}\right)_{\mu}}^{CD}.&
\ee
\end{subequations}
In section \ref{SymmAct}, I defined a derivation to be equivariant if it maps covariant objects to covariant objects under a symmetry transformation. Since $e^{-H}\underline{\delta} e^H$ (\ref{susy1}) transforms in the vielbein frame, the induced action of a diffeomorphism $\vp$ is by the Lie group action of $O\left(\vp,e^H\right)\in SO(56)$. This is further constrained to $O\left(\vp,e^H\right)\in SU(8)/\Z_2$ due to the invariance of the tensors $\Omega$ and $Q$. The transformation of $e^{-H}\underline{\delta} e^H$ that is induced by $x'=\vp(x)$ hence has the form
\be\label{tensor1}
e^{-H'}\underline{\delta}' e^{H'} 
&=&
 O^{-1}\left(\vp,e^H\right)\cdot
e^{-H}\underline{\delta} e^H
\cdot  O\left(\vp,e^H\right).
\ee
It is important to observe that the equivariant derivation $\underline{\delta}$ does not have the same explicit form in the original and the transformed frame in general, i.e. for a general $\mathfrak{u}_8$ representation $\chi$
\beg
\underline{\delta}\chi^\gamma&\neq&\underline{\delta}'\chi^\gamma.
\eeg
This is in complete analogy to the one-form valued connection $\nabla$ (\ref{Conn56}) acting on $\mathfrak{so}_{(d-1,1)}$ representations $\psi$: even if the indices of the cotangent space are contracted with the basis forms $dx^\mu$
\beg
\nabla\psi^\gamma &=& d\psi^\gamma -\delta_{\hat{\omega}}\psi^\gamma,
\eeg
it is only the Cartan differential $d$ that is invariant under a symmetry transformation, which is of course the reason why $d$ is not equivariant in general.\\

The same argumentation that allows to decompose a general connection $\nabla$ acting on $\mathfrak{so}_{(d-1,1)}$ representations $\psi$ into a Cartan differential $d$ and a one-form valued $\mathfrak{so}_{(d-1,1)}$ action $\delta_{\hat{\omega}}$, can be applied to the case of the equivariant derivation $\underline{\delta}$. Hence, I can always decompose an equivariant derivation $\underline{\delta}$ acting on $\mathfrak{u}_8$ representations $\chi$ in the following way
\be\label{deltaDefi}
  \underline{\delta}\chi^\gamma &=& \delta\chi^\gamma  -\delta_{\hat{\Lambda}}\chi^\gamma\\
    \text{with}\quad \hat{\Lambda} &:=&{\Lambda_A}^B \left({\left.\hat{M}\right.^A}_B +\frac18\delta^A_B\hat{T}\right)\nn\\
    \text{and}\quad \delta\chi^\gamma &=&\delta'\chi^\gamma.\nn
  \ee
In a next step, I will fix the $\mathfrak{u}_8$ valued parameters $\hat{\Lambda}$ by the action of $\delta$ on the vielbein $e^H$. There is a subtlety to observe: I have defined the equivariant derivation $\underline{\delta}$ on arbitrary $\mathfrak{u}_8$ representations $\chi$. The vielbein matrix $e^H$ transforms by a multiplication with $SU(8)/\Z_2$ from the right and with $E_{7(7)}$ from the left. Hence it is not clear a priori whether both indices of the vielbein matrix are affected by the $\mathfrak{u}_8$ action.\\

To answer this question, recall that the $\mathfrak{u}_8$ action $\hat{\Lambda}$ is introduced in order to guarantee covariance (\ref{tensor1}). In other words, $\hat{\Lambda}$ compensates the effect of the ``naked'' derivation $\delta$ on the transformation matrices, which are the Jacobian and the compensating Lorentz rotation $O\left(\vp,e^H\right)$.\\

If the action of $\delta$ on these matrices is trivial, then it is not necessary to add a compensating $\hat{\Lambda}$ transformation. In section \ref{VectE4}, I proved that the general diffeomorphisms preserving the tensors $\Omega$ and $Q$ form a subgroup of the affine group $A(56)$ (\ref{affine2}), whose associated Jacobians are constant matrices. Hence, the definition of a derivation (\ref{Deriv1}) implies that the action of $\delta$ on the Jacobian is trivial. The compensating Lorentz rotation $O\left(\vp,e^H\right)$ depends on the $x$-dependent vielbein $e^H$, however. Therefore, it is necessary and sufficient to define the action of $\hat{\Lambda}$ on the vielbein by a right action alone.\\

Following the conventions from section \ref{prim2}, I obtain for the $\mathfrak{u}_8$ action on the vielbein $e^H$  (with the commutation relation (\ref{MZCR}) after fixing the norm of the $\mathfrak{u}_1$ generator $\hat{T}$ by $[\hat{T},\hat{Z}_{CD}] = -2\hat{Z}_{CD}$):
\be\label{LWi}
\delta_{\hat{\Lambda}}\left(e^H\right)_\mu{}^{CD}
&=&
-2\left(e^H\right)_\mu{}^{A[C}{\Lambda_A}^{D]}.
\ee
Due to the standard relations between the vielbein and its inverse in the holomorphic frame 
\beg
{\left(e^{-H}\right)_{AB}}^\mu {\left(e^{H}\right)_{\mu}}^{CD}&=& \delta_{AB}^{CD}\\
{\left(e^{-H}\right)^{AB}}^\mu {\left(e^{H}\right)_{\mu}}^{CD}&=&0,
\eeg
the equation (\ref{LWi}) leads to the identities for derivations of the vielbein (\ref{susy1})
\beg
{\left(e^{-H}\right)_{AB}}^\mu \underline{\delta}{\left(e^{H}\right)_{\mu}}^{CD}
&=&
{\left(e^{-H}\right)_{AB}}^\mu \delta{\left(e^{H}\right)_{\mu}}^{CD}
-2\delta_{[A}^{[C}{\Lambda_{B]}}^{D]}
\\
{\left(e^{-H}\right)^{AB}}^\mu \underline{\delta}{\left(e^{H}\right)_{\mu}}^{CD}&=& 
{\left(e^{-H}\right)^{AB}}^\mu \delta{\left(e^{H}\right)_{\mu}}^{CD}.
\eeg
Since $\delta$ is a derivation, $e^{-H}\delta e^H$ is an $\mathfrak{e}_{7(7)}$ valued Maurer--Cartan form. This allows to make use of the analysis of the Maurer--Cartan form $e^{-H}d e^H$ of section \ref{cII}, which shows that I can without loss of generality identify $\hat{\Lambda}$ (\ref{deltaDefi}) with the compact part of $e^{-H}\delta e^H$ by
\be\label{susy2}
{\big.\delta}_{[A}^{[C}{{\big.\Lambda\big.}_{B]}}^{D]}&:=&\frac{1}{2}{\left(e^{-H}\right)_{AB}}^\mu \delta{\left(e^{H}\right)_{\mu}}^{CD}.
\ee
Furthermore, the non-compact part of the Maurer--Cartan form $e^{-H}\delta e^H$ is restricted to $\mathfrak{e}_{7(7)}\ominus \mathfrak{su}_8$, which is the totally antisymmetric four-tensor representation of $\mathfrak{su}_8$. Adopting the choice (\ref{susy2}), the equivariant supersymmetry variations $\underline{\delta}$ of the $d=56$ vielbein $e^H$ have the particularly nice form
\begin{subequations}\label{susy3}
\be
{\left(e^{-H}\right)_{AB}}^\mu \underline{\delta}{\left(e^{H}\right)_{\mu}}^{CD}&=& 0\\
{\left(e^{-H}\right)^{AB}}^\mu \underline{\delta}{\left(e^{H}\right)_{\mu}}^{CD}&=& {\left(e^{-H}\right)^{[AB}}^\mu \delta{\left(e^{H}\right)_{\mu}}^{CD]}.
\ee
\end{subequations}

\subsection{Fermions}\label{fermi}
Imposing the invariance of the tensors $\Omega$ and $Q$ reduced the covariance group $G$ in the vielbein frame from $SO(56)$ to $SU(8)/\Z_2$. The covering group $Spin(56)$ is hence restricted to the simply connected group $SU(8)$.\\

In section \ref{Representation}, I have explained that it is consistent for fermions $\psi$ to transform as representations of the double cover $\tilde{G}$ of the covariance group $G$, because all physical expectation values are of even degree in $\psi$. In most physical theories, fermions $\psi$ are associated to the representations of $\tilde{G}$ that are not representations of $G$.\\

For the case of the exceptional geometry defined in this chapter, the compact group is 
\beg
G&=&SU(8)/\Z_2\,\subset\, SO(56).
\eeg 
The representations of its double cover $SU(8)$ that are not $G$ representations, are characterized by the fact that they have an \textbf{odd} number of $\mathfrak{su}_8$ indices. Therefore, it is natural to associate fermions to these $SU(8)$ representations, which leads with $A,B,C=1,\dots,8$ to the definitions of the Gra\ss mann odd holomorphic fields
\begin{subequations}\label{ecDefi}
\be
\eps &=&\eps^A \hat{Z}_A\\
\chi &=&\chi^{ABC}\hat{Z}_{ABC}\quad \text{with }\chi^{ABC}\,=\,\chi^{[ABC]}.
\ee
\end{subequations}
The generators $\hat{Z}_A$ and $\hat{Z}_{ABC}$ span the \textbf{complex} vector spaces $\mathbf{8}_\C$ and $\mathbf{56}_\C$. In section \ref{chiral}, I will explain the way these holomorphic fermions $\eps$ and $\chi$ are related to \textbf{real} sections of the spin bundle over the $(d=60)$-dimensional Lorentzian manifold $(\cM^{60},\,\Omega,\,Q)$ with preserved tensors $\Omega$ and $Q$. \\

The antiholomorphic fields $\bar{\eps}$ and $\bar{\chi}$ follow by complex conjugation. For consistency with the $\mathfrak{su}_8$ action, it is natural to raise the index of the corresponding abstract generators $\hat{\bar{Z}}$ in analogy to the definition (\ref{lower6},\,\ref{ZDefi2})
\begin{subequations}\label{ZbarDefi}
\be
\hat{\bar{Z}}^A&:=& \delta^{AD}\overline{\hat{Z}_D}\\
\hat{\bar{Z}}^{ABC}&:=& \delta^{AD}\delta^{BE}\delta^{CF}\overline{\hat{Z}_{DEF}}.
\ee
\end{subequations}
This is an $SU(8)$-covariant definition on the level of the abstract generators. A comment concerning the Kronecker $\delta$ is necessary. At first, recall that only in the vector field representation $\mathbf{R}$, the lowering of an even number of $SU(8)$ indices is related to the symplectic form $\Omega$ by the definition (\ref{lower6}). In the equations (\ref{MRquer}), I have lifted this definition to the level of the abstract generators. The holomorphic generators $\hat{Z}$ (\ref{ecDefi}) with an odd number of indices do not have a vector field representation $\mathbf{R}$, however. This is due to the fact that the fields $\eps$ and $\chi$ are not sections of a tensor product of the (co)tangent bundle. Their definition involves the spin bundle. Therefore, there does not exist a natural definition of the antiholomorphic generators $\hat{\bar{Z}}$ with an odd number of $SU(8)$ indices that is canonically induced by the vector field representation $\mathbf{R}$.\\

Nevertheless, the definitions (\ref{ZbarDefi}) of $\hat{\bar{Z}}$ on the level of abstract generators is not completely arbitrary. A tensor product of two generators $\hat{Z}$ with an odd number of $SU(8)$ indices maps to vectors $\hat{Z}$ with an even number of indices, e.g.
\be\label{tensor2}
\hat{Z}_A\otimes \hat{Z}_B &=& \hat{Z}_{[AB]}+ \text{symmetric part}.
\ee
For $\hat{Z}_{[AB]}$ (\ref{defiSo7P2}), there is a vector field representation $\mathbf{R}$ (\ref{ZDefi2}) and the canonical isomorphism (\ref{lower6}) provided by the symplectic form $\Omega$ has to be applied. Therefore, the definiton of lowering an odd number of $SU(8)$ indices may only differ from (\ref{ZbarDefi}) by a sign.\footnote{At this point, it is important to keep in mind the discussion of section \ref{holom1}: I explained the different natural conventions of symplectic and of complex geometry, where the definition of the complex conjugated coordinate $\bar{z}$ is concerned. Since I chose the one of complex geometry (\ref{lower5}), the symmetric tensor product $\hat{Z}_A\otimes \hat{\bar{Z}}{}^B$ has to be related to the hermitean form $i\hat{M}$ of the abstract unitary generators $\hat{M}$ for consistency with complex conjugation (\ref{MquerM}). }\\

Due to the relation of the tensor product of odd generators $\hat{Z}_A$ and $\hat{Z}_B$ to $\hat{Z}_{AB}$ (\ref{tensor2}), the commutor of $\hat{M}$ with $\hat{Z}_{AB}$ (\ref{MZCR}) induces
\begin{subequations}\label{MZCR5}
\be\label{MZCR3}
\left[{\left.\hat{M}\right.^{P}}_{Q},\hat{Z}_{A}\right]
&=& -\left(\delta_{A}^{P}\hat{Z}_{Q}-\frac{1}{8}\delta_Q^P\hat{Z}_{A}\right)
\\
\left[{\left.\hat{M}\right.^{P}}_{Q},\hat{Z}_{ABC}\right]
&=& -3\left(\delta_{[C}^{P}\hat{Z}_{AB]Q}-\frac{1}{8}\delta_Q^P\hat{Z}_{ABC}\right)\label{MZCR4}.
\ee
\end{subequations}
These commutation relations allow to apply the definition of an $\mathfrak{su}_8$ action on a physical field from section \ref{prim2} to the holomorphic fields $\eps$ and $\chi$, which is the last ingredient to specify the supersymmetry variations (\ref{susy3}) of the vielbein $e^H$ in $56$ dimensions.

\subsection{The supersymmetry variations in $d=56$}\label{super2}
Since I focussed on the bosonic degrees of freedom of the $d=60$ vielbein that correspond to the $E_{7(7)}$ valued submatrix $e^H$, the supersymmetry variation $\underline{\delta}$ (\ref{Deriv1}) of all bosonic degrees of freedom under consideration are comprised by the relations (\ref{susy3})
\beg
{\left(e^{-H}\right)_{AB}}^\mu \underline{\delta}{\left(e^{H}\right)_{\mu}}^{CD}&=& 0\\
{\left(e^{-H}\right)^{AB}}^\mu \underline{\delta}{\left(e^{H}\right)_{\mu}}^{CD}&=& {\left(e^{-H}\right)^{[AB}}^\mu \delta{\left(e^{H}\right)_{\mu}}^{CD]},
\eeg
where the objects in the last line parametrize $\mathfrak{e}_{7(7)}\ominus \mathfrak{su}_{8}$. This $\mathfrak{su}_8$ representation space is spanned by the generators $\hat{S}_{ABCD}$ that are subject to the reality constraint (\ref{RquerR})
\beg
\overline{\hat{S}^{ABCD}} &=& \frac{1}{4!}\e^{ABCDEFGH}\hat{S}_{EFGH}.
\eeg
This circumstance has to be taken into account when relating the variation of the vielbein $e^H$ to the fermions $\eps$ and $\chi$ (\ref{ecDefi}). Thus, it is natural to define
\beg
{\left(e^{-H}\right)^{AB}}^\mu \underline{\delta}{\left(e^{H}\right)_{\mu}}^{CD}&:=& \eps^{[A}\chi^{BCD]} + \frac{1}{4!}\e^{ABCDEFGH}\bar{\eps}_E\bar{\chi}_{FGH}
\eeg
such that a contraction of this equation with $\e_{ABCDEFGH}$ is equivalent to a complex conjugation.\\

Since it is essential for a physical symmetry to map the set of solutions of the equations of motion to itself, I have to define an action of the variation $\underline{\delta}$ (\ref{Deriv1}) on the fermions $\eps$ and $\chi$. To establish the connection to supergravity, I will interpret $\eps$ as the parameter of the supersymmetry transformation. Therefore, I can without loss of generality set
\be
\underline{\delta}\!\eps^C&:=&0.
\label{eVari}
\ee
It is important to observe a difference to the standard definition of supergravity theories: I have to use the equivarant variation $\underline{\delta}$ instead of the invariant one $\delta$ (\ref{deltaDefi}), because otherwise, the relation (\ref{eVari}) would not be preserved under an action induced by a diffeomorphism of the exceptional geometry.\\

In particular, the equation (\ref{eVari}) together with the decomposition (\ref{deltaDefi}) of the equivariant derivation $\underline{\delta}$ into an invariant variation $\delta$ and a $\mathfrak{u}_8$ action implies that $\delta\! \eps^C$ does not vanish:
\be\label{eVari2}
\delta\! \eps^C &\stackrel{(\ref{deltaDefi})}{=}& +\delta_{\hat{\Lambda}}\!\eps^C\nn\\
&=&
 {\Lambda_A}^C\!\eps^A.
\ee
The action of $\hat{\Lambda}$ on the $\mathfrak{u}_8$ representation $\eps$ follows the same rules that were used to obtain (\ref{LWi}). The fact $\delta\! \eps^C\neq 0$ may be important for the evaluation of the supersymmetry algebra, whose general form is provided by (\ref{SUSY0}). As I stated in the introduction, this is beyond the scope of my thesis, however.\\

To complete the set of supersymmetry variations, I have to define an action of $\underline{\delta}$ on $\chi_{ABC}$. Since $\eps$ is the transformation parameter of a symmetry algebra, $\underline{\delta}\chi_{ABC}$ has to be linear in $\eps$. By $SU(8)$-covariance, the canonical choice is
\be\label{chiVari}
\underline{\delta}\chi^{ABC} &=& 
\bar{\nabla}^{[AB}\!\eps^{C]},
\ee
where $\bar{\nabla}^{AB}$ is the connection $\nabla$ (\ref{conn9}) in the antiholomorphic vielbein frame $\underline{d\bar{z}}_{AB}$ of the cotangent space that I also used in section \ref{cII}. The variation of the antiholomorphic section $\bar{\chi}$ hence contains the connection $\nabla$ in the holomorphic vielbein frame $\underline{dz}^{AB}$:
\beg
\underline{\delta}\bar{\chi}_{ABC} &=& 
\nabla_{[AB}\bar{\eps}_{C]}.
\eeg
To conclude, I repeat all independent supersymmetry variations:
\be\label{varii}
{\left(e^{-H}\right)_{AB}}^\mu \underline{\delta}{\left(e^{H}\right)_{\mu}}^{CD}&=& 
0
\nn\\
{\left(e^{-H}\right)^{AB}}^\mu \underline{\delta}{\left(e^{H}\right)_{\mu}}^{CD}&=& \eps^{[A}\chi^{BCD]} + \frac{1}{4!}\e^{ABCDEFGH}\bar{\eps}_E\bar{\chi}_{FGH}
\nn\\
\underline{\delta}\eps^{C} &=& 0
\nn\\
\underline{\delta}\chi^{ABC} 
&=& 
\bar{\nabla}^{[AB}\!\eps^{C]}.
\ee
The remaining degrees of freedom $c_1,c_2$ in the connection $\nabla$ (\ref{omegaDefiN}) and the normalization constants $\tau_2,\tau_7$ will be fixed by a comparison to the supersymmetry variations of $d=11$ supergravity. To do this, I have to perform an $\mathfrak{so}_7$ decomposition of the $\mathfrak{su}_8$ representations similar to the one of section \ref{boson1}.

\subsection{Decomposition into $\mathfrak{so}_7$ representations}\label{Deco}
I want to start with the variation of the bosonic degrees of freedom $e^H$ (\ref{varii}), in which I can without loss of generality substitute $\delta$ for the equivariant derivation $\underline{\delta}$ (\ref{susy3})
\be\label{bos3}
{\left(e^{-H}\right)^{AB}}^\mu \delta{\left(e^{H}\right)_{\mu}}^{CD}
=\,\,
\eps^{[A}\chi^{BCD]} + \frac{1}{4!}\e^{ABCDEFGH}\bar{\eps}_E\bar{\chi}_{FGH}.
\ee
The $E_{7(7)}$ valued matrix $e^H$ in this relation was decomposed into $\mathfrak{gl}_7$ representation by (\ref{cosetE73})
\beg
e^H&=:&e^{{h_a}^b{\left.\hat{M}\right.^a}_b}e^{A_{abc}\hat{E}^{abc}}e^{A_{a_1\dots a_6}\hat{E}^{a_1\dots a_6}}.
\eeg
It is important that the $\mathfrak{gl}_7$ representations $A_{abc}$ and $A_{a_1\dots a_6}$ in $e^H$ have only been used in the vielbein frame so far. Therefore, I can without loss of generality define the following objects with the rescaled vielbein $e$ (\ref{Weyl})
\begin{subequations}\label{vSym7}
\be
A_{m_1\dots m_3} &:=& {e_{m_1}}^{a_1}\cdots{e_{m_3}}^{a_3}A_{a_1\dots a_3},\\
A_{m_1\dots m_6} &:=& {e_{m_1}}^{a_1}\cdots{e_{m_6}}^{a_6}A_{a_1\dots a_6}.
\ee
\end{subequations}
I have shown in section \ref{boson1} that it is the rescaled vielbein $e$ and not $\tilde{e}$ that is identified with the corresponding subvielbein of $d=11$ supergravity. Therefore, this definition (\ref{vSym7}) is consistent with the standard definition of $d=11$ supergravity in which the potentials $A$ are considered as three- and six-form potentials, respectively that have a natural definition in the coordinate induced frame. From the point of view of exceptional geometry, the bosonic degrees of freedom $A_{abc}$ and $A_{a_1\dots a_6}$ do not correspond to three- or six-forms and hence, the statement (\ref{vSym7}) indeed is a definition.\\

Since $\delta$ also is a derivation, I can use the evaluation of the Maurer--Cartan form $e^{-H}de^H$ from section \ref{cII} in terms of $\mathfrak{gl}_7$ representations. A comparison of the abbreviations (\ref{vSym}) with the definition (\ref{vSym7}) while keeping in mind the rescaling of the vielbein (\ref{Weyl}) then leads to the equation
\be\label{bos4}
{\left(e^{-H}\right)^{AB}}^\mu \delta{\left(e^{H}\right)_{\mu}}^{CD}
&=&
-\frac{3}{4}
{\G^g}^{[AB}{\G_h}^{CD]}
{e_g}^m\delta {e_m}^h
\nn\\
&&
-\frac{3}{\tau_2}
{\G^{[a_1a_2}}^{[AB}{\G^{a_3]}}^{CD]}
{e_{a_1}}^{m_1}\cdots{e_{a_3}}^{m_3}
\delta A_{m_1\dots m_3}
\nn\\
&&
+\frac{i}{20\tau_2^2}\e^{a_1\dots a_6c}
{\G_{bc}}^{[AB}{\G^{b}}^{CD]}
{e_{a_1}}^{m_1}\cdots{e_{a_6}}^{m_6}
\nn\\
&&
\left(\delta A_{m_1\dots m_6}
-20A_{[m_1\dots m_3}\delta A_{m_4\dots m_6]}
\right).
\ee
Together with equation (\ref{bos3}), this defines the supersymmetry variation of the vielbein $e$ (\ref{Weyl}) as well as the ones of the $\mathfrak{gl}_7$ representations $A_{abc}$ and $A_{a_1\dots a_6}$. It is important to use the derivation $\delta$ at this point, because the $\mathfrak{u}_8$ action of the equivariant derivation $\underline{\delta}$ would not close on $\mathfrak{gl}_7$ representations.\\

The variations of the $\mathfrak{gl}_7$ representations can be cast in a particularly simple form by defining
\be\label{psichi}
(\chi_h)^{C} &:=& \frac{i}{9}\left(\delta_h^g\delta^C_D +\frac{1}{8}{{{\G_h}^g}_D}^C\right){\G_g}_{AB}\chi^{ABD}.
\ee
with $g,h=1,\dots,7$ and $A,B,C=1,\dots,8$. Since the degrees of freedom of $\chi^{ABC}$ and $(\chi_h)^{C}$ match
\beg
\binom{8}{3}\,=&56&=\,7\cdot 8, 
\eeg
it is possible to invert this relation. A short calculation reveals
\be\label{chiDefi}
\chi^{ABC} &=& 3!i{\G^f}^{[AB}(\chi_f)^{C]}.
\ee
The relation for the antiholomorphic fields $\bar{\chi}$ follows by complex conjugation keeping in mind that the Clifford matrices $\G^f$ are purely imaginary
\beg
\bar{\chi}_{ABC} &=& 3!i{\G^f}_{[AB}(\bar{\chi}_f)_{C]}.
\eeg
I want to stress that this is a mere relabelling of the degrees of freedom. Hence, the $SU(8)$-covariance is not broken, it is only obscured in complete analogy to the $E_{7(7)}$-covariance of $e^H$ in the $\mathfrak{gl}_7$ decomposition of section \ref{VielbE}.\\

Substituting the relation (\ref{chiDefi}) in (\ref{bos3}) allows to separate the linearly independent $\mathfrak{so}_7$ representations in equation (\ref{bos4}) with the $\G$-matrix relations from appendix \ref{SomeG2}:
\begin{subequations}\label{psiDefi}
\be
{e_g}^m\delta {\left.e\right._m}^h&=& 
i\!\eps^C{\G^h}_{CD}(\chi_g)^D + \text{c.c.}
\\
{e_{a_1}}^{m_1}\cdots{e_{a_3}}^{m_3}
\delta A_{m_1\dots m_3}
&=&
-\frac{i\tau_2}{8}\!\eps^C{\G_{[a_1a_2}}_{CD}(\chi_{a_3]})^D + \text{c.c.}
\\
{e_{a_1}}^{m_1}\cdots{e_{a_6}}^{m_6}
\delta A_{m_1\dots m_6}
&=&
  -\frac{i\tau_2^2}{48}\!\eps^C{\G_{[a_1\dots a_5}}_{CD}(\chi_{a_6]})^D + \text{c.c.}
 \\
  &&+20{e_{a_1}}^{m_1}\cdots{e_{a_6}}^{m_6}
A_{[m_1\dots m_3}\delta A_{m_4\dots m_6]}\nn.
 \ee
\end{subequations}
These relations already look quite similar to the ones (\ref{Trafo1}) of $d=11$ supergravity, in which the $\mathfrak{so}_7$-covariance is obvious. Having fixed the normalization of $A_{a_1\dots a_6}$ in (\ref{tau3fix}), the $\mathfrak{so}_7$-covariance is only enhanced to $\mathfrak{su}_8$ for the present ratio of prefactors
\beg
-1:\frac{\tau_2}{8}:\frac{\tau_2^2}{48}.
\eeg

Before I continue with the supersymmetry variation of the fermion $\chi$, I want to fix the $\mathfrak{u}_8$ action $\hat{\Lambda}$ (\ref{deltaDefi}) that distinguishes the standard supersymmetry variation $\delta$ from the covariant one $\underline{\delta}$. In equation (\ref{susy2}), I determined $\hat{\Lambda}$ in terms of the compact part of the Maurer--Cartan form $e^{-H}\delta e^H$. Hence, I can use the results from section \ref{cII} and obtain
\beg
{\Lambda_A}^B 
&=&
 \frac{1}{8}{e_g}^m\delta {e_m}^h  {{{\G^g}_h}_A}^B
+\frac{1}{2\tau_2}{{\G^{a_1\dots a_3}}_A}^B
{e_{a_1}}^{m_1}\cdots{e_{a_3}}^{m_3}
\delta A_{m_1\dots m_3}
\\
&&
-\frac{i}{20\tau_2^2}\e^{a_1\dots a_6c}{{\G_{c}}_A}^B
{e_{a_1}}^{m_1}\cdots{e_{a_6}}^{m_6}
\left(\delta A_{m_1\dots m_6} -20A_{[m_1\dots m_3}\delta A_{m_4\dots m_6]}\right).
\nn
\eeg

A substitution with the explicit form (\ref{psiDefi}) of the variations of the $\mathfrak{gl}_7$ representations finally leads to
\be\label{Lambexpl}
{\Lambda_A}^B 
&=&
 \left(\frac{1}{8}i\!\eps^C{\G^h}_{CD}(\chi_g)^D {{{\G^g}_h}_A}^B
-\frac{i}{8}{{\G_{c}}_A}^B
\!\eps^C{\G^{bc}}_{CD}(\chi_{b})^D \right)+\text{c.c.}
\nn\\
&&
-i\left(\frac{1}{16}{{\G^{a_1\dots a_3}}_A}^B
\!\eps^C{\G_{a_1a_2}}_{CD}(\chi_{a_3})^D
+\text{c.c.}\right).
\ee
Since ${\Lambda_A}^B$ is a part of the equivariant derivation $\underline{\delta}$, it does not transform as an $SU(8)$ tensor under the induced action of a diffeomorphism. This is due to the same reasoning showing that the Christoffel symbols or the spin connection are no tensors, either.\\

The final variation to discuss is the one of the holomorphic fermion $\chi$ (\ref{varii})
\beg
\underline{\delta}\chi^{ABC} 
&=&
\bar{\nabla}^{[AB}\!\eps^{C]}.
\eeg
The first observation is that the $\mathfrak{u}_8$ action $\hat{\Lambda}$ on $\chi$ is not trivial. With its definition (\ref{deltaDefi}), I obtain in analogy to the computation of (\ref{LWi})
\beg
\delta\chi^{ABC} 
&=&
\bar{\nabla}^{[AB}\!\eps^{C]} +\delta_{\hat{\Lambda}}\chi^{ABC} \\
&=&
\bar{\nabla}^{[AB}\!\eps^{C]} +3{\Lambda_D}^{[A}\chi^{BC]D}.
\eeg
Together with the explicit form of ${\Lambda_A}^B$ (\ref{Lambexpl}), it is obvious that the $\mathfrak{u}_8$ action $\hat{\Lambda}$ results in a quadratic expression in $\chi$. Hence, I drop it for the present calculation, because I have also neglected higher order terms in fermions in the definition of $d=11$ supergravity in section \ref{SUGRA1}.\\

To compare the remaining part to $d=11$ supergravity, I can without loss of generality truncate the $49$ additional dimensions of the exceptional geometry as I have explained in section \ref{boson1}. Furthermore, it will prove convenient to also substitute the definition of $(\chi_h)^{C}$ (\ref{chiDefi}) in the supersymmetry variation. 
Then, I obtain after some algebraic manipulations with the formul\ae{} from appendix \ref{SomeG2} and with the connection $\nabla$ (\ref{omegaDefiN}) and the abbreviation $\gamma:=\frac{c_1}{c_2}-\frac{1}{8}$:
\be\label{chigl}
\underline{\delta}(\chi_d)^{C}
 &=&
-\frac{\tau_7}{96}e^{-\frac72\sigma}\left[
e^{-\frac{7c_2\gamma}{2}\sigma}\p_d\left(e^{\frac{7c_2\gamma}{2}\sigma}\eps^C\right)
\right.\nn\\
&&+\left\{
\frac{1}{4}\left[
{\left(v^{\text{min}}_d\right)_{e}}^{f}
-3c_2{\left(v_e\right)_{d}}^{f}
\right]
 {{{\G^{e}}_f}^C}_D
  \right.\nn\\
&&
-\frac{1}{\tau_2}\left[
\left(1-\frac{c_2}{2}\right)\left(v_d\right)_{a_1 \dots a_3}
-3c_2\left(v_{a_3}\right)_{a_1 a_2 d}
\right]
{{\G^{a_1\dots a_3}}^{C}}_D
\nn\\
&&-\frac{ic_2}{16\tau_2}F_{aa_1\dots a_3}
   {\e_{bcd}}^{a_1a_2a_3a}{{\G^{bc}}^{C}}_D 
\nn\\
&&-\frac{i}{10\tau_2^2}\left[
\left(1-\frac{c_2}{2}+\frac{4c_2\gamma}{3}\right)\delta^g_c\delta^a_d
+\frac{c_2}{2}\delta^g_d\delta^a_c
\right]
\left(v_a\right)_{a_1\dots a_6}\e^{a_1\dots a_6c}
{{\G_{g}}^{C}}_D
\nn\\
&&
+c_2\left(\gamma+\frac{3}{4}\right)
\left[
{\left(v_e\right)_{d}}^{e}
 \delta_D^{C}
-\frac{2}{\tau_2}{\left(v_c\right)_{a_1  a_2}}^c
\left(
{{{\G^{a_1 a_2}}_d}^{C}}_D
+4\delta^{a_2}_d{{\G^{a_1}}^{C}}_{D}
\right)
\right.
\nn\\
&&\left.\left.\left.
+\frac{i}{15\tau_2^2}\left(v_a\right)_{a_1\dots a_6}
\e^{a_1\dots a_6c}
\left({{{\G_{cd}}^a}^{C}}_D 
+2\eta_{cd}{{\G^{a}}^{C}}_D
\right)
\right]
\right\}
\eps^D
\right].
\ee
I want to emphasize that the equation (\ref{chigl}) provides the general form of a connection $\nabla$ in the exceptional geometry that is linear in derivatives and that only depends on vielbein degrees of freedom. In particular, the constants $c_1,c_2\in \R$ uniquely fix the connection with these properties in $56$ dimensions as explained in section \ref{LCcE}.\\

I have only neglected the dependence on the additional $49$ dimensions in the equation (\ref{chigl}), because these do not correspond to directions in $d=11$ supergravity. The constants $\tau_2$ and $\tau_7$ are normalization constants that are related to the $Gl(7)$ decomposition of the $E_{7(7)}$ valued vielbein $e^H$: $\tau_2$ fixes the norm of the generators $\hat{E}^{abc}$ (\ref{defiSo7},\,\ref{EDefi}) that is associated to the field $A_{abc}$, whereas $\tau_7$ scales the relation between the real coordinate chart $(x,p)$ and the holomorphic one $z^{AB}$ (\ref{zNeu3}).\\

For a comparison to the supersymmetry variations of $d=11$ supergravity, I adopt the following choice
\begin{subequations}\label{konstfix}
\be
\tau_7&:=&-96,\\
c_1 &:=& -\frac{5}{12},\\
c_2&:=&\frac{2}{3},\\
\Rightarrow \quad \gamma &=& -\frac{3}{4}.
\nn
\ee
\end{subequations}
Then the variation (\ref{chigl}) of the holomorphic field $\chi$ takes the nice form
\be\label{chigl2}
e^{\frac74\sigma}\underline{\delta}(\chi_d)^{C}
 &=&
\p_d\left(e^{-\frac{7}{4}\sigma}\eps^C\right)
+\frac{1}{4}
{\omega_{de}}^{f}
 {{{\G^{e}}_f}^C}_D
 e^{-\frac74\sigma}
\eps^D
\nn\\
&&
+\frac{1}{12\tau_2}
F_{a_1 \dots a_4}
\left(
{{{\G^{a_1\dots a_4}}_d}^{C}}_D 
   -8\delta_d^{a_1}{{\G^{a_2\dots a_4}}^{C}}_D
   \right) e^{-\frac74\sigma}\eps^D
\nn\\
&&-\frac{i}{210\tau_2^2}F
{{\G_{d}}^{C}}_D
 e^{-\frac74\sigma}\eps^D.
\ee
The variation of the antiholomorphic field $\bar{\chi}$ follows from complex conjugation. Before I relate these fields $(\chi,\bar{\chi})$ to the real gravitino $\psi$ in the next section, I want to add some comments concerning the choice of constants (\ref{konstfix}).
\begin{itemize}
	\item The general coordinate invariance of $d=11$ supergravity reduces to $\Diff(7)$ for the fields under consideration. Therefore, all derivatives of the three- and six-form potentials $A$ have to be $\Diff(7)$-covariant $SO(7)$ tensors. It is obvious from the relation (\ref{chigl}) that this is only possible, if the constants are fixed by (\ref{konstfix}). 
	\item A joint realization with the conformal subgroup of diffeomorphisms as introduced in section \ref{secD} cannot be applied to this theory. I proved in section \ref{VectF} that this subgroup does not preserve the quartic tensor $Q$, which directly implies that it would not respect the restriction of the vielbein matrix $e^H$ to be $E_{7(7)}$ valued. 
	\item Finally, the resulting connection $\nabla$ with the conventions (\ref{konstfix}) does not coincide with the projection (\ref{projCon2}) of the Levi--Civita connection $\nabla^{(LC)}$ on $\mathfrak{su}_8$ as explained in section \ref{LCcE}. Therefore, the procedure of Borisov \& Ogievetsky is not even related to this $\nabla$ in an indirect way, i.e. by an application to the four-dimensional part of the $60$-dimensional geometry with $\Diff(4)$ symmetry.
\end{itemize}

\subsection{Comparison to $d=11$ supergravity}\label{chiral}
In section \ref{Comp1}, I have identified the bosonic fields of $d=11$ supergravity that correspond to the subsector of the exceptional geometry in $d=60$ which is parametrized by the subvielbein $e^H$ in $56$ dimensions. These fields are the unrestricted $d=7$ subvielbein of the $11$-dimensional one ${e_m}^a={E_m}^a$, the part of the three-form potential parametrized by $A_{abc}$ and the dual six-form potential $A_{a_1\dots a_6}$ with the indices ranging over $4,\dots,10$. The latter was related to the part of the three-form potential $A_{\alpha\beta\gamma}$ along the remaining four directions $\alpha,\beta,\gamma=0,\dots,3$ by the duality relation (\ref{4Form23a}). For the comparison to the subsector $e^H$ of the exceptional geometry, I have set the remaining bosonic degrees of freedom to zero. Therefore, the relevant supersymmetry transformations of the bosons in $d=11$ supergravity (\ref{Trafo1}) have the form (with $\psi_g={e_g}^m\psi_m$):
\begin{subequations}\label{susyred}
\be
	{e_g}^m\delta {e_m}^h &=& i\bar{\ep}\tilde{\G}^h\psi_g,\\
	{e_{a_1}}^{m_1}\cdots {e_{a_3}}^{m_3}	\delta A_{m_1\dots m_3} &=& -\frac{3i}{2}\bar{\ep}\tilde{\G}_{[a_1a_2}\psi_{a_3]}.
\ee
\end{subequations}
Since the identification of the degrees of freedom of the six-form potential $A_{a_1\dots a_6}$ in seven dimensions with the ones of the three-form potential $A_{\alpha\beta\gamma}$ relies on an equation of motion, the supersymmetry variation of $A_{a_1\dots a_6}$ cannot be deduced easily from the ones of the $d=11$ supergravity in the standard form (\ref{Trafo1}).\footnote{As soon as the dynamics of the complete theory in the sixty-dimensional exceptional geometry are established, the action of $\delta$ on the equation of motion (\ref{4Form23a}) may provide a non-trivial consistency check for the definitions of the variations $\delta A_{\alpha\beta\gamma}$ and $\delta A_{a_1\dots a_6}$.}\\

It catches the eye in formula (\ref{susyred}) that only the part $\psi_m$ of the gravitino with $m=4,\dots,10$ is related to the degrees of freedom of the subsector of exceptional geometry parametrized by $e^H$. Hence, it is sufficient to focus on the supersymmetry variation (\ref{Trafo1}) of these components. Due to the restriction to the degrees of freedom of $e^H$, it takes the following form in the vielbein frame to lowest order in fermions:
\be\label{psiTrafo}
	\delta\psi_g &=& \partial_g\ep +\frac{1}{4}\omega_{gab}\tilde{\G}^{ab}\ep\nn\\
&&+\frac{1}{144}\left[\left(\tilde{\G}^{abcd}{}_g-8\delta_g^a\tilde{\G}^{bcd}\right)\ep F_{abcd}
+\tilde{\G}^{\alpha\beta\gamma\delta}{}_g\ep F_{\alpha\beta\gamma\delta}
\right].
\ee

As defined in section \ref{SUGRA1} and in appendix \ref{Cliff}, the real $32\times 32$ matrices $\tilde{\G}^P$ with $P=0,\dots,10$ form a Majorana representation of the Clifford algebra. Therefore, I can without loss of generality drop the complex conjugation in the definition of the Dirac conjugate $\bar{\ep}$ of a spinor $\ep$ that was used in section \ref{u1s}. In order to avoid a confusion with the complex conjugation of the holomorphic fields $\eps$ and $\chi$ from section \ref{fermi}, I substitute the following notation for $\bar{\ep}$:
\be\label{echeck}
\check{\ep}&:=&\ep^t\tilde{\G}^0.
\ee

It is important to note that choosing the Majorana representation for the Clifford matrices does not imply that the spinors $\ep$ and $\psi$ are real objects. A complex conjugation of the supersymmetry variations of bosons (\ref{susyred}) and fermions (\ref{psiTrafo}) reveals that this choice would even be inconsistent.\\

Choosing the Majorana representation only allows to rescale the fermions $\ep$ and $\psi$ in such a way that the rescaled fermions $\epr$ and $\psir$ are real objects:
\begin{subequations}\label{rescale1}
\be
\epr &:=& \ep \sqrt{i}\\
\psir_g&:=&\psi_g \sqrt{i}\\
\text{with}\quad \sqrt{i}&=& \frac{1}{\sqrt{2}}(1+i).
\nn
\ee
\end{subequations}
This rescaling was also performed by de Wit \& Nicolai in \cite{dWN86}. It only changes the prefactor of the supersymmetry tranformation of the bosons, whose manifestly real form hence is
\begin{subequations}\label{susyred2}
\be
	{e_g}^m\delta {e_m}^h &=& \check{\epr}\tilde{\G}^h\psir,\\
	{e_{a_1}}^{m_1}\cdots {e_{a_3}}^{m_3}	\delta A_{m_1\dots m_3} &=& -\frac{3}{2}\check{\epr}\tilde{\G}_{[a_1a_2}\psir_{a_3]}.
\ee
\end{subequations}
The variation of the rescaled fermion $\psir_g$ has the same form as the one of $\psi_g$ (\ref{psiTrafo}), if $\epr$ is substituted for $\ep$. \\

Next, recall that it was essential for the proof of the $SU(8)$-covariance of $d=11$ supergravity \cite{dWN86} to reduce the Lorentz gauge group $SO(10,1)$ to $SO(3,1)\times SO(7)$ by fixing an explicit ansatz for the vielbein matrix. Since the $SU(8)$-covariance was the major motivation to discuss the exceptional geometry, it is natural to perform this $4+7$ split in the present case, too. The appropriate decomposition of the eleven real matrices $\tilde{\G}_P$ has the form
\begin{subequations}\label{gammadec}
\be
\tilde{\G}_\alpha &=& \g_\alpha\otimes \id_{8}\quad \,\text{for }\alpha\,=\,0,\dots,3,\label{G4Defi}\\
\tilde{\G}_g &=& \frac{\g_5}{i}\otimes \G_g\quad \text{for }g\,=\,4,\dots,10.
\label{G5Defi}
\ee
\end{subequations}
The Clifford matrices $\g_\alpha$ in $d=4$ are real and $\g_5:=\g^0\g^1\g^2\g^3$. Therefore, I had to introduce the imaginary unit $i$ in the line (\ref{G5Defi}), because $\G_a$ is purely imaginary by definition.\footnote{More details on these Clifford matrices can be found in appendix \ref{Cliff}.}\\

The $4+7$ split of the Clifford algebra (\ref{gammadec}) induces a decomposition of its $\mathbf{32}$-dimensional representation space into the product $\mathbf{4}\otimes \mathbf{8}$. I have explained in section \ref{Comp1} that the focus of this thesis is on the seven-dimensional part. Therefore, I want to state the index corresponding to the spin representation $\mathbf{8}$ of $\mathfrak{so}_7$ explicitly. The rescaled supersymmetry parameter $\epr$ and gravitino $\psir_g$ are hence labelled by
\beg
\epr{}^C,\quad (\psir_g)^C\quad \text{with }C=1,\dots,8\text{ and }g=4,\dots,10.
\eeg
The indices corresponding to the spin representation $\mathbf{4}$ of $\mathfrak{so}_{(3,1)}$ are kept implicit. In this notation, the definition of the conjugated spinor $\check{\epr}$ (\ref{echeck}) simplifies to
\be\label{echeck2}
{\check{\epr}}^C&:=&\big({\epr}^t\big)^C\g^0\quad \text{with }C=1,\dots,8.
\ee
In particular, the transposition $t$ only affects the $\mathbf{4}$ part, which is obvious from the decomposition of $\tilde{\G}_g$ (\ref{gammadec}). The supersymmetry variations of the bosons (\ref{susyred2}) hence have the form
\begin{subequations}\label{susyred3}
\be
	{e_g}^m\delta {e_m}^h
	&=&
	 \check{\epr}^C\frac{\g_5}{i}{\G^h}_{CD}(\psir_g)^D,
	 \\
	{e_{a_1}}^{m_1}\cdots {e_{a_3}}^{m_3}\delta A_{m_1\dots m_3} 
		&=&
		 -\frac{3}{2}\check{\epr}^C{\G_{[a_1a_2}}_{CD}(\psir_{a_3]})^D,
		 \ee
		 \end{subequations}
where I have used the fact that $\g_5$ squares to the negative identity $-\id_4$ on the $\mathbf{4}$ part (\ref{gamma6})
\be\label{gamma5b}
\g_5^2 &=& -\id_4.
\ee
This fact is of crucial importance. It canonically induces an endomorphism of the $\mathbf{32}$-dimensional representation vector space with matrix representation $\g_5\otimes \id_8$. Hence, there is an endomorphism of a real vector space that squares to the negative identity. This defines a complex structure.\\

Following the standard procedure that is for example performed in Hitchin's article \cite{H00}, I define the holomorphic objects by
\begin{subequations}\label{rescale2}
\be
{\epsr}^C &:=& \frac{1}{2}\left(\id_4 -i\g_5\right){\epr}^C \\
(\chir_g)^C&:=&\frac{1}{2}\left(\id_4 -i\g_5\right)(\psir_g)^C.
\ee
\end{subequations}
With the simple calculation
\beg
\g_5\left(\id_4 -i\g_5\right) \,\stackrel{(\ref{gamma5b})}{=}\,\g_5 +i\id_4\,\,=\,\,i\left(\id_4 -i\g_5\right),
\eeg
I can evaluate the endomorphism $\g_5$ on these holomorphic fields (\ref{rescale2}) to
\begin{subequations}\label{IAct}
\be
\g_5{\epsr}^C &=& i{\epsr}^C, \\
\g_5(\chir_g)^C&=&i(\chir_g)^C.
\ee
\end{subequations}
Hence, I can without loss of generality restrict the real $\mathbf{32}$-dimensional representation space to a complex $\mathbf{8}_\C$-dimensional vector space, on which the action of $SU(8)$ is well-defined. Due to the interpretation of $\g_5$ as an imaginary unit, this representation of $SU(8)$ is also referred to as chiral $SU(8)$ \cite{dWN86}.\newpage

A comment should be added concerning the position of the spinor indices. Cremmer and Julia \cite{CJ79} have already observed that lowered and raised indices are equivalent before the action of the chiral $SU(8)$ is introduced. Furthermore, they mentioned that the canonically associated complex version $\check{\epsr}$ of the Dirac conjugate of a real spinor $\check{\epr}$ also is holomorphic. This fact directly follows from the calculation
\beg
{\check{\epsr}}^C &\stackrel{(\ref{echeck2})}{=}&\big({\epsr}^t\big)^C\g^0\\
&\stackrel{(\ref{rescale2})}{=}&\frac{1}{2}\left(\left(\id_4 -i\g_5\right){\epr}^C\right)^t\g^0\\
&=&
\frac{1}{2}\big({\epr}^t\big)^C\left(\id_4 +i\g_5\right)\g^0\\
&=&
\frac{1}{2}\big({\epr}^t\big)^C\g^0\left(\id_4 -i\g_5\right)\\ 
&\stackrel{(\ref{echeck2})}{=}&
\frac{1}{2}{\check{\epr}}^C\left(\id_4 -i\g_5\right),
\eeg
which uses $\{\g^0,\g_5\}=0$ and the antisymmetry of $\g_5$. Therefore, it is consistent and necessary to denote the conjugated spinor $\check{\epsr}{}^C$ of the holomorphic one $\epsr{}^C$ with raised indices, too.\\

The corresponding antiholomorphic spinor follows from complex conjugation and lowering the index with the Kronecker $\delta$ as in equation (\ref{ZbarDefi}). The antiholomorphic supersymmetry transformation parameter $\bar{\epsr}$ is hence associated to the real parameter $\epr$ by
\be\label{complcon}
\bar{\epsr}_C&:=& \delta_{CD}\overline{{\epsr}^D}\\
&\stackrel{(\ref{rescale2})}{=}&\frac{1}{2}\left(\id_4 +i\g_5\right){\epr}_D,\nn
\ee
keeping in mind that the position of the index of the real spinor $\epr$ was arbitrary. As a last step, I perform a chiral rescaling and multiply by the determinant of the vielbein (\ref{Weyl}) in complete analogy to \cite{dWN86}:
\begin{subequations}\label{rescale3}
\be
{\eps}^C &:=& \sqrt{-\g_5}e^{+\frac{7}{4}\sigma}{\epsr}^C \\
(\chi_g)^C&:=&\sqrt{-\g_5}e^{-\frac{7}{4}\sigma}(\chir_g)^C\\
\text{with}\quad \sqrt{-\g_5}&:=& \frac{1}{\sqrt{2}}\left(\id_4 -\g_5\right).
\nn
\ee
\end{subequations}
The conjugate spinor transforms with the same matrix $\sqrt{-\g_5}$ from the right.\\

Finally, this shows that the supersymmetry variation of the siebenbein $e$ in $d=11$ supergravity (\ref{susyred3}) and in the exceptional geometry (\ref{psiDefi}) exactly coincide. It should be noted that there is no distinction between spinors and conjugated spinors in the subsector of exceptional geometry that corresponds to $e^H$. For the transformations in this subsector, it was sufficient to reduce the tensor product of $\mathbf{8}$ with the four-dimensional space $\mathbf{4}$ to the real sixteen-dimensional space $\mathbf{8}_\C$ with the complex structure $\g_5$.\newpage

Furthermore, the supersymmetry variations of the three-form potential $A_{abc}$ (\ref{susyred3},\,\ref{psiDefi}) in both theories can be identified, if the normalization constant $\tau_2$ of the generator $\hat{E}^{abc}$ (\ref{defiSo7},\,\ref{EDefi}) is fixed to
\be\label{tau2fix}
\tau_2&:=&12.
\ee

This was the last normalization constant remaining. Therefore, the comparison of the numerical factors in the supersymmetry variation of the fermion poses a highly non-trivial consistency check. To perform this, recall the identity $\e^{\alpha\beta\gamma\delta}\e_{\alpha\beta\gamma\delta}=-4!$ for Minkowskian signature $(-1,1,1,1)$. Then, the first order equation of motion of the three-form potential (\ref{4Form23a}) takes the form
\be\label{F7Defi3}
\e^{\alpha\beta\gamma\delta}F_{\alpha\beta\gamma\delta} &=& -\frac{4!}{7!} F
\ee
with the abbreviation $F$ introduced in (\ref{F7Defi2}). Together with the decomposition of the $\tilde{\G}$ matrices (\ref{gammadec}) and the relations (\ref{gamma6}) from appendix \ref{Cliff}, this identity transforms the supersymmetry variation of the fermion in $d=11$ supergravity (\ref{psiTrafo}) to
		 		 \beg
		\delta (\psi_g)^C &=& \partial_g\ep^C +\frac{1}{4}\omega_{gab}{{\G^{ab}}^C}_D\ep^D\\
&&
+\frac{1}{144}F_{abcd}\left({{{\G^{abcd}}_g}^C}_D-8\delta_g^a{{\G^{bcd}}^C}_D\right)\frac{\g_5}{i}\ep^D \\
&&
-\frac{4!i}{7!144}F{{\G_g}^C}_D\ep^D.
\eeg
As I mentioned before, it does not make a difference if the spinor indices are raised or lowered, before $SU(8)$-covariant objects are introduced. And this can be achieved with the redefinitions (\ref{rescale1},\,\ref{rescale2},\,\ref{rescale3}) that commute with each other, of course. Hence, the supersymmetry variations of the fermions $\psi_g$ and $\g_5\psi_g$ can be combined into a holomorphic one, which has the form
\beg
		\delta \left(e^{\frac{7}{4}\sigma}(\chi_g)^C\right) &=& \partial_g\left(e^{-\frac{7}{4}\sigma}\eps^C\right) +\frac{1}{4}\omega_{gab}{{\G^{ab}}^C}_D\eps^D\\
&&
+\frac{1}{144}F_{abcd}\left({{{\G^{abcd}}_g}^C}_D-8\delta_g^a{{\G^{bcd}}^C}_D\right)\frac{\g_5}{i}\eps^D\\
&&
-\frac{4!i}{7!144}F{{\G_g}^C}_D\eps^D.
\eeg

Together with the equation (\ref{IAct}) and the fixing of $\tau_2$ (\ref{tau2fix}), this exactly agrees with the supersymmetry variation of the fermion of exceptional geometry (\ref{chigl2}). Since I have neglected non-linear terms in fermions for this calculation, I can without loss of generality drop the term that results from a variation of $\sigma$ (\ref{Weyl},\,\ref{psiDefi}). \\

I will conclude with some remarks:
\begin{itemize}
	\item The holomorphic supersymmetry parameter $\eps^C$ does not introduce additional supercharges to the theory. The sixteen real dimensions that are spanned by this complex eight-dimensional vector, are contained in the $32$-dimensional real space that is spanned by the supercharges of supergravity.
	\item Furthermore, I want to address the different representations of $SU(8)$ that have been used in this thesis: I started with a diffeomorphism in the $56$-dimensional subsector of the exceptional geometry. Due to the restricted form of the vielbein and the arbitrarily fixed vielbein gauge, this induces an $SU(8)/\Z_2\subset SO(56)$ action on the vielbein frame. Hence, the $SU(8)/\Z_2$ action on the bosons can be presented in terms of real $56\times 56$ matrices as used for the explicit form of the vielbein in section \ref{Matrix}. In particular, there is no necessity to introduce an imaginary unit $i$ for the bosons. I only introduced it in order to make the notation less cumbersome. I want to stress that there is no complex structure on the underlying manifold.\\
	
	The situation is different for the fermions. The endomorphism $\g_5\otimes\id_8$ of the $\mathbf{32}$-dimensional representation space of the Clifford algebra in eleven dimensions provides a complex structure. This allowed to use the chiral representation of the covering group $SU(8)$ to define an action on the fermions.
	\item The geometric description of the fermions for this subsector of exceptional geometry follows the same line of argumentation. A fermion $\chi$ is hence defined as a section of the tensor product of a real eight-dimensional vector bundle with the spin bundle in four dimensions that provides the complex structure. To obtain a full understanding of the geometric interpretation, it is essential to discuss the complete theory in the sixty-dimensional exceptional geometry.
	\item  I want to stress that it is not necessary to describe fermions as representations of $Spin(59,1)$ or $Spin(56)$ at any stage, because there are no diffeomorphisms in the exceptional geometry that would induce an action on the vielbein frame that is not in $SU(8)$. Since the chiral $SU(8)$ action on the fermions $\eps$ and $\chi$ includes an action on the spin bundle in four dimensions, it is in fact unlikely that there is a canonical relation of $\eps$ and $\chi$ to $Spin(56)$ representations, because these would not affect the four-dimensional spin bundle in general.
\end{itemize}

\chapter{Conclusion and outlook}\label{CONCL}
West's idea \cite{W00} to link the formalism of non-linear realization to supergravity in eleven dimensions proved useful for a construction of an exceptional geometry. I showed that this formalism provides a natural tool to describe the independent degrees of freedom of the vielbein in a restricted geometrical setting.\\

In this thesis, I focussed on a $56$-dimensional subsector of the sixty-dimen- sional exceptional geometry. For the comparison with $d=11$ supergravity, it was essential to discuss $\mathfrak{u}_8$ representations as physical fields. In this setting, the class of equivariant connections $\nabla$ in $d=56$ was essentially parametrized by two constants $c_1,c_2\in \R$ (\ref{sympConE}). I used $\nabla$ to define supersymmetry transformations in $56$ dimensions that link the vielbein to fermions. After a restriction to the seven common coordinates of the $56$-dimensional subsector of exceptional geometry and $d=11$ supergravity, these supersymmetry transformations of the fields in this subsector exactly coincide with the ones of $d=11$ supergravity, if the constants $c_1,c_2$ are fixed by (\ref{konstfix}).\\

Concerning the $56$-dimensional dynamics, the fermionic Lagrangian in this subsector is fixed by $SU(8)$-covariance modulo quartic terms in $\chi$ to
\beg
\mathcal{L}&=&\chi^{ABC}\nabla^{DE}\chi^{FGH}\e_{ABCDEFGH}+ \text{c.c.}.
\eeg
From a technical point of view, it is highly probable that the same constants (\ref{konstfix}) for $c_1,c_2$ in $\nabla$ will lead to the fermionic Lagrangian of $d=11$ supergravity in a truncation to the common seven coordinates and fields. For the dynamics of the bosons, the natural candidate would be the invariant scalar $\tilde{R}$ (\ref{RicciN}) that is constructed from this connection $\nabla$ by taking the $\mathfrak{so}_{56}$ trace of its curvature tensor.\\

The final step to show that all solutions of $d=11$ supergravity are in fact special solutions of a sixty-dimensional theory that does not contain additional fields, would consist in coupling the remaining four dimensions to the $56$-dimensional subsector in a consistent way. The remaining degrees of freedom of supergravity would have to be included in the sixty-dimensional exceptional vielbein, which may involve further dualisations.\\

Another interesting aspect is that the possible appearance of non-linear terms in fermions is highly restricted. Due to $SU(8)$-covariance, the following term is the unique choice for a holomorphic contorsion tensor that is quadratic in $\chi$
\beg
{\left(K_{C_1C_2}\right)_A}^B &=& \e_{C_1C_2AD_1\dots D_5}\chi^{D_1\dots D_3}\chi^{D_4D_5B}.
\eeg
Hence, the higher order terms in fermions provide a further consistency check, if the solutions of $d=11$ supergravity \cite{CJS78} are equivalent to the solutions of a supergravity in a sixty-dimensional exceptional geometry.\\

Since the constants $c_1,c_2$ of the $56$-dimensional subsector of the theory are uniquely fixed by a comparison to $d=11$ supergravity, it would further be interesting to perform a decomposition of $\mathfrak{e}_{7(7)}$ with respect to its $\mathfrak{gl}_6\otimes \mathfrak{sl}_2$ subalgebra and compare the dynamics and the supersymmetry variations to IIB supergravity as explained in section \ref{summ1}.\\

My final comment addresses the symmetry structure of an exceptional geometry. Following the argumentation of section \ref{Deco}, I want to emphasize that the procedure of a joint realization of conformal and exceptional symmetry does not work in the present context, because a symmetric tensor is preserved. This fact immediately implies that such a Borisov \& Ogievetsky like procedure can also be excluded to work for an $E_{8(8)}$ construction on a $(248+3)$-dimensional exceptional geometry with preserved symmetric octic tensor.\\

It was only after a restriction to the seven common coordinates of the $56$-dimensional subsector of exceptional geometry and $d=11$ supergravity that the remaining $Gl(7)$ symmetry was enhanced to $\Diff(7)$.\footnote{The symmetry transformation that corresponds to the nilpotent generators $\hat{E}$ (\ref{EDefi}) is also enhanced in this restriction to the gauge symmetry of the three-form potential. This procedure was first observed by West in \cite{W00}.} Since it is clear from section \ref{summ1} that the action of the remaining generators of $\mathfrak{e}_{7(7)}$ would violate this restriction, there is no joint realization of $E_{7(7)}$ and $\Diff(7)$ with an $E_{7(7)}$ valued vielbein matrix.

\section*{Acknowledgements}
I would like to thank Hermann Nicolai for supervising my thesis and for inspiring discussions. Furthermore, I am grateful to Thibault Damour, Marc Henneaux, Axel Kleinschmidt and Peter West for clarifying comments and valuable advice.\\

During my work on this dissertation, I benefitted from the support of the Studienstiftung des deutschen Volkes, of the Albert-Einstein-Institut, Potsdam, of the International Solvay Institute, Brussels and of the Universit\'e Libre de Bruxelles.

\chapter{Appendix}\label{CHAP6}
\section{To chapter \ref{CHAP2}}
\subsection{To the proof of theorem \ref{Satz1}}\label{Pf1}
This part shows that there is a one-to-one correspondence between $\Diff(d)$ and $\mathfrak{diff}_d$. Given any diffeomorphism $\vp_A\in \Diff(d)$, the equation (\ref{exp9})
\be\label{exp87}
\vp_A^\mu(x) &=& x^\mu + X_A^\mu(x) + \mathcal{O}(A^2)
\ee
uniquely prescribes a vector field $X_A\in \mathfrak{diff}_d$, because it is analytic due to the analyticity of $\vp_A$ and linear in $A$ by construction. This is the easy part. The more complicated one is to show how to construct a diffeomorphism $\vp_A$ from a given vector field $X_A$. As $X_A$ is linear in the multilabel $A$, I can extract a scalar label $t\in\R$, in the following way: For arbitrary, but fixed $t\in \R\backslash \{0\}$, it is possible to define a multilabel $A'$ such that
\be
A &=& t\cdot A'\nn\\
\Rightarrow\quad X_A &=& t X_{A'}.\label{tDefi}
\ee
Furthermore, as $\vp_A$ is analytically connected to the identity map, I can without loss of generality parametrize the slope by $t\in [0,1]$
\be\label{slope}
x^\mu(t)&:=&\vp^\mu_{tA'}(x)\\
\Rightarrow\quad x^\mu(t=0)&=&\vp^\mu_{A=0}(x)\,=\, x^\mu\nn.
\ee
With the definition (\ref{tDefi}), the equation (\ref{exp87}) has the form
\beg
x^\mu(t) &=& x^\mu + tX_{A'}^\mu(x) + \mathcal{O}(t^2)
\eeg
Since this expansion holds for every $x^\mu$ in the domain of validity of the diffeomorphism $\vp_A$, which is analytically continued to the identity map, it holds in particular for the slope $x^\mu(s)$ with $0\leq s\leq t$. This implies
\beg
x^\mu(t) &=& x^\mu(s) + (t-s)X_{A'}^\mu(x(s)) + \mathcal{O}((t-s)^2)
\eeg
and by differentiation for all $s\in [0, t]$
\begin{subequations}\label{Diff2}
\be\label{DiffEQ}
\left.\frac{\p x^\mu(w)}{\p w}\right|_{w=s} &=& X_{A'}^\mu(x(s)),\\
x^\mu(t=0)&=& x^\mu.\label{DiffEQ2}
\ee
\end{subequations}
This ordinary differential equation exactly is the definition of an integral curve to a vector field \cite{Wa83}. As an ordinary differential equation of first order with specified initial condition has a unique solution, the question of finding a unique diffeomorphism $\vp_A$ for a vector field $X_A\in \mathfrak{diff}_d$ is equivalent to solving (\ref{Diff2}). As the vector field is analytic, the solution $x^\mu(t)$ will allow a Taylor expansion about $t=0$.
\beg
x^\mu(t)&=& \sum\limits_{n=0}^\infty \frac{1}{n!} t^n\left.\left(\frac{\p}{\p s}\right)^n x^\mu(s)\right|_{s=0}\\
&=&
 x^\mu +\sum\limits_{n=1}^\infty \frac{1}{n!} t^n\left.\left(\frac{\p}{\p s}\right)^{n-1}\left(\left.\frac{\p x^\mu(w)}{\p w}\right|_{w=s}\right) \right|_{s=0}
 \\
&\stackrel{(\ref{DiffEQ})}{=}&
 x^\mu +\sum\limits_{n=1}^\infty \frac{1}{n!} t^n\left.\left(\frac{\p}{\p s}\right)^{n-1}X_{A'}^\mu(x(s)) \right|_{s=0}
\eeg
As the dependence of $X_{A'}$ on $s$ is only implicit via $x(s)$, one can replace all $s$-derivatives by $x$-derivatives with the help of the chain rule
\beg
\frac{\p}{\p s}X_{A'}^\mu(x(s)) &=& \frac{\p x^\nu(s)}{\p s} \left.\frac{\p}{\p x^\nu}X_{A'}^\mu\right|_{x=x(s)}
\\
&\stackrel{(\ref{DiffEQ})}{=}&\left.X_{A'}^\nu \p_\nu X_{A'}^\mu\right|_{x=x(s)}.
\eeg
Repeating this procedure yields for $n>0$
\beg
\left(\frac{\p}{\p s}\right)^{n-1} X_{A'}^\mu(x(s)) &=& \left(X_{A'}^\nu(x)\p_\nu\right)^{n-1}X_{A'}^\mu(x(s)),
\eeg
which leads to
\beg
x^\mu(t)&=& x^\mu + \sum\limits_{n=1}^\infty \frac{1}{n!}t^n\left.\left(X_{A'}^\nu(x)\p_\nu\right)^{n-1}X_{A'}^\mu\right|_{x=x(0)} \\
&\stackrel{(\ref{DiffEQ2})}{=}&\exp\left(t X_{A'}^\nu(x)\p_\nu\right)x^\mu \\
&\stackrel{(\ref{tDefi})}{=}&\exp\left( X_{A}^\nu(x)\p_\nu\right)x^\mu.
\eeg
With the identity (\ref{slope}), I finally obtain the formula (\ref{Diffeom2})
\beg
\vp_A^\mu(x) &=&\exp (X_A^\nu(x)\p_\nu) x^\mu.
\eeg
If I had started with an arbitrary $X_A\in \mathfrak{diff}_d$, the theorem of Picard-Lindel\"of implies that there exists $t>0$ such that the initial value problem (\ref{Diff2}) has a unique solution. This proves the one-to-one correspondence between $\Diff(d)$ and $\mathfrak{diff}_d$. It also explains the subtlety with the domain of validity, I mentioned at the end of section \ref{Diffeom}: given a vector field $X_A\in \mathfrak{diff}_d$, there is an open set $U\subset \R^d$ on which there is a unique solution and on which the Taylor expansion (\ref{Diffeom2}) is valid. This solution does not have to be globally well-defined, in general.\qed

\subsection{Isometries}\label{Isometry}
The constraint of preserved Minkowski metric $\vp_A^* \eta= \eta$ restricts $\Diff(d)$ to the finite dimensional subgroup of Poincar\'e transformations. To prove this, it is without loss of generality sufficient to consider the infinitesimal version of the isometry constraint (\ref{Isometry1}) with the expansion for a general diffeomorphism (\ref{exp9}):
\be\label{om5}
0&=& \p_\mu X_A^{\sigma_1}(x)\eta_{\sigma_1\nu} +\p_\nu X_A^{\sigma_2}(x)\eta_{\sigma_2\mu} +\mathcal{O}(A^2).
\ee
With the abbreviation $X_\nu :=X_A^{\sigma_1}(x)\eta_{\sigma_1\nu}$ this reads for constant $\eta_{\mu\nu}$
\beg
0&=& \p_\mu X_\nu +\p_\nu X_{\mu} +\mathcal{O}(A^2).
\eeg
Acting with a second differential operator yields
\beg
0&=& \p_\sigma\p_\mu X_\nu +\p_\sigma\p_\nu X_{\mu} +\mathcal{O}(A^2)\\
0&=& \p_\mu\p_\nu X_\sigma +\p_\mu\p_\sigma X_{\nu} +\mathcal{O}(A^2)\\
0&=& \p_\nu\p_\sigma X_\mu +\p_\nu\p_\mu X_{\sigma} +\mathcal{O}(A^2).
\eeg
Adding the first two equations and subtracting the last one provides
\beg
0&=& 2\p_\sigma\p_\mu X_\nu +\mathcal{O}(A^2).
\eeg
This implies that only polynomials of degree $n=0$ and $n=1$ may contribute to the series (\ref{poly6}), which restricts the degrees of freedom of the vector fields to the ones of the affine group. Imposing (\ref{om5}) on top restricts to the Poincar\'e group.

\subsection{The integral curve to the conformal vectorfield $X^{\mathfrak{c}}_a$}\label{V1}
In section \ref{Subalg}, I showed that the conformal algebra in $d$ dimensions is the same as the one of the orthogonal group in $d+2$ dimensions $\mathfrak{so}_{(d-p+1,p+1)}$, with the appropriate signatures. In this section, I will stick to Lorentzian signature $p=1$, but the argumentation is the same for any $p$. The formula (\ref{Diffeom2})
\beg
\vp_A^\mu(x) &=&\exp (X_A^\nu(x)\p_\nu) x^\mu
\eeg
defined for any vector field $X_A\in \mathfrak{diff}_d$ locally a diffeomorphism $\vp_A$. This is in particular the case for all the vector fields with generators $\hat{P}$, $\hat{K}$, $\hat{L}$ and $\hat{D}$ (\ref{PRep},\, \ref{KRep},\, \ref{LDefi},\, \ref{DDefi}). Therefore, an action of the conformal group on the $d$ coordinates $x^\mu$ is defined. Since the translation generators $\hat{P}$ are included in this group, the non-linear action on the vector space parametrized by $x^\mu$ is transitive.\footnote{As coordinates on a manifold parametrize the open set $\R^d\approx U\subset \cM^d$ (\ref{PDefi0}), this transitivity is only to be understood locally from a manifold point of view.}\newpage

On the other hand, the algebra $\mathfrak{so}_{(d-p+1,p+1)}$ has a natural action on a $(d+2)$-dimensional vector space, its vector representation $\mathbf{d+2}$. It is a linear action. Due to its orthogonality, the $d+2$-metric $\eta_{AB}$ is preserved. This allows to define a $(d+1)$-dimensional submanifold of the $(d+2)$-dimensional vector space, the so-called ``light cone''. Introducing $d+2$ coordinates 
\be\label{coordV6}
V^A=(V^\mu,V^{d+1},V^{d+2})^A,
\ee
this submanifold is parametrized by all $V$ fulfilling the following constraint\footnote{In other words, this submanifold, is the locus of the following polynomial.} (\ref{vConstr}):
\be
0&=& \eta_{AB} V^AV^B\nn\\
&=&\eta_{\mu\nu}V^\mu V^\nu +\left(V^{d+1}\right)^2-\left(V^{d+2}\right)^2\label{vConstr2}\\
&=&\eta_{\mu\nu}V^\mu V^\nu +\left(V^{d+1}-V^{d+2}\right)\left(V^{d+1}+V^{d+2}\right).
\nn
\ee
Starting from a $(d+2)$-dimensional vector space, the projective space also provides the definition of a $(d+1)$-dimensional submanifold. Due to the linear action of the orthogonal group on $\R^{d+2}$, the projective space is preserved under an action of the orthogonal group. From the definition of the light cone submanifold, it is obvious that the intersection of the two $(d+1)$-dimensional submanifolds is a $d$-dimensional one. Thus, there is a $d$-dimensional submanifold of $\R^{d+2}$ that is preserved under an action of the Lorentz group. What is more, the group action on this $d$-dimensional submanifold also is transitive.\footnote{For every $\alpha\in \R\backslash\{0\}$ and every vector $V$ fulfilling (\ref{vConstr2}), $\alpha V$ also fulfills the constraint (\ref{vConstr2}). Hence by definition, the projective equivalence class of any vector also fulfills (\ref{vConstr2}). On the other hand, assume that there were two equivalence classes $[V_1]\neq 0$ and $[V_2]\neq 0$ both fulfilling (\ref{vConstr2}) that are not linked by a group action. This would imply that both generate different orbits. Translating this statement to the algebra level, which is possible due to the linear action, this however contradicts the irreducibility of the orthogonal group. Hence, the action is indeed transitive.} \\

Hence, there are two transitive group actions on $d$-dimensional manifolds. This implies that there must be a diffeomorphism linking the two manifolds.\footnote{This can be proved by identifying one element on one manifold with one on the other. Due to the transitivity, this fixing induces an identification of all elements on the two manifolds, which allows to identify the manifolds.} Parametrizing the projective light cone by the $(d+2)$-dimensional coordinates $V$ constrained by (\ref{vConstr2}), by $\mathfrak{so}_{(d-1,1)}$-covariance this identification must be of the form
\be\label{Vers1}
x^\mu &=& \frac{V^\mu}{aV^{d+1} +bV^{d+2}},
\ee
because due to the identification (\ref{dplus2}) of the algebra generators, the $\mathfrak{so}_{(d-1,1)}$ generators do not act on the $d+1^{\text{st}}$ nor on the $d+2^{\text{nd}}$ coordinate. The constants $a,b\in \R$ are uniquely fixed by comparing the action of the translation generators $\hat{P}$ on both coordinates: On the one hand, (\ref{GLMatrix}) implies
\beg
\exp\left(c^\mu\mathbf{R}_{\hat{P}_\mu} \right){x}^\mu &=& x^\mu +c^\mu.
\eeg
On the other hand, the identification (\ref{dplus2}) defines the action of $c^\mu\hat{P}_\mu$ in the representation of linear vector fields (\ref{PRep}) acting on the $d+2$ coordinates $V$
\beg
c^\mu\mathbf{R}_{\hat{P}_\mu} 
&=&
  2c^\mu\left(\mathbf{R}_{{\left.\hat{L}\right.^{d+1}}_\mu}  -\mathbf{R}_{{\left.\hat{L}\right.^{d+2}}_\mu} \right)\\
&=& c^\mu\left(V^{d+1}\frac{\p}{\p V^\mu} -\eta_{\mu\nu}V^\nu\frac{\p}{\p V^{d+1}}  -V^{d+2}\frac{\p}{\p V^\mu} -\eta_{\mu\nu}V^\nu\frac{\p}{\p V^{d+2}} \right)
\eeg
where the minus sign in the last line is due to $\eta_{d+2,d+2}=-1$. Hence $a$ and $b$ are subject to the constraint with (\ref{Vers1})
\beg
\exp\left(c^\mu\mathbf{R}_{\hat{P}_\mu} \right)\frac{V^\mu}{aV^{d+1} +bV^{d+2}} &=& \frac{V^\mu}{aV^{d+1} +bV^{d+2}} +c^\mu
\eeg
which uniquely fixes $a=-b=1$ (\ref{vCoord}):
\be\label{vCoord3}
x^\mu &=& \frac{V^\mu}{V_{d+1}-V_{d+2}}.
\ee
Finally, this identification allows the evaluation of the conformal diffeomorphism in a closed form. Recall that the identification (\ref{dplus2}) also defines the action of $a^\mu\hat{K}_\mu$ in the representation of linear vector fields (\ref{PRep}) acting on the $d+2$ coordinates $V$:
\beg
a^\mu\mathbf{R}_{\hat{K}_\mu} 
&=&
 -2a^\mu\left(\mathbf{R}_{{\left.\hat{L}\right.^{d+1}}_\mu}  +\mathbf{R}_{{\left.\hat{L}\right.^{d+2}}_\mu} \right)\\
&=& -a^\mu\left(V^{d+1}\frac{\p}{\p V^\mu} -\eta_{\mu\nu}V^\nu\frac{\p}{\p V^{d+1}}  +V^{d+2}\frac{\p}{\p V^\mu} +\eta_{\mu\nu}V^\nu\frac{\p}{\p V^{d+2}} \right).
\eeg
The minus sign in the last line is again due to $\eta_{d+2,d+2}=-1$. A direct evaluation on the coordinates (\ref{vCoord3}) does not provide a great simplification. However, if the sign in the denominator was different, the evaluation would indeed be easy. Recalling that the constraint (\ref{vConstr2}) linked the sum of $V_{d+1}$ and $V_{d+2}$ to its difference, one is led to discuss
\be
\frac{x^\sigma}{\eta_{\mu\nu} x^\mu x^\nu}
&\stackrel{(\ref{vCoord3})}{=}&
\frac{ \frac{V^\sigma}{V^{d+1}-V^{d+2}}}{\eta_{\mu\nu}  \frac{V^\mu}{V^{d+1}-V^{d+2}}  \frac{V^\nu}{V^{d+1}-V^{d+2}}}
\nn\\
&=&\frac{V^\sigma \left(V^{d+1}-V^{d+2}\right)}{\eta_{\mu\nu}  V^\mu V^\nu}
\nn\\
&\stackrel{(\ref{vConstr2})}{=}& -\frac{V^\sigma}{V^{d+1}+V^{d+2}}.
\label{inv1}
\ee
This allows a direct evaluation with derivative operators acting on $V$:
\beg
\exp\left(\mathbf{R}_{a^\mu\hat{K}_\mu}\right)\left(\frac{x^\sigma}{\eta_{\mu\nu} x^\mu x^\nu}\right)
&=&
\exp\left(\mathbf{R}_{a^\mu\hat{K}_\mu}\right)\left(-\frac{V^{\sigma}}{V^{d+1}+V^{d+2}}\right)\\
&=&
-\frac{V^{\sigma}}{V^{d+1}+V^{d+2}} +a^\sigma\\
&=&
\frac{x^\sigma}{\eta_{\mu\nu} x^\mu x^\nu} +a^\sigma.
\eeg
Since the integral curve $\vp^{\mathfrak{c}}_{a}$ (\ref{KGen2},\,\ref{SCTcoord4}) has the form
\beg
\left(\vp^{\mathfrak{c}}_{a}\right)^\nu (x)&=&\exp\left(\mathbf{R}_{a^\mu\hat{K}_\mu}\right)x^\nu,
\eeg
I obtain the identity
\beg
\frac{\left(\vp^{\mathfrak{c}}_{a}\right)^\sigma}{ \vp^{\mathfrak{c}}_{a}\cdot \vp^{\mathfrak{c}}_{a}}&=&\frac{x^\sigma}{x\cdot x} +a^\sigma,
\eeg
where I have used the abbreviation introduced in equation (\ref{SCTcoord}). With the inversion (\ref{Invdefi})
\beg
\Inv(x)^\nu := \frac{x^\nu}{x\cdot x},
\eeg
this formula can be solved to (\ref{SCTcoord47})
\beg
\left(\vp^{\mathfrak{c}}_{a}\right)^\mu(x)
&=&
\left(\Inv\circ \vp^{\mathfrak{a}}_{(0,a)}\circ \Inv\right)^\mu (x),
\eeg
which has the explicit form (\ref{SCTcoord})
\beg
\left(\vp^{\mathfrak{c}}_{a}\right)^\mu(x)
&=&
\frac{x^\mu +a^\mu\,  x\cdot x }{1 +2 a\cdot x + a\cdot a\, x\cdot x}.
\eeg
An expansion in $a$ proves that its linear part indeed is the conformal Killing vector field $X^{\mathfrak{c}}_{a}$ (\ref{KGen}), what was expected from theorem \ref{Satz1}, of course.

\subsection{Global properties of the conformal group}\label{Global}
I have already mentioned in section \ref{Representation} that the same real Lie algebras do not imply that the corresponding Lie groups coincide. The same is true for the conformal group. In order to be a subgroup of $\Diff(d)$, every element of the orthogonal group $SO(d-p+1,p+1)$ must also uniquely correspond to a diffeomorphism $\vp_A$ that maps the open set (at least locally) to itself. This is not the case, because for even $d$, the two diagonal matrices 
\beg
(+\id_d,-\id_2),(-\id_d,+\id_2)&\in& SO(d-p+1,p+1)
\eeg
correspond to the same diffeomorphism $-\id_d$ on the $x^\mu$ coordinates by the relation (\ref{vCoord}).\footnote{For non-Euclidean signature, i.e. $p\neq 0$, they are even in the same connected component as $+\id_d$.} This implies that for even dimensions, the conformal subgroup of $\Diff(d)$ is isomorphic to $SO(d-p+1,p+1)/\Z_2$.\\

It is obvious from comparing (\ref{vCoord}) and (\ref{inv1}) that the inversion $\Inv$ (\ref{Invdefi}) corresponds to the pair of diagonal matrices $(\id_d,-1,1)$ and $(-\id_d,1,-1)$. For even dimension $d$, both matrices are of determinant $-1$ and therefore, they are not an element of $SO(d-p+1,p+1)$. As $\Inv$ has a pole, it is not an analytic map. Since $\Diff(d)$ consists of analytic maps by definition, $\Inv$ is not an element of $\Diff(d)$.\newpage

The $\Z_2$ argumentation is the same for odd spacetime dimension $d$, but in this case the inversion $\Inv$ is an element of $SO(d-p+1,p+1)/\Z_2$. Hence there is no strict isomorphism between the conformal subgroup of $\Diff(d)$ and $SO(d-p+1,p+1)/\Z_2$. This statement is independent of the signature $(d-p,p)$. There are elements like $\Inv$ in $SO(d-p+1,p+1)/\Z_2$ that cannot be embedded in $\Diff(d)$. More details on the conformal group can be found in \cite{D1}.

\subsection{Proof of Ogievetsky's theorem}\label{OgieP}
I will quote Ogievetsky's original proof here \cite{O73}. It is performed by induction over the degree $n$. The definition of the affine linear generators (\ref{PRep}) provides the starting point for $n=0$ and $n=1$\beg
A^\mu_{\mathbf{0}} \mathbf{R}_{\hat{P}_{\mathbf{0},\mu}(x)} &=& c^\mu {\mathbf{R}}_{\hat{P}_{\mu}} \,=\, c^\mu \p_\mu\\
A^\mu_{\mathbf{1}} \mathbf{R}_{\hat{P}_{\mathbf{1},\mu}(x)} &=& 
{A_\mu}^\nu {\mathbf{R}}_{{\left.\hat{M}\right.^\mu}_\nu} \,=\, {A_\mu}^\nu x^\mu\p_\nu
\eeg
with $x$ independent arbitrary parameters $c$ and $A$. It suffices to show that any vector field (\ref{P2Defi})
\beg
\mathbf{R}_{\hat{P}_{\mathbf{n},\mu}(x)}&=& \big(x^1\big)^{n_1}\cdots \big(x^d\big)^{n_d}\p_\mu
\eeg
with $n_1+n_2+\dots +n_d=n$ and $n_i,n\in \N_0$ arbitrary and $\mu=0,\dots,d-1$ arbitrary can be written as a linear combination of (commutators of) the generators $\hat{M}$ and $\hat{K}$ for $n\geq 2$, too. For every $\mu=0,\dots,d-1$, the conformal vector field $\hat{K}_\mu$ provides a particular vector field of degree $n=2$ (\ref{KRep}):
\beg
{\mathbf{R}}_{\hat{K}_\mu} &=& \eta_{\tau\nu}x^\tau\left( x^\nu  \delta_\mu^\sigma -2  \delta_\mu^\nu  x^\sigma\right)\p_\sigma. 
\eeg
These vector fields form an Abelian subalgebra (\ref{KK})
\beg
\left[{\mathbf{R}}_{\hat{K}_\mu},{\mathbf{R}}_{\hat{K}_\nu}\right] &=& 0.
\eeg
Hence, another generator of degree $n=2$ is necessary to construct vector fields of degree $n>2$. I have shown in section \ref{Subalg} that commutators of $\hat{K}$ (\ref{KRep}) with all the generators of the affine algebra $\mathfrak{a}_{d}$ apart from the symmetric generator 
\beg
{\left.\hat{S}\right.^\mu}_\nu &:=& \frac{1}{2}\left({\left.\hat{M}\right.^\mu}_\nu +\eta_{\nu\sigma}{\left.\hat{M}\right.^\sigma}_\tau \eta^{\tau\mu}\right)
\eeg
close to the conformal algebra $\mathfrak{so}_{(d-p+1,p+1)}$. Hence, it must be this generator that extends the conformal algebra to $\mathfrak{diff}_d$. Ogievetsky chose to consider the particular generator
\beg
{\mathbf{R}}_{{\left.\hat{S}\right.^m}_m} &=& x^m \p_m
\eeg
with \textbf{no summation} on the index $m$. For the rest of this proof, I introduce the following convention:
\begin{itemize}
	\item Greek indices $\mu,\nu,\dots$ and Latin ones $m,n,\dots$ have the same domain $0,\dots, d-1$.
	\item If a Greek index occurs twice, a summation is implicit. For a Latin index, this is not the case.
	\item I introduce the abbreviation $x_\mu:= \eta_{\mu\nu} x^\nu$, $x_m:= \eta_{m\nu} x^\nu$ and $\p^\mu:=\eta^{\mu\nu}\p_\nu$ et cetera. I want to emphasize again that this is a mere abbreviation. The $\eta$ (\ref{etaSkal}) in the conformal vector field $\hat{K}$ (\ref{KGen}) should not be iterpreted as a metric on a manifold but as a mere prescription how to construct a particular vector field of degree $n=2$.
		\end{itemize}
With these conventions, a short calculation with the standard rules for derivatives shows

\beg
\hat{H}_{(m,q)}&:=&\left[{\mathbf{R}}_{{\left.\hat{S}\right.^m}_m},{\mathbf{R}}_{\hat{K}_q}\right]\\
&=& \left[x^m \p_m,x_\nu\left( x^\nu  \delta_q^{\sigma} -2  \delta_q^{\nu}  x^\sigma\right)\p_\sigma\right]\\
&=& 2x^m x_m  \p_q -2 \delta_q^m x_q x^\sigma \p_\sigma -2x^m x_q \p_m\\
 &&-\delta_q^m x^\nu x_\nu\p_m +2x_q x^m \p_m\\
 &=& 2x^m x_m  \p_q -2\delta_q^m x_q x^\sigma \p_\sigma 
 -\delta_q^m x^\nu x_\nu\p_m .
\eeg
Hence, $\hat{H}_{(m,q)}$ is another vector field of degree $n=2$, but it is still not general. Next, evaluate another commutator, where there is no summation implicit for the indices again
\beg
\hat{W}_{(m,q)}
&:=&
\left[{\mathbf{R}}_{{\left.\hat{L}\right.^q}_m}
,\hat{H}_{(m,q)}\right]\\
&=&
\frac{1}{2}\left[x^q \p_m -x_m\p^q,2x^m x_m  \p_q -\delta_q^m \left(2x_q x^\sigma \p_\sigma 
 + x^\nu x_\nu\p_m \right)\right]\\
&=&
\left[x^q \p_m -x_m\p^q,x^m x_m  \p_q \right]\\
&=&
2x^q x_m  \p_q -\delta^q_m x_m x^m\p_q -x^mx_m\p_m\\
\Rightarrow\quad \sum\limits_{q=0}^{d-1}\hat{W}_{(m,q)}  &=& 2 x_m  x^\sigma\p_\sigma  -(d+1)x^mx_m\p_m.
\eeg
For every $m,q=0,\dots, d-1$, the vector fields $\mathbf{R}_{\hat{K}_q}$, $\hat{H}_{(m,q)}$ and $\hat{W}_{(m,q)}$ are contained in the closure of the two algebras. Hence, their linear combination is contained, too:
\beg
\hat{X}_{(m)}
&:=&-\frac{1}{d}\Big(\mathbf{R}_{\hat{K}_m}+\sum\limits_{q=0}^{d-1}\hat{W}_{(m,q)} 
-\frac{1}{2}\sum\limits_{q=0}^{d-1}\left(1-\delta^q_{ m}\right)\hat{H}_{(q,m)} \Big)\\
&=&
-\frac{1}{d}\Big(x_\nu\left( x^\nu  \p_m -2  \delta_m^{\nu}  x^\sigma\p_\sigma\right) 
+2 x_m  x^\sigma\p_\sigma  -(d+1)x^mx_m\p_m
\\
&&
-\sum\limits_{q=0}^{d-1}\left(1-\delta^q_{m}\right)x^q x_q\p_m \Big)\\
&=&
x^m x_m\p_m.
\eeg
To sum up, for $m\neq q$, the following vector fields are contained in the closure
\beg
\hat{H}_{(m,q)}&=&2x^m x_m  \p_q, \\
\hat{X}_{(m)}&=& x^m x_m\p_m,\\
\hat{W}_{(m,q)}+\hat{X}_{(m)}&=&2x^q x_m  \p_q
\eeg
Commuting the last one again with $q\neq n\neq m$ leads to
\beg
\hat{Z}_{(n,m,q)}
&:=&
\frac{1}{2}\left[{\mathbf{R}}_{{\left.\hat{M}\right.^n}_q}
,\hat{W}_{(m,q)}+\hat{X}_{(m)}\right]\\
&=&
\left[x^n\p_q,x^q x_m  \p_q\right]\\
&=&
x^n x_m \p_q.
\eeg
Since upper indices differ from lowered indices just by a minus sign, this implies that all vector fields $\mathbf{R}_{\hat{P}_{\mathbf{n},q}(x)}$ of degree $n=2$ also are in the closure:\\ for $q\neq n\neq m$ and $m\neq q$, $\hat{Z}$ contains them all, the case of exactly two equal indices is covered by either $\hat{H}$ or $\hat{W}+\hat{X}$ and the one with all three indices equal by $\hat{X}$.\\

Commuting two different vector fields $\mathbf{R}_{\hat{P}_{\mathbf{n},\mu}(x)}$ of degree $n=2$ will in general result in a vector field $\mathbf{R}_{\hat{P}_{\mathbf{n},\mu}(x)}$ of degree $n=3$. This is the step of induction: assume that the general vector field
\beg
\mathbf{R}_{\hat{P}_{\mathbf{n},\mu}(x)}&=& \left(x^1\right)^{n_1}\cdots \left(x^d\right)^{n_d}\p_\mu
\eeg
of degree $n=n_1+n_2+\dots +n_d$ is contained in the closure. Without loss of generality, assume $n\geq 2$. Then the vector fields
\beg
\mathbf{R}_{\hat{P}^{(m)}_{\mathbf{n+1},\mu}(x)} &:=&\left[\hat{X}_{(m)},\mathbf{R}_{\hat{P}_{\mathbf{n},\mu}(x)}\right]\\
&=& \left(n_m-2\delta_{\mu }^m\right)x_m\mathbf{R}_{\hat{P}_{\mathbf{n},\mu}(x)}
\eeg
for $m=0,\dots,d-1$ are of degree $n+1$ and they are contained in the closure. Unless the prefactor vanishes, this is the general form of a vector field of degree $n+1$, which would complete the proof by induction.\\

Given $\hat{P}_{\mathbf{n},\mu}(x)$, there are two possibilities to obtain $n_m-2\delta_{\mu }^m=0$:
\begin{enumerate}
	\item $n_m=0$ and $m\neq \mu$
	\item $n_m=2$ and $m=\mu$
\end{enumerate}
For the first case, recall that $n\geq 2$. Hence there is an index $k\neq m$ with $n_k\neq 0$ and $n_k\neq 1$ if $k=\mu$. Then, the commutator 
\beg
\left[x^m x^k\p_k,\mathbf{R}_{\hat{P}_{\mathbf{n},\mu}(x)}\right]
&=& \left(n_k-\delta^k_{\mu }\right)x^m\mathbf{R}_{\hat{P}_{\mathbf{n},\mu}(x)}
\eeg
maps to $x^m\mathbf{R}_{\hat{P}_{\mathbf{n},\mu}(x)}$ with a non-zero coefficient.\\

For the second case and for arbitrary $k\neq m=\mu$, the commutator 
\be\label{last}
\left[x^m x^k\p_m,\mathbf{R}_{\hat{P}_{\mathbf{n},m}(x)}\right]
&=&\left(n_m-1\right)x^k\mathbf{R}_{\hat{P}_{\mathbf{n},m}(x)}
\ee
does not vanish due to $n_m=2$. Thus, the last step of the proof consists of showing that for the case $n_m=2$ and $m=\mu$, a vector field with $n_m=3$ can be generated. This is shown adding a double commutator to the commutator of $\hat{M}$ with (\ref{last}) keeping in mind $k\neq m$ in (\ref{last})
\beg
&&\left[ x^m\p_k,x^k\mathbf{R}_{\hat{P}_{\mathbf{n},m}(x)}\right] -\frac{1}{4}\left[x^m x^m\p_k,\left[x^k\p_m,\mathbf{R}_{\hat{P}_{\mathbf{n},m}(x)}\right]\right]\\
&=& (n_k+1)x^m \mathbf{R}_{\hat{P}_{\mathbf{n},m}(x)} - x^k \mathbf{R}_{\hat{P}_{\mathbf{n},k}(x)}\\
&&-\frac{1}{4}\left(n_m(n_k+1)x^m\mathbf{R}_{\hat{P}_{\mathbf{n},m}(x)} -2n_m x^k\mathbf{R}_{\hat{P}_{\mathbf{n},k}(x)}\right)
\\
&\stackrel{n_m=2}{=}&\frac{1}{2}(n_k+1)x^m\mathbf{R}_{\hat{P}_{\mathbf{n},m}(x)}.
\eeg
This proves Ogievetsky's theorem.\qed

\subsection{Proof of corollary \ref{Coro2}}\label{Ogie6}
{
The proof of Ogievetsky's theorem \ref{Ogie2} consists of two parts: At first, I showed that a general vector field (\ref{P2Defi}) of polynomial degree $n=2$ is contained in the closure. Then, I used this to construct vector fields $\mathbf{R}_{\hat{P}_{\mathbf{n},\mu}}$ of arbitrary degree $n$ by induction. Since the theorem is valid for any dimension $d>1$, an application of the first part of the proof to a $w$-dimensional setting implies that all vector field of the form (\ref{P2Defi})
\be\label{P2Defi2}
\mathbf{R}_{\hat{P}_{\mathbf{n},\mu}(x)}&=& \big(x^1\big)^{n_1}\cdots \big(x^d\big)^{n_d}\p_\mu
\ee
with $n_1+\dots +n_w=2$ and $q=0,\dots, w-1$ are contained in the closure. Their commutators with the general linear generators $\hat{M}$ (\ref{PRep}) in $d$ dimensions also are in the closure. Using Latin indices for $x$ and $\p$ for the range $0,\dots, w-1$ and Greek ones for the rest $w,\dots, d-1$, the multiple commutators of 
\beg
\mathbf{R}_{{\left.\hat{M}\right.^\mu}_m} &=& x^\mu \p_m\\
\mathbf{R}_{{\left.\hat{M}\right.^n}_\nu} &=& x^n \p_\nu
\eeg
with the vector field (\ref{P2Defi2}) generates all vector fields of degree $n=2$ in $d$ dimensions:
\beg
\left[x^\mu \p_m,x^m x^p\p_q\right]&=& (1+\delta_{m}^p) x^\mu x^p\p_q\\
\left[x^q \p_\mu,x^m x^p\p_q\right]&=& - x^m x^p\p_\mu\\
\left[x^\nu \p_p,x^\mu x^p\p_q\right]&=&  x^\nu x^\mu \p_q\\
\left[x^q \p_\tau,x^\mu x^p\p_q\right] -\delta_\tau^\mu x^q x^p\p_q&=&  -x^\mu x^p\p_\tau\\
\left[x^q \p_\tau,x^\nu x^\mu\p_q\right] -2\delta_\tau^{(\nu}x^{\mu)} x^q \p_q&=&  -x^\nu x^\mu \p_\tau.
\eeg
Hence, the second part of the proof of Ogievetsky's theorem is applicable.\qed\\
Note added: The restriction $w>1$ only is necessary for this proof to work. The corollary \ref{Coro2} remains true for $w=1$ and $d>1$.
}

\section{To chapter \ref{CHAP3}}
\subsection{Clifford algebras}\label{Cliff}
For a diagonal, non-degenerate metric $\eta$ of signature $(d-p,p)$, a Clifford algebra $\mathcal{A}$ is a vector space endowed with a multiplication. It is spanned by the free algebra generated by the objects $\G_a$ with $a=1,\dots,d$ modulo the Clifford property
\be\label{Clifford}
	\left\{\G_a,\G_b\right\} &=& 2\eta_{ab}\id\\
	\text{with}\quad \left\{\G_a,\G_b\right\}&:=&\G_a\G_b + \G_b\G_a.
\ee
This implies in particular that the antisymmetric products of matrices $\G_a$ are linearly independent of the generators $\G_a$. It is conventional to introduce the following abbreviation for these objects for $m=1,\dots,d$
\be\label{AntisymG}
\G_{a_1\dots a_n}&:=&\G_{[a_1}\cdots \G_{ a_m]}.
\ee
As usual, the antisymmetrization is normalized by
\be\label{Strenght1}
\G_{[a_1}\cdots \G_{ a_m]}&=&\frac{1}{m!}\G_{a_1}\cdots \G_{ a_m} +\text{permutations}.
\ee
Due to the non-degeneracy of the metric $\eta$, the position of the indices can be adjusted as explained in section \ref{Gstructur}.\\

The definition of representations of a Lie algebra on vector spaces $V$ of dimension $n$ from section \ref{Representation0} can also be transferred to the Clifford algebra. Hence the generators $\G_a$ are presented as $n\times n$ matrices. The lowest possible $n$ depends on the dimension $d$ and the signature $(d-p,p)$ of $\eta$.\footnote{The review \cite{dW02} contains more information on this topic as well as further references.} For the eleven-dimensional flat Minkowski metric 
\beg
\eta=\diag(-1,+\id_{10}),
\eeg
the lowest dimension $n$ of the representation space $V$ is $32$. Furthermore, it is possible to choose the matrix representations of the Clifford generators $\hat{\G}^P$ with $P=0,\dots,10$ to only consist of real $32\times 32$ matrices, which is referred to as the Majorana property. For my thesis, I adopt this choice with the sign convention
\begin{subequations}\label{ConvG11}
\be
\tilde{\G}^{P_0\dots P_{10}} &=& \e^{P_0\dots P_{10}}\id_{32}\\
\text{with}\quad \e^{0\,1\,2\,3\,4\,5\,6\,7\,8\,9\,10}&:=&1.
\ee
\end{subequations}

The Clifford algebra of major importance for this thesis corresponds to the Euclidean metric $\eta$ in seven dimensions. I use the pseudo-Majorana representation that implies that all generators $\G_a$ with $a=1,\dots,7$ are presented by purely imaginary $8\times 8$ matrices. These are normalized by 
\begin{subequations}\label{ConvG7}
\be
\G^{a_1\dots a_7} &=& -i\e^{a_1\dots a_{7}}\id_{8}\\
\text{with}\quad \e^{1\,2\,3\,4\,5\,6\,7}&:=&1.
\ee
\end{subequations}
An explicit form of the real matrices $\frac{1}{i}\G^a$ is provided in appendix C of \cite{CJ79}.\newpage

Finally, I use the Majorana representation of the Clifford algebra associated to the Minkowski metric of signature $(-1,1,1,1)$. These real $4\times 4$ matrices will be denoted by $\g_\alpha$ with $\alpha=0,\dots,3$. Following the decomposition of the isometry group $SO(10,1)$ of the Minkowski metric in eleven dimensions into the product $SO(3,1)\times SO(7)$ of the isometry groups of four and seven dimensions, the real $32\times 32$ matrices $\tilde{\G}$ allow for a decomposition into the matrices $\g$ and $\G$ that is provided by the identification 
\begin{subequations}\label{gammadec2}
\be
\tilde{\G}_\alpha &=& \g_\alpha\otimes \id_{8}\quad \,\text{for }\alpha\,=\,0,\dots,3
\\
\tilde{\G}_g &=& \frac{\g_5}{i}\otimes \G_g\quad \text{for }g\,=\,4,\dots,10
\label{G5Defi2}
\ee
\end{subequations}
with the obvious change in the labeling of the seven matrices $\G_g$. The consistency of the normalizations (\ref{ConvG11}) and (\ref{ConvG7}) with the decomposition (\ref{gammadec2}) implies the normalization
\be\label{gamma5}
\g_5&:=&\g^0\g^1\g^2\g^3 \quad \text{for } \e^{0\,1\,2\,3}\,=\,1.
\ee
This leads to the identities
\begin{subequations}\label{gamma6}
\be
\g_5\e^{\alpha_1\dots \alpha_4} &=&\g^{\alpha_1\dots \alpha_4},\\
\g_5^2&=& -\id_4.
\ee
\end{subequations}
The square of $\g_5$ is evaluated with the help of the Clifford property (\ref{Clifford}) keeping in mind the Lorentzian signature of $\eta$. I want to emphasize that I had to introduce the imaginary unit $i$ in the line (\ref{G5Defi2}), because the matrices $\G_a$ are purely imaginary, whereas $\g_5$ and $\tilde{\G}$ are real matrices.\\

Of further importance for this thesis is the fact that the Clifford algebra provides a representation $\mathbf{R}$ of the orthogonal Lie algebra $\mathfrak{so}$ for the corresponding signature. In equation (\ref{RS}), I used this fact for the four-dimensional Minkowski space:
\beg
\mathbf{R}_{{\left.\hat{L}\right.^e}_f} &=& \frac{1}{4}{\g^e}_f.
\eeg
It is obvious that the matrix representation of the orthogonal algebra $\mathfrak{so}$ consists of real matrices for both Majorana and pseudo-Majorana representations of the Clifford algebra.\\

To conclude, I want to stress that I distinguished the generators of the Clifford algebr\ae{} for the different metrics under consideration by the different names $\tilde{\G}$, $\G$ and $\g$. It is furthermore conventional in supergravity to introduce ``dressed'' $\tilde{\G}_M$ matrices that correspond to the standard Clifford matrix representations $\tilde{\G}_A$ by a multiplication with the vielbein
\beg
\tilde{\G}_M &:={E_M}^A\tilde{\G}_A.
\eeg
This convention to distinguish the Clifford generators in the vielbein frame from the ones in the coordinate induced frame merely by a different naming of the indices is only used once in equation (\ref{Trafo1}). For the Clifford generators $\G$ in $d=7$, I will provide additional identities in appendix \ref{SomeG2}.

\subsection{Explicit calculation of $[\nabla_a,\nabla_b]$}\label{Expl1}
In order to construct an invariant Lagrangian, the commutator of the connection $\nabla$ (\ref{conn45}) with itself is the key ingredient as in the case of an internal symmetry. The difference in the evaluation for an external symmetry is that the algebra action of the second covariant derivative also acts on the coordinate index of the first one. Therefore, I have included the detailed computation of $[\nabla_a,\nabla_b]$ to illustrate the procedure of section \ref{prim2}. I want to start with an arbitrary physical field or section of a tensor bundle $\psi$ that I decompose along the generators $\hat{T}_\gamma$ of the corresponding representation space $V$ (\ref{psiParam}):
\beg
\psi = \psi^\gamma \hat{T}_\gamma.
\eeg
The connection $\nabla$ is defined on the coefficients $\psi^\alpha$ by (\ref{conn45})
\be\nn
\nabla_d \psi^\gamma&=& \p_d \psi^\gamma
- 
{{\omega_d}_b}^c\delta_{{\left.\hat{L}\right.^b}_c}\psi^\gamma\\
&=:&
\p_d \psi^\gamma
- 
{{\omega_d}_\beta}^\gamma\psi^\beta.\label{conn48}
\ee
Only in this computation, I will use the abbreviation defined in the second line. The notation should not be confused with the Lie group multiplication (\ref{ODefi}), which will not be used during this calculation.
The second covariant derivative then acts on the section
\beg
\xi &=& \nabla_d\psi^\gamma  \hat{T}_\gamma \otimes \hat{P}^d
\eeg
in the product bundle with the basis vectors $\hat{P}^d$ parametrizing the antifundamental representation $\overline{\mathbf{d}}$. The Lorentz action on the coefficient $\nabla_d\psi^\gamma$ follows the definition (\ref{SymmAct6})

\beg
\left(\delta_{{\left.\hat{L}\right.^e}_f} \nabla_d\psi^\gamma\right)\hat{T}_\gamma \otimes \hat{P}^d
&:=&
 -\nabla_d\psi^\gamma \mathbf{R}_{{\left.\hat{L}\right.^e}_f}\left(\hat{T}_\gamma \otimes \hat{P}^d\right)\\
&=&
-\nabla_d\psi^\gamma \mathbf{R}_{{\left.\hat{L}\right.^e}_f}\left(\hat{T}_\gamma \right)\otimes \hat{P}^d
-\hat{T}_\gamma \otimes \nabla_d\psi^\gamma \delta_f^d\hat{P}^e,
\eeg
where the last relation follows from duality and (\ref{ComRel2},\, \ref{GlAction}).\footnote{The group action on the contraction of fundamental and antifundamental representation $\hat{P}_a\hat{P}^a$ is trivial by definition. Passing to the linearized part defines the algebra action on the dual vector: $0= \mathbf{R}_{{\left.\hat{L}\right.^e}_f} (\hat{P}_d\hat{P}^d )
=(-\delta^e_d\hat{P}_f)\hat{P}^d +\hat{P}_d \mathbf{R}_{{\left.\hat{L}\right.^e}_f} (\hat{P}^d )$.\\ This implies $ \mathbf{R}_{{\left.\hat{L}\right.^e}_f}(\hat{P}^d)= +\delta^d_f\hat{P}^e$.}
 Comparing the basis vectors results in

\beg
\nabla_a\left(\nabla_d\psi^\gamma\right) 
&=&
\p_a\left(\nabla_d\psi^\gamma\right) - 
{{\omega_a}_e}^f\delta_{{\left.\hat{L}\right.^e}_f}\left(\nabla_d\psi^\gamma\right) \\
&=&
 \p_a \left(\nabla_d\psi^\gamma\right)
 -{\omega_{a\beta}}^\gamma\left(\nabla_d\psi^\beta\right)
 +{\omega_{ad}}^f\left(\nabla_f\psi^\gamma\right)\\
 &=&
  \p_a \left(\p_d \psi^\gamma -{\omega_{d\beta}}^\gamma \psi^\beta\right)
 -{\omega_{a\beta}}^\gamma\left(\p_d \psi^\beta -{\omega_{d\alpha}}^\beta \psi^\alpha\right)
 +{\omega_{ad}}^f \nabla_f\psi^\gamma.
\eeg
Hence, the commutator of two covariant derivatives on $\psi^\alpha$ has the form 
\beg
\left[\nabla_a,\nabla_d\right]\psi^\gamma 
  &=&
   \left[\p_a ,\p_d\right]^f \p_f\psi^\gamma -2\left(\p_{[a}{\omega_{d]\beta}}^\gamma\right) \psi^\beta\\
 &&
 +2{\omega_{[a|\beta|}}^\gamma{\omega_{d]\alpha}}^\beta \psi^\alpha
 +2{\omega_{[ad]}}^f \nabla_f\psi^\gamma.
    \eeg
   The first term is the commutator (\ref{Lie8}) of the vector fields $\p_a$ (\ref{ehneu}). The third term can be transformed in a more conventional form with the help of the commutation relation (\ref{ComRel1}) and of (\ref{SymmAct6},\,\ref{conn48}):
\beg
2{\omega_{[a|\beta|}}^\gamma{\omega_{d]\alpha}}^\beta \hat{T}_\gamma
&=&
-2{\omega_{[d|\alpha|}}^\beta{\omega_{a]e}}^f \mathbf{R}_{{\left.\hat{L}\right.^e}_f}\left(\hat{T}_\beta\right)\\
&=&
+2{\omega_{[d|g|}}^h{\omega_{a]e}}^f \mathbf{R}_{{\left.\hat{L}\right.^e}_f}\left(\mathbf{R}_{{\left.\hat{L}\right.^g}_h}\left(\hat{T}_\alpha\right)\right)\\
&\stackrel{(\ref{ComRel1})}{=}&
{\omega_{dg}}^h{\omega_{ae}}^f\left(\delta_f^g \mathbf{R}_{{\left.\hat{L}\right.^e}_h} -\delta^e_h\mathbf{R}_{{\left.\hat{L}\right.^g}_f}\right) \left(\hat{T}_\alpha\right)\\
&=&
2{\omega_{[a|e|}}^g{\omega_{d]g}}^f \mathbf{R}_{{\left.\hat{L}\right.^e}_f}  \left(\hat{T}_\alpha\right).
\eeg
Substituting the original expression for the abbreviation (\ref{conn48}) in all terms, I obtain the abstract equation acting on any physical field $\psi^\gamma$:
\beg
\left[\nabla_a,\nabla_d\right]
 &=&
  \left[\p_a ,\p_d\right]^f\p_f +2{\omega_{[ad]}}^f \nabla_f  \\
  &&
   -\left(2\p_{[a}{\omega_{d]e}}^f
 +2{\omega_{[a|e|}}^g{\omega_{d]g}}^f
  \right) \delta_{{\left.\hat{L}\right.^e}_f}
   \eeg
I defined $\psi^\gamma$ to transform as a Lorentz tensor under a local Poincar\'e action. The derivative $\p_f$ is not equivariant, however.  Hence, I have to reshuffle the terms to get the covariant expression (\ref{curv3}):
\be\label{curv3b}
\left[\nabla_a,\nabla_d\right]
&=&
\left(\left[ \p_a ,  \p_d \right]^f +2{\omega_{[ad]}}^f \right)\nabla_f\\
&&
-\left(2\p_{[a}  {\omega_{d]e}}^f -\left[ \p_a ,  \p_d \right]^c {\omega_{ce}}^f +2{\omega_{[a|e|}}^g {\omega_{d]g}}^f\right)\delta_{{\left.\hat{L}\right.^e}_f}. \nn
\ee

\section{To chapter \ref{CHAP4}}
\subsection{Proof of theorem \ref{prop1}}\label{prop1beweis}

{
I want to start the proof with the observation that in a local theory, a tensor can only be constructed from contracting other tensors and covariant derivatives thereof.\footnote{In particular, integrations are not admitted in a local theory.} Since the minimal connection $\nabla^{\text{min}}$ (\ref{minimalCon}) provides a covariant derivative with affine linear equivariance, I can without loss of generality focus on tensors $X(e)$ that are not of derivative form.\\

To discuss the vielbein dependence of $X(e)$, it is important to recall from equation (\ref{eTrafo2}) that the vielbein matrix $e$ does not transform as a Lorentz tensor under an affine linear diffeomorphism $\vp^{\mathfrak{a}}_{(A,c)}$ on its own:
\beg
	e'(x') &= & \left(e^{-A}\right) \cdot  e(x) \cdot U(\vp^{\mathfrak{a}}_{(A,c)},e(x)).
\eeg
Next, recall from section \ref{ART} that the fixing of the matrix representative for the vielbein in the orbit $[e]$ was arbitrary. Hence, the tensor $X$ must be independent of the fixing of the vielbein gauge, i.e. independent of the choice for the matrix form of the vielbein $e$. Therefore, the global left action decouples from the local right action. This implies that Lorentz covariance requires invariance of $X$ under a global left action by $e^{-A}$. \\

The latter is tantamount to an arbitrary $Gl(d)$ left action on an essentially arbitrary $Gl(d)$ matrix $e(x)$. Hence, invariance of $X(e)$ can only be achieved if the entire dependence of $X$ on the matrix $e$ is of the form
\beg
X(e) &=& X\left(e^{-1}(x)\cdot R(x)\cdot e(x)\right)\\
\text{with}\quad \left[R(x), e^{A}\right] &=& 0 \quad \forall\, A\in \mathfrak{gl}_d.
\eeg
The constraint on the $\mathfrak{gl}_d$ valued operation $R(x)$ in the last line follows from an evaluation of the formula (\ref{formula2}). Therefore, $R(x)$ has to be commute with all invertible matrices. Hence, it is proportional to the identity matrix. Since $e^{A}$ does not depend on $x$, the action of $R(x)$ on the variable $x$ is arbitrary. Due to the locality of the theory however, the only admissible action on the coordinates is a (multiple) derivative.\\

Furthermore, recall that global left actions also affect the coordinates and hence the derivative operators in the coordinate frame $\p_\mu$. This is the reason to use the vielbein frame $\p_a={e_a}^\mu\p_\mu$ (\ref{ehneu}) for derivative operators.\\

In order to achieve invariance under global left actions, $X$ may hence only depend on arbitrary powers of derivatives $\p_a$ in the vielbein frame acting on the vielbein $e$ in the following way:
\be\label{depend}
X(e,\p) &=& X\left( \bigcup\limits_{n=0}^\infty e^{-1}\p_{a_1}\dots \p_{a_n}e, \p_a\right).
\ee
The dependence on $\p_a$ is a symbolic notation for the fact that $X$ may only depend on derivatives of Lorentz tensors by the derivative $\p_a$ in the vielbein frame (\ref{ehneu}).\\

The right action on the vielbein matrix $e$ is $x$-dependent in general. This implies that requiring $X$ to be a Lorentz tensor restricts the dependence further. I start by discussing dependences on first order derivatives acting on the vielbein, i.e. only the $n=1$ terms in (\ref{depend}). Under the affine linear diffeomorphism $\vp^{\mathfrak{a}}_{(A,c)}$, this matrix transforms as induced by (\ref{eTrafo2}):
\beg
e^{-1}\p_{a}e &\mapsto & O^{-1}\cdot\left({e'}^{-1}\p'_{a}e'\right)\cdot O + O^{-1}\p'_a O.
\eeg
Since the compensating Lorentz action $O(x)\in SO(d-1,1)$ is $x$-dependent, $O^{-1}\p'_a O$ provides a non-vanishing contribution that is $\mathfrak{so}_{(d-1,1)}$ valued. In other words, it is antisymmetric with respect to the Minkowski metric as defined in (\ref{AntisyM}). Thus, the symmetric part of this matrix transforms as a tensor. This exactly is the object $v$ defined in (\ref{omega2}). The remaining part is $v^{\text{min}}$ (\ref{omega3}), which is antisymmetric or $\mathfrak{so}_{(d-1,1)}$ valued and transforms as a Lorentz connection.\\

Finally, for the left invariant derivative operator $\p_a$ acting on arbitrary Lorentz tensors to be equivariant under $\vp^{\mathfrak{a}}_{(A,c)}$, I have to replace it by the minimal connection $\nabla^{\text{min}}$ (\ref{minimalCon}).\footnote{The vielbein $e$ is not a Lorentz tensor (\ref{eTrafo2}). Therefore, I must not replace the partial derivative in $v$ (\ref{omega2}) by this connection, of course.} \\

A comparison with the dependence \ref{prop12} in the theorem \ref{prop1} reveals that completing the proof is equivalent to showing that it is impossible to construct any Lorentz tensor under $\vp^{\mathfrak{a}}_{(A,c)}$ from the objects with more than one derivatives acting on $e$, i.e. $n>1$ alone. Nonetheless, objects like this will show up in an evaluation of the minimal connetion $\nabla^{\text{min}}$ on the tensor $v$ (\ref{omega2}), but then they are accompanied by other terms. In analogy to the $n=1$ case, I ask if there is a choice in arranging the indices such that
\beg
O^{-1}\p'_{a_1}\dots \p'_{a_n} O
\eeg
vanishes. For the matrix representation of $O(x)\in SO(d-1,1)$, define the corresponding matrix representation of the Lie algebra element by $c\Lambda(x)\in \mathfrak{so}_{(d-1,1)}$ such that $O(x) = e^{c\Lambda(x)}$ with $c\in \R$ arbitrary. I can without loss of generality assume that $O(x)$ has general form, i.e. there is no restriction on the $x$-dependence of $O$, because the tensorial property of $X$ must hold for a general vielbein in any gauge.\footnote{nota bene: for a fixed vielbein, it is however not true that every $O(x)$ is induced by a diffeomorphism $\Diff(d)$ due to the integrability condition.} Hence, there are no restrictions on $\Lambda(x)$ either. This implies that the derivative indices $a_1,\dots, a_n$ cannot help in making the object transform as a tensor: this must be achieved by restricting the matrix form alone. Similarly to the case $n=1$, $\Lambda$ must be antisymmetric for the linear order in $c$ to vanish. However, the non-linear orders in $c$ give a non-trivial constraint in this case:
\be\label{Up}
O^{-1}\p'_{a_1}\dots \p'_{a_n} O &=& \p'_{a_1}\left(O^{-1}\p'_{a_2}\dots \p'_{a_n} O\right)\nn\\
&& +O^{-1}\p'_{a_1} O \cdot O^{-1}\p'_{a_2}\dots \p'_{a_n} O.
\ee
Taking the symmetric part for $n=2$ results in
\beg
\left(O^{-1}\p'_{(a_1} \p'_{a_2)} O\right)_{(cd)} &=& \left(O^{-1}\p'_{(a_1} O\right)_{[cf]}\eta^{fg} \left(O^{-1}\p'_{a_2)}O\right)_{[gd]}
\eeg
which does not vanish.\footnote{The antisymmetric part in $a_1a_2$ does not vanish either. For simpilicity, I have only stated the symmetric part.} As the relation (\ref{Up}) links a term with $n$ ``internal'' derivatives to terms with $n-1$ ``internal'' derivatives, this statement generalizes for all $n>1$. Hence, it is impossible to generate tensors from these terms alone and the only tensor is $v$ (\ref{omega2}) and derivatives $\nabla^{\text{min}}$ thereof, which completes the proof.\qed\\

In section \ref{gravmatter2}, I have introduced the torsion and the curvature tensors of a connection. For the case of the minimal connection $\nabla^{\text{min}}$, the theorem \ref{prop1} restricts these objects to be constructed from the symmetric tensor $v$ (\ref{omega2}). A short calculation shows that this indeed is the case:
\beg
\left[\nabla^{\text{min}}_a,\nabla^{\text{min}}_d\right]
&=&
-2{\left(v_{[a}\right)_{d]}}^f\nabla^{\text{min}}_f
+2{\left( v_{[a}\right)_{e}}^c {\left( v_{d]}\right)_{c}}^f
\delta_{{\left.\hat{L}\right.^e}_f}. \nn
\eeg
}

\subsection{Proof of (\ref{Ident56})}\label{PrIdent56}
{
In this section, I will prove that the Lorentz rotation $O(\vp^{\mathfrak{c}}_a,e)$ (\ref{zu3},\,\ref{SpinRot}), induced by the conformal diffeomorphism $\vp^{\mathfrak{c}}_a$, and the one $O(a,\bar{C})$ (\ref{conftrafo}) compensating a global left action by $e^{a^\mu\hat{K}_\mu}$ coincide.\\

To start, recall from section (\ref{suff}) that after fixing the Lorentz gauge in such a way that the unimodular vielbein $\bar{e}$ (\ref{hbarD}) is a Lorentz representation, the induced Lorentz transformation $O(\vp^{\mathfrak{c}}_a,e)$ was uniquely determined by the space-time coordinate $x^\mu$ and the constant parameter $a^\nu$ alone (\ref{SpinRot}). The Lorentz group element $O(\vp^{\mathfrak{c}}_a,e)$ corresponds via the homomorphism $\exp$ (\ref{exp}) to a Lie algebra element $\Lambda \in \mathfrak{so}_{(d-1,1)}$, whose generators are antisymmetric with respect to the Minkowski metric $\eta$ (\ref{AntisyM}). However, there is a unique way modulo a scalar function how an antisymmetric tensor can be built from two vectors. And this scalar function is uniquely determined by the following argument:\\

The concatenation or group multiplication of two consecutive conformal diffeomorphisms $\vp^{\mathfrak{c}}_a,\,\vp^{\mathfrak{c}}_b\in \Diff(d)$ implies for the induced Lorentz actions (\ref{VielbeinTrafo}) with $x'=\vp^{\mathfrak{c}}_b(x)$ and $x''=\vp^{\mathfrak{c}}_a(x')$:
\beg
O\left(\vp^{\mathfrak{c}}_a\circ \vp^{\mathfrak{c}}_b,e''\right) &=& O\left( \vp^{\mathfrak{c}}_b,e\right) \cdot O\left(\vp^{\mathfrak{c}}_a,e'\right).
\eeg
Since $K(d-1,1)$ is an Abelian subgroup (\ref{abelgroup}) of $\Diff(d)$, this equation takes the form with $O=e^\Lambda$ and $\Lambda\in \mathfrak{so}_{(d-1,1)}$
\be\label{UTrafo9}
e^{\Lambda(a+b,{x''})}&=& e^{\Lambda(b,{x})} \cdot e^{\Lambda(a,{x'})}.
\ee
As sketched before, the matrix $\Lambda(b,x)$ is antisymmetric and it also is an analytic function of the two vectors $b^\mu$ and $x^\nu$ alone. Therefore, it must have the form
\beg
\Lambda(b,{x}) &=& f(x\cdot x,\,b\cdot b,\, b\cdot x)\eta_{\mu\tau}b^\tau x^\nu {\left.\hat{L}\right.^\mu}_\nu
\eeg
with the antisymmetric Lorentz generators $\hat{L}$ (\ref{LDefi}). Due to Lorentz covariance, the scalar function $f$ must analytically depend on the three scalar parameters $x\cdot x$, $b\cdot b$  and $b\cdot x$ with the abbreviation $b\cdot x$ for the contraction with the Minkowski metric $\eta$ defined in (\ref{SCTcoord}). Finally, an iterative evaluation of (\ref{UTrafo9}) with the Baker-Campbell-Hausdorff formula \cite{J79} uniquely fixes the dependence of $f$ on these three scalar parameters as a power series.\\

Hence, the transformation (\ref{UTrafo9}) uniquely fixes the Lorentz transformation $O\left( \vp^{\mathfrak{c}}_a,e\right)$ that is induced by the conformal diffeomorphism $\vp^{\mathfrak{c}}_a$  and the vielbein $e$ in the symmetric gauge. Thus, it is is sufficient to show that the same relation (\ref{UTrafo9}) also holds for the Lorentz transformations that compensate global left actions on the conformal coset $\bar{C}$ (\ref{gaugefC}).\\

A priori, the compensating Lorentz rotation $O(a,\bar{C})$ (\ref{conftrafo}) for a global left action by $K(d-1,1)$ on the coset $\bar{C}$ may depend on all coset parameters $x,\sigma,\Phi$. The commutation relations (\ref{CFT3}) however show that in evaluating (\ref{conftrafo}), no Lorentz algebra generator $\hat{L}$ can be produced by the generators associated to the parameters $\sigma$ and $\Phi$. Hence, the compensating Lorentz group element can only depend on the two vector valued parameters $a$ and $x$.\\

Now, it is easy to conclude that the two Lorentz group objects must coincide: two consecutive global left actions of $K(d-1,1)$ with parameters $a$ and $b$ on the conformal coset (\ref{gaugefC}) induce two compensating Lorentz rotations for the coset. Since $K(d-1,1)$ is an Abelian group (\ref{abelgroup},\,\ref{KK}), their product must be the compensating Lorentz rotation of a left action with $a+b$. This implies that the condition (\ref{UTrafo9}) also is valid for the compensating Lorentz rotations of the coset. As this fixes the local Lorentz rotation uniquely, the Lie group elements must be equal
\beg
	O(\vp^{\mathfrak{c}}_a,e) &=&O(a,\bar{C}),
\eeg
if the vielbein matrix is symmetric. If another choice for the vielbein gauge was fixed, the compensating Lorentz rotation $O(\vp^{\mathfrak{c}}_a,e)$ would also depend on the unimodular vielbein $\bar{e}$ explicitly. Since $\bar{e}$ is not encoded in the conformal coset $\bar{C}$, the equality (\ref{Ident56}) does not hold for any other gauge choice. \qed
}

\section{To chapter \ref{CHAP5}}
\subsection{Symplectomorphisms}\label{symp12}
In local coordinates, a symplectic form $\Omega$ can always be written in the following way
\beg
\Omega &=& \Omega_{\mu\nu}dx^\mu\wedge dx^\nu.
\eeg
By Darboux's theorem \cite{McDS95}, it is always possible to restrict the symplectic form to have constant coefficients $\Omega_{\mu\nu}$. This is what I have done in (\ref{omr1}). Then, the constraint (\ref{om1}) can be written in coordinates for all $\vp_A\in \Diff(d)$
\be\label{om3}
\Omega_{\mu\nu} &=&\frac{\p\vp_A^{\sigma_1}}{\p x^\mu}\frac{\p\vp_A^{\sigma_2}}{\p x^\nu}\Omega_{\sigma_1\sigma_2}.
\ee
Since every element $\vp_A\in \Diff(d)$ locally is uniquely fixed by its associated vector field $X_A\in \mathfrak{diff}_d$, that are related by (\ref{exp9}) 
\beg
\vp_A^\mu(x) &=& x^\mu + X_A^\mu(x) + \mathcal{O}(A^2),
\eeg
the equation (\ref{om3}) implies (\ref{om4})
\beg
\Omega_{\mu\nu} &=&\left(\delta^{\sigma_1}_\mu + \p_\mu X_A^{\sigma_1}(x)\right) \left(\delta^{\sigma_2}_\nu + \p_\nu X_A^{\sigma_2}(x)\right)\Omega_{\sigma_1\sigma_2} +\mathcal{O}(A^2)
\nn\\
\Leftrightarrow\quad 0&=& \p_\mu X_A^{\sigma_1}(x)\Omega_{\sigma_1\nu} -\p_\nu X_A^{\sigma_2}(x)\Omega_{\sigma_2\mu} +\mathcal{O}(A^2)\\
\Leftrightarrow\quad 0&=& 2\Omega_{\sigma[\nu}\p_{\mu]} X_A^{\sigma}(x)  +\mathcal{O}(A^2).
\eeg
For constant coefficients $\Omega_{\mu\nu}$, this constraint (\ref{om4}) is equivalent to requiring that the Lie derivative of $\Omega$ along a vector field $X$ vanishes. Due to the one-to-one correspondence between $\Diff(d)$ and $\mathfrak{diff}_d$, I can drop the $\mathcal{O}(A^2)$ restriction.\\

Next, I will solve this constraint for $d=56$ by using the split (\ref{om0}). The equation (\ref{om4}) hence has the form
\beg
0&=& 2\Omega_{\alpha[\nu}\p_{\mu]} X_A^{\alpha} + 2\Omega_{\alpha+28\,[\nu}\p_{\mu]} X_A^{\alpha+28},
\eeg
which decomposes into four constraints for $\alpha,\beta,\gamma=1,\dots, 28$
\beg
0&=& 2\Omega_{\alpha[\beta}\p_{\gamma]} X_A^{\alpha} + 2\Omega_{\alpha+28\,[\beta}\p_{\gamma]} X_A^{\alpha+28}\\
0&=& 2\Omega_{\alpha[\beta}\p_{\gamma+28]} X_A^{\alpha} + 2\Omega_{\alpha+28\,[\beta}\p_{\gamma+28]} X_A^{\alpha+28}\\
0&=& 2\Omega_{\alpha[\beta+28}\p_{\gamma]} X_A^{\alpha} + 2\Omega_{\alpha+28\,[\beta+28}\p_{\gamma]} X_A^{\alpha+28}\\
0&=& 2\Omega_{\alpha[\beta+28}\p_{\gamma+28]} X_A^{\alpha} + 2\Omega_{\alpha+28\,[\beta+28}\p_{\gamma+28]} X_A^{\alpha+28}.
\eeg
These simplify with the explicit form of $\Omega$ (\ref{omr1}) to
\beg
0&=& -2\delta_{\alpha[\beta}\p_{\gamma]} X_A^{\alpha+28}\\
0&=&-\delta_{\alpha\gamma}\p_{\beta} X_A^{\alpha} -\delta_{\alpha\beta}\p_{\gamma+28} X_A^{\alpha+28}\\
0&=& \delta_{\alpha\beta}\p_{\gamma} X_A^{\alpha} +\delta_{\alpha\gamma}\p_{\beta+28} X_A^{\alpha+28}\\
0&=& 2\delta_{\alpha+28\,[\beta+28}\p_{\gamma+28]} X_A^{\alpha}.
\eeg
If I introduce the following notation
\beg
Y_\beta&:=&\delta_{\alpha\beta} X_A^{\alpha+28}\\
Y_{\beta+28}&:=&\delta_{\alpha+28\,\beta+28} X_A^{\alpha},
\eeg
the first and the last line imply that $X_A^{\alpha+28}$ and $X_A^{\alpha}$, respectively, considered as one-forms, are closed:
\beg
0&=& \p_{[\gamma}Y_{\beta]}\\
0&=& \p_{[\gamma+28}Y_{\beta+28]}.
\eeg
From section \ref{Diffeom} it follows that the domain of validity $U_\alpha\subset \cM^{56}$ of the diffeomorphisms $\vp_A\in \Diff(56)$ is simply connected. Then, Poincar\'e's lemma implies that these forms are exact with functions or zero forms $H$ and $G$:
\beg
Y_\beta&:=&\p_\beta G\\
Y_{\beta+28}&:=&\p_{\beta+28}H.
\eeg
Substituting these definitions into the remaining two equations, I obtain
\beg
0&=&-\delta_{\alpha\gamma}\p_{\beta} \delta^{\e+28\,\alpha+28}\p_{\e+28}H
-\delta_{\alpha\beta}\p_{\gamma+28} 
\delta^{\e\alpha}\p_{\e}G
\\
&=&
-\p_{\beta}\p_{\gamma+28}H
-\p_{\gamma+28} 
\p_{\beta}G
\\
0&=&-\delta_{\alpha\beta}\p_{\gamma} \delta^{\e+28\,\alpha+28}\p_{\e+28}H
-\delta_{\alpha\gamma}\p_{\beta+28} 
\delta^{\e\alpha}\p_{\e}G
\\
&=&-\p_{\gamma}\p_{\beta+28}H
-\p_{\beta+28} 
\p_{\gamma}G.
\eeg
The general solution obviously is $G=-H$ modulo affine linear contributions. Since all translational vector fields can be constructed from $G=-H$, this is no constraint on the level of the vector fields. Hence, the general solution to this set of differential equations is provided by an arbitrary function $H(x^\alpha,p_\beta)$ (\ref{om9}) with $p_\beta =\delta_{\beta\,\alpha+28}x^{\alpha+28}$ (\ref{omr2}):
\beg
X^\alpha &=&\frac{\p}{\p p_\alpha} H\\
X_\beta &=&-\frac{\p}{\p x^\beta} H.
\eeg
I want to denote this infinite dimensional subgroup of symplectomorphisms of the Lie group $\Diff(56)$ by $\Symp(56)$. If I dropped the restriction of general diffeomorphisms to $\Diff(d)$ from section \ref{Diffeom}, there would be further solutions to the constraint (\ref{om1}). In this context, $\Symp(56)$ is the subgroup of all symplectomorphisms that correspond to Hamiltonian vector fields, which is called the subgroup of Hamiltonian flows \cite{McDS95}.

\subsection{Vector fields preserving $\Omega$ and $Q$}\label{ESympm}
In section \ref{SympV1}, I discussed the symplectomorphisms $\Symp(56)$, i.e. the infinite dimensional Lie subgroup of $\Diff(56)$ preserving the symplectic form: $\vp^*\Omega=\Omega$ (\ref{om1}). I want to restrict $\Symp(56)$ further to the ones that also preserve the quartic symmetric tensor $Q$ (\ref{QDefi})
\be\label{Qinv2}
\vp^*Q &=& Q.
\ee
As explained in the sections \ref{Diffeom} and \ref{symp12}, I restricted the symplectomorphisms to Hamiltonian flows. Therefore, the constraint (\ref{Qinv2}) is equivalent to requiring (\ref{Q3})
\beg
	4Q_{\rho(\mu_1\mu_2\mu_3}\p_{\mu_4)}X^\rho &=& 0
\eeg
with a Hamiltonian vector field $X^\rho$ (\ref{om9}). In section \ref{ESy}, I stated that the general solution to this constraint is provided by the Hamiltonian (\ref{Eom10})
\beg
H_{(\Lambda,\Sigma,c)}(z,\bar{z})&:=& \frac{i\tau_7^2}{8}{\Lambda_A}^B\left(\delta_{[D_2}^{A}\delta_{D_1]}^{[C_1}\delta_{B}^{C_2]}
-\frac{1}{8}\delta_{B}^{A}\delta_{D_1D_2}^{C_1C_2}\right)z^{D_1D_2}\bar{z}_{C_1C_2}\\
&& +\frac{i\tau_7^2}{32}\Sigma^{[C_1C_2C_3C_4]}\left(\bar{z}_{C_1C_2}\bar{z}_{C_3C_4}
-\frac{1}{4!}\e_{C_1\dots C_8}z^{C_5C_6}z^{C_7C_8}\right)\nn\\
&&+\frac{i\tau_7^2}{16}c^{M_1M_2}\bar{z}_{M_1M_2} -\frac{i\tau_7^2}{16}\bar{c}_{M_1M_2}z^{M_1M_2}\nn.
\eeg
In order to show this, I proceed in the following way:
\begin{enumerate}
	\item As the tensor $Q$ in the $SU(8)$-covariant coordinates $(z^{[AB]},\bar{z}_{[CD]})$ (\ref{QDefi}) is easier to handle, I start by rewriting the constraint (\ref{Q3}) in these coordinates.
	\item Next, I show that $H_{(\Lambda,\Sigma,c)}$ (\ref{Eom10}) solves this constraint.
	\item As the constraint is $SU(8)$-covariant, it is sufficient to exclude all other quadratic Hamiltonians that correspond to the other $SU(8)$ representations.
	\item Finally, I will show that Hamiltonians of higher polynomial degrees are excluded, what is expected from Cartan's theorem \cite{C09}.
\end{enumerate}

I start by restating (\ref{Q3}) in the holomorphic coordinates $(z,\bar{z})$ with the summation introduced in section \ref{ESy}:
\beg
Q_{A_1A_2(\mu_1\mu_2\mu_3}\p_{\mu_4)}X^{A_1A_2} +{Q^{A_1A_2}}_{(\mu_1\mu_2\mu_3}\p_{\mu_4)}\bar{X}_{A_1A_2} &=& 0.
\eeg
Due to the explicit form of the quartic tensor (\ref{QDefi}), this leads to the following equations that always occur in pairs of holomorphic and antiholomorphic objects
\beg
Q_{A_1A_2(B_1B_2|B_3B_4|B_5B_6|}\p_{B_7B_8)}X^{A_1A_2} &=&0\\
\bar{Q}^{A_1A_2(B_1B_2|B_3B_4|B_5B_6|}\bar{\p}^{B_7B_8)}\bar{X}_{A_1A_2} &=& 0
\eeg
with all indices up and down, respectively. Symmetrizations are to be taken over pairs of indices. For the next set of equations, one has to pay attention to the numerical factors:
\beg
\frac{1}{4}Q_{A_1A_2B_1B_2B_3B_4B_5B_6}\bar{\p}^{B_7B_8}X^{A_1A_2} +\frac{3}{4}{Q^{A_1A_2\,B_7B_8}}_{(B_1B_2|B_3B_4}\p_{B_5B_6)}\bar{X}_{A_1A_2} 
&=&0\\
\frac{3}{4}{Q_{A_1A_2\,B_7B_8}}^{(B_1B_2|B_3B_4}\bar{\p}^{B_5B_6)}X^{A_1A_2} 
+\frac{1}{4}\bar{Q}^{A_1A_2B_1B_2B_3B_4B_5B_6}\p_{B_7B_8}\bar{X}_{A_1A_2}&=&0.
\eeg
The selfdual equation finally is
\beg
\frac{2}{4}{Q_{A_1A_2(B_5B_6}}^{B_1B_2B_3B_4}\p_{B_7B_8)}X^{A_1A_2}
+\frac{2}{4}{Q^{A_1A_2(B_1B_2}}_{B_5B_6B_7B_8}\bar{\p}^{B_3B_4)}\bar{X}_{A_1A_2}&=&0,
\eeg
where I have used the abbreviations $\p_{AB}:=\frac{\p}{\p z^{AB}}$ and $\bar{\p}^{AB}:=\frac{\p}{\p \bar{z}_{AB}}$.\\

Since I want to restrict the symplectomorphisms $\Symp(56)$ to the ones that preserve $Q$ (\ref{Qinv2}), I can without loss of generality restrict to Hamiltonian vector fields that are generated in the holomorphic frame by (\ref{zNeu3}). This implies that it is sufficient to discuss the holomorphic constraint, the antiholomorphic always follows from complex conjugation. With the explicit form for the coefficients of $Q$ (\ref{QDefi}), these equations pose restrictions on the functions $H$. The one from the first holomorphic equation is
\beg
\e_{A_1A_2(B_1B_2|B_3B_4|B_5B_6|}\p_{B_7B_8)}\bar{\p}^{A_1A_2}H &=&0.
\eeg
From the second set of equiations, I use the antiholomorphic constraint
\beg
-\frac{1}{2}
\left(\delta_{B_8A_1}^{(B_1B_2}\delta_{A_2B_7}^{B_3B_4} -\frac{1}{4}\delta_{A_1A_2}^{(B_1B_2}\delta_{B_7B_8)}^{B_3B_4}\right)
\bar{\p}^{B_5B_6)}\bar{\p}^{A_1A_2}H\nn&&\\
+\frac{1}{96}\e^{A_1A_2B_1B_2B_3B_4B_5B_6}\p_{B_7B_8}\p_{A_1A_2}H&=&0
\eeg
that is related to the holomorphic one by raising the indices with $\e$, because $\e$ is in the definition of $H_{(\Lambda,\Sigma,c)}$ (\ref{Eom10}). The selfdual equation is
\beg
-\left(\delta_{A_1A_2}^{B_4B_1}\delta_{(B_5B_6}^{B_2B_3} -\frac{1}{4}\delta_{A_1A_2}^{B_1B_2}\delta_{(B_5B_6}^{B_3B_4}\right)\p_{B_7B_8)}\bar{\p}^{A_1A_2}H
\nn&&\\
-
\left(\delta_{A_1A_2}^{B_2B_3}\delta_{(B_5B_6}^{B_4B_1} -\frac{1}{4}\delta_{A_1A_2}^{B_3B_4}\delta_{(B_5B_6}^{B_1B_2}\right)\p_{B_7B_8)}\bar{\p}^{A_1A_2}H
\nn&&\\
+\left(\delta_{B_6B_7}^{A_1A_2}\delta_{B_8B_5}^{(B_1B_2} -\frac{1}{4}\delta_{B_5B_6}^{A_1A_2}\delta_{B_7B_8}^{(B_1B_2}\right)
\bar{\p}^{B_3B_4)}\p_{A_1A_2}H\nn&&\\
+
\left(\delta_{B_8B_5}^{A_1A_2}\delta_{B_6B_7}^{A(B_1B_2}
-\frac{1}{4}\delta_{B_7B_8}^{A_1A_2}\delta_{B_5B_6}^{(B_1B_2}\right)
\bar{\p}^{B_3B_4)}\p_{A_1A_2}H
&=&0.
\eeg
This appears to be a set of very restrictive constraints, for which it is not clear that there exist non-trivial solutions at all. Before discussing possible non-linear solutions, I will first show that there is a quadratic solution. Using the ansatz $H_{(\Lambda,\Sigma,c)}$ (\ref{Eom10}) yields
\beg
&&\e_{A_1A_2(B_1B_2|B_3B_4|B_5B_6|} \delta_{[B_7}^{[A_1}{\Lambda_{B_8])}}^{A_2]}
-
\frac{1}{8}\e_{A_1A_2(B_1B_2|B_3B_4|B_5B_6|} \delta_{B_7B_8)}^{A_1A_2}{\Lambda_{B}}^{B}\\
&=&
-\e_{A_2[B_7B_1B_2B_3B_4B_5B_6} {\Lambda_{B_8]}}^{A_2}
-\frac{1}{8}\e_{B_7B_8B_1B_2B_3B_4B_5B_6} {\Lambda_{B}}^{B}\\
&=&
-\frac{9}{8}\e_{[A_2B_7B_1B_2B_3B_4B_5B_6} {\Lambda_{B_8]}}^{A_2}\\
 &=&0
 \eeg
in eight dimensions. I continue with the selfdual constraint, because it also projects on the $\Lambda$ components of the ansatz $H_{(\Lambda,\Sigma,c)}$ (\ref{Eom10}). A short look at the constraint furthermore reveals that the trace part $\delta^{C_1C_2}_{D_1D_2}$ cancels. Hence, I can focus on the other part\footnote{As the constraint is linear, the trace term also fulfils the identity separately for $\Lambda\sim \delta$.} that also vanishes by itself:
 \beg
&&\left(\delta_{A_1A_2}^{B_4B_1}\delta_{(B_5B_6}^{B_2B_3} -\frac{1}{4}\delta_{A_1A_2}^{B_1B_2}\delta_{(B_5B_6}^{B_3B_4}\right)\delta_{[B_7}^{[A_1}\Lambda_{B_8])}^{A_2]}\\
&&+
\left(\delta_{A_1A_2}^{B_2B_3}\delta_{(B_5B_6}^{B_4B_1} -\frac{1}{4}\delta_{A_1A_2}^{B_3B_4}\delta_{(B_5B_6}^{B_1B_2}\right)\delta_{[B_7}^{[A_1}\Lambda_{B_8])}^{A_2]}\\
&&
-\left(\delta_{B_6B_7}^{A_1A_2}\delta_{B_8B_5}^{(B_1B_2} -\frac{1}{4}\delta_{B_5B_6}^{A_1A_2}\delta_{B_7B_8}^{(B_1B_2}\right)\delta_{[A_1}^{[B_3}\Lambda_{A_2]}^{B_4])}\\
&&-
\left(\delta_{B_8B_5}^{A_1A_2}\delta_{B_6B_7}^{A(B_1B_2} -\frac{1}{4}\delta_{B_7B_8}^{A_1A_2}\delta_{B_5B_6}^{(B_1B_2}\right)\delta_{[A_1}^{[B_3}\Lambda_{A_2]}^{B_4])}\\
&=&0.
 \eeg
Since the second constraint only involves the $\Sigma$ contribution in $H_{(\Lambda,\Sigma,c)}$ (\ref{Eom10}), this already shows that the $\Lambda$ contribution generates vector fields and hence symplectomorphisms that preserve the quartic tensor $Q$ (\ref{Q0}).\footnote{There is a different way to see this: as the parametrization of $Q$ (\ref{QDefi}) only involves $SU(8)$-invariant tensors $\delta$ and $\e$ and as $\Lambda$ corresponds to the Lie algebra $\mathfrak{su}_8$, what I show in section \ref{VectE4}, the conservation is trivial.}\\

The second constraint finally is
\beg
&&
-\frac{1}{2}
\left(\delta_{B_8A_1}^{(B_1B_2}\delta_{A_2B_7}^{B_3B_4} -\frac{1}{4}\delta_{A_1A_2}^{(B_1B_2}\delta_{B_7B_8}^{B_3B_4}\right)
\Sigma^{B_5B_6)A_1A_2}\nn\\
&&-\frac{1}{96}\e^{A_1A_2B_1B_2B_3B_4B_5B_6}
\frac{1}{4!}\e_{B_7B_8A_1A_2C_1\dots C_4}
\Sigma^{C_1C_2C_3C_4}\\
&=&
-\frac{1}{2}
\left(\delta_{B_8}^{(B_1}\delta_{B_7}^{B_4}\Sigma^{B_5B_6B_2B_3)} -\frac{1}{4}\delta_{B_7B_8}^{(B_3B_4}\Sigma^{B_5B_6B_1B_2)}\right)
\\
&&-\frac{2!6!}{4!96}\delta^{B_1B_2B_3B_4B_5B_6}_{B_7B_8C_1\dots C_4}
\Sigma^{C_1C_2C_3C_4}\\
&=&
-\frac{1}{2}\left(-1-\frac{1}{4}\right)\delta^{B_1B_2B_3B_4B_5B_6}_{B_7B_8C_1\dots C_4}
\Sigma^{C_1C_2C_3C_4}
\\
&&-\frac{5}{8}\delta^{B_1B_2B_3B_4B_5B_6}_{B_7B_8C_1\dots C_4}
\Sigma^{C_1C_2C_3C_4}\\
&=&0,
\eeg
where antisymmetrizations over pairs of indices are implicit. This proves that the ansatz $H_{(\Lambda,\Sigma,c)}$ (\ref{Eom10}) solves the constraints.\\

The third part of the proof consists in showing that no further quadratic vector fields solve this constraint. To do this, recall from the general solution for symplectomorphisms (\ref{Unity}) from section \ref{SympV1} that the general quadratic Hamiltonian is provided by the real function
\beg
H_{\text{lin}}&=& i{\Psi_{D_1D_2}}^{C_1C_2}z^{D_1D_2}\bar{z}_{C_1C_2}\\
&& +\frac{i}{2}\Pi^{C_1C_2,D_1D_2}\bar{z}_{C_1C_2}\bar{z}_{D_1D_2}
-\frac{i}{2}\bar{\Pi}_{C_1C_2,D_1D_2}z^{C_1C_2}z^{D_1D_2}\nn\\
&&+\frac{i\tau_7^2}{16}c^{M_1M_2}\bar{z}_{M_1M_2} -\frac{i\tau_7^2}{16}\bar{c}_{M_1M_2}z^{M_1M_2}\nn
\eeg
with the coefficient $\Psi$ having the appropriate reality properties such that $H_{\text{lin}}$ is real. The fact that the constraint for the conservation of the quartic tensor $Q$ is $SU(8)$-covariant\footnote{In the definition of $Q$ (\ref{QDefi}) only $SU(8)$-invariant tensors $\delta$ and $\e$ were used.} implies that it is sufficient to discuss the irreducible $SU(8)$ representations separately.\\

Next, observe that ${\Lambda_A}^B$ in the ansatz $H_{(\Lambda,\Sigma,c)}$ (\ref{Eom10}) generates the $\mathfrak{su}_8$ algebra, as I show in section \ref{VectE4}. Thus, the decomposition of the components ${\Lambda_{D_1D_2}}^{C_1C_2}$ parametrizing $\mathfrak{u}_{28}$ into irreducible $\mathfrak{su}_{8}$ representations is dictated as follows:
\beg
\mathfrak{u}_{28}&=& \mathbf{1}\oplus \mathfrak{su}_8 \oplus \mathbf{720}.
\eeg
The first equation immediately discards the singlet representation, because it corresponds to the double trace ${\Psi_{C_1C_2}}^{C_1C_2}$. Where the $\mathbf{720}$ is concerned, it is the irreducible representation of $SU(8)$ with vanishing single trace ${\Psi_{SD_2}}^{SC_2}=0$. Due to $SU(8)$-covariance of $Q$, this implies that the entire $\mathbf{720}$ is either admitted or excluded. A short check shows that the exclusion is indeed the case. The same argumentation holds for the non-compact generators. In dimensions, the decomposition (\ref{SPU}) is as follows:
\begin{center}
\begin{tabular}{cccccccc}
$\mathfrak{sp}_{56}$
&$=$& $\mathfrak{u}_{28}$& $\oplus$&$\langle\hat{S}_{(\alpha\beta)} +\text{c.c.}\rangle_\R$ & $\oplus$&$\langle i\hat{S}_{(\alpha\beta)} +\text{c.c.}\rangle_\R$\\
\\
&$=$&$\mathbf{784}$&$\oplus $&$\mathbf{406}$&$\oplus $&$\overline{\mathbf{406}}$.
\end{tabular}
\end{center}
The symmetric $\mathfrak{u}_{28}$ tensors $\hat{S}$ in $\mathbf{406}$ and $\overline{\mathbf{406}}$ further decompose under $\mathfrak{su}_{8}$
to
\begin{center}
\begin{tabular}{ccccc}
$\mathbf{406}$
&$=$& $\mathbf{70}$& $\oplus$&$ \mathbf{336}$\\
\\
&$=$&
\begin{picture}(24,6)(0,0)
\put(0,0){\line(1,0){24}}
\put(0,6){\line(1,0){24}}
\put(0,0){\line(0,1){6}}
\put(6,0){\line(0,1){6}}
\put(12,0){\line(0,1){6}}
\put(18,0){\line(0,1){6}}
\put(24,0){\line(0,1){6}}
\end{picture}
&
$\oplus $&
\begin{picture}(12,12)(0,0)
\put(0,-3){\line(1,0){12}}
\put(0,3){\line(1,0){12}}
\put(0,9){\line(1,0){12}}
\put(0,-3){\line(0,1){12}}
\put(6,-3){\line(0,1){12}}
\put(12,-3){\line(0,1){12}}
\end{picture}
\end{tabular}
\end{center}
As the second constraint links $\Pi$ to $\bar{\Pi}$ by the $\e$ tensor, one combination of the two $\mathbf{70}$ representation survives, the other one is discarded. It is obvious that the $\e$ tensor cannot link the $\mathbf{336}$ representations. Hence, the second constraint discards both of them. \\

To complete the proof, I have to show that there are no polynomials of higher degree in $(z,\bar{z})$ that preserve the completely symmetric quartic tensor. The argument is similar to the one that the group preserving a symmetric two tensor is finite dimensional.\footnote{As an example serves the Poincar\'e group, the isometry group of the Minkowski metric that I discussed in section \ref{Isometry}.} I start with the quartic tensor in the coordinate form (\ref{Q2k}).
\beg
Q&=&
-dp_m \circ dx^m \circ dp_n \circ dx^n 
\\
&&+2dp_m \circ dx^m \circ dp_{pq} \circ dx^{pq} 
+8dp_m \circ dx^{mn} \circ dp_{nq} \circ dx^{q} \nn\\
&& -\frac{\sqrt{2}}{6}\left(\e_{m_1\dots m_7}dx^{m_1}\circ dx^{m_2m_3}\circ dx^{m_4m_5}\circ dx^{m_6m_7}
\right.\nn\\
&&\left.
+ \e^{m_1\dots m_7}dp_{m_1}\circ dp_{m_2m_3}\circ dp_{m_4m_5}\circ dp_{m_6m_7}\right)\nn\\
&&
+dp_{cd}\circ dx^{cd} \circ dx^{ef}\circ dp_{ef} 
-4dx^{ab} \circ dp_{be}\circ  dx^{ef}\circ dp_{fa} 
\nn
\eeg
Consider its transformation under an infinitesimal diffeomorphism $X_A\in \mathfrak{diff}_{56}$
\be\label{quadrat}
{x'}^m &=& x^m + X_A^m(x) +\mathcal{O}(A^2)
\ee
with at least quadratic dependence of $X$ on the coordinate $x^n$. Recall that I have restricted the general diffeomorphisms $\Diff(56)$ to symplectomorphisms $\Symp(56)$ that preserve $\Omega$ (\ref{om1}). Hence, the formula (\ref{om9}) implies that the dual coordinate $p_m$ must infintesimally transform as
\beg
{p'}_m &=& p_m +Y_m(x) +\mathcal{O}(A^2)
\eeg
with $Y$ at least linear in $x^m$. Hence, in the evaluation of the pull-back, the first term in the quartic tensor will give rise to the term
\beg
2\frac{\p Y_m}{\p x^r}dx^r \circ dx^m \circ dp_n \circ dx^n. 
\eeg
A brief look at the quartic tensor $Q$ in the coordinates $(x,p)$ reveals that no other term in $Q$ gives rise to a contribution like this. Therefore, the quartic tensor cannot be invariant under a non-constant symplectomorphism that is non-linear in the $7$ coordinates $x^m$.\\

The same argumentation works for the dual coordinates $p_m$. If it was invariant under such a symplectomorphism that is at least quadratic in some other coordinates or a combination of coordinates, there must not exist a linear transformation that maps the coordinates to each other, because a concatenation would map this again to the case specified in (\ref{quadrat}). As the vector space $\mathbf{56}$ is an irreducible representation of the Lie group specified by the linear diffeomorphisms, which is $E_{7(7)}$ what I show in section \ref{VectE4}, these linear transformations do exist. This completes the proof that there are no symplectomorphisms $\Symp(56)$ that preserve $Q$ whose Hamiltonian is a polynomial of degree higher than $2$.\qed
\\

This is a very special case of Cartan's theorem \cite{C09} that lists the possible simple, infinite dimensional, non-trivial  subgroups of $\Diff(d)$: diffeomorphisms of unit determinant, symplectomorphisms and contact diffeomorphisms.\\

I conclude by mentioning that this statement does not contradict the transformation defined in \cite{GKN01}, because the one in \cite{GKN01} cannot be interpreted as a vector field: it is only possible to define a transformation preserving $Q$ that is of at least second order in derivative operators.

\subsection{Some relations for $\G$ matrices}\label{SomeG2}
For the purely imaginary $8\times 8$ Clifford matrices $\G_a$ that I defined in appendix \ref{Cliff}, the following relations can be deduced from the defining property of the Clifford algebra (\ref{Clifford}):
\begin{subequations}\label{Gnq}
\be
	{\G_a}^{AB} {\G^d}_{AB} &=& -8\delta^d_{a}
	\\
		{\G_{ab}}^{AB} {\G^{de}}_{AB} &=& 2!8\delta_{ab}^{de}
	\\
			{\G_{abc}}^{AB} {\G^{def}}_{AB} &=& -3!8\delta^{def}_{abc}.
	\ee
\end{subequations}
Small letters as indices $a,b,c,\ldots$ always have the range $1,\ldots,7$, whereas capital ones denote the matrix indices $A,B,\ldots =1,\ldots,8$. A proof by induction with (\ref{Clifford}) leads to
\beg
	\G_{a_1\dots a_n}\G_{a_{n+1}}=\G_{a_1\dots a_{n+1}} +n\G_{[a_1\dots a_{n-1}}\eta_{a_n]a_{n+1}}.
\eeg
From the definition (\ref{AntisymG}), it is obvious that a $\G$ matrix with seven indices is maximal in seven dimensions. Its normalization is (\ref{ConvG7})
\be\label{epsdefi7}
	{{\G_{a_1\dots a_7}}^A}_B &=:&-i{\id^A}_B \e_{a_1\dots a_7},
\ee
with $\e_{1234567}=1$. This leads to the matrix products
\begin{subequations}\label{G3G}
\be
	\G_{ab}\G_{xy}
	&=&
	\frac{i}{3!}\e_{ab xyrst}\G^{rst}
	  +4\delta_{xy}^{rs}\G_{r[a}\eta_{b]s}
	 +2\eta_{x[b}\eta_{a]y}\id
	 \\
	\G_{a_1\dots a_3}\G_{xy}	
		 &=&
		 \frac{i}{2!}\e_{a_1\dots a_3 xyrs}\G^{rs}
	  -6\delta_{xy}^{tu}\G_{[a_1 a_{2}|t|}\eta_{a_{3}]u}\nn\\
	 &&+3!\eta_{x[a_3}\G_{a_1}\eta_{a_{2}]y}
	 \\
	 	\G_{a_1\dots a_3}\G_{xyz}
	&=&
	\delta_{xyz}^{stu}\left(
		-i\e_{a_1\dots a_3 stu r}\G^{ r} 
			+18\delta_{a_{1}\dots a_{3}}^{f_{1}\dots f_{3}}\eta_{f_{3}t}\G_{f_1s}\eta_{f_{2}u}
	\right.\\
		&&
		\left.+\frac{3i}{2}{\e_{st[a_1 a_2}}^{ rvw}\eta_{a_{3}]u}\G_{rvw}
		 +6\eta_{sa_3}\eta_{a_{2}t}\eta_{a_{1}u}\id\right)
	\nn
\ee
\end{subequations}
with the identity matrix $\id$. Together with the $63$ linearly independent traceless $8\times 8$ matrices (\ref{Independent2}), $\id$ spans the complex vector space of all $8\times 8$ matrices. This is the core of the Fierz identity that can be proved by taking traces
\be\label{Fierz}
	8{{\G_X}^A}_B{{\G_Y}^C}_D&=& {\id^C}_B \left({{\G_X}^A}_F{{\G_Y}^F}_D\right)\\
	&&+{{\G^a}^C}_B\left({{\G_X}^A}_E{{\G_a}^E}_F{{\G_Y}^F}_D\right)\nn
	\\
	&&-\frac{1}{2!}{{\G^{ab}}^C}_B\left({{\G_X}^A}_E{{\G_{ab}}^E}_F{{\G_Y}^F}_D\right)\nn
	\\
	&&-\frac{1}{3!}{{\G^{abc}}^C}_B\left({{\G_X}^A}_E{{\G_{abc}}^E}_F{{\G_Y}^F}_D\right),\nn
\ee
where $X$ and $Y$ are short for arbitrary $d=7$ vector indices. By setting $\G_X=\G_Y=\id$ and antisymmetrizing in $[BC]$, one obtains the completeness relation\footnote{It is not important for these formal relations, if an index of a $\G$ matrix is raised or lowered. }
\be\label{complete}
2\G^{a}_{AB}\G^{a}_{CD} -\G^{ab}_{AB}\G^{ab}_{CD}&=& -16 \delta_{CD}^{AB}.
\ee
The Fierz identity (\ref{Fierz}) also provides the identity
\be\label{Finale6b}
&&{\G_{ab}}^{[AB}{\G_{cd}}^{CD]}
\\
				&=&
				-\frac{i}{3!}\e_{abcd frs}\nn
	{\G^{[f}}^{[AB}{\G^{rs]}}^{CD]}	
			-{\G_g}^{[AB}{\G_{h}}^{CD]}\left(4\delta_{[c}^g\eta_{d][b}\delta_{a]}^h +\frac{2}{3}\eta^{gh}\eta_{c[b}\eta_{a]d}\right).
\ee
Furthermore, tracing in $ac$ proves the identity
\be\label{Zerl3}
	{\G_{c}}^{[AB}{\G_{ab}}^{CD]} &=&  {\G_{[c}}^{[AB}{\G_{ab]}}^{CD]} -\frac{1}{3}\eta_{c[a}{\G_{b]f}}^{[AB}{\G^{f}}^{CD]}.
\ee

I conclude the summary of formulas for $\G$ matrices with
\begin{subequations}\label{G4G}
\be
{\G^{e}}^{[AB}{\G_{f}}^{CD]}{\G_{c}}_{AB}
&=&
-\frac{2}{3}\left( \delta^e_{c}{\G_{f}}^{CD} +\eta_{cf}{\G^{e}}^{CD} +\delta^e_f{\G_{c}}^{CD}
 \right)
\\
{\G^{j}}^{[AB}{\G_{j}}^{CD]}{\G_{c}}_{AB}
&=&-6{\G_{c}}^{CD}
\\
{\G^e}^{[AB}{\G_{f}}^{CD]}{\G_{cd}}_{AB}
&=&
\frac{2}{3}\left(
2\delta^e_{[c}{{\G_{d]f}}^{CD}} +2\eta_{f[c}{{{\G_{d]}}^e}^{CD}}
+\delta^e_{f}{{\G_{cd}}^{CD}}
 \right)\nn
 \\
 {\G^{j}}^{[AB}{\G_{j}}^{CD]}{\G_{cd}}_{AB}
&=&
2{{\G_{cd}}^{CD}}
\\
 {\G^{[a_1a_2}}^{[AB}{\G^{a_3]}}^{CD]}{\G_{c}}_{AB}
 &=&
 -2{{\G^{[a_1a_2}}^{CD}}\delta_c^{a_3]}
   \\
  {\G^{ef}}^{[AB}{\G_{f}}^{CD]}{\G_{c}}_{AB}
 &=&
 4{{{\G_{c}}^e}}^{CD}
 \\ 
  {\G^{[a_1a_2}}^{[AB}{\G^{a_3]}}^{CD]}{\G_{cd}}_{AB}
  &=&
  4\delta_{cd}^{[a_1a_2}{\G^{a_3]}}^{CD} 
 +\frac{i}{3}{\e^{a_1\dots a_3}}_{cdrs}\left.\G^{rs}\right.^{CD}
  \\
  {\G^{ef}}^{[AB}{\G_{f}}^{CD]}{\G_{cd}}_{AB}
 &=&
 8\delta_{[c}^{e}{\G_{d]}}^{CD} \\
\label{dWN1}
{\G^{[a_1a_2}}^{[AB}{\G^{a_3]}}^{C]G} &=&{\G^{[a_1}}^{[AB}{\G^{a_2a_3]}}^{C]G} \!-\!\frac{1}{3}{\G_{c}}^{[AB}{\G^{ca_1a_2a_3}}^{C]G}
\\
\label{dWN2}
{\G_{cb}}^{[AB}{\G^b}^{C]G} &=&{\G^b}^{[AB}{\G_{cb}}^{C]G} +4{\G_{c}}^{[AB}\delta^{C]G}.
\ee
\end{subequations}
Further relations in the same conventions are stated in the appendix of \cite{dWN86}.

\subsection{Action of the $\mathfrak{e}_{7(7)}$-generators on the coordinates in the $\mathfrak{gl}_7$ decomposition}\label{CRAE2}
\vspace{-0.4cm}
\beg
\left[{\left.\hat{M}\right.^{a}}_b,\hat{Q}^e\right]
&=& 
\hat{Q}^{a}\delta_b^{e}
\\
\left[{\left.\hat{M}\right.^{a}}_b,\hat{P}_e\right]
&=& 
-\hat{P}_{b}\delta_e^{a}
\\
\left[{\left.\hat{M}\right.^{a}}_b,\hat{P}_{ef}\right]
&=&
			-2\left(
\hat{P}_{b[f}\delta_{e]}^a -\frac{1}{6}\delta^{a}_b \hat{P}_{ef}
\right)
\\
\left[{\left.\hat{M}\right.^{a}}_b,\hat{Q}^{ef}\right]
&=&
		2\left(\hat{Q}^{a[f}\delta^{e]}_b  -\frac{1}{6}\delta^{a}_b\hat{Q}^{ef}\right)
		\\
\left[\hat{E}^{abc},\hat{Q}^d \right]
&=&0
\\
\left[\hat{E}^{abc},\hat{P}_d\right]
&=&
  \frac{12\sqrt{2}}{\tau_2}\hat{Q}^{[ab}\delta^{c]}_d
\\
\left[\hat{E}^{abc},\hat{P}_{ef}\right]
		&=&
	\frac{12\sqrt{2}}{\tau_2} \delta_{ef}^{[ab}\hat{Q}^{c]}
\\
\left[\hat{E}^{abc},\hat{Q}^{ef}\right]
				&=&
				\frac{2}{\tau_2}\hat{P}_{dg}
\e^{efdgabc}
\\
 \left[\hat{E}^{a_1\dots a_6},\hat{Q}^d\right]&=&0
 \\
 \left[\hat{E}^{a_1\dots a_6},\hat{P}_d\right]
 &=&
 -\frac{4 \sqrt{2}}{\tau_3} \hat{P}_{db}\e^{a_1\dots a_6b}
 \\
 \left[\hat{E}^{a_1\dots a_6},\hat{P}_{ef}\right]&=& 0\\
\left[\hat{E}^{a_1\dots a_6},\hat{Q}^{ef}\right]
 &=&
  \frac{4 \sqrt{2}}{\tau_3}\hat{Q}^{[e}\e^{f]a_1\dots a_6}
\eeg

\section*{Selbst\"andigkeitserkl\"arung}
Hiermit erkl\"are ich, die vorliegende Arbeit selbst\"andig ohne fremde Hilfe verfasst zu haben und nur die angegebene Literatur verwendet zu haben.\\
\smallskip\\
\smallskip\\
Christian Hillmann\\
Potsdam, den 24. September 2008

\end{document}